# *Highlights in*
# High-Energy Physics

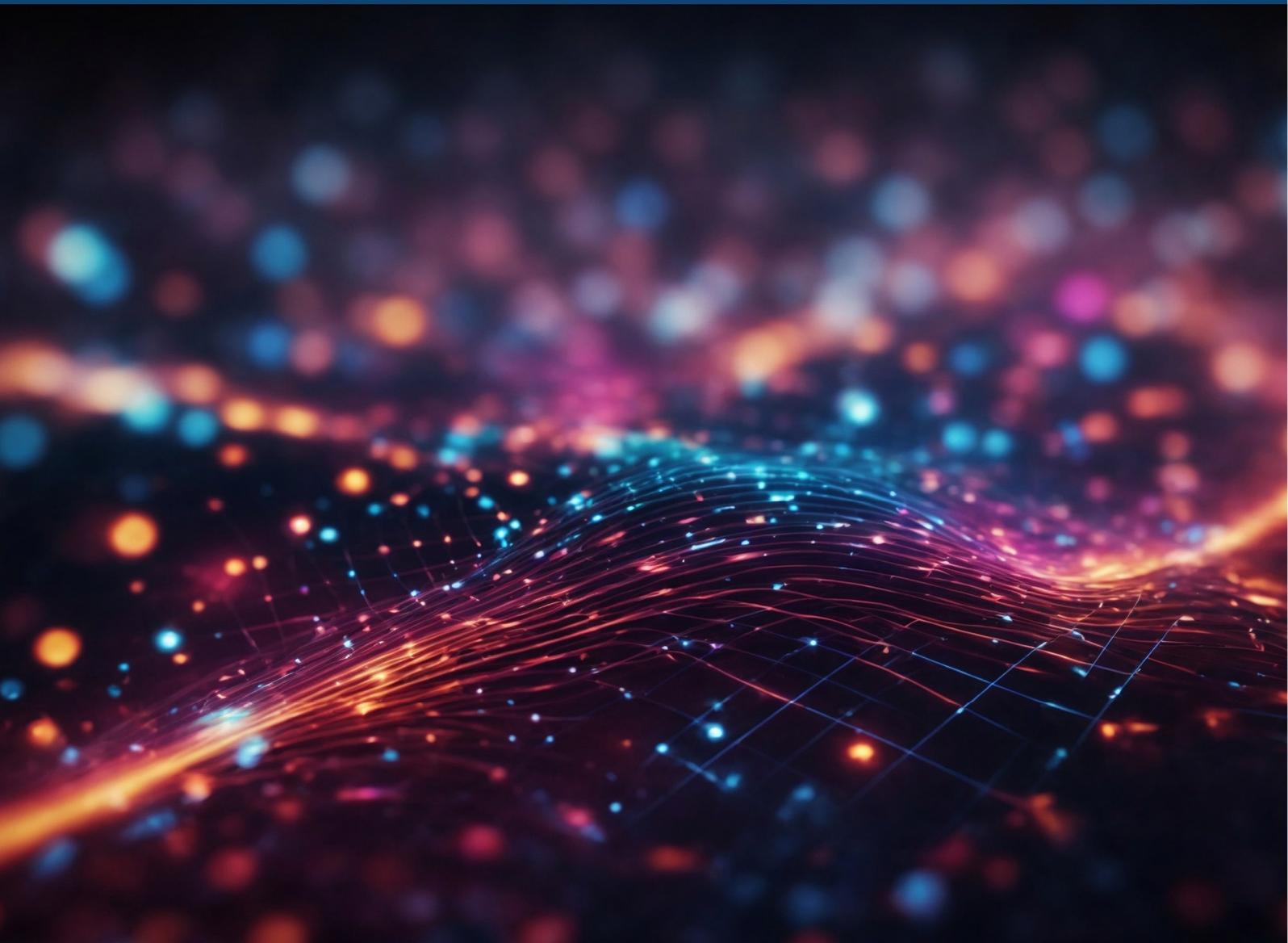

**Scilight**

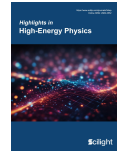

*Editorial*

# Why a New Journal in High Energy Physics

Guido Emilio Tonelli [1,2,3]

[1]   Department of Physics, University of Pisa, 56126 Pisa, Italy; guido.tonelli@pi.infn.it or guido.tonelli@cern.ch
[2]   INFN (Istituto Nazionale di Fisica Nucleare), 56127 Pisa, Italy
[3]   CERN (European Organization for Nuclear Research), 1217 Geneva, Switzerland

It is a true honor and a privilege for me to introduce you the first issue of this new journal, *Highlights in High Energy Physics* (HiHEP). It is an open access journal, publishing four issues per year, edited by Scilight Press, a rapidly growing international publishing press focused on academic, peer-reviewed works. Here (https://www.sciltp.com/journals/hihep/) you can find additional information on the new journal and on the publishing press (https://www.sciltp.com/).

Among the members of the Editorial Board (https://www.sciltp.com/journals/hihep/about/editorialTeam) that I have the honor to chair you will find many worldwide known scientists distributed with a careful balance in gender and regions

The scope of the journal is to present the most important results produced in High-Energy Physics (HEP). In particular, the main topics we would like to cover are listed here: Higgs Physics, Neutrino Physics, Physics Beyond the Standard Model, Top Quark and Electroweak Physics, Quark and Lepton Flavour Physics, Strong Interactions and Hadron Physics, Heavy ions, Astro-particle Physics and Cosmology, Dark Matter Detection, Collider Physics, Underground and Large Array Physics, Gravitation Waves astronomy, Gauge Field Theories, Quantum Gravity Theories, New Formal Theories, Accelerator Physics and Future Facilities, Present Detectors and R&D for Future Facilities, Computing, AI and Data Handling.

As you might have noticed the journal will not cover only the traditional results coming from colliders physics. We plan to host also the most exciting results in gravitational waves astronomy, or astro-particle physics since these experiments study some of the most energetic phenomena of the cosmos and their findings could be of extreme interest for all scientists. A particular attention will be dedicated also to new theories as well as to new results considered relevant for the field at large. The journal will give a particular attention to a critical discussion of the main achievements, hosting different points of view and soliciting in depth comparison of hot-from-the-press results. Main goal of the journal would be to distribute quickly, and in a critical format, the most relevant information among scientists willing to stay tuned with the latest developments. By presenting the scientific highlights it will ideally connect the various fields of activity in High-Energy Physics.

We are living a very exciting moment in HEP. Just because everything seems to fit within the "Standard Models" (SM of Particle Physics, Lambda-CDM Cosmology etc.). But we know already that this will not be the end of the story. We desperately need a new description of nature that will incorporate the impressive list of phenomena that we are not able to explain today. We don't know when and where new physics will appear, as unquestionable evidence of new particles or new interactions. It could manifest itself in an outstanding new discovery in particle physics, or in a striking new astrophysical observation. It could even be a breakthrough produced by a completely new theory.

This is why we need an exploration in all possible directions. Very likely new physics will impact several observables in various fields, and we need to react rapidly to any possible hint. To discover new physics, we need specialists in the various disciplines, but they should have a global vision on what is happening in different fields of the frontier research. To allow a more effective circulation of the information among scientists of the various disciplines we would ask authors to write their papers highlighting the physics content of their findings, while trying to avoid, whenever possible, excessively complex technical details.

Finally, another important goal of the new journal will be to offer visibility to a new generation of scientists, contributing therefore to shape the future leadership of the field. As editorial choice we would like to invite some of the youngest protagonists of our field to cover the most recent results, mostly in form of review papers or papers comparing critically results coming from different experiments or competing theoretical approaches. Too often, particularly within the large collaborations, individuals providing key contributions to outstanding analyses or even completely new results, are well known. But this is not always true outside. This journal will be the right tool to promote early career of the most promising scientists, providing them an opportunity to shine worldwide.

For this first issue we have exceptionally decided to publish the proceedings of *The Rise of Particle Physics,* a beautiful conference held in Rome on 23–24 September 2024, to mark the fiftieth anniversary of the discovery of the $J/\psi$. Going through the papers of Nobel Prize winners, like Sam Ting and Giorgio Parisi, and of other key protagonists of the so called "November Revolution", the readers will get important insights on the complex, sometime chaotic process that shaped a major milestone in establishing the Standard Model, one of the most successful theories ever. Going through memories and considerations of the heated discussions of the period preceding the discovery of the $J/\psi$ it is inevitable to acknowledge the many similarities with the present situation. Today, as in the period 1970–1974, both experimentalists and theorists are struggling to identify the right direction where to focus their attention, with the aim of producing new ideas leading to breakthrough measurements or novel theories. I really hope that the reading of this first issue of HiHEP will be inspirational for many of them.

**Conflicts of Interest**

The author declares no conflict of interest.

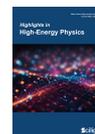

*Editorial*
# Preface

Antonio Davide Polosa

Dipartimento di FISICA, Sapienza University of Rome, 00185 Roma, Italy; antoniodavide.polosa@uniroma1.it

The conference "The Rise of Particle Physics" was conceived in late 2023 to commemorate the 50th anniversary of the discovery of the $J/\psi$ particle in 1974. Initially, we envisioned this conference as part of a series of history of physics events co-supported by the Institute of Physics (IOP); however, we soon recognized the need for a fully dedicated symposium. We applied for grants from Sapienza University of Rome and Istituto Nazionale di Fisica Nucleare (INFN), both of which decided to support our initiative.

The conference took place at Sapienza University of Rome on the 23 and 24 September 2024. More information about the event can be found on the official conference website: https://www.roma1.infn.it/conference/rise-hep-roma/

Following the conference, we began compiling and editing a document that includes (nearly) all written versions of the talks presented by the speakers. Finally, we established a collaboration with the editor Guido Tonelli with the purpose of publishing this work in the HiHep peer-reviewed journal.

The resulting document offers a conceptual journey through some of the most significant steps in the construction of the Standard Model of particle physics.

From the discovery of the $J/\psi$ particle in 1974, which triggered the "November Revolution" and solidified the Standard Model, to the development of Quantum Chromodynamics (QCD) and the experimental confirmation of weak interactions via the $W$ and $Z$ bosons, a reflection is made to trace the evolution of particle physics from a fragmented set of theoretical models to a coherent and predictive framework.

The 1970s, in particular, stand as a defining decade, transforming quantum field theory itself from a theoretical challenge (taming the infinities) into a toolkit to describe both electroweak and strong interactions.

Many of the contributions in this volume revisit that transformative period, culminating in the 2012 discovery of the Higgs boson at the Large Hadron Collider–an event that marked both a major scientific triumph and the beginning of new, enduring challenges in the field.

Beyond the theoretical breakthroughs, these proceedings highlight the role of experimental physics in the definition of the Standard Model. In addition, it is discussed how the advent of high-energy colliders, beginning with AdA and ADONE, followed by the proton-antiproton collider at CERN, shows the relation between technological innovation and fundamental research.

As we look ahead, the legacy of these discoveries continues to inspire new generations of physicists. Whether through the ongoing exploration of Beyond Standard Model physics, the search for new particles and phenomena, or the revolutionary field of gravitational wave detection, the pursuit of fundamental knowledge remains as strong as ever. It is our hope that this document not only preserves the history of these remarkable achievements but might also spark some discussions on the future directions of particle physics.

The authors thank all the contributors for their reflections, insights, and dedication to advance our understanding. May this volume serve as both a record of past achievements and a guidepost for future discoveries.

The Organizing Committee for this symposium consisted of Pia Astone, Fabio Bellini, Gianluca Cavoto, Riccardo Faccini, Stefano Giagu, Aleandro Nisati, Giulia Pancheri, Riccardo Paramatti, Antonio Davide Polosa, Shaharam Rahatlou, Paolo Valente, and Cecilia Voena.

We would like to extend a special thanks to Davide Germani for his support in editing the proceedings and to Mauro Mancini for his collaboration on various aspects related to the organization of the event.

**Conflicts of Interest**

The author declares no conflict of interest.

*Review*

# Discovery of the *J* Particle at Brookhaven National Laboratory and the Physics of Electrons and Positrons

Samuel C. C. Ting

Massachusetts Institute of Technology, Cambridge, MA 02139, USA; Samuel.Ting@cern.ch or sccting@mit.edu

**Abstract:** The discovery of the *J* particle in 1974 at Brookhaven National Laboratory led to the "November Revolution" in particle physics, fundamentally altering the Standard Model. In this article I review my experiments performed before, during, and after the "November Revolution" and their impact on modern physics.

**Keywords:** Charm; *J* Particle; AMS; L3; Gluons; Anti-matter

## 1. First experiment: Measuring the Size of the Electron (1966)

During my school years at Michigan and my years as a junior faculty member at Columbia University, I was very much interested in quantum electrodynamics (QED), particularly in various tests of QED at short distances using high-energy electron accelerators. QED, as formulated by Feynman, Schwinger, Tomonaga in 1948, assumes that electrons have no measurable radius. The theory agreed well with all experiments until the 6 GeV Cambridge Electron Accelerator (CEA) provided a most sensitive measurement of the size of the electron. At CEA, the Harvard experiment was done by the world's leading experts in the field who had spent many years to develop the technology [1]. Their results showed that the electron has a radius of ~$10^{-13}$–$10^{-14}$ cm (Figure 1). Most importantly, this experiment was independently confirmed by a group at the Cornell Electron Accelerator. Since those results touched upon the foundation of modern physics, I decided to perform an experiment with an independent method. At that time, I knew nothing about electron physics, so I received no support in the U.S. In 1965, I decided to leave Columbia University and move to the newly built 6 billion electron-volt electron accelerator (DESY) in Hamburg, Germany to re-measure the size of the electron. It was during this time at Columbia that I went to the Brandeis Summer School for Theoretical Physics and met with Luciano Maiani and have learnt a lot of physics from him, particularly the GIM mechanism [2].

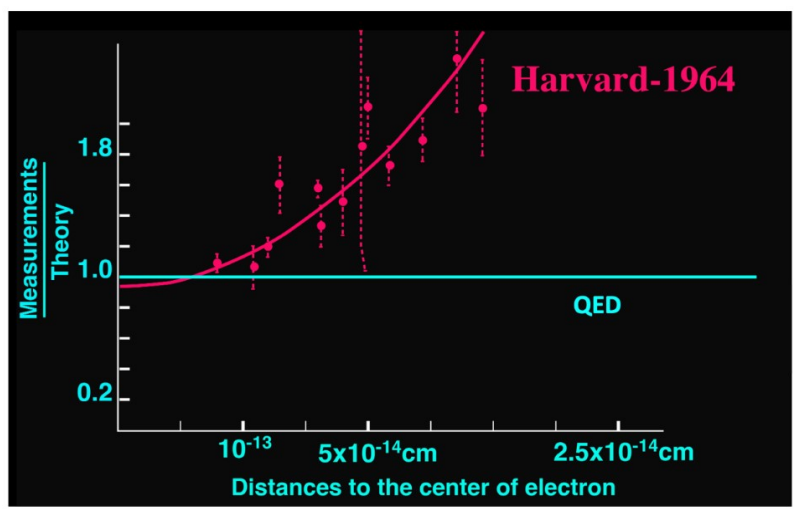

**Figure 1.** Results of the Harvard experiment showing that the electron has a radius of ~ $10^{-13}$–$10^{-14}$ cm.

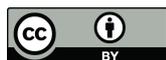

The layout of my experiment at DESY (Figure 2a) has the following unique features: use of dipole magnets and counters to measure the momentum (P); use of two Cherenkov counters separated by magnets on each arm to identify $e^\pm$, so that background $e^\pm$ produced from interactions in the first counter are swept away by the magnet and the $e^\pm$ identification of the two counters are independent; use of calorimeters to measure the energy (E); none of the detectors see the target so they are not exposed to neutron or gamma-ray backgrounds; the acceptance is defined by counters, not by the aperture of the magnet; require $E = P$ to reject large pion background. The development of this type of pair spectrometer (Figure 2b) eventually led to the $J$-Particle experiment.

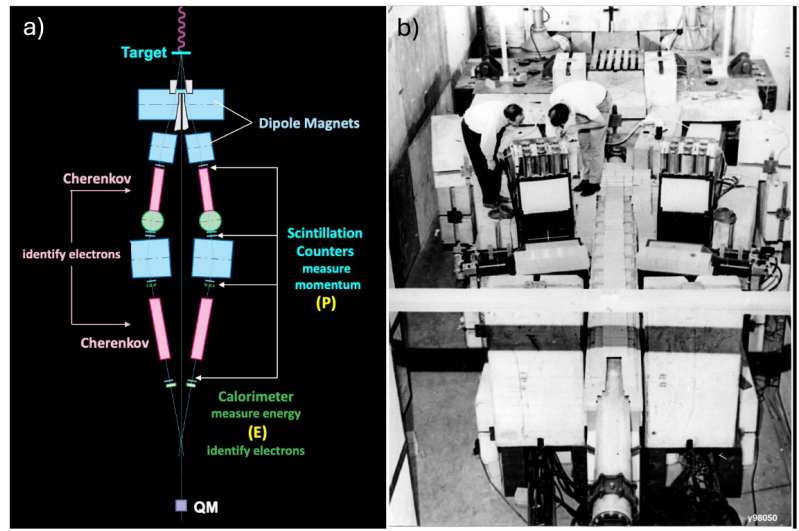

**Figure 2.** (**a**) Experimental layout of my experiment for electron size measurement at DESY; (**b**) Photo of the experiment for electron size measurement.

In 1966, after 8 months, our group completed the experiment at DESY and discovered that electron indeed has no measurable size $R_e$ <$10^{-14}$cm [3] (Figure 3). This result, which validated key aspects of QED, was first announced in 1966 at the "Rochester" conference at Berkeley (now known as the International Conference on High Energy Physics). On this occasion I met W.K.H. Panofsky, Dick Feynman, and I.I. Rabi. I maintained close contact with them for many years.

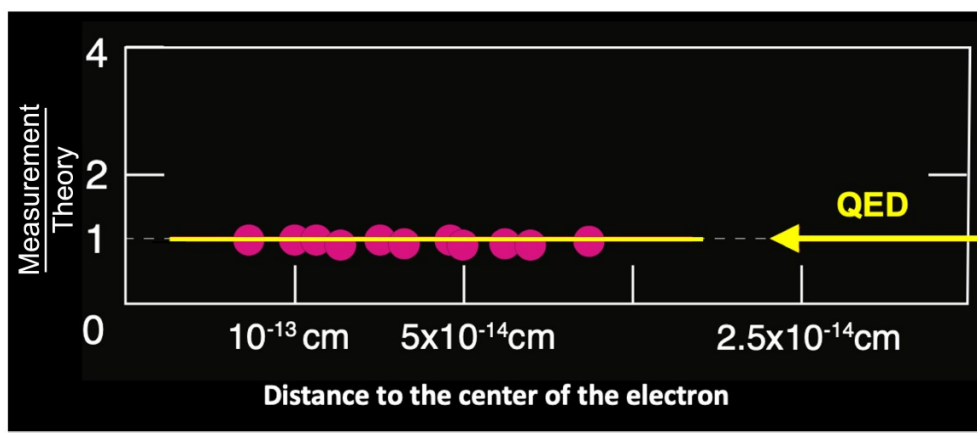

**Figure 3.** Results of our experiment showing that electron does not have measurable size up to $10^{-14}$ cm.

## 2. Studies on Photons and Heavy Photons

The QED experiment set the foundation for further studies in particle physics, showing the importance of precision measurements in particle physics. When we tuned the spectrometer magnets so that the pair mass acceptance is centered near 750 MeV, we observed a large increase in the $e^+e^-$ yield caused by an enhancement of the contribution to the $e^+e^-$ yield of the $\rho$-meson – a massive photon-like particle, which decays into $e^+e^-$ pairs [4] (Figure 4). The observation of $\rho \to e^+e^-$ decays started a series of experiments by my group on massive photon-like particles [5–10].

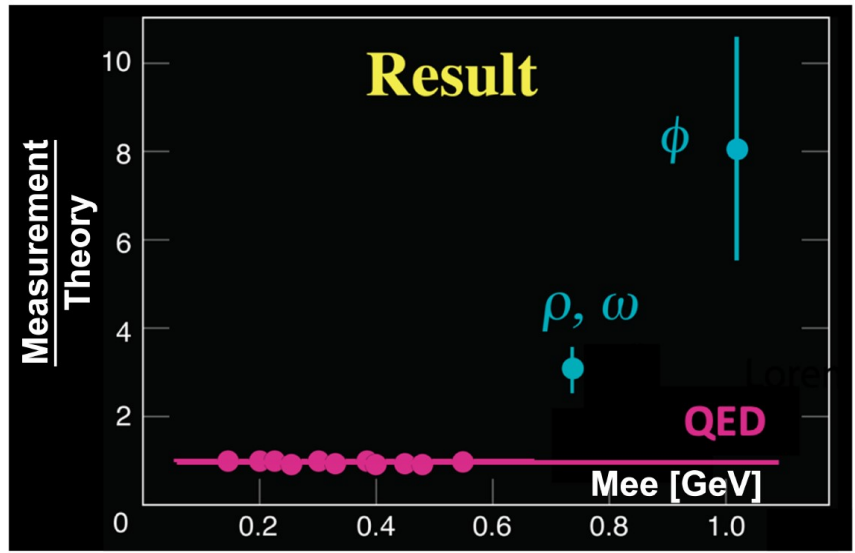

**Figure 4.** Deviation from QED due to heavy photon ($\rho$, $\omega$ and $\phi$) production.

The heavy photons $\rho$, $\omega$ and $\phi$ are resonance states of $\pi^+\pi^-$ ($\rho$), $\pi^+\pi^-\pi^0$ ($\omega$), and $K^+K^-$ or $\pi^+\pi^-\pi^0$ ($\phi$) with a rather short lifetime of typically between $10^{-24}$ and $10^{-23}$ s. They are unique in that they all have quantum numbers $J$ (spin) = 1, $C$ (charge conjugation) = $-1$, $P$ (parity) = $-1$. Thus, they are exactly like an ordinary light ray except for their heavy mass: $M_\rho = 760$ MeV, $M_\phi = 783$ MeV and $M_\phi = 1020$ MeV. Their interactions with hadrons are described by the Vector Dominance Model:

$$J_\mu(x) = \left[ \frac{m_\rho^2}{2\gamma_\rho}\rho_\mu + \frac{m_\omega^2}{2\gamma_\omega}\omega_\mu + \frac{m_\phi^2}{2\gamma_\phi}\phi_\mu \right] \tag{1}$$

To carry out these experiments accurately, we improved the detector mass resolution to $\sim$5 MeV and the background rejection to $10^8$. This allowed us to measure $\rho - \omega$ coherent interference using $\rho \to e^+e^-$ and $\omega \to e^+e^-$ decays (Figures 5 and 6) as well as forbidden $\omega \to \pi^+\pi^-$ decays, which at that time attracted significant attention [11–15] (Figures 7 and 8).

Precision measurements of the widths $\Gamma(\rho \to e^+e^-)$, $\Gamma(\omega \to e^+e^-)$ (Figures 6–8), and $\Gamma(\phi \to e^+e^-)$ (Figure 9a) resulted in verification of Weinberg's first sum rule (Figure 9b).

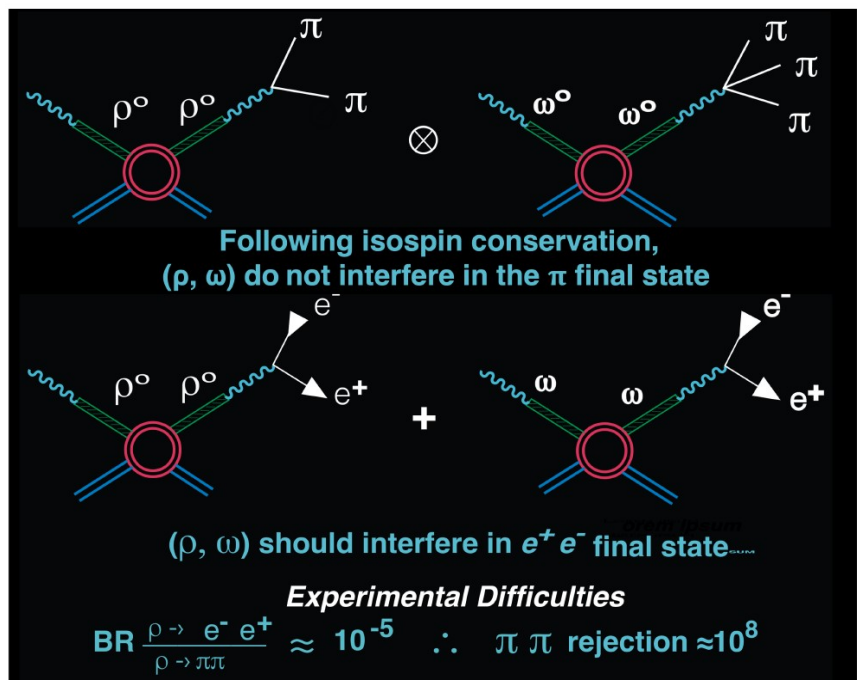

**Figure 5.** Feynman diagrams of $\rho - \omega$ coherent interference.

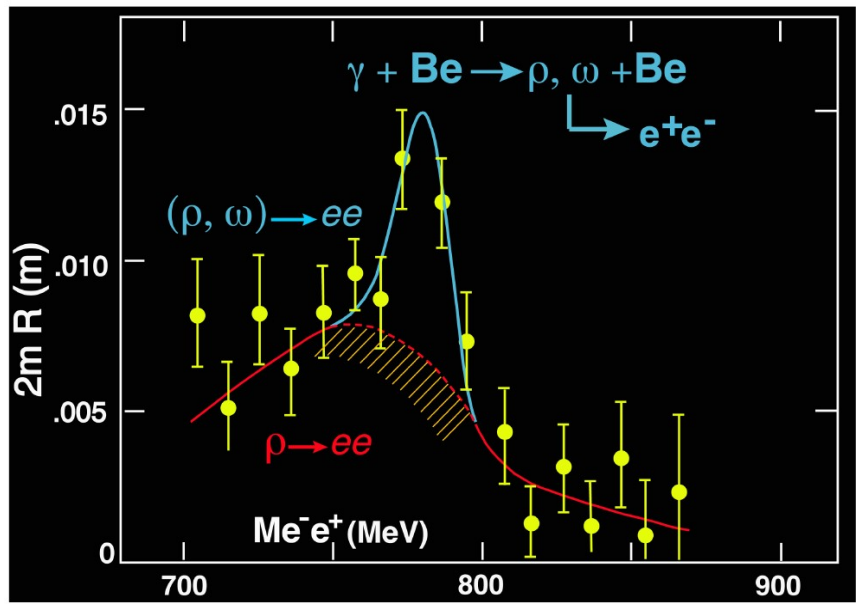

**Figure 6.** Observation of $\rho - \omega$ coherent interference in the $e^+e^-$ final state.

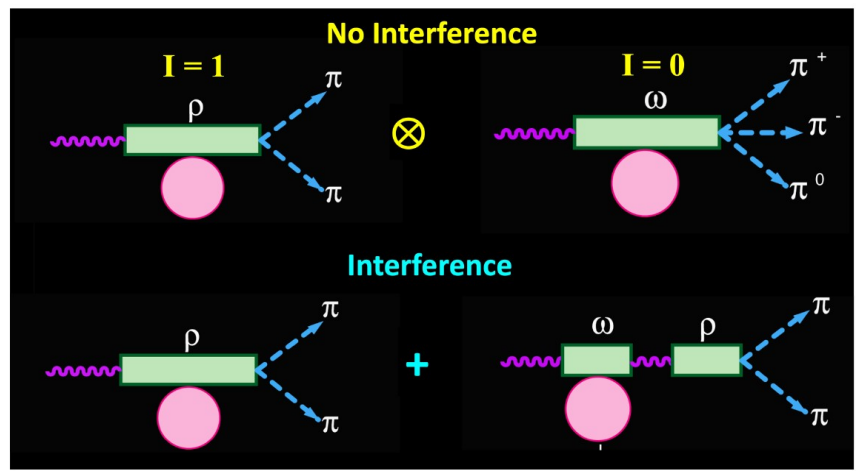

**Figure 7.** Feynman diagrams of forbidden $\omega \rightarrow \pi^+\pi^-$ decays due to isospin $I$ violation.

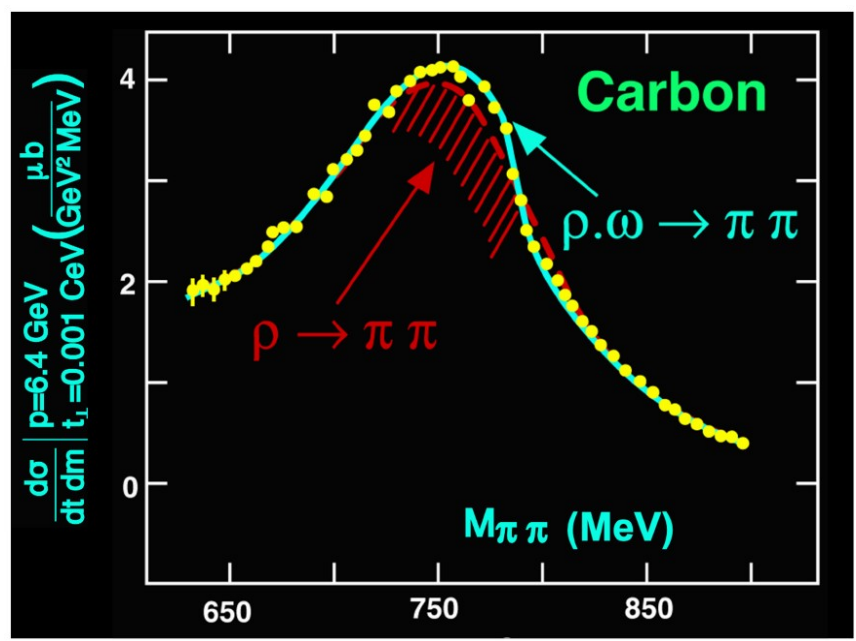

**Figure 8.** First observation of forbidden $\omega \rightarrow \pi^+\pi^-$ decays.

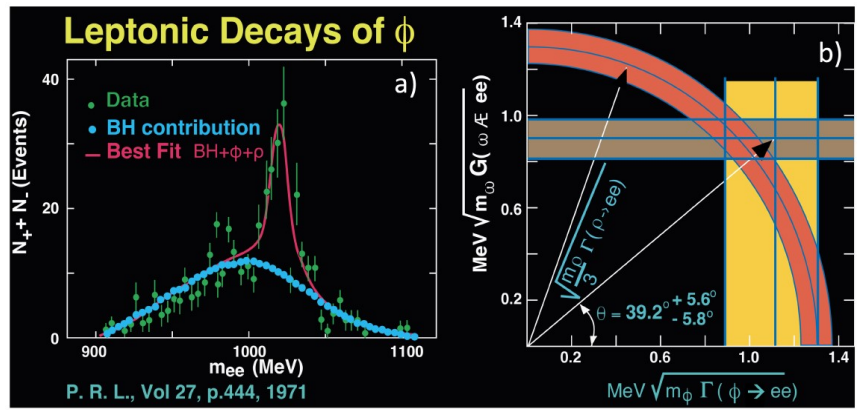

**Figure 9.** (a) invariant mass of $e^+e^-$ pairs showing the $\phi$ meson peak; (b) first validation of Weinberg's first sum rule using our data on $\Gamma(\rho \to e^+e^-)$, $\Gamma(\omega \to e^+e^-)$ and $\Gamma(\phi \to e^+e^-)$.

## 3. Discovery of the *J* Particle—The Brookhaven Experiment (1972–1974)

From previous experiments we have learned that photons and heavy photons are almost the same. They transform into each other. We can now ask a simple question: how many heavy photons exist? And what are their properties? It was inconceivable to me that there should be only three of them, and all with a mass around 1 GeV. To answer these questions, I decided to perform the first large-scale experiment to search for more heavy photons by detecting their $e^+e^-$ decay modes up to much higher mass. Figure 10 shows the photocopy of a page of the proposal E598 to Brookhaven National Laboratory. It gives the reasons I presented, in the spring of 1972, for performing an $e^+e^-$ experiment in a proton beam (Figure 11).

> The best way to search for vector mesons is through production experiments of the type p + p → V⁰ + X . The reasons are:
>           └ e⁺e⁻
>
>  (a) The V⁰ are produced via strong interactions, thus a high production cross section.
>
>  (b) One can use a high intensity, high duty cycle extracted beam.
>
>  (c) An e⁺e⁻ enhancement limits the quantum number to 1⁻, thus enabling us to avoid measurements of angular distribution of decay products.
>
>  Contrary to popular belief, the e⁺e⁻ storage ring is not the best place to look for vector mesons. In the e⁺e⁻ storage ring, the energy is well-defined. A systematic search for heavier mesons requires a continuous variation and monitoring of the energy of the two colliding beams—a difficult task requiring almost infinite machine time. Storage ring is best suited to perform detailed studies of vector meson parameters once they have been found.

**Figure 10.** Page 4 of proposal E598 submitted to Brookhaven National Laboratory early in 1972 and approved in May of the same year.

From our experience at DESY, we felt the best way to build an electron-pair spectrometer that could handle high intensities with high background rejection and at the same time have a large mass acceptance and a good mass resolution, is to repeat the concept of our spectrometer at DESY, i.e. a large double arm spectrometer and with all the detectors behind the magnets so they would not view the target directly and are not exposed to neutrons and gamma rays. To obtain the best mass resolution, we used magnets to bend the particles vertically for momentum measurement, while measuring production angles in the horizontal plane. Figure 12 shows the layout of the spectrometer and related detectors.

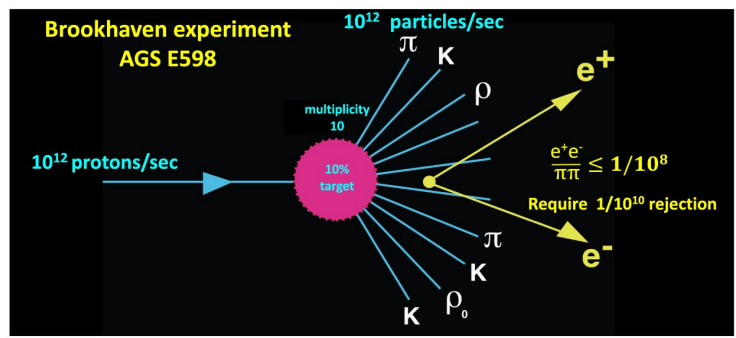

**Figure 11.** Concept of the AGS Experiment E598. The extracted beam of $10^{12}$ protons/s interact with a 10% target. The multiplicity is 10, resulting in $10^{12}$ particles/s from the target volume. The ratio $e^+e^-/\pi^+\pi^-$ is less than $1/10^8$, so a percent accuracy measurement requires $1/10^{10}$ rejection.

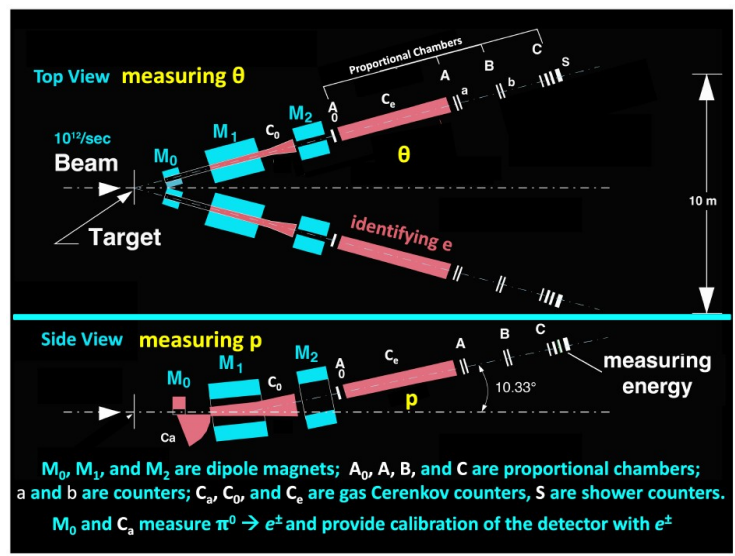

**Figure 12.** Layout of the AGS Experiment E598, which is an upgraded precision version of the DESY experiment.

The main features of the spectrometer are the following:

(1) Shielding. Shielding the detector and the control room from $10^{12}$ particles per second generated in the experimental area was of the utmost importance. The total shielding used was approximately (a) 10,000 tons of concrete, (b) 100 tons of lead, (c) 5 tons of uranium, (d) 5 tons of soap – placed on top of $C_0$, between $M_1$ and $M_2$ and around the front of $C_e$ to stop soft neutrons (Figure 13).

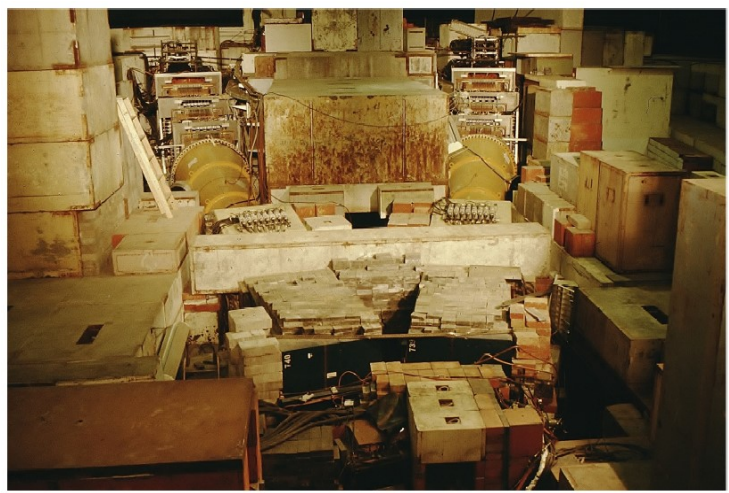

**Figure 13.** Shielding arrangement with roof open.

(2) The target. The target consists of nine pieces of $1.78$ mm thick beryllium, each separated by $7.5$ cm so that particles produced in one piece and accepted by the spectrometer do not pass through the next piece (Figure 14). This arrangement rejects accidental pairs by requiring both tracks come from the same origin.

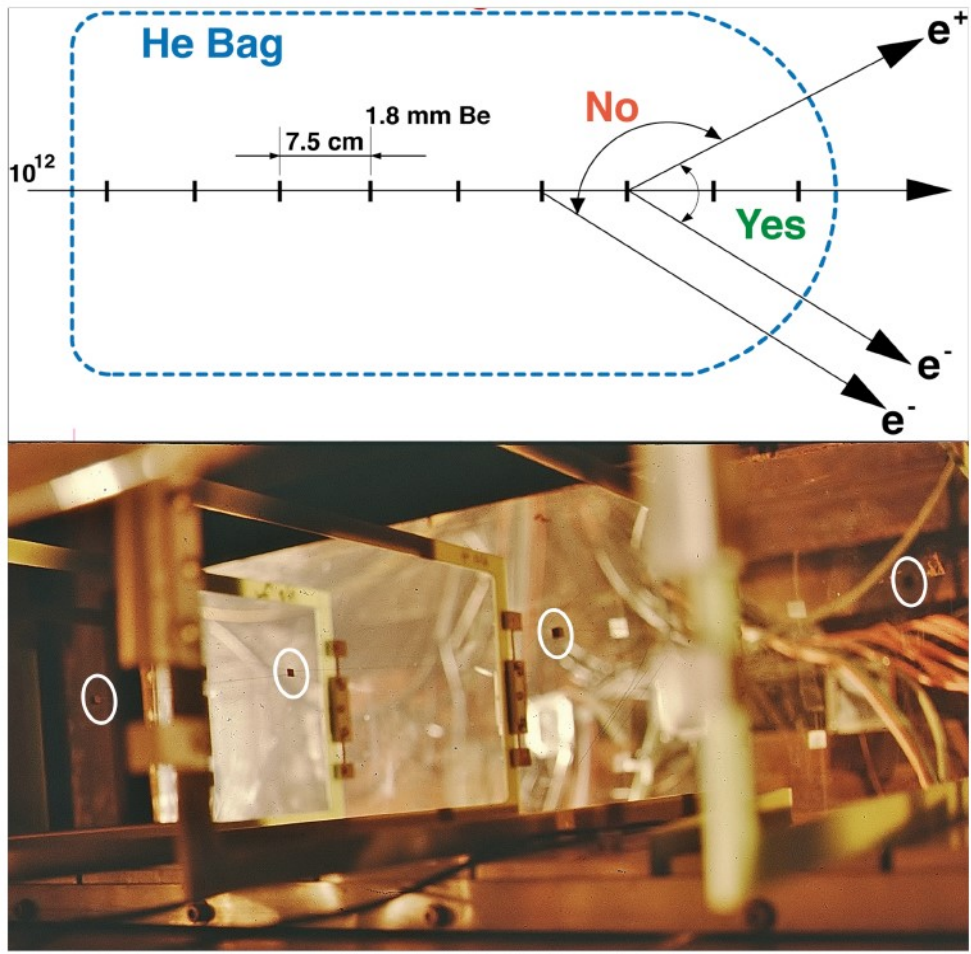

**Figure 14.** Nine separate targets to reduce the background.

(3) The magnet system. The magnetic field is measured with $3-\mathrm{D}$ Hall probe in $10^5$ points. The bending power of the dipole magnets $M_0$, $M_1$ and $M_2$ are such that none of the counters sees the target directly (Figure 15). The detector is smaller than the aperture of the magnets, so the detector itself defines the acceptance. Calibration of the detectors with pure electron beam produced in the target from $\pi^0 \rightarrow e^+e^-\gamma$ was performed (Figure 16).

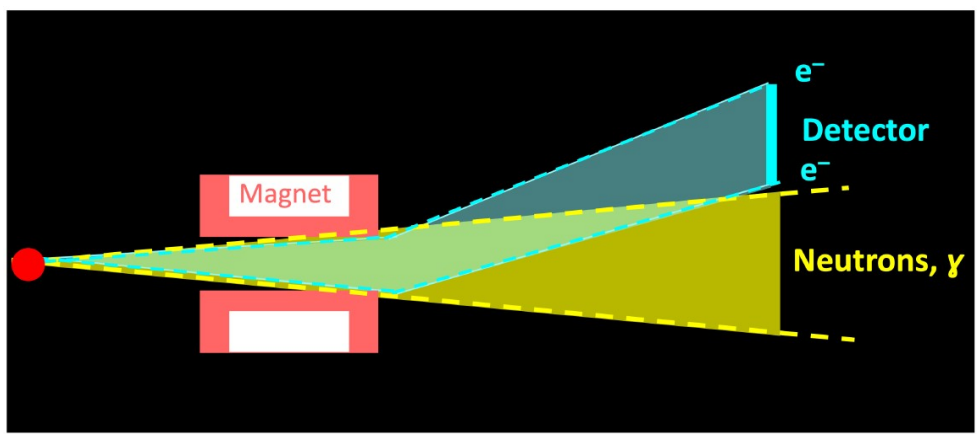

**Figure 15.** The magnets bend charged particles to an angle such that the detectors are not exposed to photons or neutrons from the target.

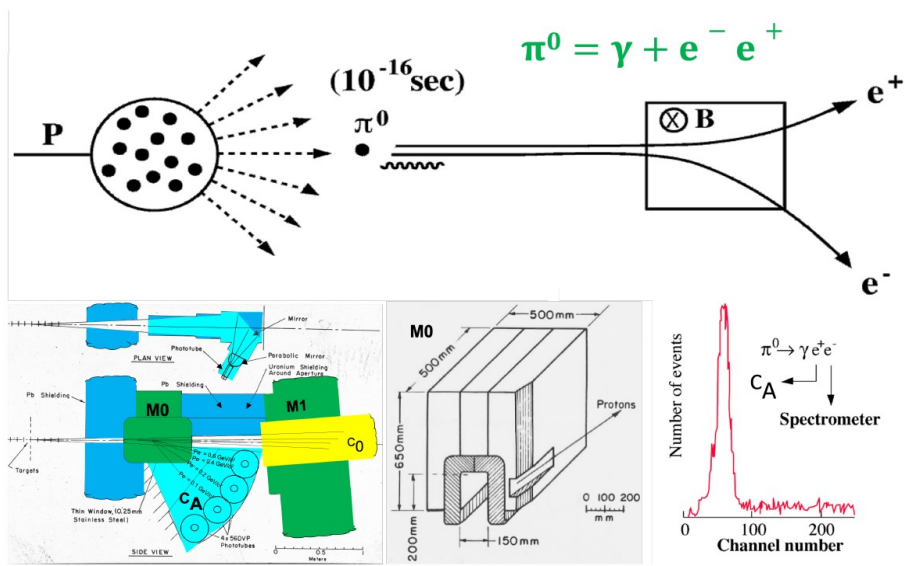

**Figure 16.** Detector calibration with a pure electron beam by placing a specially designed magnet $M_0$ close to the target followed by a special Cherenkov counter, $C_A$, to detect positrons from $\pi^0 \to \gamma e^+ e^-$ ensuring the electron entering the spectrometer.

(4) The position detectors. $A_0$, $A$, $B$ and $C$ are multiwire proportional chambers designed by the late Professor U.J. Becker. They consist of more than 8000 very fine, 20 µm, gold-plated wires, 2 mm apart, each with its own readout chain. Chambers $A$, $B$ and $C$ have wire planes rotated $60°$ with respect to each other, so that for a given hit, the sum of distances to the wire planes is a constant—a unique feature for sorting out multi-hit events and rejecting backgrounds (Figure 17). The chambers were operating at low voltage and with a special gas mixture such that they were able to operate at a rate of 20 MHz, and were also able to sort out as many as eight particles simultaneously in each arm.

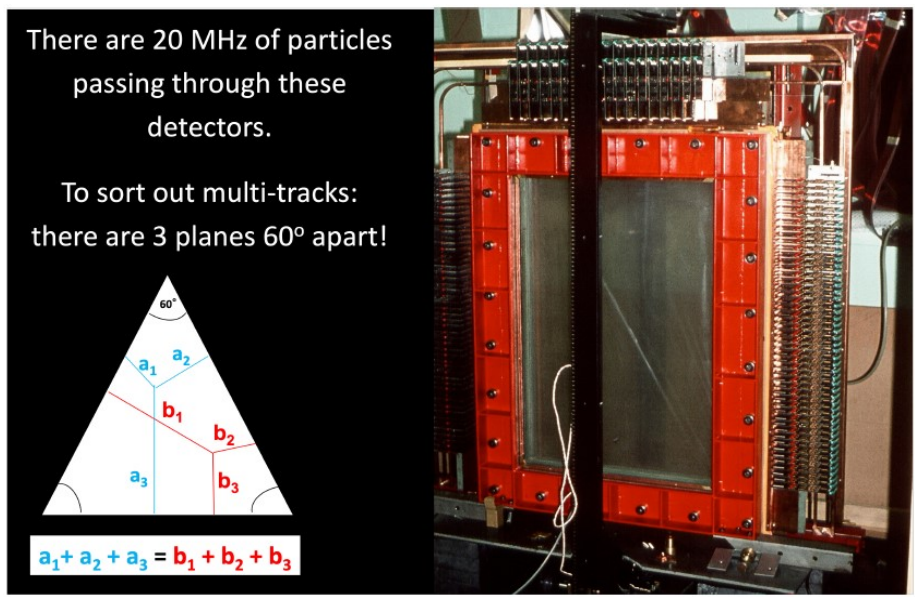

**Figure 17.** Precision position detectors, which were designed by the late Professor UJ Becker. The chamber, shown on the right, is on display in Smithsonian Institution in Washington, DC after completion of the experiment.

(5) The $\pi - e$ separation was achieved by four extremely sensitive Cherenkov Counters $C_0$, $C_E$ (Figure 18), which were designed by M. Vivargent, J. J. Aubert and myself and manufactured at LAPP, Annecy, France. Figure 19 shows a photo of J.J. Aubert in the control room.

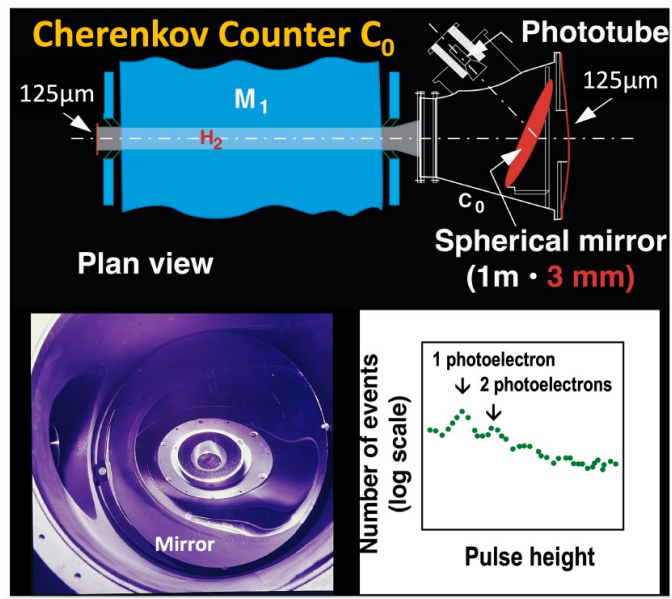

**Figure 18.** The $\pi - e$ separation was achieved by four extremely sensitive Cherenkov Counters $C_o$, $C_e$.

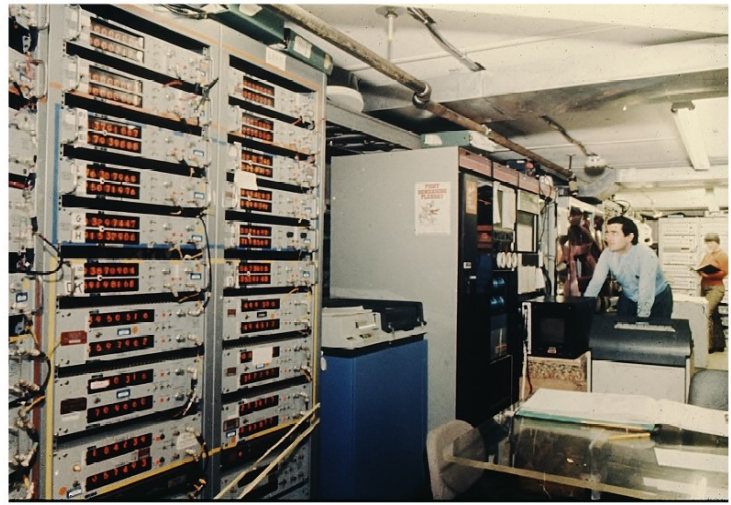

**Figure 19.** J. J. Aubert, Professor of Physics, University of Marseille, Director-General, IN2P3, France.

In the early summer of 1974 we took some data in the high-mass region of 4–5 GeV. However, analysis of the data showed very few electron-positron pairs.

By the end of August 1974 we tuned the magnets to accept an effective mass of 2.5–4.0 GeV. Immediately we saw clean, real, electron pairs. But most surprising of all is that most of the $e^+e^-$ pairs peaked narrowly at 3.1 GeV (Figure 20a). A more detailed analysis showed that the width was less than 5 MeV.

To make sure the peak we observed was a real effect and not due to the instrumentation bias, we have performed several experimental checks on our data and on the data analysis. The most important one was to collect another set of data with the magnet current lowered by 10%. This has the effect of moving the particles into different parts of the detector. If the peak is false, it will shift away. The fact that the peak remained fixed at 3.1 GeV (Figure 20b) showed right away that a real particle had been discovered [16,17].

I was considering announcing our results during the retirement ceremony for V. F. Weisskopf, who had helped us a great deal during the course of many of our experiments. This ceremony was to be held on 17 and 18 October 1974. I postponed the announcement for two reasons.

First, there were speculations on high-mass $e^+e^-$ pair production from proton-proton collisions as coming from a two-step process: $p + N \rightarrow \pi + \ldots$, where the pion undergoes a second collision $\pi + N \rightarrow e^+e^- + \ldots$ This could be checked by a measurement based on target thickness. The yield from a two-step process would increase quadratically with target thickness, whereas for a one-step process the yield increases linearly. This was quickly done.

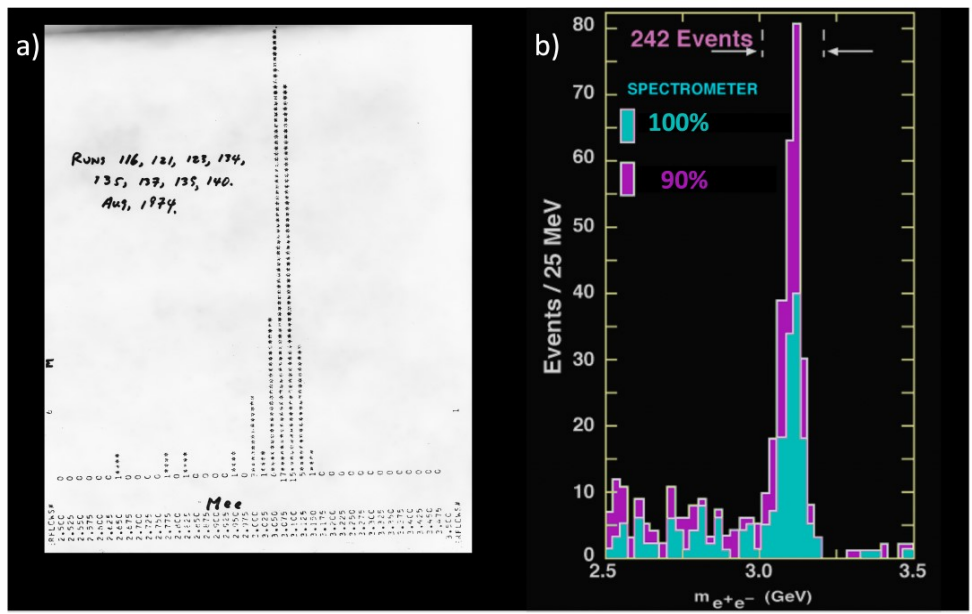

**Figure 20.** (**a**) First observation of the $J$ particle peak in August 1974. (**b**) Stability of the peak position against the change of magnetic field strength.

Second, we realized that there were earlier Brookhaven measurements [18] of direct production of muons and pions in nucleon-nucleon collisions which gave the $\mu/\pi$ ratio as $10^{-4}$, a mysterious ratio that seemed not to change from 2000 GeV of lab energy at the ISR down to 30 GeV.

This value was an order of magnitude larger than expected in terms of the three known vector mesons, $\rho$, $\omega$ and $\phi$, which, at that time, were the only possible "intermediaries" between the strong and electromagnetic interactions. We then added the $J$ meson to the three and found that the linear combination of the four vector mesons could not explain the $\mu/\pi$ ratio either.

This I took as an indication that something exciting might be just around the corner, so I decided that we would make a direct measurement of this number. Since we could not measure the $\mu/\pi$ ratio with our spectrometer, we decided to look into the possibility of investigating the $e^-/\pi^-$ ratio. On Thursday, 7 November, we made a major change in the spectrometer (Figure 21) to start the new experiment to search for more particles.

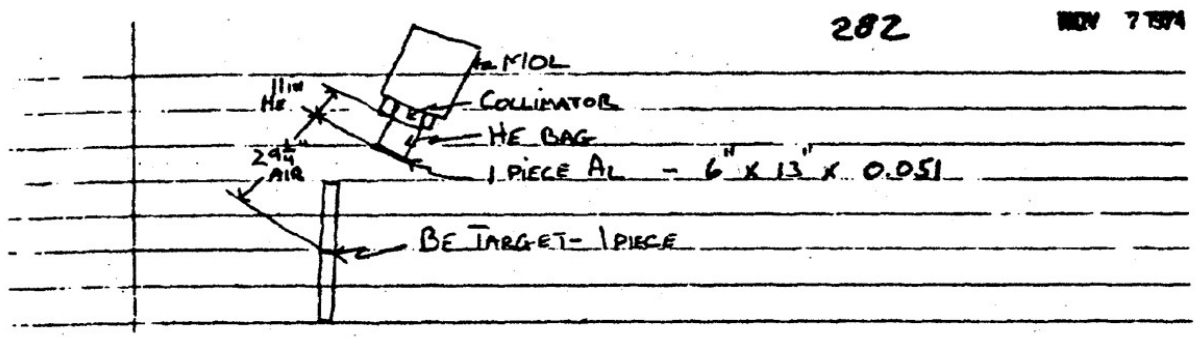

**Figure 21.** Aluminum foil arrangement in front of magnet $M_0$ in our new experiment to determine the $e/\pi$ ratio. The converter was used to determine the electron background yield.

On 6 November I paid a visit to G. Trigg, Editor of *Physical Review Letters*, to find out if the rules for publication without refereeing had been changed. Following that visit, I wrote a simple draft of a letter which emphasized only the discovery of $J$ particle and the checks we made on the data. The group photo (Figure 22) was taken when the paper was accepted for publication. On 11 November we telephoned G. Bellettini, the Director of Frascati Laboratory, informing him of our results [16,17,19]. At Frascati they started a search on 13 November, and called us back on 15 November to tell us excitedly that they had also seen the $J$ signal. They were able to publish their results [20] in the same issue of *Physical Review Letters* as ours and the results from SLAC [21] (Figure 23). This discovery was widely discussed in the media [22] (Figure 24). The impact of these papers on our understanding of particle physics is known as the "November Revolution".

The properties of the $J$ particle are truly unique: its lifetime is $10,000$ times longer than other hadronic particles. The significance of this is similar to suddenly discovering, in a remote region of the Earth, a village where people live to be, instead of 100 years old, about 1 million years old; its transitions spectrum is similar to positronium (Figure 25). This implies the existence of a new kind of matter made out of a new kind of quark-antiquark.

Many accelerators were built to study the detailed properties of this particle (Figure 26). Continuous, 40-year long studies were performed at the Beijing Electron-Positron Collider (Figures 27 and 28) where 30 new hadrons have been discovered from charmed meson production and decays by the BES detectors (Figure 29).

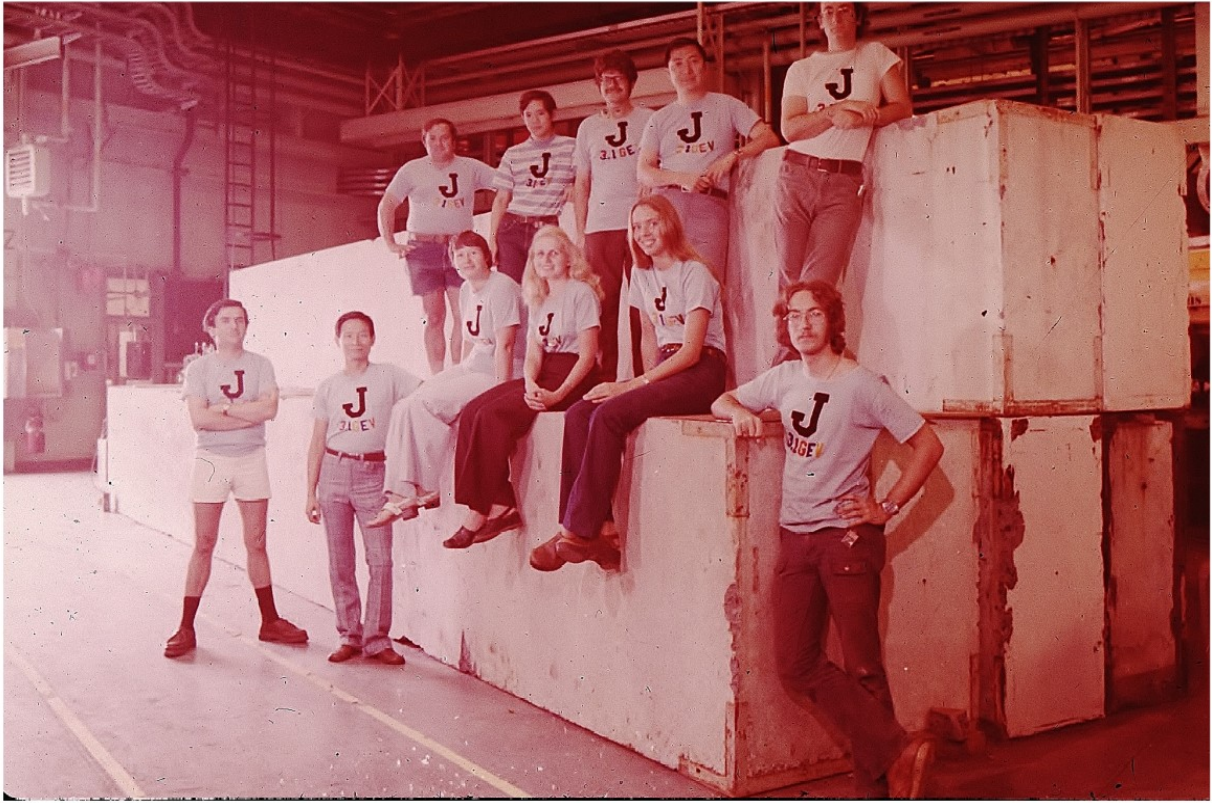

**Figure 22.** Members of the J-Particle Group.

**Figure 23.** The "November Revolution" – papers on a narrow hadronic resonance with a mass of 3.1 GeV published in the December 1974 issue of *Physical Review Letters*.

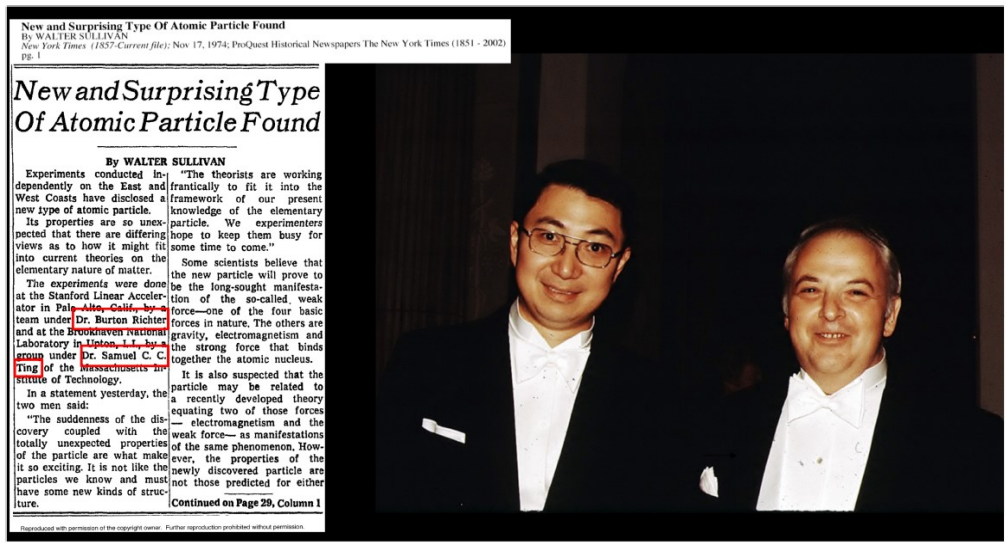

**Figure 24.** (**left**) Article about discovery of a new form of matter in New York Times [22]; (**right**) Myself and Professor B.Richter in Stockholm two years later.

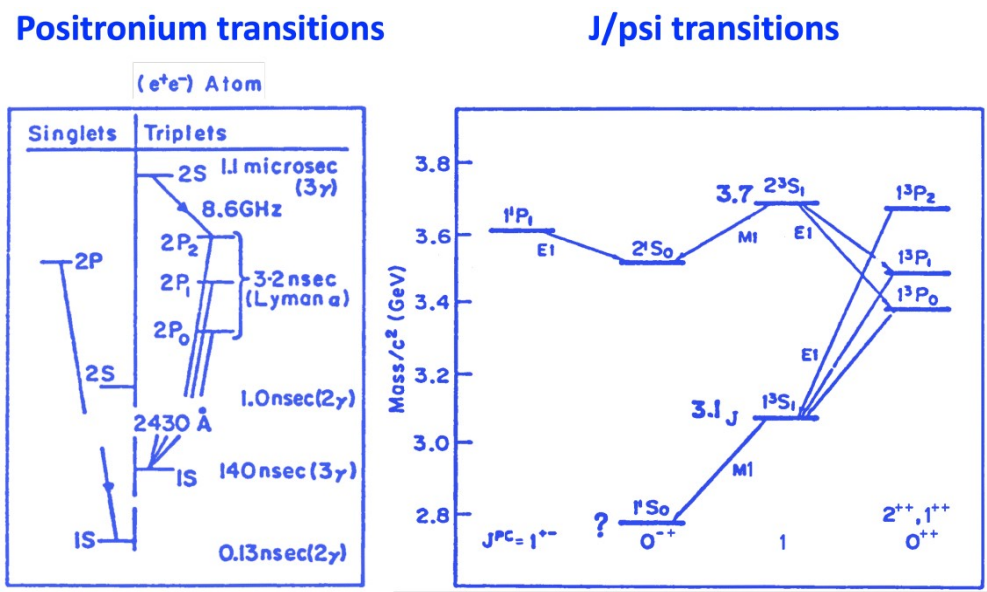

**Figure 25.** The transitions spectrum of the $J$ particle is similar to positronium. This implies the existence of a new kind of matter made out of a new kind of quark-antiquark.

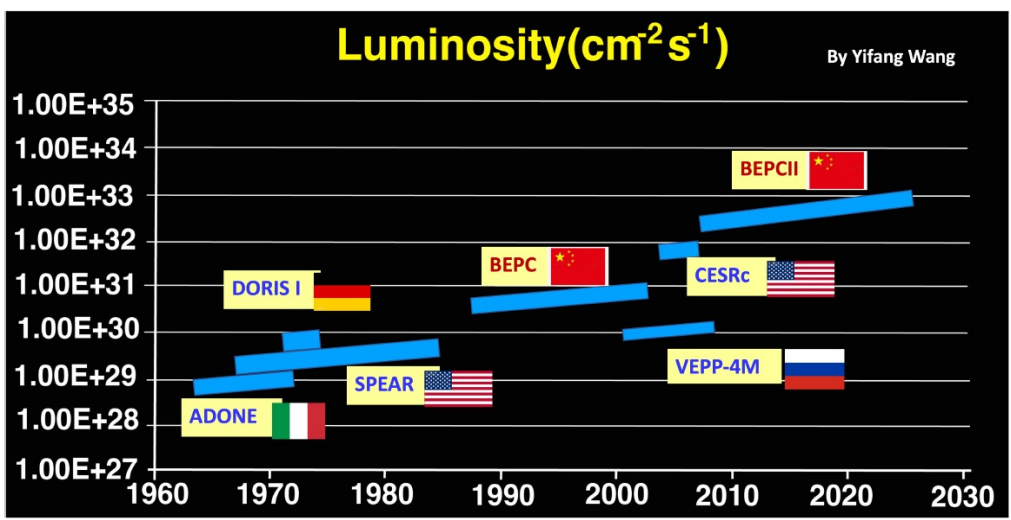

**Figure 26.** World tau-charm factories and their integral luminosities over time.

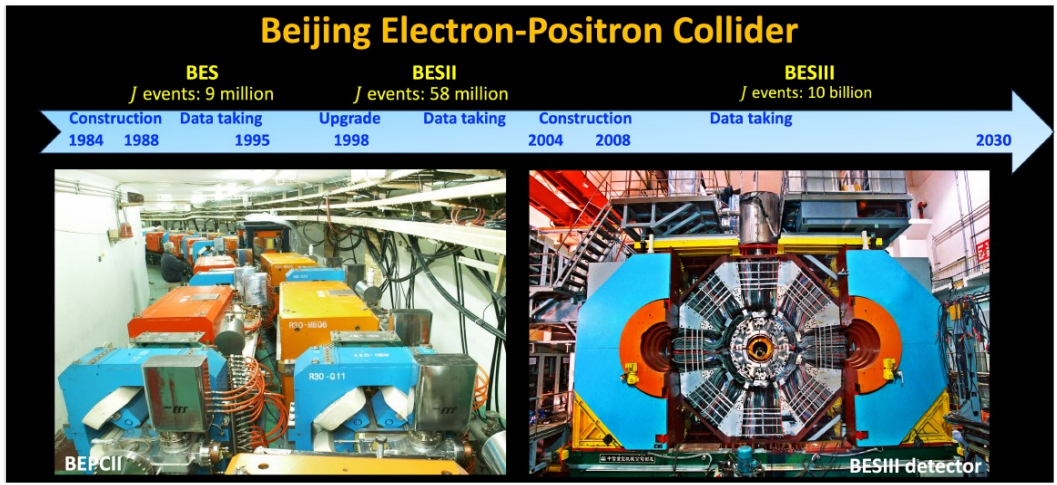

**Figure 27.** Beijing Electron-Positron Collider, BEPC and the BES detector running for 40 years at BEPC.

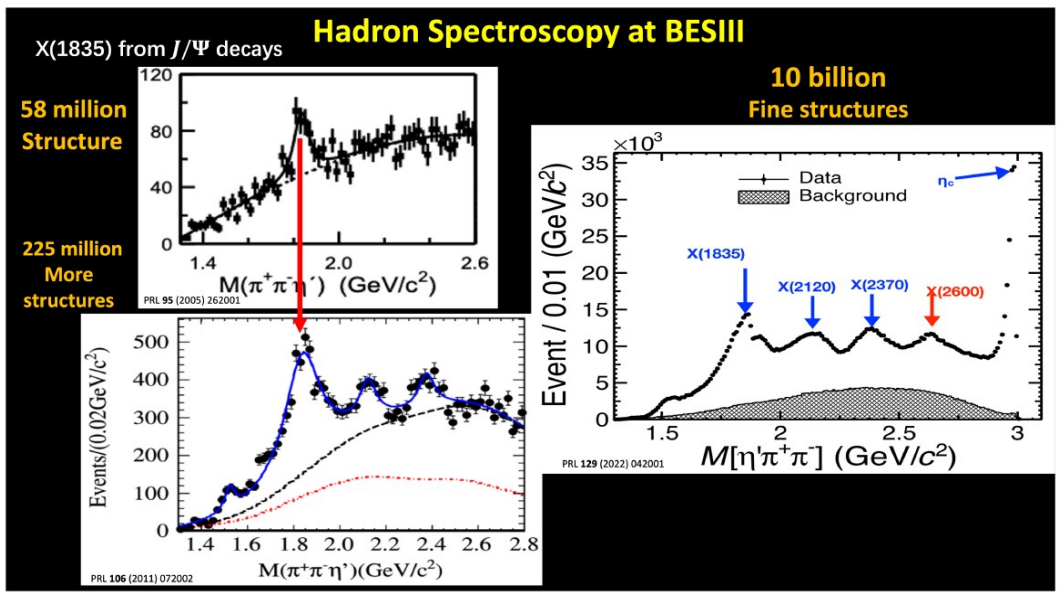

**Figure 28.** Hadron spectroscopy with the BES detectors at BEPC.

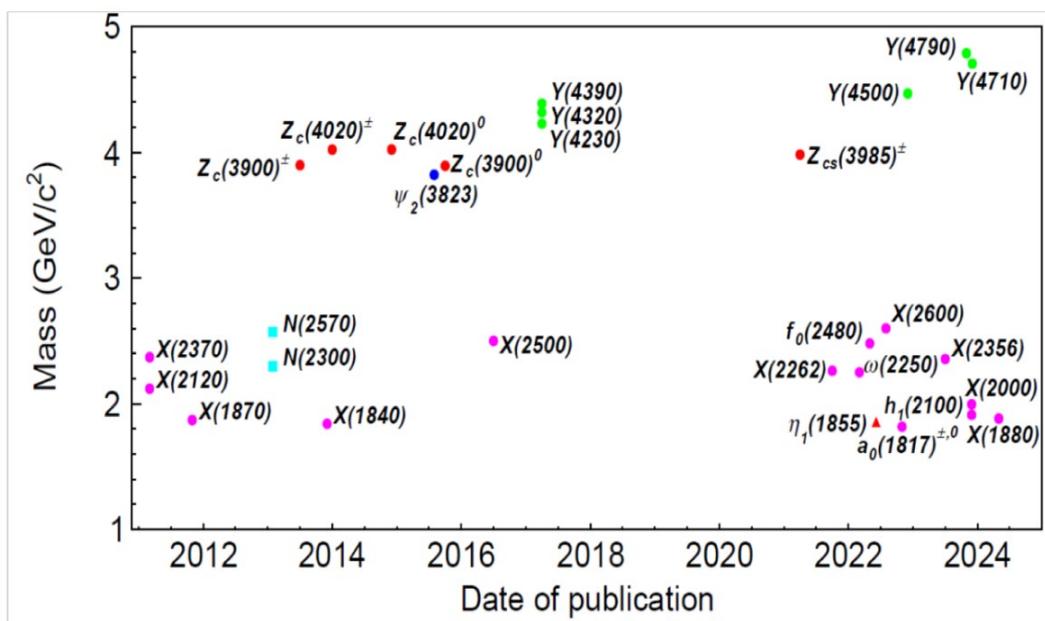

**Figure 29.** 30 new hadrons were discovered by the BES detectors from charmed meson production and decays [23].

## 4. MARK–J Experiment at DESY

In 1976 our group, in collaboration with institutes from the Europe and Asia, submitted to DESY a proposal for the MARK–J detector to measure $e^+e^-$ reactions at high energies, eventually up to $E_{cm} = 46$ GeV. The detector was designed to cover approximately $4\pi$ sr solid angle, and to measure and distinguish hadrons, electrons, neutral particles and muons. The proposal was promptly accepted. With this detector we planned to do a wide range of studies including measurements of interference effects between weak and electromagnetic interactions, look for structures in the total hadronic cross section, searches for new quarks, vector mesons and heavy leptons, study the structure of hadronic jets, etc.

The experiment, running at the PETRA $e^+e^-$ collider, produced important results on the interference effects between weak and electromagnetic interactions by studying the charge asymmetry in the reaction $e^+e^- \rightarrow \mu^+\mu^-$ [24,25] (Figure 30a), clearly showing the contribution of the weak current, long before the discovery of the $Z^0$ boson at CERN [26]. This was the earliest confirmation of electroweak theory, which provided the first opportunity to distinguish between the Standard Model [27–29] and other models that yield indistinguishable predictions for low-energy, low-momentum-transfer experiments (Figure 30b).

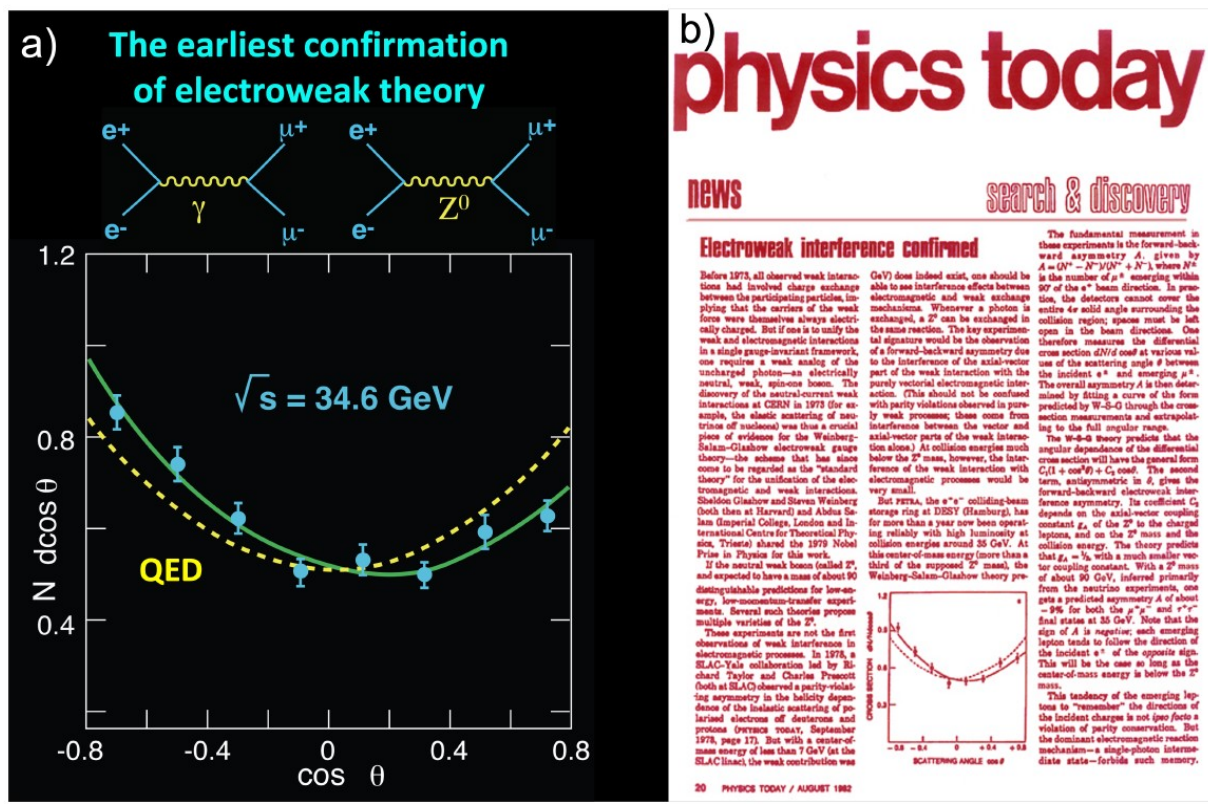

**Figure 30.** (**a**) Results on forward-backward asymmetry in the reaction $e^+e^- \rightarrow \mu^+\mu^-$ showing the contribution of the $Z^0$ boson. (**b**) Article in the August 1982 issue of *Physics Today* devoted to the observation of electroweak interference [26].

Most importantly, the experiment analyzed the properties of three-jet events, which led to the discovery of gluons [30]—the carriers of the strong force that "glue" quarks together into protons, neutrons and other particles known collectively as hadrons. The three-jet topology was clearly visible and interpreted as $q\bar{q}$ gluon bremsstrahlung (Figure 31a). MARK–J experiment was the first to report statistically significant evidence of the three-jet event pattern [31]. This observation was later confirmed by other PETRA experiments. This observation is the key in establishing the theory of the strong force [32], known as quantum chromodynamics or QCD (Figure 31b).

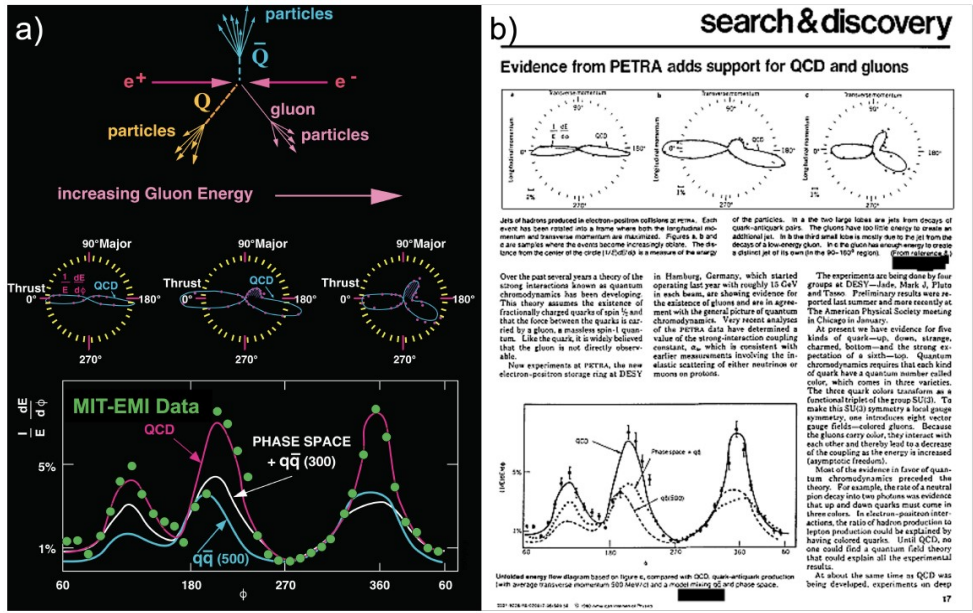

**Figure 31.** (**a**) Angular distribution of three-jet events showing bremsstrahlung emission of gluons. (**b**) Article in February 1980 issue of *Physics Today* showing the MARK–J results to the discovery of gluons [32].

## 5. L3 Experiment at CERN (1982–2003)

We spent 20 years, 1982–2003, building and operating the L3 experiment (Figure 32) at the electron-positron collider LEP at CERN. The experiment was designed to study $e^+e^-$ collisions in the center-of-mass energies ranging from 90 to 200 GeV with emphasis on high resolution energy measurements of electrons, photons and muons. It was an effort involving a worldwide collaboration of 600 physicists from 20 countries. It was the first large-scale scientific collaboration between the United States, China, the Soviet Union, India, East and West Germany.

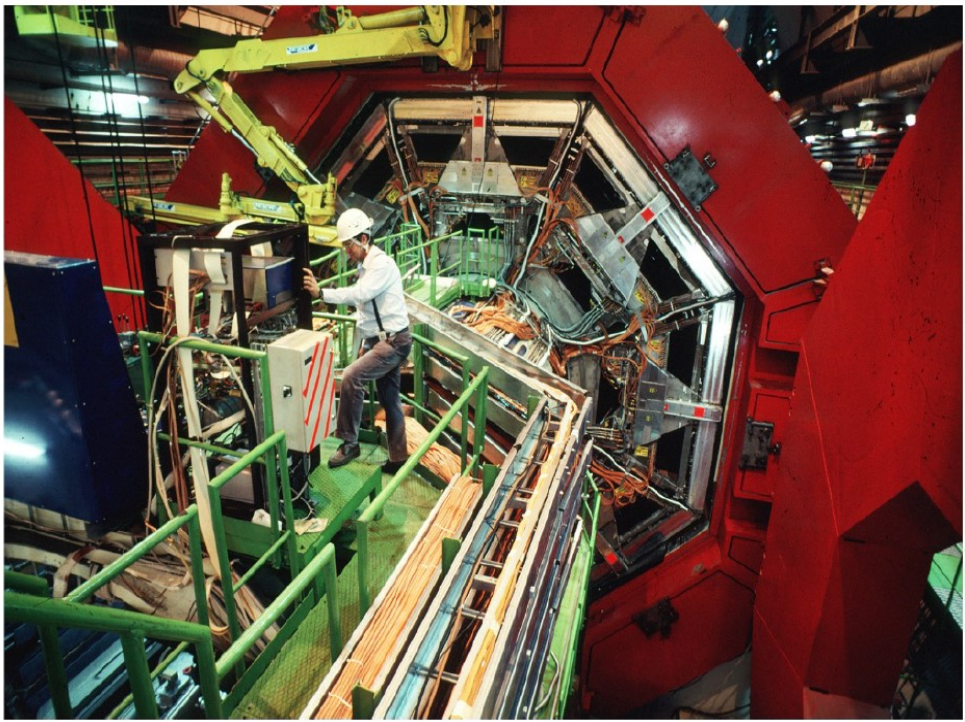

**Figure 32.** L3 detector at LEP.

The L3 detector conceptually differs from a standard $e^+e^-$ collider detector by its emphasis on high resolution measurements of leptons, photons and jets. This is implemented in the experimental setup by an accurate tracking system, a high-resolution muon spectrometer, a precision electromagnetic calorimeter as well as a $4\pi$ fine-grain hadron calorimetry. All the detectors were installed within a 10 thousand ton magnet providing a 0.5 T field. We have chosen a relatively low field in a large volume to optimize the muon momentum resolution, which improves linearly with the field but quadratically with the track length. High resolution is essential for detecting rare new phenomena with sufficient signal-to-noise ratio; identifying exclusive and inclusive final states and rejecting backgrounds; and analyzing final state properties by measuring particle energy, momentum and reconstructing mass spectra. The construction of the experiment took eight years from its conception to the beginning of data taking in summer 1989.

Our research program at LEP was very broad: precision measurements of the $Z$ boson properties (mass, width and decay channels); determination of the number of light neutrino families; measurements of the electroweak force ($\alpha(Q^2)$, $\sin\theta_W$, gauge couplings, ...); direct Higgs boson searches and Higgs mass constraints from precision electroweak measurements; QCD tests (evolution of the strong coupling constant, $\alpha_s(\sqrt{s})$, and structure of gluon jets); physics of photon final states; study of two photon interactions; testing and constraining many theories beyond the Standard Model, including supersymmetry (SUSY) and other models based on exotic particles.

We have published over 300 papers in Physics Letters. Our most notable results include studies of the energy evolution of the strong [33] (Figure 33a) and electromagnetic [34] (Figure 33b) coupling constants, as well as the model-independent determination of the number of light neutrinos [35] (Figure 34). All the results agreed with the standard model [36]. That is rather unfortunate, because when an experiment agrees with the model, what you learn is limited. When an experiment disagrees with a model, you learn much more, obviously. After 20 years of precision measurements at LEP, we have found that the electron still has no measurable size, its radius is less than $10^{-17}$ centimeters.

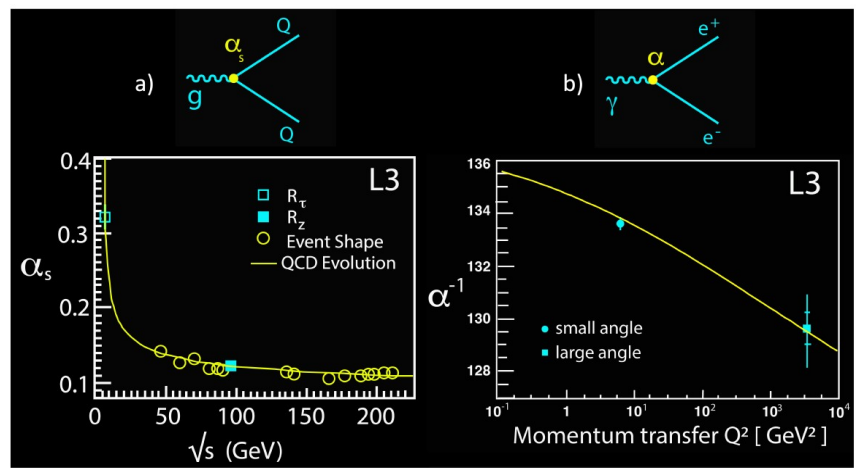

**Figure 33.** L3 experimental results: (**a**) dependence of the strong coupling constant, $\alpha_s$, on center-of-mass energy $\sqrt{s}$; (**b**) dependence of the electromagnetic fine structure constant, $\alpha$, on momentum transfer $Q^2$.

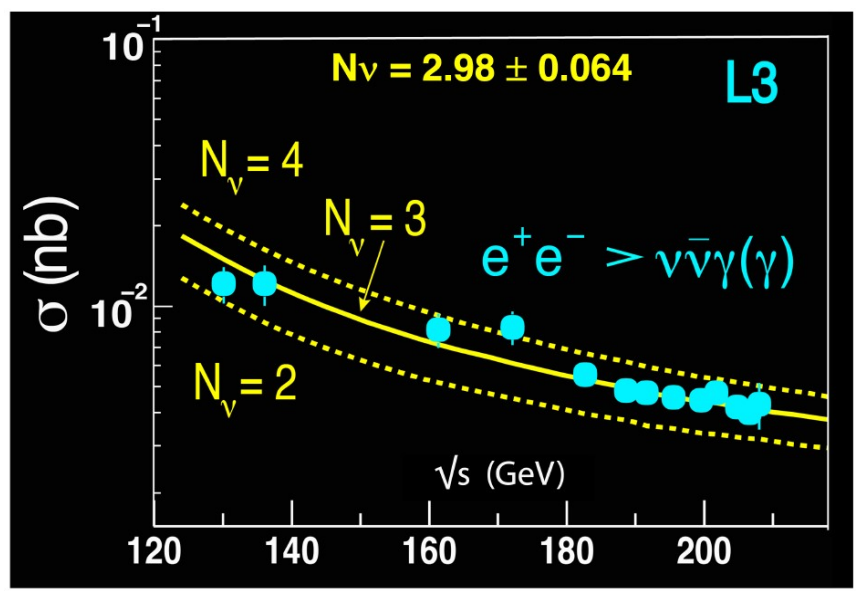

**Figure 34.** L3 experimental results: model independent determination of the number of light neutrino species using the reaction $e + e^- \rightarrow \nu\bar{\nu}\gamma$.

## 6. Alpha Magnetic Spectrometer (AMS)

The conceptual idea of AMS came to my mind when I was thinking about matter-antimatter asymmetry in the universe and related theoretical explanations. If the theory is wrong and antimatter exists in space then it can be detected experimentally. Despite having no prior experience of doing experiments in space, I decided that this is what I should do next: to look for antimatter, to study the origin of cosmic rays with a precision magnetic spectrometer in space. This is the origin of AMS.

Since charged cosmic rays have mass, they are absorbed by the 100 km of Earth's atmosphere, therefore the properties (such as charge sign or momentum) of charged cosmic rays cannot be studied on the ground. The detector must be placed in space, it must be lightweight, without sizable external magnetic field, performing well in the harsh space environment.

The AMS detector (Figure 35) consists of a permanent magnet with $1.4\,\mathrm{kG}$ field; nine planes of precision silicon tracker to measure the particle momentum, charge and sign; a transition radiation detector (TRD) to differentiate $e^{\pm}$ from protons; four planes of time-of-flight (TOF) counters to measure the particle direction, charge, and velocity; an array of anticoincidence counters to reject particles entering the detector from the side; a ring imaging Cherenkov detector to measure particle charge and velocity; and a 3D electromagnetic calorimeter (ECAL) to measure energy and directions of electrons, positrons, and photons. The AMS Collaboration includes 47 universities and research institutes from 14 countries. NASA has organized an excellent AMS Project Office (APO) to ensure that the experiment is built according to space requirements and safety specifications. All the detectors of AMS were constructed in Europe and Asia, assembled at CERN and tested at the European Space Agency Test Facility in the Netherlands.

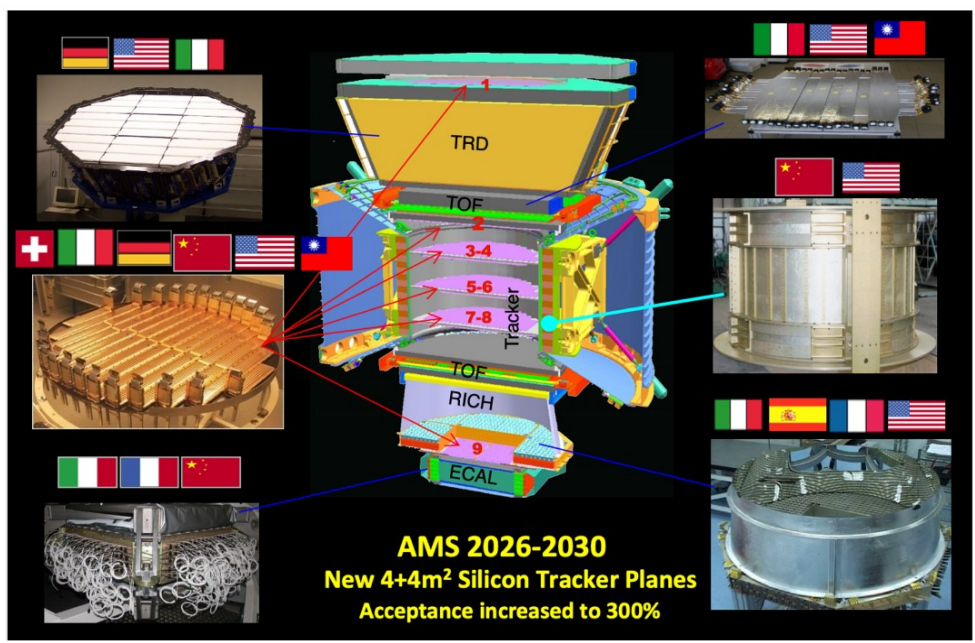

**Figure 35.** Layout of the AMS experiment showing the countries which participated in the construction of individual detectors.

As a magnetic spectrometer, AMS is unique in its exploration of a new and exciting frontier in physics research. Following a 16-year period of construction and testing and a precursor flight on the Space Shuttle in 1998 [37], AMS was installed on the International Space Station, ISS, (Figure 36) on 19 May 2011 to conduct a long duration mission of fundamental physics research in space. Its main physics objectives are the understanding of dark matter and complex antimatter in the cosmos, studies of the properties of primary and secondary cosmic rays as well as the search for new, unexpected phenomena [38]. The orders of magnitude improvement in accuracy over previous measurements is due to its precision, long exposure time in space, large acceptance, built-in redundancy and thorough calibration. In 2026 we will upgrade AMS with a new, $4 + 4\,\mathrm{m}^2$ silicon layer on top of the detector to increase the acceptance to $300\%$.

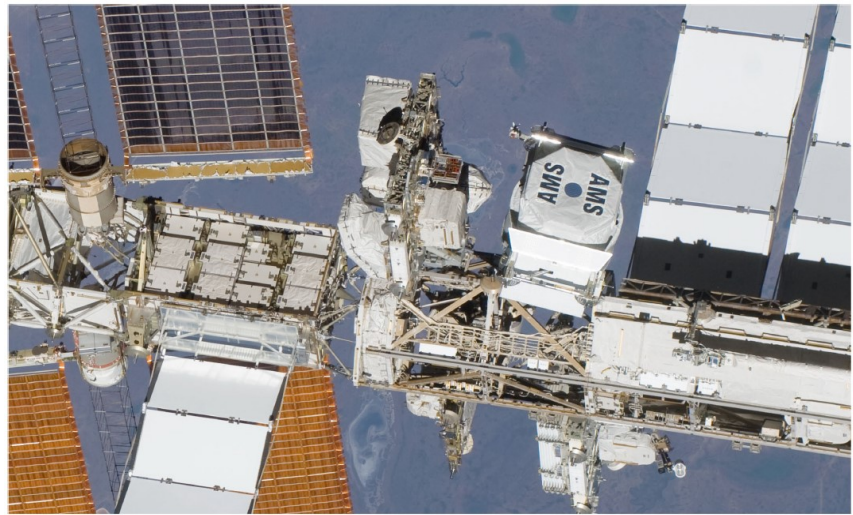

**Figure 36.** AMS on the International Space Station.

Studies of light cosmic ray antimatter species, such as positrons, antiprotons, and antideuterons, are crucial for the understanding of new phenomena in the cosmos, since the yield of these particles from traditional cosmic ray collisions is small. So far none of our results are what was expected when we first started the experiment. We found that the positron spectrum, based on 4.2 million events, exhibits complex energy dependence. Unexpectedly, after rising from ∼25 to ∼300 GeV, the spectrum suddenly cuts off and decreases quickly with energy. This means that there is a finite energy cutoff, which is measured to be ∼800 GeV. Significance of this measurement is $4.8\sigma$ (Figure 37). This complex behavior of the positron spectrum is consistent with the existence of a new source of high energy positrons with a characteristic cutoff energy, whether of dark matter or other new astrophysical origin. It is not consistent with the exclusive secondary production of positrons in collisions of cosmic rays with the interstellar media (Figures 37 and 38a). With more data collected through 2030 with the upgraded detector (Figure 38b) we will be able to provide definitive answers concerning the physics nature of the positron source term.

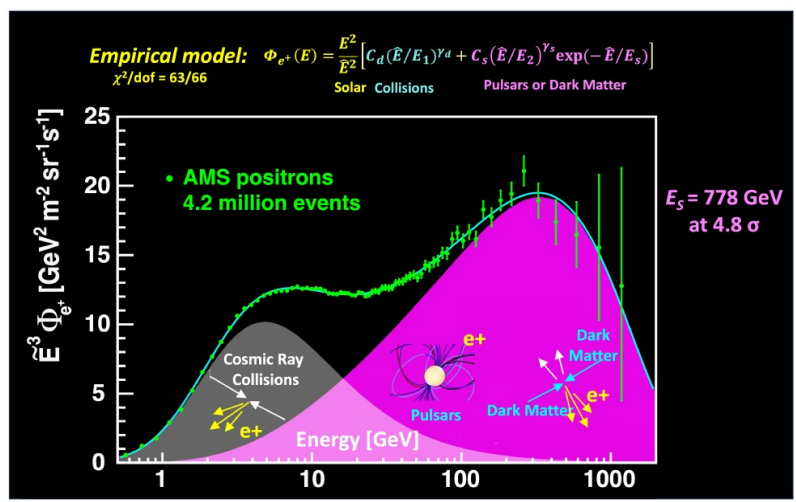

**Figure 37.** The positron flux is the sum of low energy part from cosmic ray collisions plus a high-energy term from pulsars or dark matter with a cutoff energy. The empirical formula (shown on top), which includes both cosmic ray collisions and new source term with an exponential cutoff is represented by a light blue line.

For electrons, we have measured their spectrum from very low energy to a few TeV based on 62 million electrons.

Note that the contribution of cosmic ray collisions to the electron flux is negligible. We have found that the spectrum can be described by two power law functions and the same source term observed in the positron spectrum (Figure 39). This is consistent with the dark matter annihilation which produces equal amounts of high energy electrons and positrons. We determined the significance of the positron source term in the electron spectrum to be $2.6\sigma$ (99% CL) at present. With the AMS upgrade, we will determine the significance of the charge symmetric source term to $4\sigma$.

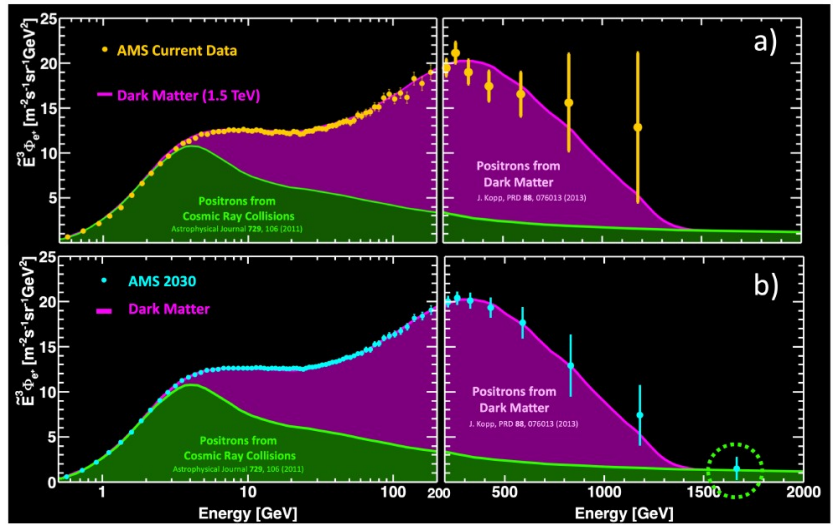

**Figure 38.** (**a**) Comparison of the AMS data with predictions of a dark matter model with $M_{DM} = 1.5\,\text{TeV}$. (**b**) The projection of AMS measurements to 2030 shows that we will not only improve the accuracy of current measurements but also provide a data point above the dark matter mass, where the contribution of cosmic ray collisions dominates.

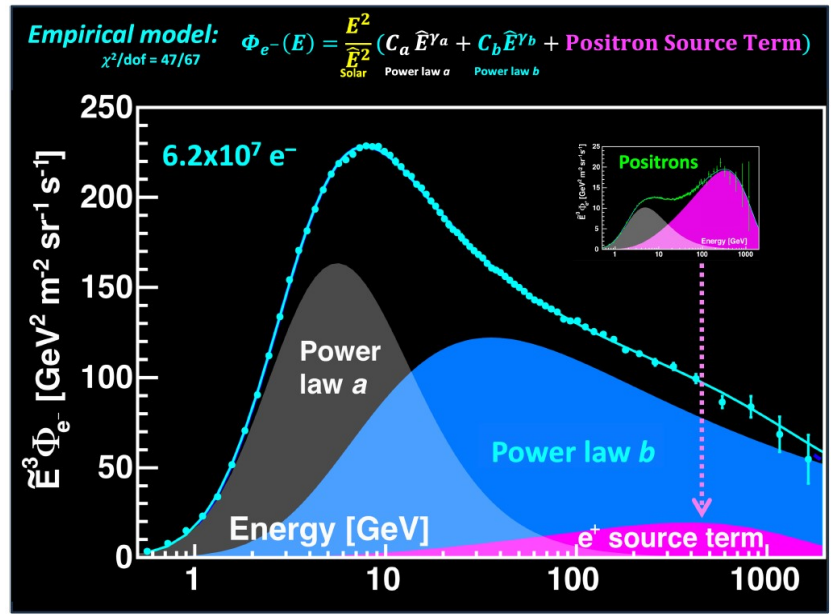

**Figure 39.** The electron spectrum with the fit results showing that the charge symmetric measured positron source term (from Figure 37) is needed to describe the behavior of the spectrum at high energies. The empirical formula (shown on top), which includes two power law functions and the positron source term with an exponential cutoff is represented by a light blue line.

For antiprotons we found that the spectrum is identical to the positron spectrum above 60 GeV (Figure 40). Indeed the ratio of the positron-to-antiproton fluxes above 60 GeV was determined to be $1.98 \pm 0.03\,(\text{stat}) \pm 0.05\,(\text{syst})$. This identical behavior of positrons and antiprotons excludes the pulsar origin of positrons.

For cosmic ray nuclei, we have measured the spectra of many different types, from helium up to iron, as a function of rigidity (i.e. momentum per unit charge). Before AMS, there were very limited measurements of cosmic rays with 30% or larger errors. Now, we can study cosmic rays with an accuracy of 1%. For each cosmic ray element we have collected tens of millions of events with energies up to multi-trillion electron volts.

Before AMS, cosmic rays were believed to have two groups. The first group are primary cosmic rays (helium, carbon, oxygen, ...) which are produced from nuclear fusion in stars and accelerated by supernova explosions. Then there is another group of secondary cosmic rays (lithium, beryllium, boron, ...) which are produced from the collision of primary cosmic rays with the interstellar media. Unexpectedly, AMS discovered that at high rigidities, primary cosmic rays actually have two classes (light and heavy nuclei), each class contains elements with unique but identical rigidity dependence. AMS also found that the secondary cosmic rays have their own rigidity dependence also with two classes (light and heavy nuclei) which is very different from the two classes of primary cosmic rays (see Figures 41 and 42). These phenomena were not predicted.

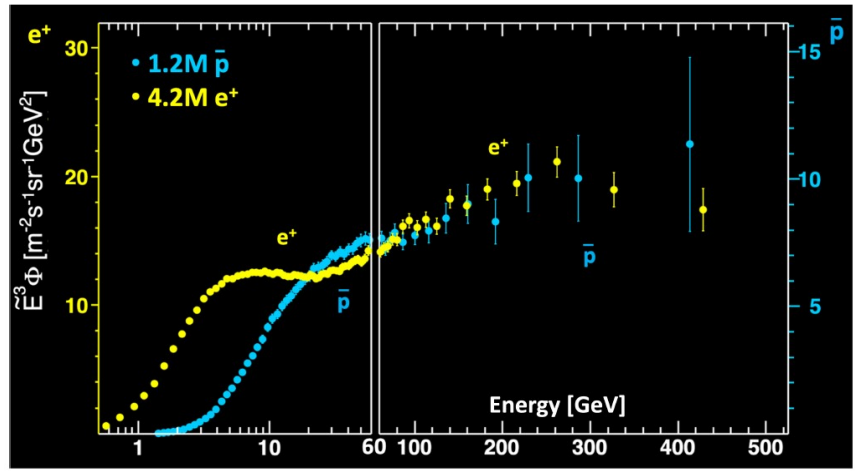

**Figure 40.** The antiproton spectrum (blue data points, right axis) and the positron spectrum (yellow data points, left axis) show identical behavior above 60 GeV.

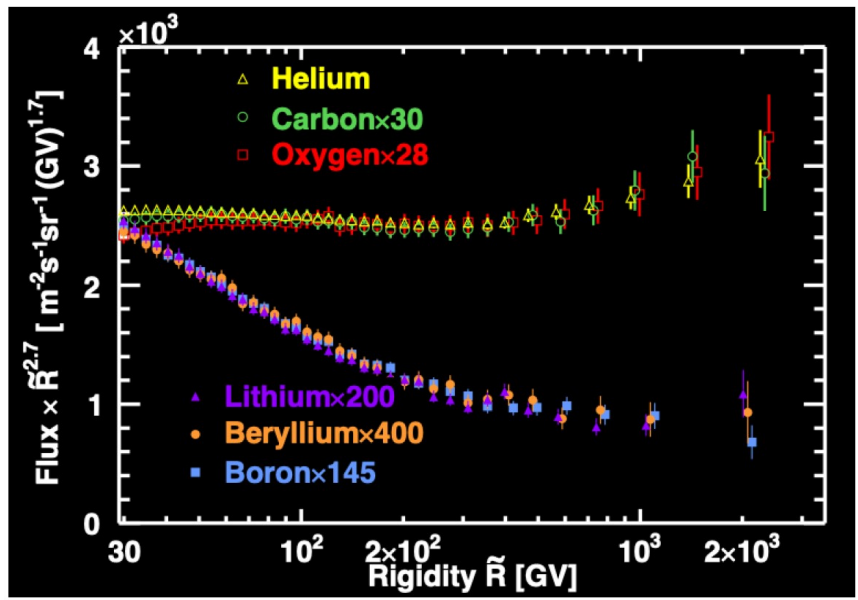

**Figure 41.** Class of light nuclei: $2 \leq Z \leq 8$ He-C-O primaries compared with Li-Be-B secondaries.

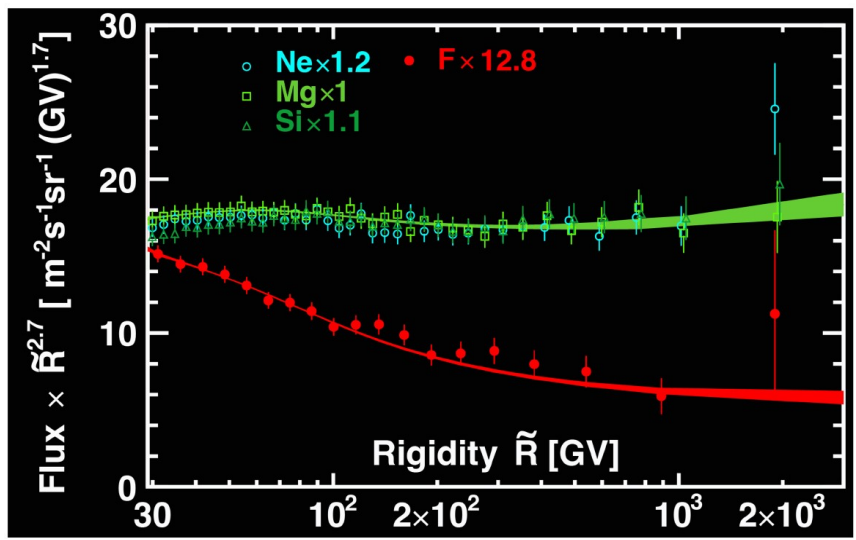

**Figure 42.** Class of heavier nuclei: $9 \leq Z \leq 14$ Ne-Mg-Si primaries compared with F secondaries.

Surprisingly, we found that any cosmic nuclei flux can be described as a linear combination of the corresponding primary and secondary fluxes (Figure 43). This is an important observation as it allows to determine for every cosmic nuclei the amount of its primary component at the source in a model-independent way. Most interesting, we found that the traditional primary cosmic rays He, C, S, Ne and Mg all have sizable secondary components. We continue these studies aiming to definitively determine the nature of all high-energy cosmic rays from $Z = 1$ to $Z = 26$ and beyond.

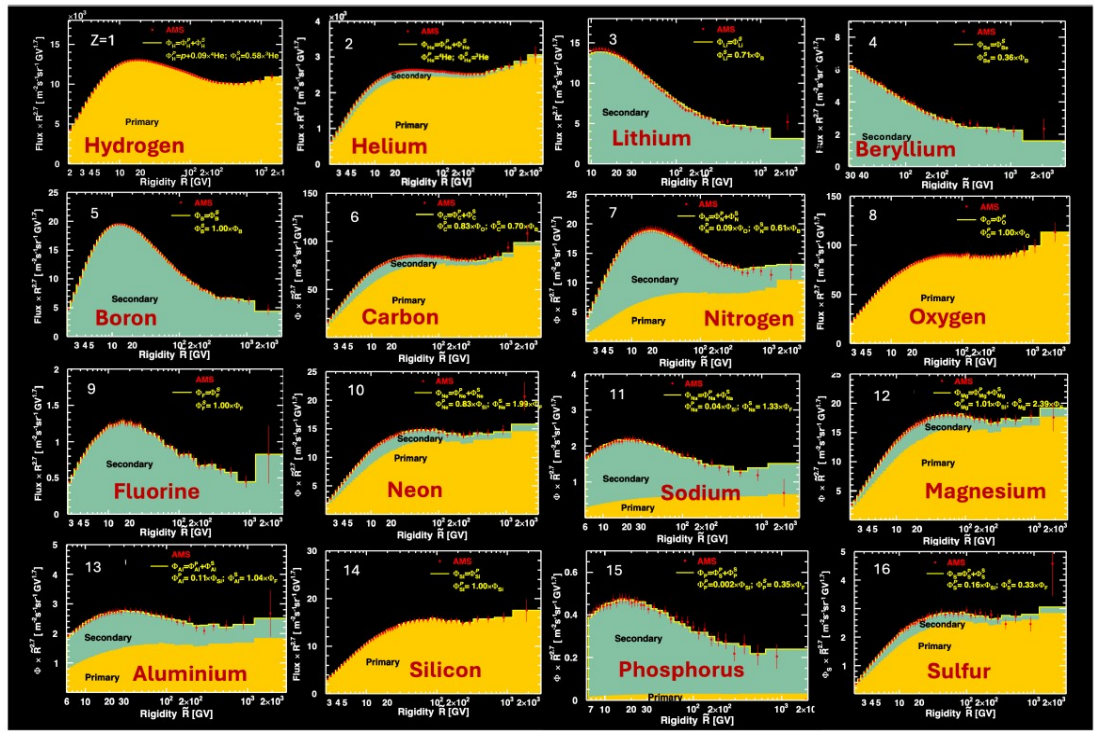

**Figure 43.** The fluxes of all cosmic nuclei from $Z = 1$ to $Z = 16$. In each plot the contributions of the primary and secondary components are indicated by the yellow and green shading, respectively.

One of the main physics research topics in the last half a century is the search for the explanation of the absence of heavy antimatter (known as baryogenesis). Baryogenesis requires strong CP symmetry breaking and a finite lifetime of the proton. To date, despite the efforts of many outstanding experiments, no evidence for strong CP symmetry breaking nor for proton decay has been found. Therefore, the observation of heavy antimatter events is of great importance. If the universe had come from the Big Bang, there should be equal amounts of matter and antimatter at the beginning. Now the universe is 14 billion years old, where is the other half of the universe made out of antimatter?

AMS is a unique precision magnetic spectrometer with large acceptance and long exposure time. We are studying anti-matter in space with highest priority and we have begun to see the anti-helium candidates. The observation of anti-helium events is the first step for AMS to study heavy antimatter. We need to collect more data and investigate higher $Z$ nuclei to see how many kinds of heavy antimatter nuclei we can find.

The latest AMS results on the fluxes of electrons, positrons, protons, antiprotons, and primary and secondary nuclei provide precise and unexpected information. The accuracy and characteristics of the data, simultaneously from many different types of cosmic rays, provide unique input to the understanding of cosmic ray production and propagation.

With data taking through the lifetime of the Space Station, we will explore the physics of complex anti-matter (anti-He, anti-C, ...), the physics of dark matter (anti-deuterons, anti-protons and positrons), the physics of cosmic ray nuclei including isotopes and high-$Z$ cosmic-rays for which there is only very limited data below 35 GeV/n, as well as the physics of the heliosphere. AMS is exploring uncharted territory and opening new domains of research. There is no plan by any country to launch another magnetic spectrometer into space. It is most important that we ensure that AMS data are precise and cover the highest energies and the highest $Z$ continuously over an extended period of time.

Space is the ultimate laboratory. Space provides particles with much higher energies than accelerators. AMS provides a first step in uncovering the mysteries of cosmic rays. The unexpected nature of the AMS results requires a new and comprehensive astrophysical model of the cosmos.

**Acknowledgments**

I want to acknowledge Luciano Maiani and Antonio Polosa for inviting me to the symposium "The Rise of Particle Physics" in Rome. I also want to thank M. Capell, A. Kounine, and Z.Weng for helping me to prepare the manuscript.

**Conflicts of Interest**

The author declares no conflict of interest.

**References**

1. Blumenthal, R.B.; Ehn, D.C.; Faissler, W.L.; et al. Wide-angle electron-pair production. *Phys. Rev.* **1966**, *144*, 1199.
2. Glashow, S.L.; Iliopoulos, J.; Maiani, L. Weak interactions with lepton-hadron symmetry. *Phys. Rev. D* **1970**, *2*, 1285.
3. Asbury, J.G.; Bertram, W.K.; Becker, U.; et al. Validity of Quantum Electrodynamics at Small Distances. *Phys. Rev. Lett.* **1967**, *18*, 65.
4. Asbury, J.G.; Becker, U.; Bertram, W.K.; et al. Leptonic Decays of Vector Mesons: The Branching Ratio of the Electron-Positron Decay Mode of the Rho Meson. *Phys. Rev. Lett.* **1967**, *19*, 869.
5. Alvensleben, H.; Becker, U.; Bertram, W.K.; et al. Validity of quantum electrodynamics at extremely small distances. *Phys. Rev. Lett.* **1968**, *21*, 1501.
6. Alvensleben, H.; Becker, U.; Bertram, W.K.; et al. Photoproduction of neutral RHO mesons from complex nuclei. *Phys. Rev. Lett.* **1970**, *24*, 786.
7. Alvensleben, H.; Becker, U.; Bertram, W.K.; et al. Observation of Coherent Interference Pattern between $\rho$ and $\omega$ Decays. *Phys. Rev. Lett.* **1970**, *25*, 1373.
8. Alvensleben, H.; Becker, U.; Chen, M.; et al. Determination of the Photoproduction Phase of $\rho^0$ Mesons. *Phys. Rev. Lett.* **1970**, *25*, 1377.
9. Alvensleben, H.; Becker, U.; Busza, W.; et al. Determination of the Photoproduction Phase of $\varphi$ Mesons. *Phys. Rev. Lett.* **1971**, *27*, 444.
10. Alvensleben, H.; Becker, U.; Busza, W.; et al. Precise Determination of $\rho - \omega$ Interference Parameters from Photoproduction of Vector Mesons Off Nucleon and Nuclei. *Phys. Rev. Lett.* **1971**, *27*, 888.
11. Glashow, S.L. Is Isotopic Spin a Good Quantum Number for the New Isobars? *Phys. Rev. Lett.* **1961**, *7*, 469.
12. Bernstein, J.; Feinberg, G. Elcetromagnetic mixing effects in elementary-particle physics. *Il Nuovo C.* **1962**, *25*, 1343.
13. Gourdin, M.; Stodolsky, L.; Renard, F.M. Electromagnetic mixing of $\rho$, $\omega$ mesons. *Phys. Lett. B* **1969**, *30*, 347.
14. Sachs, R.G.; Willemsen, J.F. Two-Pion Decay Mode of the $\omega$ and $\rho - \omega$ Mixing. *Phys. Rev. D* **1970**, *2*, 133.
15. Lütjens, G.; Steinberger, J. Compilation of Results on the Two-Pion Decay of the $\omega$. *Phys. Rev. Lett.* **1964**, *12*, 517.
16. Ting, S.C.C. Particle Discovery at Brookhaven. *Science* **1975**, *189*, 750. https://www.science.org/doi/10.1126/science.189.4205.750.a.
17. Deutsch, M. Particle Discovery at Brookhaven. *Science* **1975**, *189*, 750–816. https://www.science.org/doi/10.1126/science.189.4205.750.b.
18. Leipuner, L.B.; Larsen, R.C.; Smith, L.W.; et al. Production of prompt muons by the interaction of 28-GeV protons. *Phys. Rev. Lett.* **1975**, *34*, 103.
19. Aubert, J.J.; Becker, U.; Biggs, P.J.; et al. Experimental observation of a heavy particle $J$. *Phys. Rev. Lett.* **1974**, *33*, 1404.
20. Bacci, C.; Celio, R.B.; Berna-Rodini, M.; et al. Preliminary result of frascati (ADONE) on the nature of a new 3.1-GeV particle produced in $e^+e^-$ annihilation. *Phys. Rev. Lett.* **1974**, *33*, 1408.
21. Augustin, J.E.; Boyarski, A.M.; Breidenbach, M.; et al. Discovery of a Narrow Resonance in $e^+e^-$ Annihilation. *Phys. Rev. Lett.* **1974**, *33*, 1406.
22. Sullivan, W. New and Surprising Type Of Atomic Particle Found. *The New York Times*, 17 November 1974, p. 1.
23. Wang, Y.F. Private communication.
24. Barber, D.P.; Becker, U.; Benda, H.; et al. Physics with high energy electron-positron colliding beams with the MARK J detector. *Phys. Rep.* **1980**, *63*, 337.
25. Adeva, B.; Barber, D.P.; Becker, U.; et al. Measurement of Charge Asymmetry in $e^+e^- \longrightarrow \mu^+\mu^-$. *Phys. Rev. Lett.* **1982**, *48*, 1701.
26. Schwarzschild, B.M. Polarized plasmas may prove useful for fusion reactors. *Phys. Today* **1982**, *35*, 19.
27. Glashow, S.L. Partial-symmetries of weak interactions. *Nucl. Phys.* **1961**, *22*, 579.
28. Weinberg, S. A Model of Leptons. *Phys. Rev. Lett.* **1967**, *19*, 1264.
29. Salam, A. *Elementary Particle Physics, Eighth Nobel Symposium*; Svartholm, N., Ed.; Almquvist and Wiksell: Stockholm, Sweden, 1968; p. 367.
30. Barber, D.P.; Becker, U.; Benda, H.; et al. Discovery of three-jet events and a test of quantum chromodynamics at petra. *Phys. Rev. Lett.* **1979**, *43*, 830.

31. Schopper, H. Presented at the International Conference on Multihadron Physics, Goa, India, October 1979.

32. Lubkin, G.B. Evidence from PETRA adds support for QCD and gluons. *Phys. Today* **1980**, *33*, 17.

33. Acciarri, M.; Achard, P.; Adriani, O.; et al. Measurement of the running of the fine-structure constant. *Phys. Lett. B* **2000**, *476*, 40.

34. Achard, P.; Adriani, O.; Aguilar-Benitez, M.; et al. Determination of $a_s$ from hadronic event shapes in $e^+e^-$ annihilation at $192 \leqslant s \leqslant 208$ GeV. *Phys. Lett. B* **2002**, *536*, 217.

35. Achard, P.; Adriani, O.; Aguilar-Benitez, M.; et al. Single- and multi-photon events with missing energy in $e^+e^-$ collisions at LEP. *Phys. Lett. B* **2004**, *587*, 16.

36. L3 Collaboration. Results from the L3 experiments at LEP. *Phys. Rep.* **1993**, *236*, 1–146.

37. Aguilar, M.; Alcaraz, J.; Allaby, J.; et al. The Alpha Magnetic Spectrometer (AMS) on the International Space Station: Part I—Results from the test flight on the space shuttle. *Phys. Rep.* **2002**, *366*, 331.

38. Aguilar, M.; Cavasonza, L.A.; Ambrosi, G.; et al. The Alpha Magnetic Spectrometer (AMS) on the international space station: Part II—Results from the first seven years. *Phys. Rep.* **2021**, *894*, 1.

*Opinion*

# The Standard Model Yesterday, Today and Tomorrow

## Howard Georgi

Department of Physics, Harvard University, Cambridge, MA 02138, USA; hgeorgi@fas.harvard.edu

**Abstract:** The 1970s was the decade of the Standard Model! This decade which began with quantum field theory in disarray ended with a practical set of QFT tools for calculations for strong interactions at high and low energies and for electroweak interactions at all energies. After briefly remembering how we got there, I celebrate the remarkable achievements of particle physics in the 1970s and comment on where we go from here.

I would like to thank the organizers for inviting me to this amazing Symposium. Who would not want to talk about the Standard Model in the Eternal City? And for me, being something of a homebody, it is a wonderful opportunity for me to see old friends from the heroic period when the Standard Model was built.

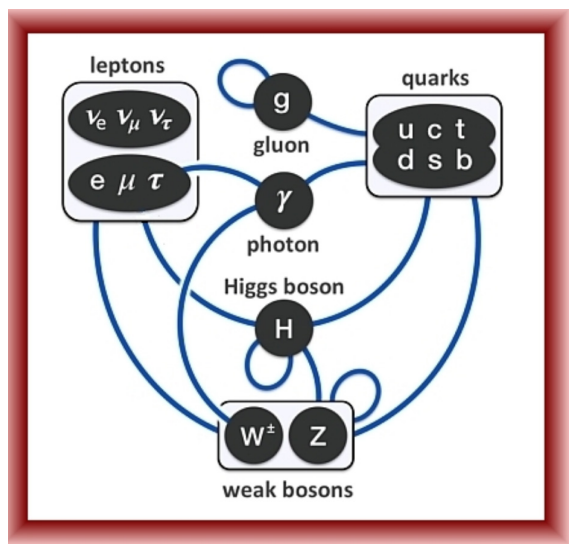

johncarlosbaez.wordpress.com

What is the Standard Model? This is my favorite cartoon version, but of course the Standard Model is much more than that. This doesn't do justice to the neutrino sector. And indeed, we don't yet have a Standard Model of neutrinos. And the Standard Model is more than a clever picture or a list of fields and interactions. The development of the Standard Model went hand in hand with a change in the way we look at high-energy physics. This is a revolution that is well worth celebrating. Perhaps that is what the organizers had in mind with the title "The Rise of Particle Physics".

But I don't think the word "Rise" quite captures the unique feel of the decade of the '70s. To show you what I mean I want to go back another 20 years and remember some of the amazing particle physics in the 1950s and 1960s and try to understand how what we are celebrating is different. Here are a few of many highlights from the particle physics timeline.

I was too young in the 1950s to understand what was going on, but I had the great pleasure of working with Bram Pais twenty years later and heard a lot of stories. Just imagine the long list of issues that were under discussion in those years: the $\pi^\circ$, the renormalisation group, the strong focusing, the bubble chamber, associated production, strangeness charge and isospin, Yang-Mills theory, the K-long K-short puzzle, the anti-proton and the anti-neutron, the detection of neutrinos and anti-neutrinos, parity violation, V-A, structure of the weak interactions, the prediction of two neutrino species, Regge poles. The list could continue with chiral symmetry breaking and the pion, macroscopic quantum interference, the Z and the weak mixing angle, the Goldstone bosons, the 8-fold way, the nucleon resonance, the evidence for two neutrinos, the $\Omega^-$, flavour mixing, the quarks and the Higgs.

By the mid 1960s, I started to understand bits and pieces. SU(3) was one of the reasons I fell in love with our tiny world of elementary particles. I remember being excited by Jim Cronin's colloquium at Harvard soon after the discovery of CP violation. Not to mention the suggestion of charm, quark color, current algebra calculations. I was a beginning graduate student when the PRL containing Weinberg's "Model of Leptons" arrived in my mailbox [1]. Among the many issues under discussion in those years I have to mention also the solar neutrino problem, the dual resonance model, scaling and the parton model, chiral anomalies, the observation of partons, the operator product expansion. The last few years of the 1960s were certainly nibbling on the edges of the Standard Model.

So why were particle physicists depressed??

The reason, I think, is that these fantastic experimental discoveries and theoretical breakthroughs created as many puzzles and frustrations as they solved. Why no Flavor Changing Neutral Currents? Why approximate SU(3) and almost exact isospin? Since the quarks are partons, why don't we see them if they are almost free? How can the parton model work for electron scattering and fail for $e^+e^-$ annihilation? Why don't we see fractional charges? Why do quark masses look so different in current algebra and constituent quarks? Why does not the $\eta'$ look like a Goldstone boson? Why is CP violation so small? The list could go on and on!

Among the many frustrations was the ad-hoc character of current algebra. And even if it had been more systematic, it was still just an application of unexplained symmetry that did not address the fundamental problems of strong interaction. We just had no good way of dealing with the strong dynamics. The problem was certainly not lack of experimental results. By the standards of the day, there was a huge amount of data and it was neatly packaged by the particle data group into our particle data books. But it was not obvious what to do with it. We lacked the theoretical tools to fit the data into a useful calculational scheme.

And of course the biggest elephant in the room was that the renormalizability of Weinberg's model of leptons was just a speculation.

Steve was an honest man, and he did not claim too much. "*Is this model renormalizable? We usually do not expect non-Abelian gauge theories to be renormalizable if the vector-meson mass is not zero, but our Z and W mesons get their mass from the spontaneous breaking of the symmetry, not from a mass term put in at the beginning. Indeed, the model Lagrangian we start from is probably renormalizable, so the question is whether this renormalizability is lost in the reordering of the perturbation theory implied by our redefinition of the fields*" [1].

He very clearly identified the issues. Massless Yang-Mills looked renormalizable but there wasn't really a proof. And the Higgs mechanism added an additional layer of uncertainty. I am pretty sure that this is why the "Model of Leptons" paper did not make a big splash when it first appeared. I was still a baby at the time, but I remember looking at it when it arrived in my mail box and like most everyone else (including, I think, Steve himself), I ignored it because I didn't know what to make of it. It didn't look renormalizable to me. This is also why I was quite annoyed in 2017 when there were conferences about the 50th anniversary of the Standard Model. That was just nonsense.

This shows up spectacularly in a plot of the citations to Steve's paper by year, Figure 1. Almost nobody cared for a few years.

The end of the 1970s was a depressing time for quantum field theory. Julian Schwinger, who was one of the creators of QED and legendary for his deep understanding of all of physics, had given up on QFT. When I took his course at Harvard in 1966–1967, he taught his alternative "source theory". He didn't even mention Yang-Mills. Amazingly, virtually all these puzzles and frustrations that had accumulated over two decades were swept away in a few years in the early 1970s.

I think that this was not a "rise" but an "eruption". It was as if the previous two decades of particle physics had built up an enormous pressure for change and the crucial results and ideas exploded out in many different ways. Some old puzzles remained and some new ones were created. The eruption was dramatic and transformed particle physics: the result was the Standard Model. This is what I think we are celebrating today. It led to some Nobel Prizes, Figure 2.

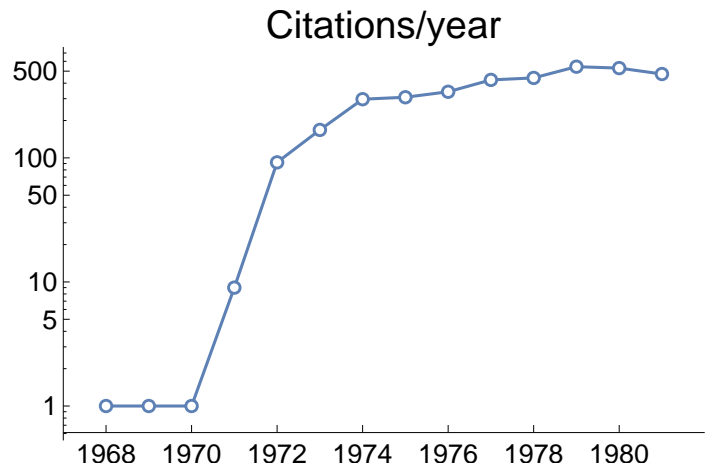

**Figure 1.** Plot of the citations/year of Steven Weinberg paper "Model of Leptons".

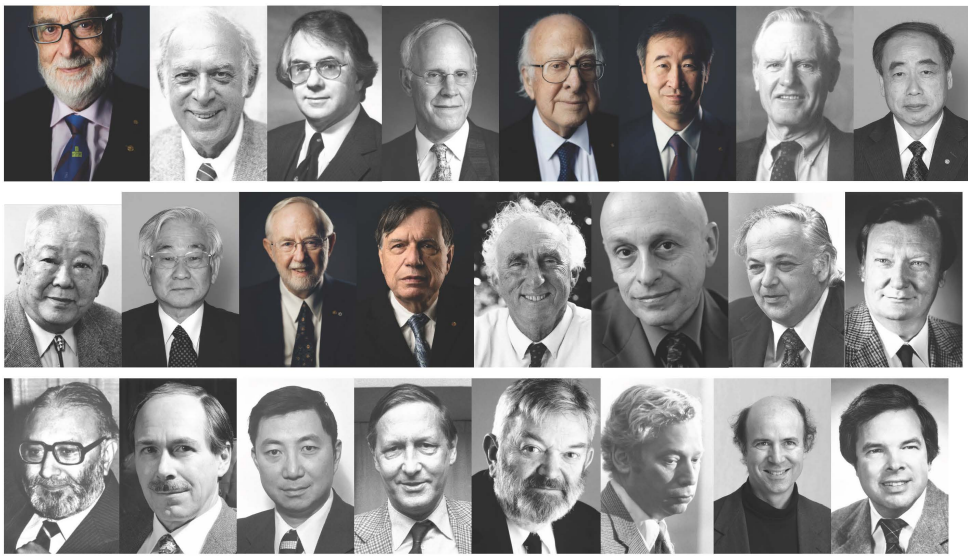

**Figure 2.** Nobel Prize winners that contributed significantly to the Standard Model.

I now want to talk briefly about some of the components of this revolution. You have already heard about an important one from Sam. And you will hear more about some of the rest in subsequent talks. But I want to try to give a sense of how they fit together into the Standard Model.

The first piece in the puzzle was the Glashow-Iliopoulos-Maiani mechanism [2]. GIM identified the relation between the symmetries of the strong and the weak interactions that was responsible for the observed suppression of flavor changing neutral current effects. If the breaking of a chiral flavor symmetry is proportional to the mass matrix that defines what the flavors mean, then quark mixing for quarks of a given charge can be transformed away. While they only discuss four flavours, their mechanism was obviously extendable to any number of flavors. While this was brilliant, and critical for the success of the Standard Model, it is possibly the single most annoying piece of physics I know. What the GIM mechanism tells us is that we have absolutely no idea what flavor really is. Nothing distinguishes the different flavors of the same charge except their masses. We all know in our hearts that this can't be true and our experimental friends have been searching for a violation ever since without definitive results.

In my view, the real birth of the Standard Model was 't Hooft's demonstration of renormalizability [3]. I love the concise abstract of this paper because it captures in three short sentences the enormity of what he did, Figure 3. He opened the door to a "large" (in fact infinite) new set of QFT models. I frequently thank my lucky stars that this happened just as I was beginning my first postdoc because I had a beautiful shiny new theoretical playground to explore, including, of course, a version of Weinberg's model of leptons. It was a fun time for model building because all the models were new—and because there were a relatively small number of model builders and a finite number of models. It was easy to understand everything that was going on. This has long since ceased to be even imaginable. 't Hooft's work grew out of the amazing general work on QFT done by him and Veltman [4]. It is also important to acknowledge the contributions of Lee and Zinn-Justin [5–8] and others in making this accessible to the community.

## RENORMALIZABLE LAGRANGIANS FOR MASSIVE YANG-MILLS FIELDS

### G.'t HOOFT

*Our result is a large set of different models with massive, charged or neutral, spin one bosons, photons, and massive scalar particles. Due to the local symmetry our models are renormalizable, causal, and unitary. They all contain a small number of independent physical parameters.*

**Figure 3.** The fundamental paper of Gerard 't Hooft.

The next piece to fall into place was dimensional transmutation. I am referring, of course, to the classic paper by Coleman and Erick Weinberg, "Radiative Corrections as the Origin of Spontaneous Symmetry Breaking" [9] This was an enormously influential paper on a not very interesting subject. This paper is a true classic. It did much more than to explain how renormalization converts a dimensionless parameter into a dimensional parameter and to show how to calculate the famous Coleman-Weinberg potential. It was a handbook on modern quantum field theory, a sort of "Well-Tempered Clavier" for QFT. It is worth reading every few years to see the work of the master Sidney Coleman at the height of his powers.

Motivated by experiments in quasi-elastic neutrino proton scattering that did not observe neutral currents, Glashow and I constructed a model based on the simple group SO(3) without a Z [10]. It is really ugly with heavy leptons and no explanation of universality, but we liked the fact that there was only one coupling and the fact that electric charge was quantized. We thought that this trivial algebraic quantization was different from Dirac's quantization in the presence of magnetic monopoles. In retrospect, it is clear that we should have thought harder about it. I think we did this at the end of 1971, but for some reason we didn't submit it until March 1972, by which time Steve Weinberg had done something that was also wrong, but was a little more interesting, Figure 4.

### Mixing Angle in Renormalizable Theories
### of Weak and Electromagnetic Interactions*

#### Steven Weinberg

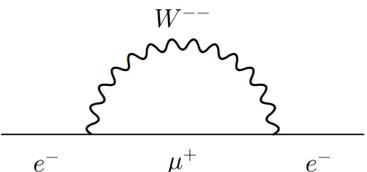

**Figure 4.** Steven Weinberg paper, 15 April 1972.

I call this Weinberg's second model of leptons [11]. His group was SU(3) cross SU(3) but it is easier to explain with a single SU(3) as shown. This is a real unified theory that has extra interactions with a very large VEV that break SU(3) down to SU(2) cross U(1) and leaves just the Standard Model of leptons at low energy. This super-strong symmetry breaking is a feature of all subsequent unified theories. The reason that Weinberg's version was more complicated was that he thought he might have a finite and calculable electron-muon mass ratio. At the time this didn't seem as ridiculous as it does now. But he got the renormalizations wrong and the ratio actually turns out to be infinite and thus a free parameter after renormalization. Glashow and I figured out how to fix the issue and the result was totally uninteresting because there were many parameters [12].

Weinberg's model wasn't renormalizable anyway because it had SU(3) gauge anomalies. 't Hooft had stated the importance of this in his original paper and Bouchiat, Iliopoulos and Meyer took it seriously and noticed that SU(2) and U(1) anomalies would cancel between leptons and quarks for three colours [13].

The next crucial piece of the model to appear was asymptotic freedom [14,15]. I have nothing to add about priority, 't Hooft, and all that. But I will simply say that this was amazing and occupied much of my time for many years after the discovery. It is clear, though, that all the authors assumed (just as I and everyone else did at that time) that in a realistic color SU(3) theory of the strong interactions, the gauge symmetry would somehow have to be broken dynamically to give mass to the gluons.

I will say more about the Pati-Salam idea of lepton number as as a 4th colour [16] in a few minutes, Figure 5. The original papers are very frustrating and difficult to read, mostly because Salam disliked the idea of fractionally charged quarks and he insisted on breaking the color symmetry, mixing with the SU(2)s to produce integrally charged quarks. Even back in 1972, they should have known that this was not consistent with what was observed in deep inelastic scattering, for example. But if you just don't break the gauge symmetry, the model shows how beautifully the quarks and the leptons fit together. If they hadn't done this crazy symmetry breaking, I think this would have been the first model with electric charge quantized in the right way with fractionally charged quarks.

$$\mathrm{SU(2)}_L \times \mathrm{SU(2)}_R \times \mathrm{SU(4)}$$

$$SU(2)_L \left\{ \overbrace{\begin{pmatrix} u_{rL} & u_{gL} & u_{bL} & \nu_L \\ d_{rL} & d_{gL} & d_{bL} & e_L^- \end{pmatrix}}^{SU(4)} \right.$$

$$SU(2)_R \left\{ \overbrace{\begin{pmatrix} u_{rR} & u_{gR} & u_{bR} & \nu_R \\ d_{rR} & d_{gR} & d_{bR} & e_R^- \end{pmatrix}}^{SU(4)} \right.$$

**Figure 5.** Pati-Salam: lepton number as a 4th colour.

The next really important piece was confinement, Figure 6. People had been looking for fractional charges for many decades when quarks were proposed, and while I have not tried to do an exhaustive search, it would astonish me if the idea of quark confinement had not been discussed extensively long before asymptotic freedom.

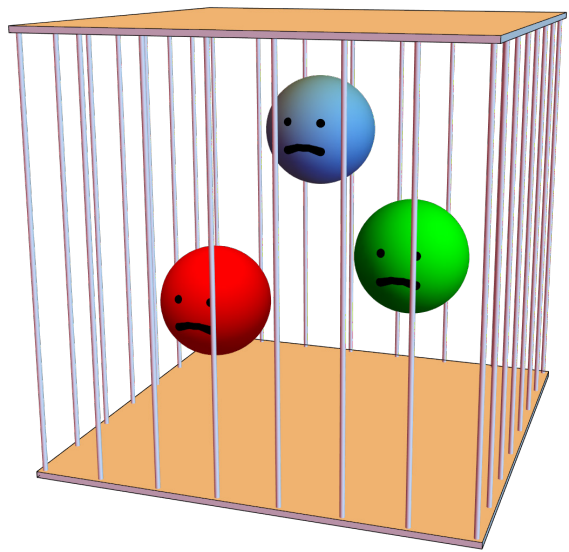

**Figure 6.** Quarks confinement.

What was really new and revolutionary was the idea that leaving the color gauge symmetry unbroken might kill two birds with one stone by confining both quarks and massless gluons so that no massless particles appear as physical states, but fractionally charged quarks could still look like real particles inside hadrons, making sense of Feynman's parton picture. As far as I know, the first mention of this in the published literature is by Weinberg [17] but I think that several other people came to similar conclusions at about the same time.

Confinement allowed Glashow and me to think about unifying SU(3) with SU(2) cross U(1). It was obvious how to do this if you thought properly about Pati-Salam and I quickly constructed first the SO(10) theory [18] and then SU(5) [19]. One of my few regrets is that I did not discuss SO(10) in a footnote of the SU(5) paper, although I gave talks about it. This allowed Fritsch and Minkowski to find it later independently [20]. We didn't know quite what to do with the coupling constants but Glashow realized that we could suppress proton decay adequately with superstrong breaking at a huge scale. The gorgeous fit of the quarks and leptons and gauge bosons into these unified groups helped to make the Standard Model gauge group look more reasonable.

The large mass scale and large renormalizations in GUTs forced us to think more carefully about the renormalization group in the presence of broken symmetry. A few months after SU(5), Helen Quinn, Steve Weinberg and I [21] understood that the only practical way to deal with this was effective field theory - to calculate in the broken theory ignoring the heavy particles and to put the symmetry in as a boundary condition at the high scale. GUTs forced us into the first full effective field theory calculation in which the process of renormalization was taken apart into the two different steps of matching couplings at the boundaries between different effective theories and running in the effective theory between the boundaries.

1974 was an absolutely crazy time because there was so much going on all over the world trying to turn the prescient guess about unbroken asymptotically free color SU(3) into testable predictions. Deep-inelastic electron and neutrino scattering and $e^+e^-$ annihilation were two of the primary pressure points where theory and experiment could interact. In deep inelastic scattering, it looked like the renormalization group improved parton model was working beautifully. This helped to confirm the SU(2) × U(1) model and pin down the weak mixing angle. But for many of us, the reverse logic was equally important. By this time, we were starting to believe in SU(2) × U(1), so the success of the parton model description definitely helped convince us that color and quarks were real and that we could calculate.

But the success of the parton model in deep inelastic scattering just heightened the puzzle of R - the ratio of hadrons to muons - in $e^+e^-$, Figure 7. Very soon after asymptotic freedom Appelquist and I [22] and independently Zee [23] had done the rather simple calculation of the color radiative corrections to the free-quark result for $R$ in the short distance limit. This gives a modest enhancement - but is not even close to explaining the data. It was becoming clear that something more was going on—and it was becoming obvious to us that it was charm.

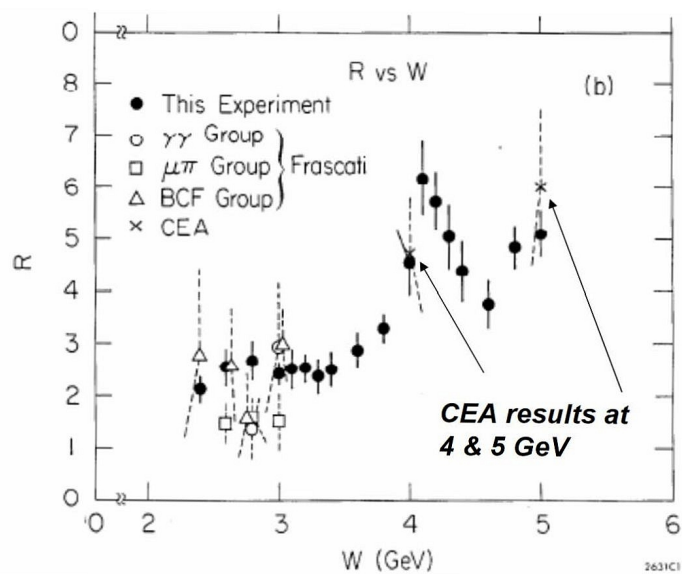

SLAC-PUB-1520, LBL-3621; January 1975

**Figure 7.** The puzzle of R, the ratio of hadrons to muons in $e^+e^-$ collisions.

The week before the $\psi$ was discovered at SPEAR, Burt Richter was at Harvard as a Loeb lecturer, giving talks on his theory that the reason for the apparently steadily rising $R$ was that the electron was a hadron a small part of the time and we were seeing the constant cross-section of the hadronic component. We had a fancy lunch in the department library to celebrate his lectures, and there Appelquist and Politzer suggested that he look for narrow states. He did not appear to take this suggestion seriously.

Shortly after Richter's visit, the November Revolution changed the particle physics world dramatically [24,25]. We have heard about this from Sam, but let me just remind you that of the 8 theory papers that appeared in Physical

Review Letters in the first issue devoted to theory after the discovery, most were totally nutty, including several by brilliant physicists who should have known better. I will just briefly show the abstracts:

Are the New Particles Baryon-Antibaryon Nuclei? - Alfred S. Goldhaber and Maurice Goldhaber - Baryon-antibaryon bound states and resonances could account for the new particles, as well as narrow states near nucleon-antinucleon threshold, which were reported earlier.

These are terrific physicists but energy scales are all wrong.

Interpretation of a Narrow Resonance in $e^+e^-$ Annihilation - Julian Schwinger - A previously published unified theory of electromagnetic and weak interactions proposed a mixing between two types of unit-spin mesons, one of which would have precisely the characteristics of the newly discovered neutral resonance at 3.1 GeV. With this interpretation, a substantial fraction of the small hadronic decay rate can be accounted for. It is also remarked that other long-lived particles should exist in order to complete the analogy with $\rho^0$, $\omega$, and $\phi$.

Ditto

Possible Explanation of the New Resonance in $e^+e^-$ Annihilation—S. Borchardt, V. S. Mathur, and S. Okubo—We propose that the recently discovered resonance in $e^+e^-$ annihilation is a member of the 15⊕1 dimensional representation of the SU(4) group. This hypothesis is consistent with the various experimental features reported for the resonance. In addition, we make a prediction for the masses of the charmed vector mesons belonging to the same representation.

Mentions charm but completely misses the point. SU(4) is useless except for keeping track of the charm quantum number

Model with Three Charmed Quarks—R. Michael Barnett—The spectroscopy and weak couplings of a quark model with three charmed quarks are discussed in the context of recent results from Brookhaven National Laboratory, Stanford Linear Accelerator Center, and Fermi National Accelerator Laboratory.

I have no idea what he was thinking.

Possible Interactions of the $J$ Particle—H. T. Nieh, Tai Tsun Wu, and Chen Ning Yang—We discuss some possible interaction schemes for the newly discovered particle $J$ and their experimental implications, as well as the possible existence of two $J^0$s like the $K_S - K_L$ case. Of particular interest is the case where the $J$ particle has strong interactions with the hadrons. In this case $J$ can be produced by associated production in hadron-hadron collisions and also singly in relative abundance in $ep$ and $\mu p$ collisions.

Again great physicists, really confused.

Is Bound Charm Found? - A. De Rújula and S. L. Glashow - We argue that the newly discovered narrow resonance at 3.1 GeV is a $^3S_1$ bound state of charmed quarks and we show the consistency of this interpretation with known meson systematics. The crucial test of this notion is the existence of charmed hadrons near 2 GeV.

Exactly right.

Remarks on the New Resonances at 3.1 and 3.7 GeV - C. G. Callan, R. L. Kingsley, S. B. Treiman, F. Wilczek, and A. Zee - This is a collection of comments which may be useful in the search for an understanding of the recently discovered narrow resonances at 3.1 and 3.7 GeV.

They should have been brave and committed to charm!

Heavy Quarks and $e^+e^-$ annihilation—Thomas Appelquist and H. David Politzer—The effects of new, heavy quarks are examined in a colored quark-gluon model. The $e^+e^-$ total cross section scales for energies far above any quark mass. However, it is much greater than the scaling prediction in a domain about the nominal two-heavy-quark threshold, despite $e^+e^-$ being a weak-coupling problem above 2 GeV. We expect spikes at the low end of this domain and a broad enhancement at the upper end.

Brilliant prediction—sadly submitted too late! If they had submitted this a few weeks earlier it might have been a Nobel Prize.

The confusions here by great physicists are an indication of just how revolutionary the idea of charmonium was at the time. It seems obvious now, but it certainly wasn't then.

The discovery of charmonium should have convinced everyone. But the month after the discovery I attended a conference at the University of Miami and I was astonished to find that most people we met at the conference were not convinced, and that many were convinced that the charm explanation was ruled out because charmed particles had not been seen. This situation changed surprisingly slowly over the next year and a half. I think the hang-up was that many physicists could not wrap their heads around what a huge difference the large mass of the $c$ quark would make for the properties of charmed mesons and baryons. Consider, for example, this entry from the stable particle section of the 1974 particle data book, Figure 8.

## Stable Particle Table *(cont'd)*

| Particle | $I^G(J^P)C_n$ | Mass (MeV) Mass² (GeV)² | Mean life (sec) cτ (cm) | Mode | Fraction[a] | p or $p_{max}$[b] (MeV/c) |
|---|---|---|---|---|---|---|
| $K^0$ | $\frac{1}{2}(0^-)$ | 497.70 ±0.13 S=1.1* m²=0.248 | 50% $K_{Short}$, 50% $K_{Long}$ | | | |
| $K_S^0$ | $\frac{1}{2}(0^-)$ | 0.886×10⁻¹⁰ ±.007 S=2.4* cτ=2.66 | $\pi^+\pi^-$ | ( 68.77±0.26)% S=1.1* | 206 |
| | | | | $\pi^0\pi^0$ | ( 31.23 )% | 209 |
| | | | | $\mu^+\mu^-$ | ( <0.3 )10⁻⁶ | 225 |
| | | | | $e^+e^-$ | ( <35 )10⁻⁵ | |
| | | | | $\pi^+\pi^-\gamma$ | c( 2.0±0.4 )10⁻³ | 206 |
| | | | | $\gamma\gamma$ | ( <0.4 )10⁻³ | 249 |
| $K_L^0$ | $\frac{1}{2}(0^-)$ | 5.179×10⁻⁸ ±0.040 cτ=1553 | $\pi^0\pi^0\pi^0$ | ( 21.3±0.6 )% S=1.1* | 139 |
| | | | | $\pi^+\pi^-\pi^0$ | ( 11.9±0.4 )% S=2.2* | 133 |
| | | | | $\mu\pi\nu$ | ( 27.5±0.5 )% S=1.1* | 216 |
| | | $m_{K_L}-m_{K_S}=0.5403×10^{10} ℏ \text{ sec}^{-1}$ ±0.0035 | | $\pi e\nu$ | ( 39.0±0.6 )% S=1.1* | 229 |
| | | | | $\pi e\gamma\nu$ | c( 1.3±0.8 )% | 229 |
| | | | | $\pi^+\pi^-$ | ( 0.177±0.018)% S=4.9* | 206 |
| | | | | $\pi^0\pi^0$ | ( 0.093±0.019)% S=1.5* | 209 |
| | | | | $\pi^+\pi^-\gamma$ | c( <0.4 )10⁻³ | 206 |
| | | | | $\pi^0\gamma\gamma$ | ( <2.4 )10⁻⁴ | 231 |
| | | | | $\gamma\gamma$ | ( 4.9±0.4 )10⁻⁴ | 249 |
| | | | | $e\mu$ | ( <1.6 )10⁻⁹ | 238 |
| | | | | $\mu^+\mu^-$ | i( <1.6 )10⁻⁸ | 225 |
| | | | | $e^+e^-$ | ( <1.6 )10⁻⁹ | 249 |
| | | | | $e^+e^-\nu$ | ( <2.8 )10⁻⁵ | 249 |

**Figure 8.** Stable Particle Table in 1974 Particle Data Group.

The corresponding table for $D^0$ today runs to over 10 pages. The particle physics world had really changed! Meanwhile, from 1974 to 1978, the development of the details of the Standard Model continued at a rapid pace. Wilson constructed lattice gauge theories. Kobayashi and Maskawa extended GIM to six quarks and predicted CP violation. SLAC found P-wave charmonium states, silencing many of the skeptics. 't Hooft and Polyakov showed that unified theories have monopole solitons, connecting algebraic charge quantization in unified theories with Dirac's. Witten constructed an effective field theory of low-energy effects of charm. Charmed particle masses were calculated at Cornell and Harvard. Perl discovered the tau lepton. Cornell and CERN groups did systematic charmonium calculations. Goldhaber discovered charmed particles just where they were predicted. 't Hooft showed how instantons solve the chiral U(1) problem. Witten constructed an effective theory of weak interactions. Politzer and Altarelli-Parisi developed the QCD parton model. The Upsilon was discovered at FNAL. Peccei and Quinn discovered their symmetry in a 2-Higgs model, while Gilman and Wise estimated epsilon prime over epsilon in the KM model. In 1978, Prescott and Taylor found parity violation in deep inelastic electron-deuteron scattering, vanquishing the last challenge to the Standard Model from atomic parity violation experiments. Weinberg and Wilczek established the properties of axions in the Peccei-Quinn model. Experimenters at DESY found evidence for gluons in hadronic jets. Dimopoulos and Susskind and Weinberg showed that Technicolor could break the electroweak symmetry dynamically. Weinberg capped a remarkable decade with his Phenomenological Lagrangians paper that established the Effective Chiral Theory.

Thus the 1970s was the decade of the Standard Model. This decade which began with quantum field theory in disarray ended with a practical set of QFT tools for calculations for strong interactions at high and low energies and for electroweak interactions at all energies. After the Prescott-Taylor experiment, the gauge symmetry of the Standard Model was firmly established and the focus shifted to the physics of the symmetry breaking and to radiative corrections. In the 1990s, the t quark was discovered at Fermilab and LEP pinned down the properties of the Z and this left very little wiggle room for symmetry breaking by any mechanism other than the vacuum expectation value of a scalar Higgs particle. Nevertheless, until the Higgs was discovered in 2012, I was hoping against hope that we had missed something and that something very different would appear. Even though it appeared where it had to appear based on previous data and radiative corrections, its existence still amazes me. It remains unique - the one example of its kind. Since it stubbornly refuses to disappear, I very much hope that we can find something about it that suggests where to look next.

The current status of the Standard Model can be summed up in this wonderful cartoon by Randall Munroe about unexpected experimental results, Figure 9.

Our hero suggests betting that any interesting result is wrong! Of course I am not being totally serious here. I hope that the g-2 anomaly, for example, holds up. That would be a problem for the Standard Model. And problems is what we need! We have to flesh out the neutrino sector. Maybe that will generate problems for us. But at the moment we have lots of puzzles but no problems. We have issues like dark matter that may be a problem for particle physics, but may be something else entirely. We have no problems like the inconsistencies in particle physics itself that led to the Standard Model. I had hoped to give some kind of an overview of the approaches to our puzzles. But it turns out that this is Impossible.

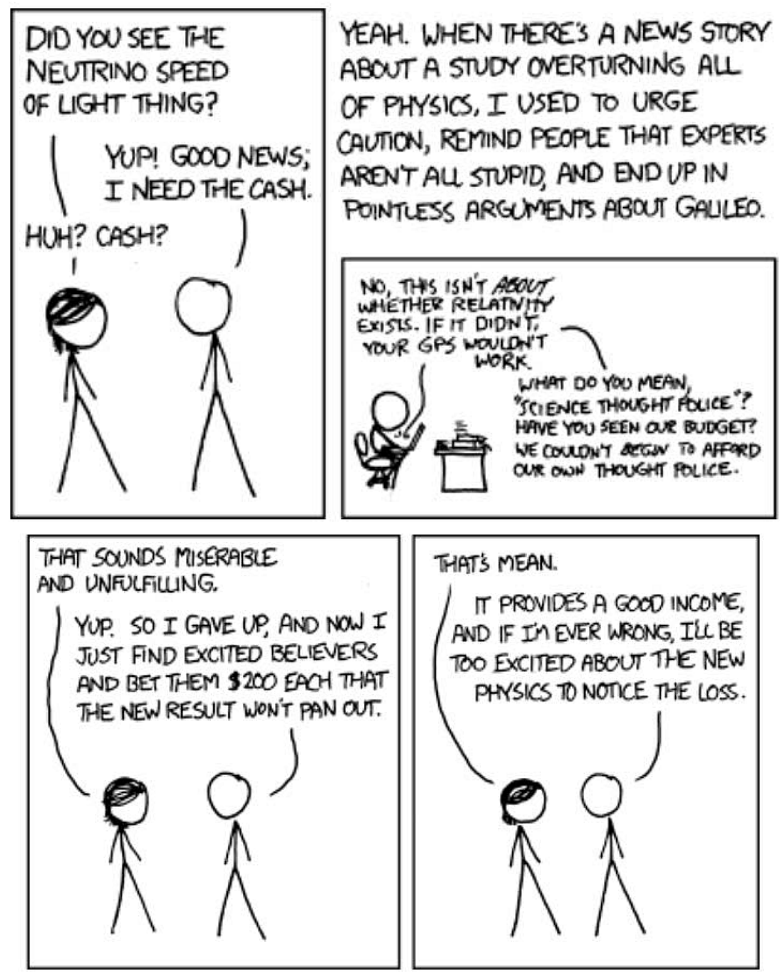

**Figure 9.** Betting that any interesting new physics result is wrong!

I used Inspire to count the number of papers that mention the Standard Model as a function of time, Figure 10. While there is a kink in the exponential growth after the 1980s, we are still in a period of doubling every 10 years. There is just too much to even think about, let alone summarize. Because I am at Harvard I will just mention one currently popular approach to our puzzles.

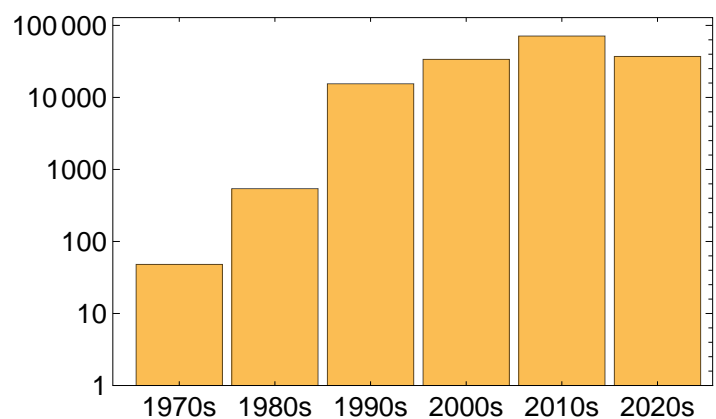

**Figure 10.** Number of papers mentioning the Standard Model as a function of time.

The Swampland conjectures [26], as I understand them which is not very well, are a set of constraints on effective field theories motivated by string theory and quantum gravity.

The swamplanders draw amusing pictures. Only if a theory satisfies their conjectures, we are asked to believe, will it poke out of the mess of the swamp onto the high land of the string landscape.

They claim to be able to draw testable conclusions for example about neutrino masses. I think the idea is that a theory needs enough light degrees of freedom or it will be pulled into the swamp. This does not seem too restrictive to me. I assume that if neutrinos are Majorana, or too heavy, they will just tell us that there are additional light particles that we haven't seen. The swampland conjectures reminded me of the time back in the early '90s when my late friend Nathan Isgur used to talk about the "brown muck of low energy hadron physics". Though it has very little to do with my talk, I can't resist showing a drawing done by the multitalented Michael Peskin for a Heavy Quark meeting, Figure 11.

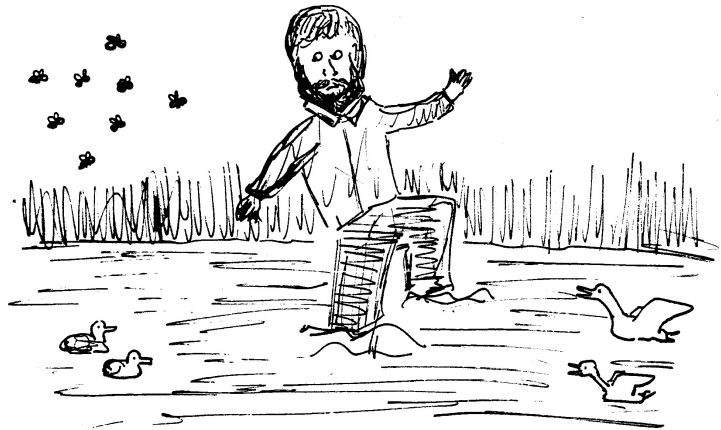

**Figure 11.** H. Georgi escapes from hadron dynamics (Original drawing by Michael Peskin).

Now that is a "swamp"! Let me close by briefly discussing one puzzle that I think has been solved by my grand-student Matt Schwartz and his students [27]. They find that the universe will last for at least $10^{65}$ years. So at least we will have plenty of time to solve the other puzzles of the Standard Model. Personally, I plan to spend my time in the swamp.

**Conflicts of Interest**

The author declares no conflict of interest.

**References**

1. Weinberg, S. A Model of Leptons. *Phys. Rev. Lett.* **1967**, *19*, 1264–1266. https://doi.org/10.1103/PhysRevLett.19.1264.

2. Glashow, S.L.; Iliopoulos, J.; Maiani, L. Weak Interactions with Lepton-Hadron Symmetry. *Phys. Rev. D* **1970**, *2*, 1285–1292. https://doi.org/10.1103/PhysRevD.2.1285.

3. Hooft, G. Renormalizable Lagrangians for massive Yang-Mills fields. *Nucl. Phys. B* **1971**, *35*, 167–188. https://doi.org/10.1016/0550-3213(71)90139-8.

4. Hooft, G.; Veltman, M.J.G. Regularization and renormalization of gauge fields. *Nucl. Phys. B* **1972**, *44*, 189–213. https://doi.org/10.1016/0550-3213(72)90279-9.

5. Lee, B.W.; Zinn-Justin, J. Spontaneously Broken Gauge Symmetries. I. Preliminaries. *Phys. Rev. D* **1972**, *5*, 3121–3137. https://doi.org/10.1103/PhysRevD.5.3121.

6. Lee, B.W.; Zinn-Justin, J. Spontaneously Broken Gauge Symmetries. II. Perturbation Theory and Renormalization. *Phys. Rev. D* **1973**, *8*, 4654. https://doi.org/10.1103/PhysRevD.5.3137.

7. Lee, B.W.; Zinn-Justin, J. Spontaneously Broken Gauge Symmetries. III. Equivalence. *Phys. Rev. D* **1972**, *5*, 3155–3160. https://doi.org/10.1103/PhysRevD.5.3155.

8. Lee, B.W.; Zinn-Justin, J. Spontaneously Broken Gauge Symmetries. IV. General Gauge Formulation. *Phys. Rev. D* **1973**, *7*, 1049–1056. https://doi.org/10.1103/PhysRevD.7.1049.

9. Coleman, S.R.; Weinberg, E.J. Radiative Corrections as the Origin of Spontaneous Symmetry Breaking. *Phys. Rev. D* **1973**, *7*, 1888–1910. https://doi.org/10.1103/PhysRevD.7.1888.

10. Georgi, H.; Glashow, S.L. Unified Weak and Electromagnetic Interactions without Neutral Currents. *Phys. Rev. Lett.* **1972**, *28*, 1494. https://doi.org/10.1103/PhysRevLett.28.1494.

11. Weinberg, S. Mixing Angle in Renormalizable Theories of Weak and Electromagnetic Interactions. *Phys. Rev. D* **1972**, *5*, 1962–1967. https://doi.org/10.1103/PhysRevD.5.1962.

12. Georgi, H.; Glashow, S.L. Attempts to Calculate the Electron Mass. *Phys. Rev. D* **1973**, *7*, 2457–2463. https://doi.org/10.1103/PhysRevD.7.2457.

13. Bouchiat, C.; Iliopoulos, J.; Meyer, P. An anomaly-free version of Weinberg's model. *Phys. Lett. B* **1972**, *38*, 519–523. https://doi.org/10.1016/0370-2693(72)90532-1.

14. Politzer; D, H. Reliable Perturbative Results for Strong Interactions? *Phys. Rev. Lett.* **1973**, *30*, 1346–1349. https://doi.org/10.1103/PhysRevLett.30.1346.

15. Gross, D.J.; Wilczek, F. Ultraviolet Behavior of Non-Abelian Gauge Theories. *Phys. Rev. Lett.* **1973**, *30*, 1343–1346. https://doi.org/10.1103/PhysRevLett.30.1343.

16. Pati, J.C.; Salam, A. Lepton number as the fourth "color". *Phys. Rev. D* **1974**, *10*, 275–289. https://doi.org/10.1103/PhysRevD.10.275.

17. Weinberg, S. Non-Abelian Gauge Theories of the Strong Interactions. *Phys. Rev. Lett.* **1973**, *31*, 494–497. https://doi.org/10.1103/PhysRevLett.31.494.

18. Georgi, H. The State of the Art—Gauge Theories. In Proceedings of the 1974 Meeting of the Division of Particles and Fields of the APS, Williamsburg, VA, USA, 5–7 September 1974.

19. Georgi, H.; Glashow, S.L. Unity of All Elementary-Particle Forces. *Phys. Rev. Lett.* **1974**, *32*, 438–441. https://doi.org/10.1103/PhysRevLett.32.438.

20. Fritzsch, H.; Minkowski, P. Unified interactions of leptons and hadrons. *Ann. Phys.* **1975**, *93*, 193–266. https://doi.org/10.1016/0003-4916(75)90211-0.

21. Georgi, H.; Quinn, H.R.; Weinberg, S. Hierarchy of Interactions in Unified Gauge Theories. *Phys. Rev. Lett.* **1974**, *33*, 451–454. https://doi.org/10.1103/PhysRevLett.33.451.

22. Appelquist, T.; Georgi, H. $e^+e^-$ Annihilation in Gauge Theories of Strong Interactions. *Phys. Rev. D* **1973**, *8*, 4000–4002. https://doi.org/10.1103/PhysRevD.8.4000.

23. Zee, A. Electron-Positron Annihilation in Stagnant Field Theories. *Phys. Rev. D* **1973**, *8*, 4038–4041. https://doi.org/10.1103/PhysRevD.8.4038.

24. Aubert, J.J.; Becker, U.; Biggs, P.J.; et al. Experimental Observation of a Heavy Particle *J*. *Phys. Rev. Lett.* **1974**, *33*, 1404–1406. https://doi.org/10.1103/PhysRevLett.33.1404.

25. Augustin, J.-E.; Boyarski, A.M.; Breidenbach, M.; et al. Discovery of a Narrow Resonance in $e^+e^-$ Annihilation. *Phys. Rev. Lett.* **1974**, *33*, 1406–1408. https://doi.org/10.1103/PhysRevLett.33.1406.

26. Vafa, C. The String Landscape and the Swampland. *arXiv* **2005**, arXiv:hep-th/0509212.

27. Andreassen, A.; Frost, W.; Schartz, M.D. Scale-invariant instantons and the complete lifetime of the Standard Model. *Phys. Rev. D* **2018**, *97*, 056006. https://doi.org/10.1103/PhysRevD.97.056006.

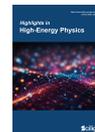

*Opinion*

# The Rise of Gauge Theories: From Many Models to One Theory

Jean Iliopoulos

Laboratoire de Physique de l'Ecole Normale Supérieure, ENS-PSL, CNRS, Sorbonne Université, Université Paris Cité, 75005 Paris, France; ilio@phys.ens.fr

**Abstract:** By a worth noting coincidence, the year 2024 marks two anniversaries: The 70th anniversary of CERN and the 50th anniversary of the $J/\Psi$ discovery. In this talk I will present very briefly the significance of these anniversaries [1] . CERN's founding fathers were among Europe's leading scientists. CERN became the most successful International Institution. Scientists built the Scientific European Union long before politicians thought of the Economic European Union. If the Scientific Institution appears to be more solid than the Economic one, it is because its foundations are made with ideas. Regarding the second anniversary, I want to argue that $J/\Psi$ was not just a new resonance; we knew already a large number of them. It was not even only the first indirect evidence for the existence of a new quark flavor. It was all that, but it was also much more. It was the final proof which convinced the large majority of our community that we were witnessing a radical change of paradigm in our understanding of microscopic physics. From phenomenological models and specific theories, each one applied to a restricted set of experimental data, we had to think in terms of a fundamental theory of universal validity. From many models to the STANDARD THEORY of particle physics. For most of us this transition was a revelation, for some others it was a painful experience. It is appropriate to combine it with CERN's anniversary because the experimental verification of the Standard Model started at CERN with the neutral currents and was completed also at CERN with the BEH scalar boson. It is this story that I will attempt to narrate in this note, although I do not consider myself to be an unbiased observer.

**Keywords:** gauge theories; charm; standard model

## 1. The Origins

If I had to assign a year to the birth of experimental particle physics I would choose 1950, the discovery of the neutral pion at the Berkeley electron synchrotron. It is the first "elementary" particle discovered with an accelerator (the charged pion was discovered in 1947 in cosmic rays). The existence of $\pi^0$ was predicted in 1938 by N. Kemmer who wrote the first isospin invariant theory for the nuclear forces (Kemmer has not received the appropriate recognition for his groundbreaking work. His equations for the pion-nucleon interaction can be found in every textbook, but his name is rarely mentioned). So, $\pi^0$ is the first particle whose existence was predicted by an argument based on an internal symmetry and also the first to be discovered in an accelerator. Since that time accelerators became the main engines of discovery. With their use the field expanded very rapidly and today the "Table of Elementary Particles" has several hundred entries, although we know that very few among them are "elementary".

During the second half of the 20th century experimental particle physics has followed a monotonically rising trajectory, in contrast, that of its theoretical counterpart was more circuitous. Modern theoretical high energy physics has a precise date of birth: 2 June 1947, the date of the Shelter Island Conference. The most important contributions which were presented in that conference were not theoretical breakthroughs but two experimental results: the non-zero values of the Lamb shift and of the electron anomalous magnetic moment. Both these results were important because, for the first time, they showed, beyond any possible doubt, that the predictions based on the Dirac equation for the electron are only approximately correct. This in turn motivated a serious study of quantum field theory. In the following months R.P. Feynman and J.S. Schwinger, independently, using apparently different formulations, set up the program for the renormalised perturbation expansion of Quantum Electrodynamics and Schwinger gave the first calculation of $g-2$. As it turned out, similar results were obtained independently in Japan by Sin-Itiro Tomonaga, who obtained also the first complete calculation of the Lamb shift. The equivalence of all these approaches was formally shown by F. Dyson in 1948. Rare are the examples in physics in which so much progress was accomplished in such a short time. The agreement between theory and experiment in the phenomena of quantum electrodynamics is impressive.

The natural next step was to apply this approach to the other two interactions, to wit the strong and the weak ones. The former were represented by the isospin invariant pion-nucleon interaction and it was shown that the renormalisation program applied to it. However, the results were useless because the effective pion-nucleon coupling constant turned out to be very large, on the order of 10, making the power series expansion meaningless. The dynamics appears to be dominated by phenomena such as resonance production, which cannot be described by ordinary perturbation theory. There remained the weak interactions. Naturally, they are much weaker, but now a new problem appeared: the renormalisation program is a quite complex process and applies only to a small number of quantum field theories. The Fermi theory, which describes very well the low energy weak interactions, is not one of them.

This double failure soon tarnished the glory of quantum field theory. The disappointment was such that the subject was not even taught in many universities. By the late fifties the theoretical high energy physics landscape was fragmented in many disconnected domains, having no common trends and often ignoring each other. For strong interaction processes the main approach was based on the assumed analytic properties of the $S$-matrix elements, but we had also several simple models, none with any solid theoretical basis, each one applied to a particular corner of phase space. For weak interactions the Fermi theory proved to be a very good phenomenological model, but it had no logical justification and no obvious connection with anything else. The most important progress in our understanding of Nature's fundamental laws came from the application of symmetry principles and not from dynamical calculations. Quantum field theory was noticeable essentially by its absence. A totally marginal subject confined to very few precision calculations in quantum electrodynamics. Many physicists had only vague and often erroneous ideas about it and, to a certain extent, this misunderstanding has survived even today. The "Dark Ages" would last two decades.

## 2. The Secret Road to the New Theory

It is usually said that progress in science occurs when an unexpected experimental result contradicts the current theoretical beliefs. This forces scientists to change their ideas and leads to a new theory. This has often been the case in the past, but the revolution we are going to describe here had a theoretical, rather an aesthetic motivation. It was a triumph of abstract theoretical thought which brought geometry into physics. This "secret road" has been long and circuitous with no discernible well defined direction. Many a time it gave the impression of leading to a dead end. In fact it was a long series of isolated and mostly confidential contributions and many important ideas had to be rediscovered again and again. Several milestones did not seem to point to a single path and most of them went unnoticed when they were first proposed. The pioneers were often unaware of each other's work and it is only now that we can see a coherent picture. A strict chronological order is not possible and I choose instead to group together a few important contributions which became the main ingredients of the Standard Model. Notice that most of them had motivations unrelated to their final application. The construction of the Standard Model used many concepts and techniques of quantum field theory. A short list includes:

- Gauge invariance.
- Renormalisation, in particular for gauge invariant field theories.
- Renormalisation and symmetry, including the phenomenon of anomalies in the conservation of symmetry currents.
- The renormalisation group and the property of asymptotic freedom.
- Higher internal symmetries and the quark model.

The phenomenon of spontaneous symmetry breaking, in particular in the presence of long range gauge forces. I had no time to present the full story of all these discoveries in Rome and a short account can be found in reference [1]. I gave only a discussion of the concept of gauge invariance which goes back to classical electrodynamics. I do not know who was the first to remark that the dynamical system described by the components of the electric and magnetic fields $\boldsymbol{E}$ and $\boldsymbol{B}$ – that is, six degrees of freedom in our counting—was in fact redundant because some of the equations do not involve any time derivatives and should be considered as constraints. It seems that the first person who attempted to reduce the redundancy was C.F. Gauss who, in some manuscript notes in 1835, introduced the concept of the "vector potential" $A$. It was further developed by several authors and was fully written by G. Kirchhoff in 1857, following earlier work, in particular by F. Neumann. The components of the electric and magnetic fields could be expressed in terms of the vector and scalar potentials, thus reducing the number of degrees of freedom from six to four. It was soon noticed that it still carried redundant variables and several "gauge conditions" were used. The condition, which in modern notation is written as $\partial_\mu A^\mu = 0$, was proposed by the Danish mathematical physicist L.V. Lorenz in 1867. It seems that Maxwell favored the condition $\nabla \cdot A(x) = 0$ which today we call *"the Coulomb gauge"*. Using Maxwell's equations we can immediately see the redundancy of the system $(\Phi, A)$ because the equation for $\Phi$ does not involve any time derivative. Lorenz arrived to this conclusion because he had a formulation of classical electrodynamics equivalent to Maxwell's.

In this context, an interesting story is the following: Around the years 1840 F.E. Neumann and, independently, W.E. Weber, studied the interaction between two closed electric circuits carrying currents $I$ and $I'$, respectively. Their methods and physical assumptions were different and so were the expressions they obtained for the magnetic interaction energy between the two circuits. However, they both seemed to fit Ampère's measurements. In the years after 1870, H.L.F. von Helmholtz criticised and compared the two expressions. In particular, he noticed that Neumann's and Weber's formulae for the elementary magnetic interaction energy, differ by a quantity which can be expressed as a multiple of a perfect differential, so, they give the same result when integrated over the closed circuits. In our present terminology he showed that the two expressions are *gauge equivalent*. He even went a step further: he generalised the two results by exhibiting a one-parameter family of expressions for the magnetic energy which interpolate between those of Neumann and Weber. It was the first example of a *family of gauges*. By the end of the century H.A. Lorentz published a book and some encyclopedia articles with the full classical electromagnetic theory. The invariance under gauge transformations of the vector and scalar potentials is an integral part of it.

At the beginning of the 20th century the development of the general theory of relativity offered a new paradigm for a gauge theory and triggered several attempts to unify electromagnetism and gravitation. In reference [1] I mention the work of G. Nordström, as well as that of T. Kaluza and O. B. Klein, which is used today in supergravity and superstring theories. An independent approach is due to H.K.H. Weyl in 1919. It is mostly known for his unsuccessful attempt to enlarge the gauge symmetry of general relativity to include invariance under local scale transformations of the metric: $g \rightarrow e^{2\lambda(x)}g$, with $\lambda(x)$ an arbitrary function of the space-time point $x$. He called the resulting invariance *eichinvarianz* in German, which was translated in English as *gauge invariance*. Weyl was the first to understand that the classical electromagnetic theory was mathematically incomplete. People had discovered by trial and error its property of gauge invariance, but the underlying global symmetry was missing. Weyl proposed to identify it with scale transformations. It was a wrong physical answer to a correct mathematical question. The laws of physics are in no way invariant under global scale transformations. As it turned out, the correct answer was found in the framework of quantum mechanics.

The first person who understood that the invariance of quantum mechanics under phase transformations of the wave function could be the missing global symmetry of electromagnetism, was V.A. Fock. In a paper published in 1926, just after Schrödinger wrote his equation, he showed that promoting the invariance from global to local, is equivalent to writing the equation for an electron in an external electromagnetic field. The extension of this simple idea to non-abelian internal symmetries – in particular the $SU(2)$ isospin transformations developed by Heisenberg and Kemmer between 1932 and 1938 – took many years and followed a very complicated and counter-intuitive way [1]. It was finally achieved by C.N. Yang and R.L. Mills in 1954. Since that time gauge theories became part of high energy physics.

## 3. The Standard Model–A Theoretical Speculation

The entire theoretical scheme, which became the Standard Theory of particle physics, was fully written in 1973. It is a gauge theory based on the group $SU(3) \times SU(2) \times U(1)$, spontaneously broken to $SU(3) \times U(1)_{em}$. It describes all phenomena we observe in our experiments. Yet, it was not generally accepted. For most physicists it was a wild theoretical speculation with no connection to the real world. The main reason for this negative attitude was the mistrust towards quantum field theory, but it is also true that the model seemed to make many strange predictions with little, if any, experimental support. Let me mention some of them.

- The model predicted the existence of 12 vector bosons. But only one, the photon, was known! Three, $(W^\pm, Z^0)$ were predicted to be very heavy (For the 1973 physicists a mass of 100 GeV was essentially infinite!) and the other eight, the gluons, were declared *unobservable* by a strange property of *confinement*.
- The model predicted the existence of a scalar boson, the BEH, with unknown mass. For many physicists it was

a heresy coming after the assumed triumph of the $V - A$ theory of weak interactions.

- Neutral currents were predicted but the obvious ones, the $K^0 \rightarrow \mu^+ + \mu^-$ decay, were excluded. Gargamelle had established the existence of strangeness conserving neutral currents but not everybody was convinced (I won some bottles of very fine wine by betting for neutral currents, in particular against Jack Steinberger). Furthermore, the possible existence of weak neutral currents had been envisaged before the formulation of gauge theories, so their existence was not considered as a decisive proof of the new theory.

- Probably the most "extravagant" prediction was that of the charmed quark, implying the existence of an entire new hadronic world of charmed particles. For most people the arguments were not considered serious. I still remember the objections: some obscure higher order effects – triangle diagrams for the anomalies, or square diagrams for the absence of flavor violating neutral currents – would dictate the structure of the world? Totally absurd! In retrospect, I think that the large majority of physicists rejected this particular prediction because it went against the prevailing philosophy of compartmentalisation of high energy physics. Theoretical arguments motivated by properties of weak interactions were not admissible to make predictions on hadronic physics, a domain reserved exclusively to strong interactions (*Sutor, ne supra crepidam or, Let the cobbler stick to his last.*)

- The QCD prediction for the ratio $R(Q^2) = \frac{\sigma(e^+ + e^- \rightarrow \text{hadrons})}{\sigma(e^+ + e^- \rightarrow \mu^+ + \mu^-)}$ seemed to be in violent contradiction with experiment.

## 4. The London Conference

This brings me to the 17th International Conference on High Energy Physics held in London in July 1974. It can be viewed as the last Conference of the Dark Ages and the first Conference of the New Era. Several old and new results were announced in this Conference. I will mention only those which are important for our story.

- D.C. Cundy gave the plenary report on neutrino physics. Obviously, the Gargamelle results were the central point. I remember him saying: *"Those who have bet on neutral currents, now is the time to pay and collect."*

- B. Richter gave the plenary report on $e^+e^- \rightarrow$ hadrons. In Figure 1 I show the graph he presented.

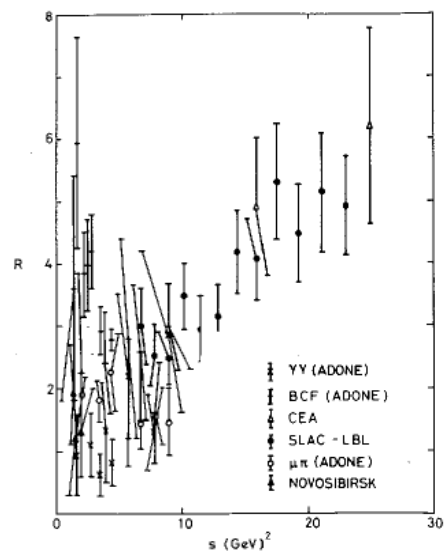

**Figure 1.** A compilation of all early measurements of the ratio $R$, as presented in the 1974 London International Conference on High Energy Physics by Burton Richter.

I remind you that the QCD prediction was that $R$ should approach the value $R = 2$ (the sum of the electric charges squared of $u$, $d$ and $s$ quarks) from above. No such behavior is visible in Figure 1.

- In my plenary report on gauge theories at the same Conference I said: *"... the hadron production cross section, which absolutely refuses to fall, creates a serious problem. The best explanation may be that we are observing the opening of the charmed thresholds, in which case everything fits together very nicely."* The addition of a charmed quark would add an extra 4/3 to $R$ (By a numerical accident, the data of Figure 1 contain also the production of the $\tau$ lepton which was not known at the time). A. Salam, who was the Chairman of the Conference, had ready a bottle of a fine Bordeaux red wine to reward speakers who finish on time. Naturally I was late and I said something like: *"I know I am about to loose my bottle, but I am ready to bet now a* *whole case that, if the weak interaction sessions of this Conference were dominated by the discovery of the neutral currents, the entire next Conference will be dominated by the discovery of the charmed particles."* This convinced Salam to give me the bottle which I opened immediately, poured myself a generous libation, offered the rest to those sitting in the first row, and drank "to charm!".

## 5. The Charming Theory of the New Particles

In November 1974 both Brookhaven and Stanford published their results. SPEAR decided to sweep the region above 3 GeV in fine steps of 1 MeV. To their great surprise they obtained a totally different picture, Figure 2.

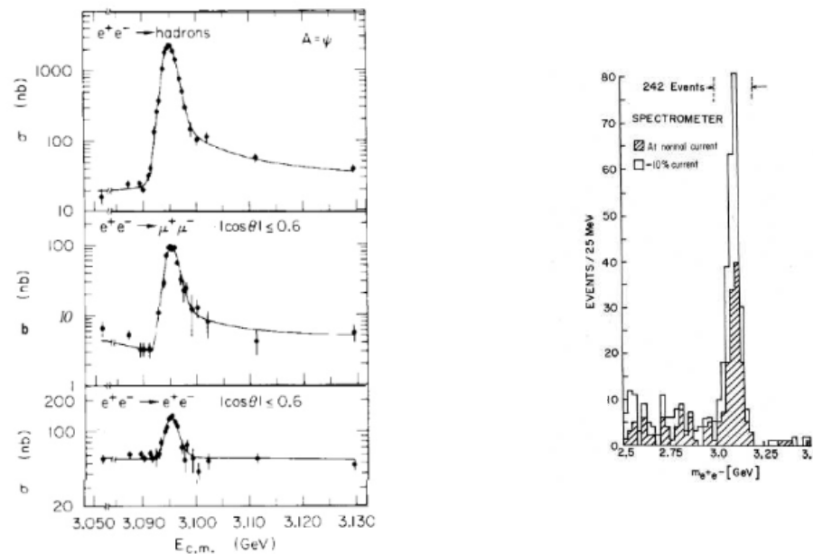

**Figure 2.** The discovery of the $J/\Psi$ meson in Nov. 1974 independently by SPEAR (**left**) and AGS (**right**). Both exhibit peaks in the oppositely charged dielectron mass spectrum consistent with the $J/\Psi$ mass at 3.1 GeV. This result was also confirmed by the Frascati group.

I was in Paris at the end of 1974 when I received a telephone call from A. Lagarrigue inviting me to an informal meeting to discuss important results from SPEAR. G. 't Hooft was visiting Ecole Normale and I took him along. In those days news did not travel with the speed of internet. B. Jean-Marie had come from Stanford with the results. They were indeed very impressive. A 3 GeV hadron, decaying into pions, with a width of less than 100 keV? Incredible! Lagarrigue asked me what I thought of it and I confess I was bewildered. It was 't Hooft who first gave me the explanation. It is simple to understand in QCD. Let us consider the series of $1^-$ mesons: $\rho$ is a bound state of a quark-antiquark pair of the first family. Its mass lays well above the $2\pi$ mass and its width is very large $\sim 147$ MeV. The $\phi$ is an $s\bar{s}$ bound state with a mass of 1020 MeV. It lays barely above the threshold of a $K\bar{K}$ pair, yet its branching ratio to $K\bar{K}$ is 83% despite the fact that the phase space is tiny. The pure pionic partial width is only 650 keV. In the old days we had a rule, called the OZI (Okubo-Zweig-Iizuka) rule, one of those empirical rules of the dark ages with no real theoretical justification. It stated that in a quark-antiquark bound state, the decay modes requiring the annihilation of the initial $q\bar{q}$ pair, were highly suppressed (It was called "the re-arrangement model" by H. Rubinstein). Let us come now to $J/\Psi$ and assume it is a $c\bar{c}$ bound state. The $0^-$ mesons are supposed to be the pseudo-Goldstone bosons of spontaneously broken chiral symmetry. The latter is very good for the first family, questionable for the $s$ quark, and very poor for charm. Therefore we expect the charmed $0^-$ mesons $D$ to be quite heavy and the mass of $J/\Psi$ to lay below the $D\bar{D}$ threshold. As a result $J/\Psi$ decays mainly into pions. The decay amplitude for a $1^-$ meson goes through three gluons, so the width is proportional to $\alpha_s^3$. Between 1 and 3 GeV $\alpha_s$ has dropped by a factor of two, so we expect the $J/\Psi$ width to be 8 times smaller than the $\phi$ pionic width. It is precisely what is found experimentally. As I said, I first heard this argument from 't Hooft, but later I found it in papers by Appelquist and Politzer as well as De Rujula and Glashow. The first has a submission date of November 19. This makes me believe it was found before the experimental discovery.

Within a year the entire region between 3 and 5 GeV was studied in detail, see Figure 3.

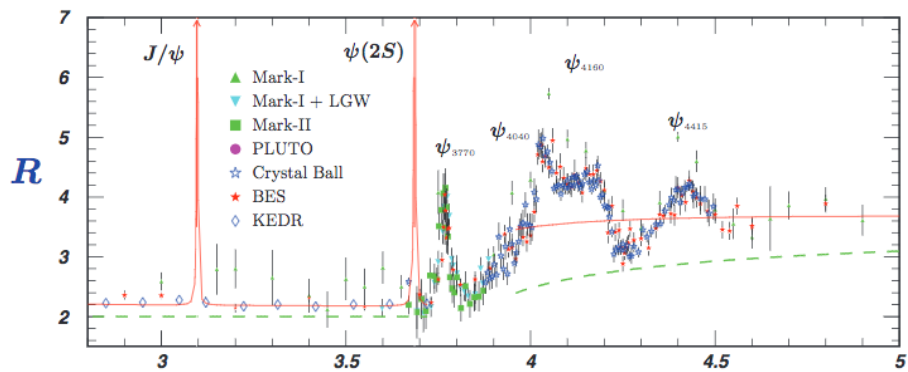

**Figure 3.** The value of $R$ for energies between 3 and 5 GeV.

As expected, there are broader resonances with masses $\geq 4$ GeV which are those lying above the $D\bar{D}$ threshold. Lo and behold, the particles with naked charm were found among the decay products of these resonances. It was in 1976. In the meantime a rich charmonium spectroscopy (I believe that the term 'charmonium" was coined by A. De Rujula) was discovered in full agreement with the theoretical predictions. Although nobody paid the bet I offered in the 1974 Conference, the entire 1976 one was indeed dominated by charmed particles and gauge theories. The report in this Conference by A. De Rujula had the title: *"Theoretical basis of the new particles"*. The first sentence is *"I review the four-quark standard gauge field theory of weak, electromagnetic and strong interactions"*. Now we talk about "the four-quark standard gauge field theory". The phase transition from Many Models to One Theory was complete. The order parameter has been the fraction of physicists who changed their views: a small minority before 1974 to the large majority after 1976. The complete verification of the theory took many more years and many great discoveries, but the mood of the community had changed. The following discoveries of the vector bosons, the $b$ and $t$ quarks which complete the third family of the $\tau$ lepton, the gluon jets and the BEH scalar as well as the very good general fit using all available data, were no more great surprises, they were expected. THE STANDARD MODEL had become THE STANDARD THEORY.

## 6. From Dream to Expectation

Feynman has said that progress in physics is to prove yourself wrong as soon as possible. For half a century now we have not been able to prove the Standard Theory is wrong. It has passed successfully all tests and all its predictions have been brilliantly verified. What comes next?

In 2011 the European Physical Society awarded to Glashow, Maiani and myself the High Energy Physics Prize. At this occasion we were invited to speak at the European Conference in Grenoble. The title of my talk was "Following the Path of Charm" and I tried to argue that precision measurements at a certain energy scale allow us to make predictions at some higher scale. Let me start from an expanded version of a plot showing the value of $R$ from low energies up to and above the $Z^0$ mass, Figure 4.

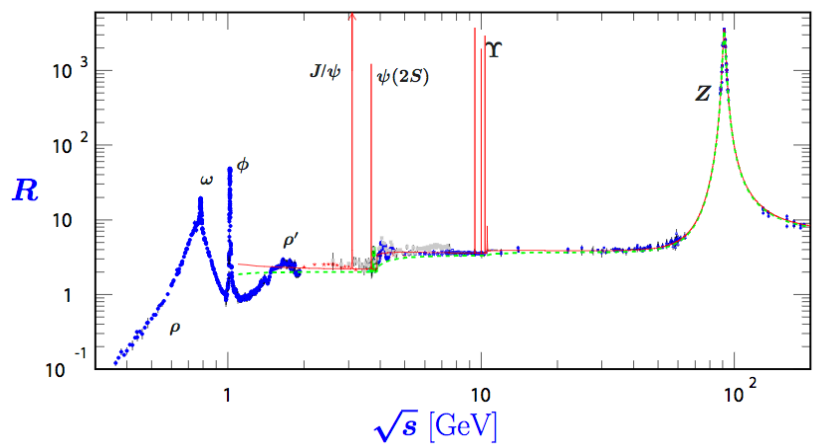

**Figure 4.** The ratio $R$ from low energies, up to and above the $Z$ mass. The green curve is the parton model prediction and the red one includes QCD corrections. Remarkable agreement.

What is most remarkable is the precision with which theoretical predictions fit the experimental data in most of the energy interval. The only regions in which there is no agreement are the one at low energy below 1 GeV, in which the QCD is in the strong coupling regime, and small, very localised regions signaling the thresholds for the production of new species of hadrons, charm or *b*. Outside those regions, our theory gives an excellent fit, although it is just the result of one loop perturbation expansion. I concluded that, for reasons that are not fully understood, perturbation theory is reliable outside the regions of strong interactions.

I tried to use this fact in order to make predictions regarding the multi-hundred GeV scale which was expected to be explored by LHC. The argument was based on the low energy data which favored a Higgs particle of relatively low mass, $\leq 200$ GeV. It went roughly as follows: The existing data are compatible with the Standard Theory only if the Higgs is light. Therefore, if we find a very heavy Higgs we must also find new interactions which invalidate the Standard Theory calculations. If on the other hand we find a light Higgs, we must find new interactions which stabilise its mass at this low value. I thought that both possibilities were good news for LHC. In my talk at EPS in 2011 I said:

> *"I want to exploit this experimental fact* [the validity of perturbation theory] *and argue that the available precision tests of the Standard Model allow us to claim with confidence that new physics is present at the TeV scale and the LHC can, probably, discover it. The argument assumes the validity of perturbation theory and it will fail if the latter fails. But, as we just saw, perturbation theory breaks down only when strong interactions become important. But new strong interactions imply new physics".*

My conclusion was that, for LHC, which was about to start operating, new physics was around the corner!
Today we know that LHC found no corner!
But I secretly believe the argument is correct, only the corner is a bit further down.
Although I will not see it, I am confident some among our young colleagues will find it.

## Conflicts of Interest

The author declares no conflict of interest.

## Reference

1. Iliopoulos J. From Many Models to ONE THEORY. *arXiv* **2025**, http://arxiv.org/abs/2501.10233.

*Review*

# From Charm to CP Violation

Luciano Maiani [1,2]

[1] Sapienza University of Rome and INFN , Piazzale Aldo Moro 2, I-00185 Rome, Italy; luciano.maiani@cern.ch
[2] CERN , 1211 Geneva, Switzerland

**Abstract:** The paper provides a personal recollection of key developments in high-energy physics, from the quark model to CP violation and the November revolution. It highlights how pivotal theoretical breakthroughs, as the GIM mechanism and the formulation of the Standard Model, have been deeply intertwined with experimental discoveries shaping modern particle physics. Lastly a few suggestions are given to orient the search for a more fundamental theory beyond the Standard Model.

**Keywords:** quark model; charm; GIM; standard model; CP violation

## 1. Late 1960: Hopes to a Get a Fundamental Theory of Particle Interactions

The years 1960's saw an impressive progress of the theoretical understanding of particle physics, that led to the hope to arrive at a fundamental theory of all particle interactions.

- Abundant spectroscopic data supported the Gell-Mann-Zweig hypothesis [1,2] that hadrons are all composite states of elementary quarks in three flavours ($u$, $d$, $s$): baryons made by three quarks, e.g., $p = (uud)$, $n = (udd)$, etc. and mesons by quark antiquark pairs, e.g., $\pi^+ = u\bar{d}$, $\pi^0 = (u\bar{u} - d\bar{d})/\sqrt{2}$, etc.;
- The Cabibbo Theory [3] described in a simple and elegant way the Weak Interactions of hadrons and leptons, in agreement with the $V - A$ [4–6], current × current Fermi theory and the universality principle. In quark language:

$$\mathcal{L}_F = \frac{G}{\sqrt{2}} J^\lambda J_\lambda$$
$$J^\lambda = \bar{\nu}_e \gamma^\lambda (1 - \gamma_5) e + \bar{\nu}_\mu \gamma^\lambda (1 - \gamma_5)\mu + \bar{u}\gamma^\lambda(1-\gamma_5)d_C \tag{1}$$
$$d_C = (\cos\theta_C \, d + \sin\theta_C \, s)$$

with $G$ the Fermi constant and $\theta_C$ a new universal constant known as the Cabibbo angle.

There were clouds, however, pointing out that something was missing:

- Do quark clash with Fermi-Dirac statistics? with first ideas about color by Han and Nambu [7];
- The structure of basic strong interactions was not clear: mediated by *abelian gluons*? determined by *Veneziano duality* [8]?
- Fermi theory is not renormalizable. Is it mediated by a $W^\pm$ boson? Are divergences for hadron weak decays cured by form-factors?

A new line of investigations was opened by J. Schwinger [9], suggesting the unification of Weak and Electromagnetic interactions in a Yang-Mills, non-abelian, gauge theory. The idea was followed by a sequence of bright papers:

- the theory of Glashow (1961) [10], based on the gauge group: $SU(2)_L \otimes U(1)_Y$;
- the Brout-Englert-Higgs Mechanism [11,12], proposing spontaneous symmetry breaking of the gauge symmetry to give a mass to the W boson(1965);
- the model by Weinberg [13] and Salam[14], incorporating the Brout-Englert-Higgs idea in Glashow's model (1967).

The unified theories, however, could be applied to lepton interactions only: embedding the Cabibbo Theory in $SU(2)_L \otimes U(1)_Y$ would produce flavour-changing neutral currents to first order in the Fermi constant, in blatant contradiction with data. Does unification work for leptons only? Or, are flavour-changing neutral currents suppressed by form factors ?

## 2. The $G\Lambda^2$ Puzzle, 1968

The discussion on higher order weak interactions was opened in 1968 by a calculation by Boris Ioffe and Evgeny Shabalin [15], indicating that $\Delta S = \pm 1$ neutral currents and $\Delta S = 2$ amplitudes would result from higher order weak interactions, even in a theory with one charged $W$ boson only. Amplitudes, see Figure 1, were found to be divergent, of order $G(G\Lambda^2)$, and in disagreement with experiments, unless limited by an ultraviolet cut-off $\Lambda = 3$–4 GeV (from $\Delta m_K$);

The result was based on current algebra commutators: it shows that hadron form factors are irrelevant. Current commutators imply hard constituents.

Similar results were found by R. Marshak and collaborators [16] and by F. Low [17] and the exceedingly small value of the cut-off raised a wide discussion.

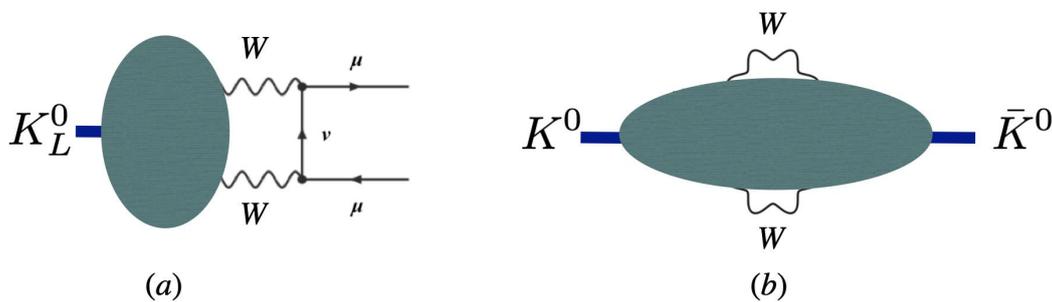

**Figure 1.** Decays and mixing in the Ko system resulting from higher order weak corrections.

Attempts were made during 1968-69 to make the amplitude more convergent:

- Introducing more than one Intermediate Vector Boson(Gell-Mann, Low, Kroll, Ruderman) [18]: far too many were needed;
- Introducing negative metrics (ghost) states (T.D.Lee and G.C. Wick), of mass = $\Lambda$ [19];
- Another line was to cancel the quadratic divergence, in correspondence to a specific value of the Cabibbo angle, i.e., "computing" the Cabibbo angle (Gatto, Sartori, Tonin,Cabibbo, Maiani) [20,21];
- In this context, it was realised that quadratic divergent amplitudes at order $G\Lambda^2$ would also arise, in the intermediate vector boson theory, with potential violations of strong interaction symmetries (parity, isospin, SU(3) and strangeness). However, with chiral $SU(3)_L \otimes SU(3)_R$ breaking described by a $\mathbf{3} \otimes \bar{\mathbf{3}}$.

  … but the small cutoff in the $G(G\Lambda^2)$ terms still called for an explanation.

## 3. A Personal Recollection

The Ioffe-Shabalin problem was still on the table in November 1969, when I moved to Harvard and met with John Iliopoulos, at work with Shelly Glashow on the $G(G\Lambda^2)$ corrections [22]. We discussed for long, usually two of us arguing against the one at the blackboard, apparently getting nowhere. But during our discussions a change in paradigm occurred. Previous works had been done in the framework of the algebra of currents, Figure 1, but slowly we began to phrase more and more our discussion in terms of quarks.

In quark language, the Ioffe-Shabalin problem is represented by the box diagram in Figure 2a. The divergent amplitude is proportional to the product of the couplings of quarks d and s to the u quark, as required by the Cabibbo theory. By January 1970 we got convinced that we had to modify the weak interaction theory. Once we realised that, the solution was just under our eyes. A fourth quark of charge $+2/3$, called the charm quark, had been introduced by Bjorken and Glashow (and others), for entirely different reasons. In the weak interaction, the charm quark is coupled to $s_C$, the quark left out in the Cabibbo theory, represented by the orthogonal combination to $d_C$ of Equation (1). The exchange of a c-quark, Figure 2b, cancels the singularity and produces an amplitude of order $G[G(m_c^2 - m_u^2)]$, the GIM mechanism [23].

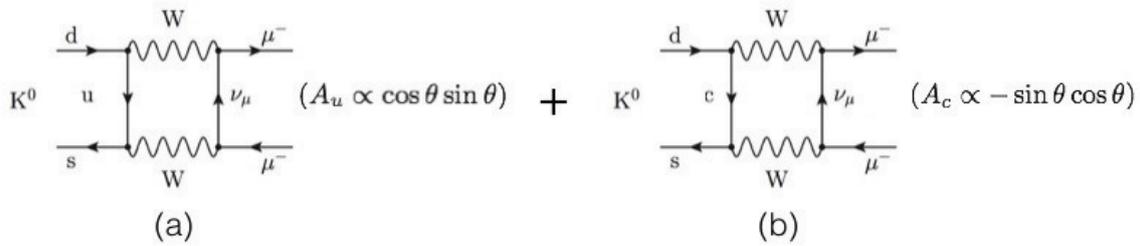

**Figure 2.** GIM mechanism for $K^0 \to \mu^+ \mu^-$.

With two generations, Cabibbo weak mixing $d_C = (\cos\theta \, d + \sin\theta \, s)$ is replaced by a unitary $2 \times 2$ matrix $U$

$$U = \begin{pmatrix} \cos\theta & \sin\theta \\ -\sin\theta & \cos\theta \end{pmatrix} \tag{2}$$

Charged currents in four-flavor space $(u, c, d, s)$ are given by the matrices $C$ and $C^\dagger$:

$$C = \begin{pmatrix} 0 & U \\ 0 & 0 \end{pmatrix}; \; C_3 = [C, C^\dagger] = \begin{pmatrix} UU^\dagger & 0 \\ 0 & -U^\dagger U \end{pmatrix} = \begin{pmatrix} \mathbf{1} & 0 \\ 0 & -\mathbf{1} \end{pmatrix} \tag{3}$$

The neutral current generator, $C_3$, is flavour diagonal. Thus, a unified gauge theory including charged and neutral vector bosons, as in Refs. [10,13,14], with flavour conserving neutral currents is possible [23].

The observed strangeness changing processes appear to one loop (the first weak interaction loop ever computed), are finite and determined by the mass difference $m_c - m_u$. Ioffe's cutoff becomes the prediction: $m_c \sim 1.5$ GeV.

A detailed study made later by B. W. Lee and M. K. Gaillard, in the Glashow-Weinberg-Salam theory with two generations of quark and leptons, confirmed the charm quark mass prediction [24].

## 4. Predictions and Facts Following GIM

**Quark-Lepton Symmetry.** The fourth quark restores a full quark-lepton symmetry. In fact, quark-lepton symmetry is mandatory for the cancellation of the Adler-Bell-Jackiw anomalies.

Charm is needed, fractionally charged, color triplet quarks are necessary, as shown by Bouchiat, Iliopolos and Meyer in [25], removing the last obstacle towards a renormalizable $SU(2)_L \otimes U(1)_Y$ theory,

**Neutrino neutral current processes.** Flavor diagonal, neutral current processes are predicted in Yang-Mills theory, to order $G[C, C^\dagger]$; in the unified theory, they appear in lowest order, mediated by $Z^0$.

**1973.** The Gargamelle Collaboration led by the French physicist Andrè Lagarrigue, operating the heavy liquid, large bubble chamber exposed to the high energy CERN beam, observes muonless neutrino events: events in which a hadronic jet is produced without an energetic muon track, interpreted as the neutral current process:

$$\nu_\mu + N \to \nu_\mu + \text{hadrons} \tag{4}$$

The Collaboration observed also events with an isolated electron track, interpreted as:

$$\nu_\mu + e \to \nu_\mu + e. \tag{5}$$

**1974.** The neutrino experiment in Fermilab led by Carlo Rubbia announced the observation of two-muon neutrino events. It was the first positive indication, at least in the Western countries, of the existence of a new quark flavour produced off the nucleons in the $\nu_\mu \to \mu$ transition (the first muon in the final state) and stable enough to decay semileptonically (the second muon). In quark language:

$$\nu + d \to c + \mu^- \to s + \nu_\mu + \mu^+ + \mu^- \tag{6}$$

(Cabibbo suppressed, rate $\propto \sin^2\theta_C$).

**The Search for Charmed Particles.** In 1970 there was no experimental evidence of weakly decaying hadrons beyond the lowest lying strange baryons and mesons. GIM explanation: ... *Suppose they are all relatively heavy, say 2 GeV. ...will decay rapidly ($10^{-13}$ sec) by weak interactions....into a very wide variety of uncharmed final states ?..are copiously produced only in associated production, such events will necessarily be of very complex topology...Charmed mesons could easily have escaped notice* [23].

In fact, in 1971, K. Niu and collaborators observed kinks in cosmic ray emulsion events, indicating unstable particles with lifetimes of order of $10^{-12}$ to $10^{-13}$ sec [26], values that were in the right ballpark for charmed particles. They were indeed identified, in Japan, with the weakly decaying $p'$ particle of the extended Sakata model. Following the discovery of two neutrino flavours at Brookhaven (Lederman, Schwarz ad Steinberger, 1962), the Sakata model, with elementary constituents the observed baryons ($p$, $n$, $\Lambda$), had been extended to ($p$, $n$, $p'$, $\Lambda$), to restore hadron-lepton symmetry. In the Sakata model translation of the Cabibbo theory, $p'$ was coupled weakly to $n' = -\sin\theta_C\ n + \cos\theta_C\ \Lambda$.). Cosmic rays event, unfortunately, were not paid much attention at the time and the Niu events went essentially unnoticed in western countries.

## 5. The November Revolution

The November Revolution was opened by the simultaneous and independent discovery of a new mesonic particle, at Brookhaven (November 12) and SLAC [27,28] (13 November), with mass 3.098 GeV and width 93 keV, decaying into $e^+e^-$, $\mu^+\mu^-$ and hadrons. The authors named this particle $J$ and $\Psi$ respectively, hence the name $J/\Psi$ adopted since then.

A week later, a Frascati collaboration at the electron-positron collider Adone confirmed the $J/\Psi$ particle [29].

Remarkably, the three papers followed each other so closely that they could appear in the same issue of the Physical Review Letters.

The news arrived in our group in Roma precisely in the day fixed by Nicola Cabibbo for a briefing in Frascati to convince them to search for a narrow width, charm-anticharm vector meson in the Adone energy range(A more detailed account of the events in Roma following $J/\Psi$ discovery are found in [22].).

Already in Harvard we had guessed that this resonance was probably below threshold for decay into a pair of charmed mesons, indeed the $\phi$ meson is almost in this condition with respect to the $K\bar{K}$ pair. However, the $\phi$ example led to guess a $J/\Psi$ width in the order of 1 MeV, 10 times the observed value.

In Roma, we noted that a $Z^0$ of 3 GeV mass coupled with the Fermi constant would decay with about the observed rate. Also, in the form of a confidential rumour, we got from Frascati that the $\mu^+\mu^-$ events could exhibit a backward-forward asymmetry (The initially found asymmetry has very slowly disappeared with the increase of statistics.), a signal of parity violation. All that convinced us to write a quick paper with the proposal of a light $Z^0$ [30].

Unfortunately for us, the Harvard group was light years ahead. Following previous work by Applequist and Politzer [31], De Rujula and Glashow could explain the large difference of $\phi$ and $J/\Psi$ widths as due to the strong dependence of the QCD constant from the difference in momentum scale of the two decays and use the smallness of the width to support the charm-anticharm interpretation of the $J/\Psi$ [32] (see Alvaro De Rujula's talk at this Conference).

In these days, things were moving fast. Our paper was received (by Nuovo Cimento Letters) on November 20, De Rujula and Glashow (by Phys. Rev. Letters) on November 27. Few days later, I was reached in Trieste by a phone call by Nicola Cabibbo with the news of the discovery of $\psi'$ at SLAC (November 25), which definitely supported the hadronic nature of the $J/\Psi$.

In the same days, Greco and Dominguez [33] advanced the proposal of a $c\bar{c}$ resonance, based on an estimate of the rise of the $e^+e^- \to$ hadrons cross section above the $J/\Psi$ mass, based on duality arguments (see Mario Greco's talk at this Conference). This interpretation was questionable due to the presence of problematic $e - \mu$ events. Only in 1977 M. Perl and coll. could prove that these events are due to a new sequential lepton, the $\tau$ lepton [34], whose contribution has to be subtracted to obtain the correct $c\bar{c}$ contribution.

In 1976, the $D^0$ meson, the lightest weakly decaying charmed meson, was discovered by the Mark I detector (SLAC).

In the same year L. Lederman and coll. discovered $\Upsilon = b\bar{b}$ evidence of the 3rd generation together with the $\tau$ lepton.

The observation of the top quark at Fermilab (1994) completed the third generation. Since then we have seen no evidence for any further elementary constituent.

## 6. CP Violation in K Decays

With 4 quarks in 2 doublets (N=2 quark generations with 2N, up and down, lefthanded flavours) it was shown in the GIM paper that the weak coupling matrix U can always be made real [23]. Already commited to a new quark, we did not investigate what would happen with additional quark generations.

In 1973, Kobayashi and Maskawa, explored the possibility of CP violation with more quark generations and discovered that one complex phase is allowed with three quark generations, offering a possible Weak Interaction source of the observed CP violation [35]. In general, for N generations and 2N left-handed fields, one may have a number of irreducible phases given by:

$$N_{\text{phases}} = \frac{(N-1)(N-2)}{2} \tag{7}$$

vanishing in the cases of Cabibbo Theory and GIM. The consequences of KM observation have been studied in 1976 by myself and by Pakvasa and Sugawara, after discovery of the $\tau$ lepton and of the b quark [36,37].

**A few formulae.** CP violation in K decays is well described by the *milliweak interaction* based on the Cabibbo-Kobayash-Maskawa mixing matrix (Closely related parameters $\bar{\rho}$ and $\bar{\eta}$ are frequently used in the literature, with:

$$\rho + i\eta = \frac{(\bar{\rho} + i\bar{\eta})\sqrt{1 - A^2\lambda^4}}{\sqrt{1 - \lambda^2}[1 - A^2\lambda^4(\bar{\rho} + i\bar{\eta})]} = (\bar{\rho} + i\bar{\eta})(1 + \mathcal{O}(\lambda^2)) \tag{8}$$

In the Wolfenstein parametrisation [38], valid to order $(\sin\theta_C)^3$:

$$U_{CKM} = \begin{pmatrix} 1 - \frac{1}{2}\lambda^2 & \lambda & A\lambda^3(\rho - i\eta) \\ -\lambda & 1 - \frac{1}{2}\lambda^2 & A\lambda^2 \\ A\lambda^3[1 - (\rho + i\eta)] & -A\lambda^2 & 1 \end{pmatrix} \tag{9}$$

with:

$$\lambda = \sin\theta_C = 0.2253 \pm 0.0007; A = 0.808^{+0.022}_{-0.015}; \rho = 0.132 \pm 0.018, \ \eta = 0.341 \pm 0.013 \tag{10}$$

CP violating $K^0 \to 2\pi$ decays are determined by two different mechanisms: $K_1 - K_2$ mixing (parameter $\epsilon$) and CP-violation in non-leptonic Hamiltonian (parameter $\epsilon'$). One defines:

$$K_1 = \frac{K^0 - \bar{K}^0}{\sqrt{2}} \ (CP = +1); \ K_2 = \frac{K^0 + \bar{K}^0}{\sqrt{2}} \ (CP = -1) \tag{11}$$

and the mass eigenstates are (CPT symmetry assumed):

$$K_S = N(K_1 + \epsilon K_2), \ K_L = N(K_2 + \epsilon K_1) \tag{12}$$

- $\epsilon$ arises from the $K^0 - \bar{K}^0$ mixing amplitude, which is $2^{nd}$ order in the Weak Interactions, and it could receive a CP-violating contribution from a Superweak Interaction.
- $\epsilon'$ competes with non leptonic interactions of order $G$. In the Superweak theory $\epsilon' = 0$, and $\epsilon'/\epsilon \neq 0$ indicates a milliweak nature of CP violation.

One defines two CP violating parameters

$$\eta_{+-} = \frac{A(K_L \to \pi^+\pi^-)}{A(K_S \to \pi^+\pi^-)}; \ \eta_{+-} = \epsilon + \frac{<K_2|H_{NL}|\pi^+\pi^->}{<K_1|H_{NL}|\pi^+\pi^->} = \epsilon + \epsilon'$$

$$\eta_{00} = \frac{A(K_L \to \pi^0\pi^0)}{A(K_S \to \pi^0\pi^0)}; \ \eta_{00} = \epsilon + \frac{<K_2|H_{NL}|\pi^0\pi^0>}{<K_1|H_{NL}|\pi^0\pi^0>} = \epsilon - 2\epsilon'$$

and the ratio $\epsilon'/\epsilon$ can be obtained experimentally from the ratio

$$R = |\frac{\eta_{00}}{\eta_{+-}}|^2 = 1 - 6 \ \mathcal{R}e(\epsilon'/\epsilon) \tag{13}$$

**CKM and $\epsilon$.** In the Standard Theory $\epsilon$ arises from the familiar box diagrams Figure 3, via $d - t$ and $s - t$ couplings in the KM matrix, expected to be small: a milliweak effect, even with a not-so-small CP violating phase.

**$\epsilon'/\epsilon$I: Theory.** CP violation is in the operator $\mathcal{O}_t$, the $t - \bar{t}$ penguin, Figure 4, which gives an imaginary contribution to the decay amplitude (with respect to the CP conserving amplitude of order $\lambda$):

$$Im(\mathcal{O}_t)/\lambda = (-\eta A^2\lambda^4) \ (\bar{s}_L\gamma_\mu t^A d_L)(\bar{q}_R\gamma^\mu t^A q_R) \tag{14}$$

$$\eta A^2\lambda^4 \sim 0.5 \cdot 10^{-3} \tag{15}$$

leading to a theoretical prediction $\epsilon'/\epsilon = \mathcal{O}(10^{-3})$.

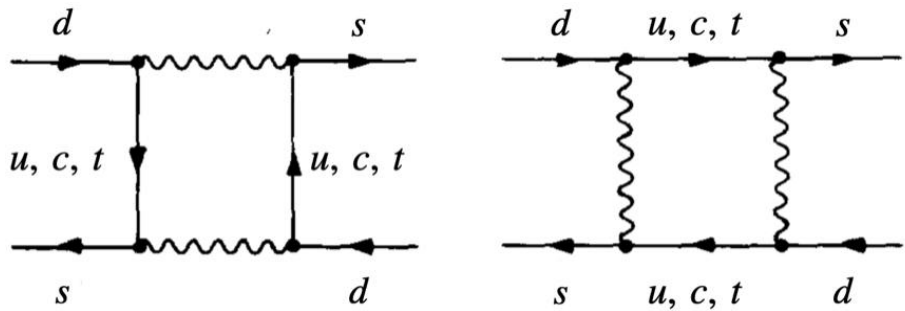

**Figure 3.** GIM mechanism for $K^0 - \bar{K}^0$ mixing.

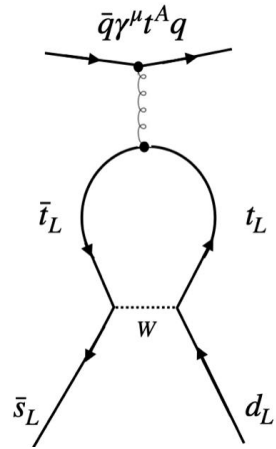

**Figure 4.** Penguin diagram yielding an imaginary contribution to the decay amplitude.

After several controversial results, a non vanishing value of $\epsilon'/\epsilon$ was established in 2001 by the NA48 (CERN) and KTeV (FermiLab) Collaborations, with the observation of the double ratio Equation (13):

$$\epsilon'/\epsilon \cdot 10^3 = 1.53 \pm 0.26 \text{ (NA48)}; \quad 2.07 \pm 0.28 \text{ (KTeV)}$$

$$[\text{PdG}(2024) : 1.66 \pm 0.23] \tag{16}$$

## 7. The Electric Dipole Moment of the Neutron

Milliweak theories of CP violation are in danger to contradict the existing experimental limits to the electric dipole moment (e.d.m.) of the neutron, a T and P violating, $\Delta S = 0$, effect.

In a milliweak theory, we would expect a neutron e.d.m. of the order: $d_e(\mathrm{n}) < 0.18 \; 10^{-25} \; e \cdot \mathrm{cm}$

$$d_e(\mathrm{n}) = e \cdot r_P \cdot (G M_P^2)|\epsilon| \sim 10^{-21} \; e \cdot \mathrm{cm} \tag{17}$$

($r_P$ and $M_P$ the nucleon's radius and mass), vs. the present PdG upper limit: $d_e(\mathrm{n}) < 0.18 \; 10^{-25} \; e \cdot \mathrm{cm}$.

The Standard Theory evades the limit at one loop level, since, e.g., for the d-quark, the correction is determined by the real product $V_{dq}(V_{qd})^*$, which brings the previous estimate to $\sim 10^{-24} \; e \cdot \mathrm{cm}$ [36].

Further investigations [39] led to further reductions of the estimate and later to the conclusion that the two loop contribution to $d_e(\mathrm{n})$ vanishes also to two loop order [40], bringing the estimate below $\sim 10^{-30} \; e \cdot \mathrm{cm}$.

## 8. CP Violation in B Decays

- 1986. I. Bigi and A. Sanda predict direct CP violation in B decay [41].
- 2001, Belle and BaBar discover CP violating mixing effects in B-decays.
- New impetus in theory calculations of $V_{ub}$, $V_{cb}$, $f_B$, ... from particle phenomenology and Lattice QCD.

We report in Figure 5 examples of lattice QCD determination of the CKM parameter $V_{ub}$ and $V_{cb}$ by the $UT_{fit}$ Lattice Collaboration [42] from electroweak effects, including CP violation in K and B decays.

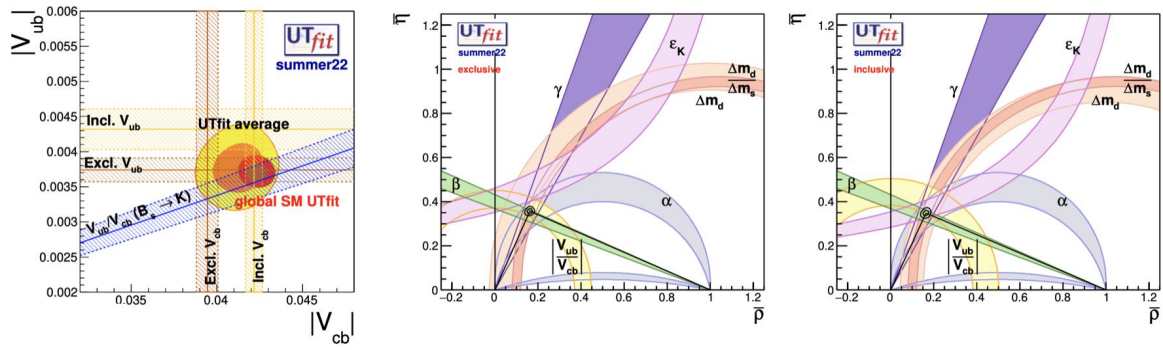

**Figure 5.** Examples of lattice QCD determination of the CKM parameter $V_{ub}$, $V_{cb}$.

## 9. The Special Role of Massive Quarks in QCD

- QCD is asymptotically free. Quarks carry color, associated to $SU(3)_{col}$ and flavour, associated to $S(3)_{flavour}$, and are confined inside color singlet hadrons. The momentum scale which marks the transition from non-perturbative to perturbative QCD is called $\Lambda_{QCD}$. Numerically: $\Lambda_{QCD} \simeq 250$ MeV.

- The decrease of the colour coupling $\alpha_S(q^2)$ has been observed at LEP and LHC and is consistent with the value $\alpha_S(q^2 = M_Z^2) = 0.1185$.

- Quarks $u$ and $d$ have small masses, $m_q \simeq \Lambda_{QCD}$, and fall in the color strong interaction regime. The strange quark is marginal, but from charm onward we are in the heavy quark region ($m_Q >> \Lambda_{QCD}$). Inclusive semileptonic decays are calculable like deep inelastic processes.

- Bound states involve short distance forces, implying a calculable spectrum of charmonia and bottomonia, as first observed by Politzer and Appelquist [31].

- Heavy quark $c\bar{c}$ or $b\bar{b}$ pairs inside hadrons are not easily created or destroyed. A hadron decaying into $J/\Psi$ or $\Upsilon$+ light hadrons, most likely contains a valence $c\bar{c}$ or $b\bar{b}$ pair: heavy-quark counting is possible.

Parton model and perturbative QCD have been used in the years 1980s to compute inclusive semileptonic widths and the energy spectra of the charged lepton in charmed and beauty quarks, see e.g., [43–46] and have provided a valuable method to determine the CKM couplings $V_{ub}$, $V_{cb}$ from inclusive semileptonic decays of charm and beauty.

The special role of heavy quarks as indicators of multiquark structures, mentioned in the last bullet, was recognised in the years 2000, in connection with the observation of so-called *unanticipated charmonia* discussed in the next Section.

## 10. Unanticipated Charmonia: X, Y, Z and More

Unanticipated, hidden charm/beauty resonances not fitting in predicted charmonium/bottomonium spectra have been observed, classified initially as **X**, **Y** and **Z** particles.

- X, e.g., $X(3872)$ (BELLE, BaBar, 2003): neutral, typically seen in $\Psi + 2\pi$, positive parity: $J^{PC} = 0^{++}$, $1^{++}$, $2^{++}$.

- Y, e.g., $Y(4260)$ (BaBar, 2005): neutral, seen in $e^+e^-$ annihilation with Initial State Radiation (ISR): $e^+e^- \to e^+e^- + \gamma_{ISR} \to Y + \gamma_{ISR}$, therefore $J^{PC} = 1^{--}$.

- Z, e.g., $Z(4430)$ (BELLE, 2007; confirmed by LHCB, 2014): typically $J^{PC} = 1^{+-}$, charged or neutral; mostly seen to decay in $\Psi + \pi$ ($Z(3900)$), (BESIII 2013), and $h_c(1P) + \pi$ ($Z(4020)$), (BESIII, 2013). 4 valence quarks manifest in the charged $Z$: ($c\bar{c}u\bar{d}$). $Z_b$ observed ($b\bar{b}u\bar{d}$).

Figure 6 reports, in black, a recent determination of the spectrum of charmonia, Ref. [47,48], using the Cornell potential [49–51], see also [52]. In red, the lowest lying unexpected charmonia.

Unexpected, electrically neutral states differ from charmonia also by their decay modes, e.g., $X(3872) \to J/\Psi + \rho^0/\omega^0$ with a substantial violation of isospin symmetry, not expected for a pure $c\bar{c}$ bound state.

We know by now about one hundred meson resonances that contain two quark pairs, tetraquarks ($c\bar{c}q\bar{q}$ or $cc\bar{q}\bar{q}$), and a few baryons with ($c\bar{c}qqq$) composition, pentaquarks: an entire family of new states is showing up.

Multiquark hidden charm hadrons, in hadron colliders, originate mostly from the decays of mesons and baryons containing b-quark, via the weak decay $b \to c + (\bar{c}s)$. For $B^+$ decay see Figure 7, taken from [53]. A similar diagram for $\Lambda_b$ gives rise to pentaquark production: $\Lambda_b \to K + \mathcal{P} \to K + J/\Psi + p$.

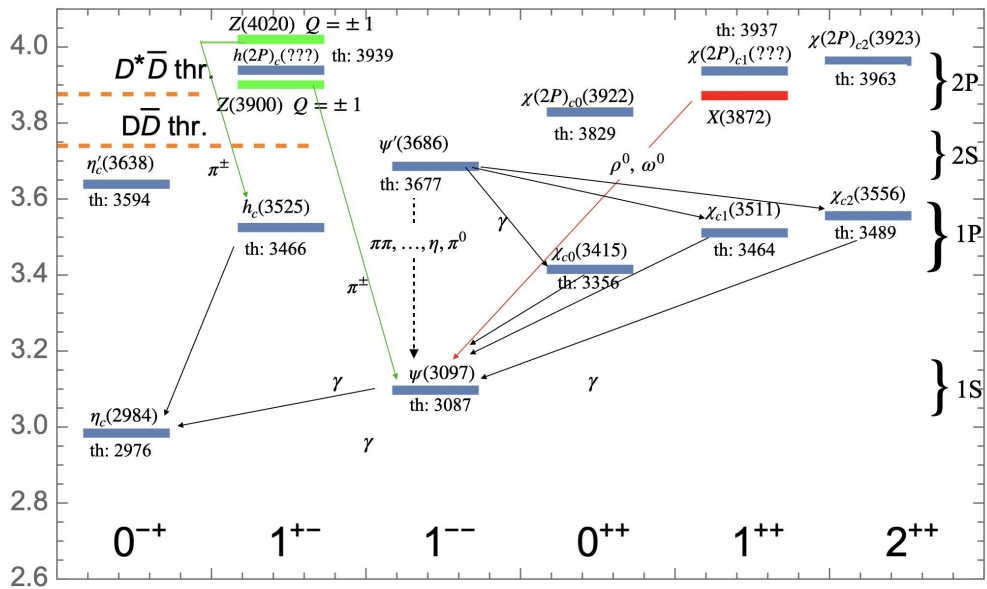

**Figure 6.** Predicted and observed charmonia, $S1,2$ and $P1,2$ states in (black). In red the first discovered unanticipated charmonia. Figure from Ref. [47] .

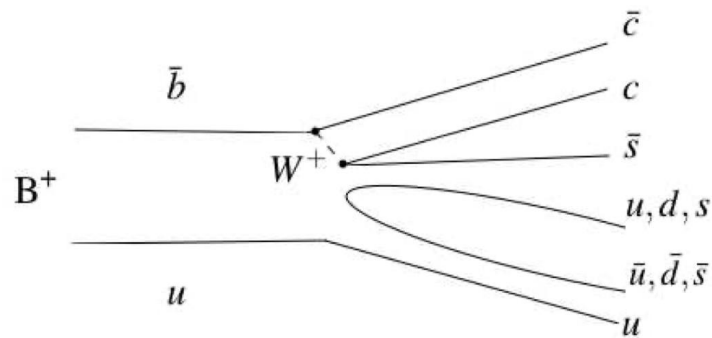

**Figure 7.** Quark diagram for $B^+ \to K^+ + X$, with $X = (c\bar{c}q\bar{q}')$. Figure from [53].

The challenge, after X, Y and Z particle discovery, is to reconcile their structure with what we know about the binding of the classical mesons ($q\bar{q}$) and baryons ($qqq$) by QCD interactions.

There is no consensus yet. A few alternatives still under discussion can be described as follows, see also [54].

- The first guess advanced for $X(3872)$ was that of a *hadron molecule*, pioneered by E. Braaten [55]: a $(D\bar{D}^* +$ C-conjugate) meson pair bound by the same nuclear forces that bind nucleons in atomic nuclei. The rationale was the closeness of $X$ mass to the $D\bar{D}^*$ threshold, reminding the deuteron $pn$ bound state, see [56] for a more recent review. In the first papers, pion exchange was considered to provide the binding force, but this is incompatible with exotic mesons such as $J/\Psi - \phi$ or $Z_{cs}$, which cannot be bound by pion exchange. A recently advanced hypothesis [57] to connect exotic hadrons to the known mesons is that *contact interactions*, described by a chiral Effective Field Theory, produce exotic hadrons as real or virtual poles in the scattering amplitude of meson pairs.

- A different scheme is provided by the *compact diquark-antidiquark* picture proposed in 2005 and further specified in 2014 [58,59]. It describes the exotic hadrons as multiquark states bound by QCD forces, in addition to but independent from $q\bar{q}$ mesons or $qqq$ baryons. Closeness to meson pair thresholds of the lowest lying exotic hadron masses would be the obvious consequence of the fact that these exotic hadrons are made of the same quarks as Gell-Mann Zweig mesons and baryons.

- Models based on pure QCD interactions and specific to hidden charm or hidden beauty exotic hadron, called Hadrocharmonia, have been considered by Voloshin and coll [60] and by Braaten and coll [61], in which a color octet, heavy quark pair is formed by QCD forces, with color being shielded by a cloud of light quarks and antiquarks. They are dominated by QCD interactions, and are, in fact, simple variations of the last mentioned scheme.

Exotic $SU(3)$ flavour multiplets, with a characteristic scale of symmetry breaking is a distinctive prediction of compact tetraquarks. The newly found $J - \phi$ and $J - K$ exotics, Figure 8, fit into nonets with $X(3872)$, $Z_c(3900)$, $Z_c(4020)$.

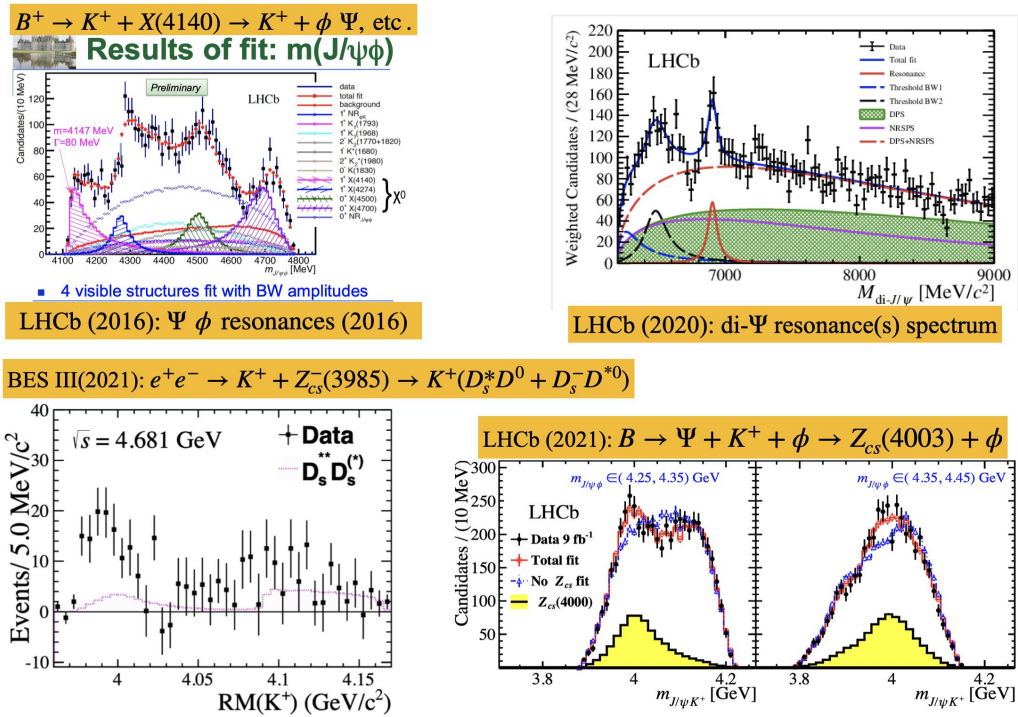

**Figure 8.** The new wave of multiquark states discovered by LHCb and BES III, 2016-2021. All can be described as multiquark states bound by QCD forces, $J/\Psi - \phi$ and $J/\Psi K$ states fit into SU(3) flavour nonets with $X(3872)$, $Z_C(3900)$ and $Z_c(4020)$.

Much remains to be done, to produce more precise data and to search for still missing particles, to complete the flavour multiples required by QCD bound, multiquark Exotics.

Many questions still remain unanswered. Among the missing particles:

- $X(3872)^+$: the I=1 partner of X(3872), with decays into: $X^+ \to J/\psi \, \rho^\pm \to J/\psi \, \pi^+\pi^0$. $X^+$ is expected to be produced in B non leptonic decays within the bounds [62]

$$0.057 < \frac{\Gamma(B^0 \to K^+ X^- \to K^+ \psi \, \pi^0\pi^-)}{\Gamma(B^0 \to K^0 X(3872) \to K^0\psi \, \pi^+\pi^-)} < 0.50 \qquad (18)$$

- Is $X(3872)$ split into two lines?
- Double charm or beauty spectrum to be explored: can we find $\mathcal{T}_{cc}^{++}(?)$, $\mathcal{T}_{bb}^-$ ?;
- For several missing states, we have predicted mass and decay modes, see e.g., [63];
- Are Exotic hadrons produced in hadron collisions at large $p_T$, besides $X(3872)$?.

These questions will require more luminosity, better energy definition, detectors with exceptional qualities... a lot of work... and will keep us busy for quite a while.

**Conflicts of Interest**

The author declares no conflict of interest.

**References**

1. Gell-Mann, M. A Schematic Model of Baryons and Mesons. *Phys. Lett.* **1964**, *8*, 214–215.
2. Zweig, G. *An Su(3) Model For Strong Interaction Symmetry And Its Breaking*; CERN-TH-412; CERN: Genève, Switzerland, 1964.
3. Cabibbo, N. Unitary Symmetry and Leptonic Decays. *Phys. Rev. Lett.* **1963**, *10*, 531.
4. Sudarshan, E.C.G.; Marshak, R.E. Chirality Invariance and the Universal Fermi Interaction. *Phys. Rev.* **1958**, *109*, 1860.
5. Feynmann, R.P.; Gell-Mann, M. Theory of the Fermi Interaction. *Phys. Rev.* **1958**, *109*, 193.
6. Gerstein, S.S.; Zeldovich, Y.B. *ZhETE* 1955, 29, 698. [JETP **2**, 576 (1956)].

7. Han, M.Y.; Nambu, Y. Three-Triplet Model with Double SU(3) Symmetry. *Phys. Rev.* **1965**, *139*, B1006.

8. Veneziano, G. Construction of a crossing - symmetric, Regge behaved amplitude for linearly rising trajectories. *Nuovo Cim. A* **1968**, *57*, 190.

9. Schwinger, J. A Theory of the Fundamental Interactions. *Ann. Phys.* **1958**, *2*, 407.

10. Glashow, S.L. Partial Symmetries of Weak Interactions. *Nucl. Phys.* **1961**, *22*, 579–588.

11. Englert, F.; Brout, R. Broken Symmetry and the Mass of Gauge Vector Mesons. *Phys. Rev. Lett.* **1964**, *13*, 321.

12. Higgs, P.W. Spontaneous Symmetry Breakdown without Massless Bosons. *Phys. Rev.* **1966**, *145*, 1156.

13. Weinberg, S. A Model of Leptons. *Phys. Rev. Lett.* **1967**, *19*, 1264–1266.

14. Salam, A. *Elementary Particle Theory*; Svartholm, N., Ed.; Almquist and Wiksell: Stockholm, Sweden, 1968; pp. 367–377.

15. Ioffe, B.L.; Shabalin, E.P. Neutral Currents and the Applicability Limit of the Theory of Weak Interactions. *Yad. Fiz.* **1967**, *6*, 828.

16. Mohapatra, R.N.; Rao, J.S.; Marshak, R.E. Second-Order Weak Processes and Weak-Interaction Cutoff. *Phys. Rev.* **1968**, *171*, 1502.

17. Low, F.E. Difficulties of the Theory of Weak Interactions. *Comments Nucl. Part. Phys.* **1968**, *2*, 33.

18. Gell-Mann, M.; Goldberger, M.L.; Kroll, N.M.; et al. Amelioration of Divergence Difficulties in the Theory of Weak Interactions. *Phys. Rev.* **1969**, *179*, 1518.

19. Lee, T.D.; Wick, G.C. Negative Metric and the Unitarity of the S Matrix. *Nucl. Phys. B* **1969**, *9*, 209–243

20. Gatto, R.; Sartori, G.; Tonin, M. Weak Selfmasses, Cabibbo Angle, and Broken SU(2) ⊗ SU(2). *Phys. Lett. B* **1968**, *28*, 128–130.

21. Cabibbo, N.; Maiani, L. Origin of the Weak-Interaction Angle. II. *Phys. Rev. D* **1970**, *1*, 707–718.

22. Maiani, L.; Bonolis, L. The Charm of Theoretical Physics (1958–1993). *Eur. Phys. J. H* **2017**, *42*, 611–661.

23. Glashow, S.L.; Iliopoulos, J.; Maiani, L. Weak Interactions with Lepton-Hadron Symmetry. *Phys. Rev. D* **1970**, *2*, 1285.

24. Gaillard, M.K.; Lee, B.W. Rare Decay Modes of the K-Mesons in Gauge Theories. *Phys. Rev. D* **1974**, *10*, 897.

25. Bouchiat, C.; Iliopoulos, J.; Meyer, P. An Anomaly Free Version of Weinberg's Model. *Phys. Lett. B* **1972**, *38*, 519–523

26. Niu, K.; Mikumo, E.; Maeda, Y. A possible decay in flight of a new type particle. *Prog. Theor. Phys.* **1971**, *46*, 1644–1646

27. Aubert, J.J.; Becker, U.; Biggs, P.J.; et al. Experimental Observation of a Heavy Particle *J*. *Phys. Rev. Lett.* **1974**, *33*, 1404–1406.

28. Augustin, J.E.; Boyarski, A.M.; Breidenbach, M.; et al. Discovery of a Narrow Resonance in $e^+e^-$ Annihilation. *Phys. Rev. Lett.* **1974**, *33*, 1406–1408.

29. Bacci, C.; Celio, R.B.; Berna-Rodini, M.; et al. Preliminary Result of Frascati (ADONE) on the Nature of a New 3.1-GeV Particle Produced in $e^+e^-$ Annihilation. *Phys. Rev. Lett.* **1974**, *33*, 1408. Erratum in *Phys. Rev. Lett.* **1974**, *33*, 1649.

30. Altarelli, G.; Cabibbo, N.; Petronzio, R.; et al. Is the 3104-MeV Vector Meson the $\phi_c$ or the $W_0$? *Lett. Nuovo Cim.* **1974**, *11*, 609.

31. Appelquist, T.; Politzer, H.D. Orthocharmonium and $e^+e^-$ Annihilation. *Phys. Rev. Lett.* **1975**, *34*, 43.

32. De Rujula, A.; Glashow, S.L. Is Bound Charm Found? *Phys. Rev. Lett.* **1975**, *4*, 46–49.

33. Dominguez, C.A.; Greco, M. Charm, Evdm and Narrow Resonances in $e^+e^-$ Annihilation. *Lett. Nuovo Cim.* **1975**, *12*, 439.

34. Perl, M.L. The Discovery of the Tau Lepton. SLAC-PUB-5937. In Proceedings of the Third International Symposium on the History of Particle Physics: The Rise of the Standard Model, Stanford, CA, USA, 24–27 June 1992.

35. Kobayashi, M.; Maskawa, T. *CP*-Violation in the Renormalizable Theory of Weak Interaction. *Prog. Theor. Phys.* **1973**, *49*, 652.

36. Maiani, L. *CP* violation in purely lefthanded weak interactions. *Phys. Lett. B* **1976**, *62*, 183.

37. Pakvasa, S.; Sugawara, H. *CP* violation in the six-quark model. *Phys. Rev. D* **1976**, *14*, 305.

38. Wolfenstein, L. Parametrization of the Kobayashi-Maskawa Matrix. *Phys. Rev. Lett.* **1983**, *51*, 1945.

39. Ellis, J.; Gaillard, M.K.; Nanopoulos, D.V. Lefthanded Currents and CP Violation. *Nucl. Phys. B* **1976**, 109, 213–243.

40. Shabalin, E.P. The Electric Dipole Moment of The Neutron in a Gauge Theory. *Sov. Phys. Usp.* **1983**, *26*, 297.

41. Bigi, I.I.; Sanda, A.I. *CP* Violation in Heavy Flavor Decays: Predictions and Search Strategies. *Nucl. Phys. B* **1987**, *281*, 41–71.

42. Bona, M.; Ciuchini, M.; Derkach, D.; et al. New UTfit Analysis of the Unitarity Triangle in the Cabibbo-Kobayashi-Maskawa scheme. *Rend. Lincei Sci. Fis. Nat.* **2023**, *34*, 37–57.

43. Cabibbo, N.; Corbo, G.; Maiani, L. Lepton Spectrum in Semileptonic Charm Decay. *Nucl. Phys. B* **1979**, *155*, 93.

44. Altarelli, G.; Cabibbo, N.; Corbo, G.; et al. Leptonic Decay of Heavy Flavors: A Theoretical Update. *Nucl. Phys. B* **1982**, *208*, 365.

45. Grinstein, B.; Wise, M.B.; Isgur, N. Weak Mixing Angles from Semileptonic Decays Using the Quark Model. *Phys. Rev. Lett.* **1986**, *56*, 298.

46. Altomari, T.; Wolfenstein, L. Comment on Weak Mixing Angles from Semileptonic Decays in the Quark Model. *Phys. Rev. Lett.* **1987**, *58*, 1583.

47. Maiani, L.; Benhar, O. *Relativistic Quantum Mechanics: An Introduction to Relativistic Quantum Fields*; CRC Press: Boca Raton, FL, USA, 2016,

48. Soni, N.R.; Joshi, B.R.; Shah, R.P.; et al. $Q\bar{Q}$ ( $Q \in \{b,c\}$ ) spectroscopy using the Cornell potential. *Eur. Phys. J. C* **2018**, *78*, 592.

49. Eichten,E.; Gottfried, K.; Kinoshita, T.; et al. The Spectrum of Charmonium. *Phys. Rev. Lett.* **1975**, *34*, 369; Erratum in *Phys. Rev. Lett.* **1976**, *36*, 1276.

50. Eichten, E.; Gottfried, K.; Kinoshita, T.; et al. Charmonium: The Model. *Phys. Rev. D* **1978**, *17*, 3090; Erratum in *Phys. Rev. D* **1980**, *21*, 313.

51. Eichten, E.; Gottfried, K.; Kinoshita, T.; et al. Charmonium: Comparison with Experiment. *Phys. Rev. D* **1980**, *21*, 203.

52. Brambilla, N.; Krämer, M.; Mussa, R.; et al. Heavy Quarkonium Physics. *arXiv* **2004**, arXiv:hep-ph/0412158.

53. Bigi, I.; Maiani, L.; Piccinini, F.; et al. Four-quark mesons in non-leptonic $B$ decays: Could they resolve some old puzzles? *Phys. Rev. D* **2005**, *72*, 114016.

54. Ali, A.; Maiani, L.; Polosa, A.D. *Multiquark Hadrons*; Cambridge University Press: Cambridge, UK, 2019.

55. Braaten, E.; Kusunoki, M. Low-energy universality and the new charmonium resonance at 3870-MeV. *Phys. Rev. D* **2004**, *69*, 074005.

56. Guo, F.K.; Hanhart, C.; Meißner, U.G.; et al. Hadronic molecules. *Rev. Mod. Phys.* **2018**, *90*, 015004; Erratum in *Rev. Mod. Phys.* **2022**, *94*, 029901.

57. Zhang, Z.H.; Ji, T.; Dong, X.K.; et al. Predicting isovector charmonium-like states from X(3872) properties. *arXiv* **2024**, arXiv:2404.11215.

58. Maiani, L.; Piccinini, F.; Polosa, A.D.; et al. Diquark-antidiquarks with hidden or open charm and the nature of X(3872). *Phys. Rev. D* **2005**, *71*, 014028,

59. Maiani, L.; Piccinini, F.; Polosa, A.D.; et al. The Z(4430) and a New Paradigm for Spin Interactions in Tetraquarks *Phys. Rev. D* **2014**, *89*, 114010,

60. Dubynskiy, S.; Voloshin, M.B. Hadro-Charmonium. *Phys. Lett. B* **2008**, *666*, 344.

61. Braaten, E.; Langmack, C.; Smith, D.H. Selection Rules for Hadronic Transitions of XYZ Mesons. *Phys. Rev. Lett.* **2014**, *112*, 222001.

62. Maiani, L.; Polosa, A.D.; Riquer, V. $X(3872)$ tetraquarks in $B$ and $B_s$ decays. *Phys. Rev. D* **2020**, *102*, 034017.

63. Maiani, L.; Polosa, A.D.; Riquer, V. Open charm tetraquarks in broken SU(3)F symmetry. *Phys. Rev. D* **2024**, *110*, 034014.

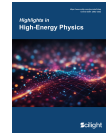

*Review*

# When the Standard Model Was Ignored

Alvaro De Rújula [1,2]

[1] Instituto de Física Teórica (UAM/CSIC), Universidad Autónoma de Madrid, 28049 Madrid, Spain
[2] Theory Division, CERN, CH 1211 Geneva, Switzerland; Alvaro.DeRujula@cern.ch

**Abstract:** It has become fashionable to celebrate anniversaries of, among other subjects, crucial steps in science. The Standard Model (SM) of particles and forces is not an exception. I shall emphasize the fact that "the community" was not always aware of the obvious: that the SM is a convincing contender for being part of "the truth"... and always was. In this writeup many figures reproduce good old times' transparencies, a touch of nostalgia.

**Keywords:** asymptotic freedom; November Revolution; charmonium spectroscopy

## 1. A Slow Start

The first "unified" gauge theory was Glashow's $SU(2) \otimes U(1)$ 1961 model of the weak and electromagnetic interactions. It became a theory with Weinberg's 1967 paper "A model of leptons", followed by Salam's 1968 conference proceeding on the same subject. Use the notation $\{\text{Year}, \text{Number of references}\}$ to recall the early success of these works: $\{67, 0\}$, $\{68, 0\}$, $\{69, 0\}$, $\{70, 1\}$, $\{71, 4\}$. The abrupt 1971 rise was due to the publication of GIM (It is unnecessary to give a GIM reference, "everybody" knows what the GIM mechanism is. The same for $SU(2) \otimes U(1)$. Citations in this paper will be sieved, but not systematic.) and its two correct guesses (that charmed quarks ought to exist and that $SU(2) \otimes U(1)$ might be renormalizable).

In a couple of extra years one can already see the start of an exponential trend: $\{72, 64\}$, $\{73, 162\}$. What had happened? As Sidney Coleman put it: *Gerard's kiss transmogrified Steve's frog into an enchanted prince*, that is 't Hooft published his proof that anomaly-free non-abelian gauge theories are renormalizable. Since Gerard did not attend the September 2024 Rome conference on *The Rise of Particle Physics* I refer to his recollection of the subject [1].

## 2. Time for Surprises

### 2.1. Discovery of Asymptotic Freedom

Renormalizable gauge theories have couplings whose strength increases at short distances. That had ceased to be true in the spring of 1973, when Gross and Wilczek (then at Princeton) and Politzer (then at Harvard) proved that non-abelian gauge theories, such as QCD, were asymptotically free [2,3]. Characteristically the papers reached PRL one week apart. But there were various precedents, the earliest one in 1970, by Khriplovich [4].

The first two phenomenological QCD papers, on the other hand, had no precedents. This time Princeton [5] was one day ahead of Harvard [6]. Both articles disregarded a (temporary [7]) lack of QCD-understanding of "Bloom-Gilman" duality and used data on the $q^2$-dependence of the magnetic form factor of the proton to study $\alpha_s$, the strong fine-structure "constant". Only in [6] the theoretical analysis and the data did agree well enough for $\Lambda^2$ to be determined: it lies between 0.2 and 1.0 GeV$^2$, see Figure 1. Breaking with the tradition of Newton and Fermi, the constant $\Lambda$ was thereafter called $\Lambda_{QCD}$ and not $\Lambda_{ADR}$. In Figure 2 this leading twist and leading log result is compared with next and next-to-next order ones, differing, as expected, by about 30%.

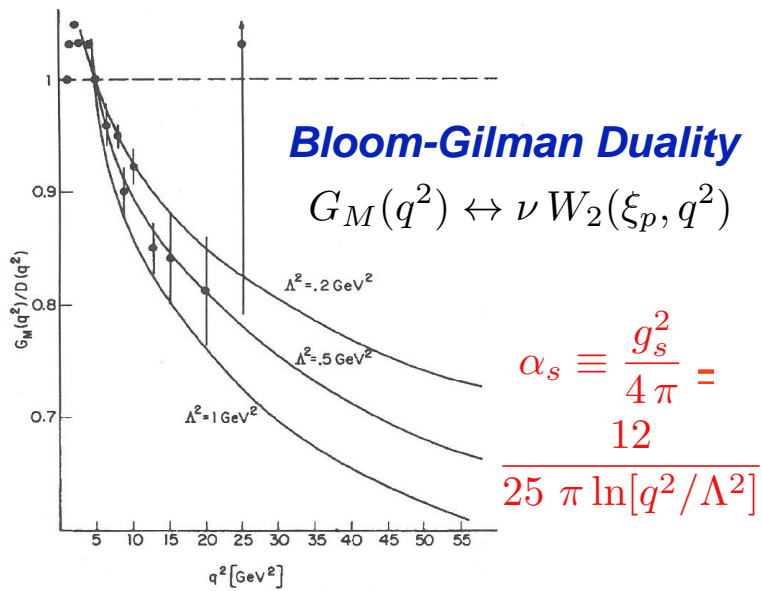

**Figure 1.** Data on the proton's magnetic form factor (normalized to a specific dipole approximation) and a QCD fit [6].

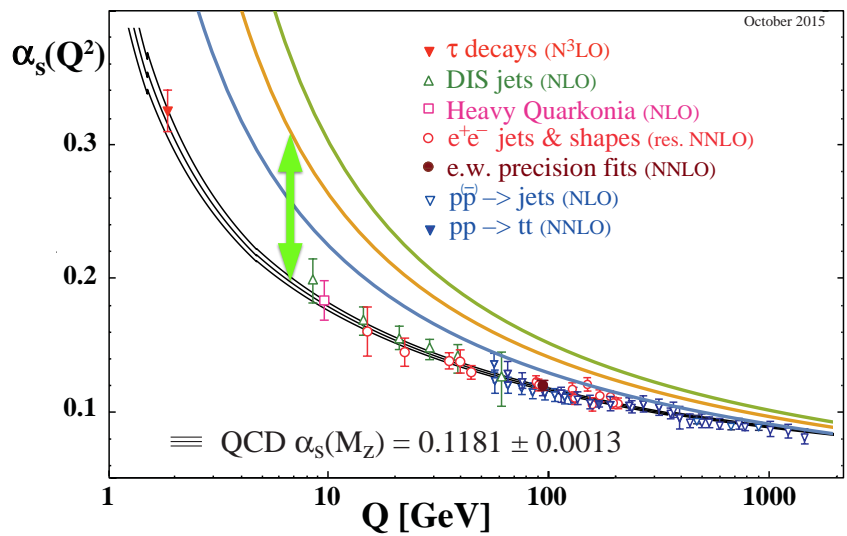

**Figure 2.** The colored lines are $\alpha_s(Q^2)$ with uncertainty estimates [6]. The rest are later results.

### 2.2. The November Revolution

A revolution took place in 1974, see Figure 3. The discovery of the $\Psi$ was announced one day after the one of the J (pronounced *Ting*, as in *Ting/Psi*). In the upper image from Augustin et al. a point is highlighted in the figure. It was this peculiarly high point, the chronicles of the time say, which allowed Richter and collaborators to catch the devil by the tail. The observed resonance has a long higher-energy tail due to initial-state radiation: $e^+ e^- \to \Psi + \gamma$ with the $\gamma$ escaping detection. To hit gold (as in a medal) it sufficed to move down in collider energy from the observed tail-point.

The symbols "# $\sigma's$?" in Figure 3 refer to the fact that nobody at the time insisted on quoting a number of standard deviations. Perhaps, as Val Telegdi used to put it, *If you need statistics to argue that you have made a discovery... perhaps you haven't*.

The J/$\Psi$ was an unprecedented hadron, as dramatized in Figure 4. Well above the vertical scale of lifetimes one finds electrons, photons, neutrinos and protons, all of them—but one species—for very good reasons. From $\sim 10^{-10}$s to $\sim 10^{-5}$s one finds particles that decay weakly. Below that, a few electromagnetically decaying ones. And at the busy horizontal band the hadronic resonances known in 1974. The J/$\Psi$ is in the middle of nowhere. The black line would be the lifetime of the $SU(2) \otimes U(1)$ $Z$, had it been that incredibly light. This misled a collection of my Italian friends [8] who proposed that interpretation, by far the least wrong of the wrong ones immediately published, not cited here out of respect for many eminent and far-out-of-the-box thinkers.

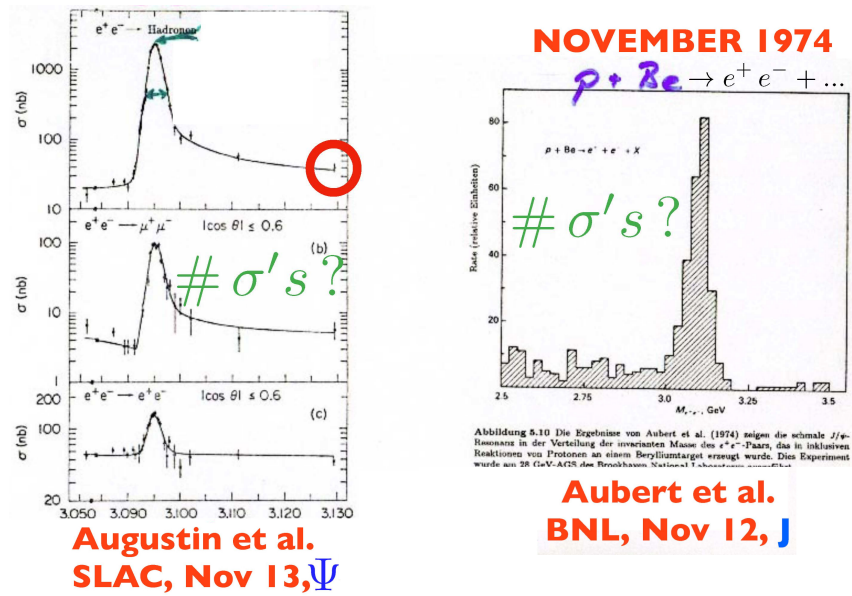

**Figure 3.** The data on the discovery of the J and the Ψ.

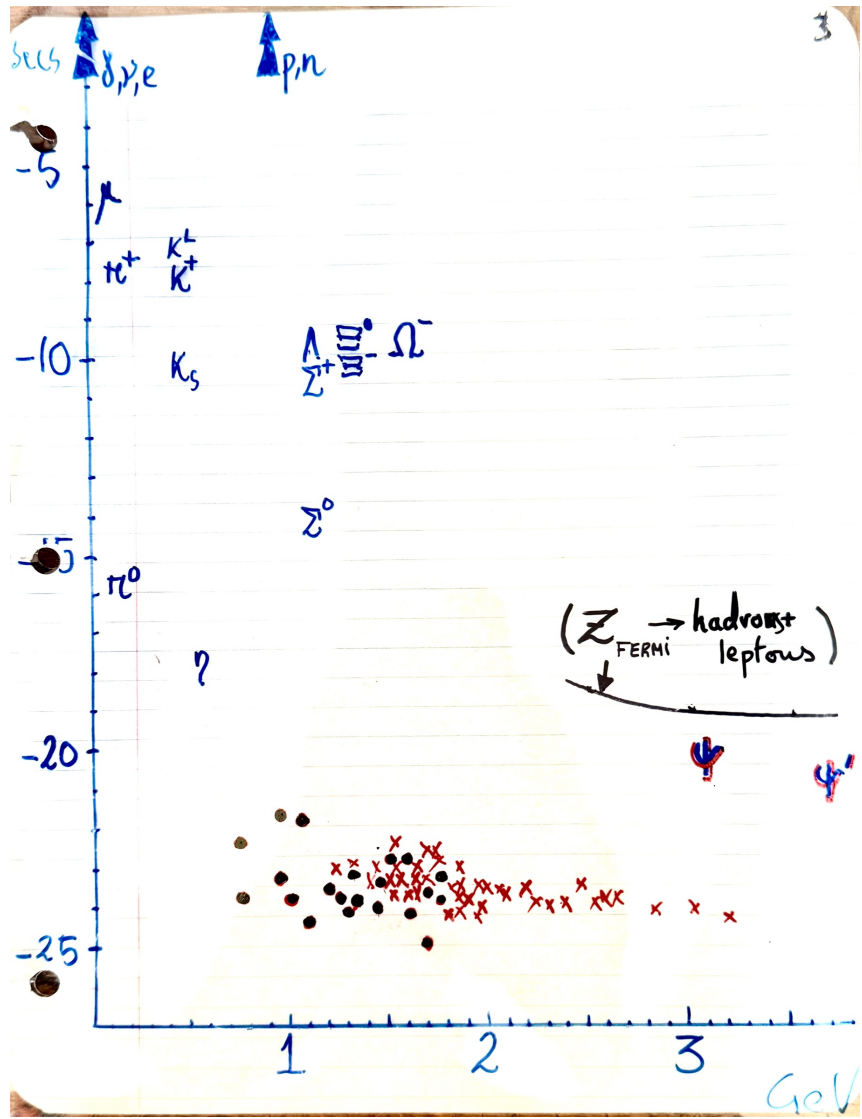

**Figure 4.** Masses and lifetimes of the particles known in 1974.

At Harvard the narrow width of the J/Ψ was not a surprise. Appelquist and Politzer, before the discovery, were already thinking about Charmonium, but they were hesitant because they could not believe their results, then based on a too Coulomb-like potential. But they knew [9], as Glashow and I also figured out a couple of offices away, that this composite vector boson decayed mainly into three gluons [10]. *Abusus not tollit usum,* Glashow and I compared the 3 gluon decay of the J/Ψ to the analogous one of the φ to pions, relating them by evolving $α_s$ as in Figure 2. Lo and behold our predicted hadronic width of the J/Ψ is the currently measured one (We also correctly predicted the mass of the $D^*$ and the existence of the Ψ', discovered before our paper was published.).

Two crucial ingredients in the understanding of charmonium are charmed quarks (As it is unknown, quarks were invented by André Petermann, see Who invented quarks?) and color. But, who invented quark color? This may constitute another surprise, west of the Iron Curtain. *Three identical quarks cannot form an antisymmetric S-state. In order to realize an antisymmetric orbital S-state, it is necessary for the quark to have an additional quantum number,* wrote Boris Struminsky (Greenberg and Han and Nambu are often credited in the West, but their colors are not the currently accepted ones.) in 1965 [11,12].

## 3. Fun Times

For the believers in the then non-standard Standard model the mid 70s were a great time to work. Competitors were very limited in number. The ideas to develop were occasionally very simple. One example: the charmonium Harvard crowd gathered for an entire night at my home—with my guest Éduard Brézin cooking delicious crêpes—to write a paper on predicting the realm of Charmonium Spectroscopy [13]. Our friends from Cornell wrote an article on the same subject [14]. It was more scholarly than ours, they had borrowed from Ken Wilson a program to implement the "Cornell potential".

The predictions are shown in Figure 5. After a couple of years finding limits below the predicted ones on the atom-like radiative decays, the experimentalists at DESY observed a line. The SLAC group found several transitions and announced their results a couple of weeks later, not quoting the DESY ones. This time a referee obliged them to give proper reference. Long live charmonium.

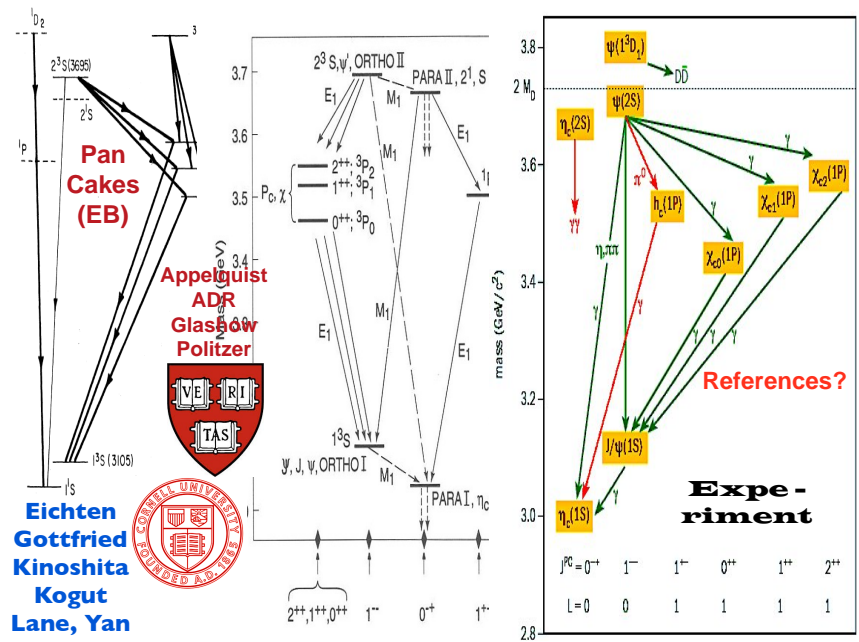

**Figure 5.** Charmonium spectroscopy. The data providers (right) did not refer to theory.

The SLAC data on the ratio R are shown in Figure 6. In a paper whose title *Finding fancy flavors counting colored quarks* was duly censored by PRL, Georgi and I made a "standard" analysis of the data. We found its best interpolation to $σ(e^+e^- \rightarrow \text{quarks} + \text{leptons})$ ("dual" to the resonances), including the charmed quark (and perhaps others) and a heavy lepton, all of them then hypothetical. We extended this analytical function to the spacelike domain, and compared it there with the perturbative-QCD predictions. We concluded that *the "old" theory with no charm is excluded, the standard model with charm is acceptable if heavy leptons are produced and six quark models are viable if no heavy leptons are produced* [15].

Times change. If a similarly novel analysis with a solid theoretical basis and statistical significance aplenty was made today on, say, LHC data, "the community" would conclude that a crucial discovery had been made. At the time only Martin Perl, inching his way towards discovering the $\tau$, thanked us for the encouragement . Our work was thereafter totally forgotten (except at ITEP [16]).

In 1975 Glashow, Georgi and I wrote a paper [17] starting with the words *Once upon a time,* to refer to pre-standard model physics as surpassed. We introduced a QCD-improved constituent quark model with "hyperfine" mass splittings due to one-gluon exchange. That provided a successful description of all S-wave hadrons made of $u$, $d$ and $s$ quarks, explaining, for instance, the $\Sigma^0/\Lambda$ mass difference, large though both particles have the same spin and $uds$ constituency. With only one extra parameter (the charmed quark mass implied by the $c\bar{c}$ states) we predicted, correctly, the masses of all S-wave singly charmed mesons and baryons. Much later lattice gauge theories caught up with us, see Figure 7.

Still in 1975 Samios and collaborators [18, 19]took an impressive bubble chamber photo of the process $\nu_\mu + p \rightarrow \mu^- \Sigma_c^{++}, \Sigma_c^{++} \rightarrow \pi^+ \Lambda_c^+, \Lambda_c^+ \rightarrow \pi^+ \pi^+ \pi^- \Lambda$. The masses of $\Sigma_c^{++}$ and $\Lambda_c^+$ were precisely the ones we predicted. For once, the experimentalist said so.

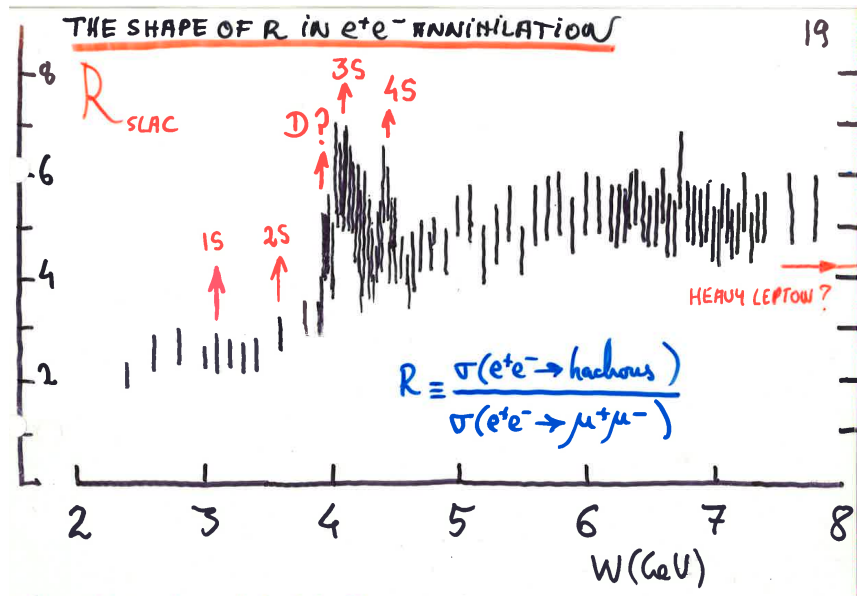

**Figure 6.** The well-know ratio R, measured at SLAC.

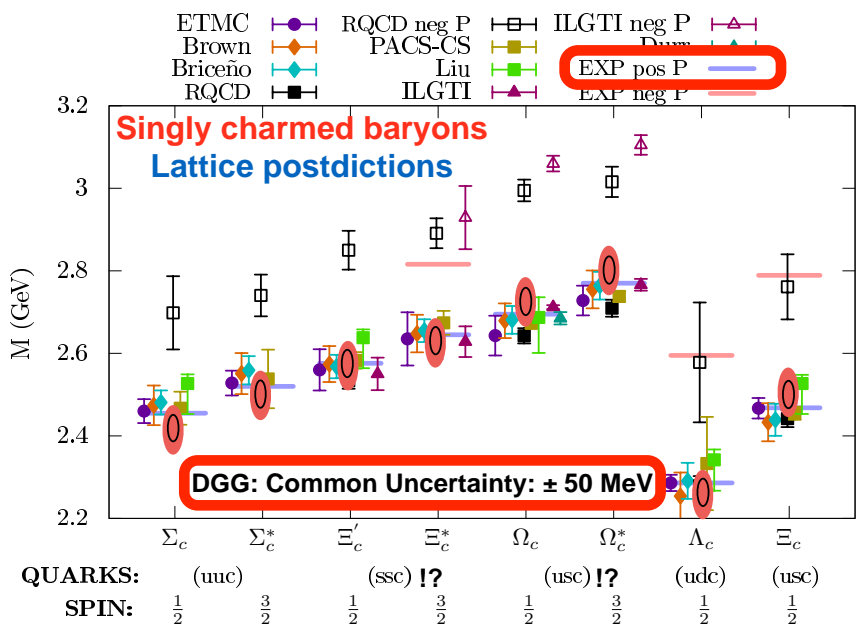

**Figure 7.** Masses of the singly charmed positive parity baryons compared to the data, the predictions in [17] and a collection lattice results.

## 4. Becoming Standard?

Given the many successes of the Standard Model one would have thought that by 1976 it would have been generally accepted. Not so. In July I gave a talk to an extremely well attended conference in Tbilisi [20]. I could not resist showing the twelve established quarks, as in Figure 8, and choosing for them the colors of the Spanish republican flag. Since Georgia was then part of the Soviet Union this triggered a standing ovation. Even a collection of quark pins for children became a must.

At Tbilisi intriguing new SLAC data on $e^+e^-$ annihilation to states containing kaons were presented, see Figure 11.

The experimentalists suspected that their results had to do with charm (a good fraction of events contained kaons) but did not understand them. To do it, one had to believe our predictions for the masses and decays of charmed mesons, as in Figure 9. Notice that the decay $D^{0*} \to D^+\pi^-$ (or its charge conjugate) is forbidden, while in all other $D^* \to D\pi$ decays the pion has a very small kinetic energy. In Figure 10 we see how, at a collider energy just above the $D^*D$ production threshold, a slow $\pi$ may help fake a recoiling mass $M \approx m(D^*)$.

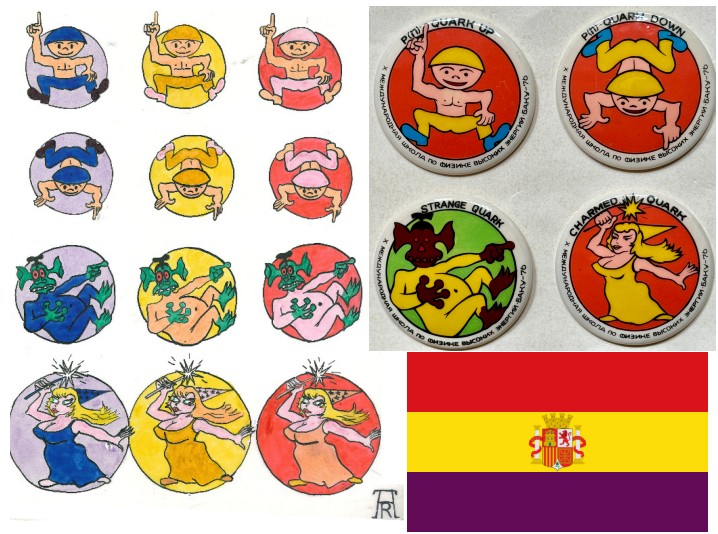

**Figure 8.** Three colors, four quarks and their pins in Cyrillic.

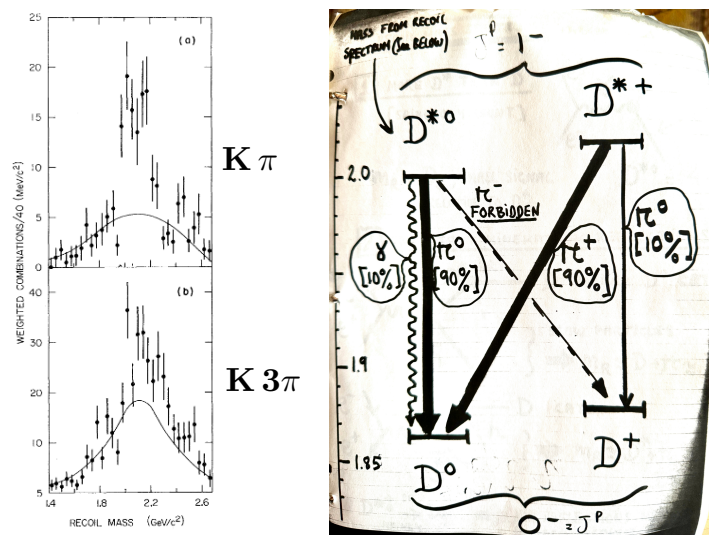

**Figure 9.** Invariant mass distributions of recoils against $K\pi$ and $K\pi\pi$ and the peculiarities of $D^* \to D$ decays.

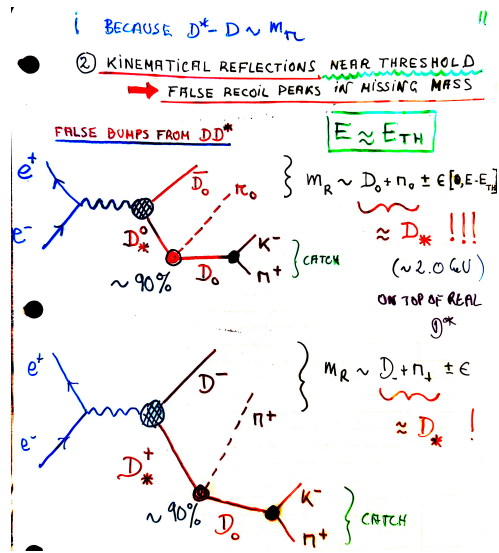

**Figure 10.** Origin of the "fakes" in the invariant mass of the ensembles of particles recoiling against $K^-\pi^+$.

Based on the above considerations we [21] could make a description (not meant to be a fit) of the observed recoiling mass spectra, with only one tuned parameter ($b$ in $\exp[-(bE)]$ describing the suppression of $D^*D$ production at a collider's energy not far above threshold. The results, shown in Figure 11, must have convinced the experimentalists, for they eventually made a correct analysis and published the discovery of charm.

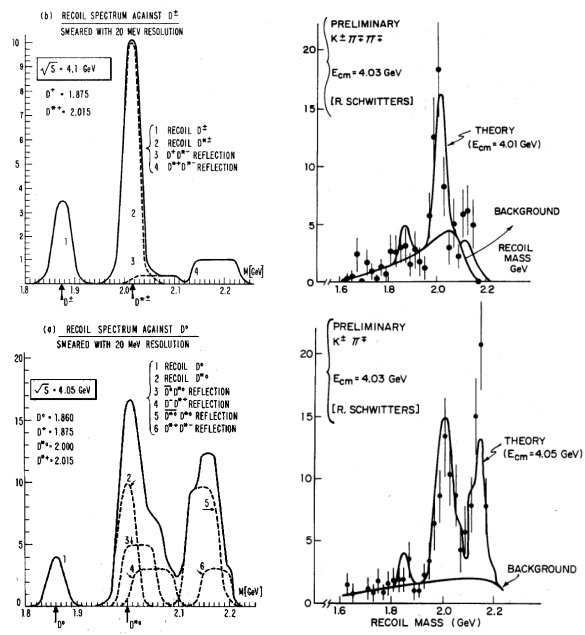

**Figure 11.** Recoil mass spectra in $D^*D$ production [21].

## 5. The Higgs boson

Most scientists, often justifiably, feel under-cited. We all contribute to this. There are a lot of possible citations that I skipped. Citations are peculiar. One example which I shall not cite: perhaps the most cited article on the search for "the Higgs" is one that recommends not to waste time looking for it.

The discovery of the Higgs definitely established the most standard Standard Model. One aspect of it had to do with peculiar citations. In Figure 12, I reproduce, with extra commentary, a transparency shown by Joe Incandela in his Higgs discovery talk. It specifies that, in the search for $H \to Z\,Z^* \to 4$ leptons, an analysis was used accumulating the event-by-event likelihood ratio of it being signal or background. In tiny lettering and not spoken by Incandela there was an arXiv reference to [22] where the method was introduced for possible neutral objects of spin-parity $0^+, 0^-, 1^-, 1^+$ and $2^+$. In the next transparency Incandela showed that this analysis contributed $3.2\sigma$ to the discovery of a $0^+$ object, which, combined with the analysis of $H \to \gamma\gamma$, resulted in a $5\sigma$ result. Quasi-standing ovation.

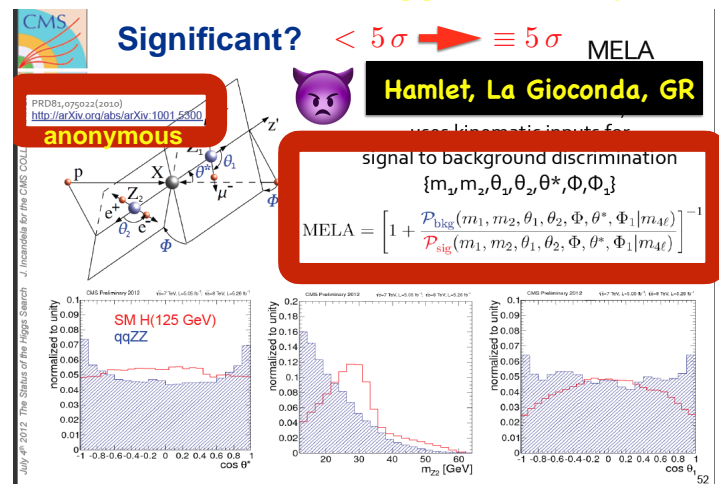

**Figure 12.** A transparency by Joseph Incandela .

Without the mentioned analysis CMS could not have reached the sacred $5\sigma$ at the time ATLAS did. ATLAS did not use this methodology, was it luckier? In Figure 12 I added "Hamlet, La Gioconda, General Relativity". Why? No doubt because "CMS" considered that, as in the cases just mentioned, who the authors of [22] were... was so well known that it would be offensive to believe that the audience did not know.

## 6. Conclusions

The interplay between theory, phenomenology and experiment in the development of the Standard Model was a blessing for those who participated. This being a compte rendu of a half-hour talk I have left a lot out. I wrote and cited more extensively on this subject in [23,24].

### Conflicts of Interest

The author declares no conflict of interest.

## References

1. 't Hooft, G. The discovery of the renormalizability of non-Abelian gauge theories. *Subnucl. Ser.* **2002**, *38*, 656.
2. Gross, D.J.; Wilczek, F. Ultraviolet behavior of non-abelian gauge theories. *Phys. Rev. Lett.* **1973**, *30*, 1343.
3. Politzer, H.D. Reliable Perturbative Results for Strong Interactions? *Phys. Rev. Lett.* **1973**, *30*, 1346.
4. Khriplovich, I.B. Green's functions in theories with non-Abelian gauge group. *Sov. J. Nucl.Phys.* **1970**, *10*, 235.
5. Gross, D.J.; Treiman, S.B. Hadronic form factors in asymptotically free field theories. *Phys. Rev. Lett.* **1974**, *32*, 1145.
6. De Rújula, A. Proton Magnetic Form Factor in Asymptotically Free Field Theories. *Phys. Rev. Lett.* **1974**, *32*, 1143.
7. De Rújula, A.; Georgi, H.; Politzer, H.D. Demythification of electroproduction local duality and precocious scaling. *Ann. Phys.* **1977**, *103*, 315.
8. Altarelli, G.; Cabibbo, N.; Petronzio, R.; et al. Is the 3104 MeV vector meson the phisub(c) or the $W_0$. *Lett. Nuovo Cim.* **1974**, *11*, 609.
9. Appelquist, T.; Politzer, H.D. Heavy Quarks and $e^+e^-$ Annihilation. *Phys. Rev. Lett.* **1975**, *34*, 43.
10. De Rújula, A.; Glashow, S.L. Is bound charm found? *Phys. Rev. Lett.* **1975**, *34*, 46.
11. Struminsky, B.V. Magnetic moments of barions in the quark model. *JINR* **1965**, JINR-Preprint P-1939
12. Bogoliubov, N.N.; Struminsky, B.V.; Tavkhelidze, A.N. *On Composite Models in the Theory of Elementary Particles*; JINR Pub.: Dubna, Russia, 1965.
13. Appelquist, T.; De Rújula, A.; Glashow, S.L.; et al. Spectroscopy of the New Mesons. *Phys. Rev. Lett.* **1975**, *34*, 365.
14. Eichten, E.; Gottfried, K.; Kinoshita, T.; et al. Spectrum of Charmed Quark-Antiquark Bound States. *Phys. Rev. Lett.* **1975**, *34*, 369.
15. De Rujula, A.; Georgi, H. Counting quarks in $e^+e^-$ annihilation. *Phys. Rev.* **1976**, *D13*, 1296.
16. Shifman, M.A.; Vainshtein, A.I.; Zakharov, V.I. QCD and resonance physics. applications. *Nucl. Phys.* **1979**, *B147*, 385.
17. De Rújula, A.; Georgi, H.; Glashow, S.L. Hadron masses in a gauge theory. *Phys. Rev.* **1975**, *D12*, 147.
18. Cazzoli, E.G.; Cnops, A.M.; Connolly, P.L.; et al. Evidence for $\Delta S = -\Delta Q$ currents or Charmed-Baryon production by neutrinos. *Phys. Rev. Lett.* **1975**, *34*, 1125.

19. Samios, N. *History of Original Ideas and Basic Discoveries in Particle Physics*; Newman, H.B., Ypsilantis, T., Eds.; Plenum Press: New York, NY, USA, 1996.

20. De Rújula, A. Theoretical Basis of the New Particles. In Proceedings of the XVIII International Conference on High-Energy Physics, Tbilisi, Georgia, 15–21 July 1976.

21. De Rújula, A.; Georgi, H.; Glashow, S.L. Charm spectroscopy via electron-positron annihilation.. *Phys. Rev. Lett.* **1976**, *37*, 785.

22. De Rújula, A.; Lykken, J.; Pierini, M.; et al. Higgs boson look-alikes at the LHC. *Phys. Rev.* **2010**, *D82*, 013003.

23. De Rújula, A. QCD, from its inception to its stubbornly unsolved problems. *Int. J. Mod. Phys.* **2019**, *A34*, 1930015.

24. De Rújula, A. *50 Years of Yang-Mills Theory*; 't Hooft, G., Ed.; World Scientific: Singapore, 2005.

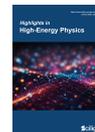

*Review*

# The Discovery of the *W* and *Z* Bosons at the CERN Proton-Antiproton Collider

Luigi Di Lella

Physics Department, University of Mainz, 55122 Mainz, Germany; Luigi.Di.Lella@cern.ch

**Abstract:** This article describes the scientific and technical achievements that led to the discovery of the weak intermediate vector bosons, $W^{\pm}$ and $Z$, from the original proposal to modify an existing high-energy proton accelerator into a proton-antiproton collider and its implementation at CERN, to the design, construction and operation of the detectors which provided the first evidence for the production and decay of these two fundamental particles.

**Keywords:** UA1; UA2; W and Z discovery

## 1. Introduction

The first experimental evidence for the existence of the weak neutral $Z$ boson, as predicted by the electro-weak theory formulated in the 1960s, was obtained in 1973 with the discovery of the Neutral Currents (NC). Those were neutrino interactions yielding only visible hadrons as final products. They were explained as resulting from the process $\nu_{\mu}(\bar{\nu}_{\mu}) + N \rightarrow \nu_{\mu}(\bar{\nu}_{\mu}) + $ hadrons, mediated by the exchange of a massive virtual particle, the neutral mediator of the weak interaction, the $Z^{o}$. Within the framework of the electroweak theory, the measurement of the cross-section ratio between the NC and the CC (Charged – Current) interactions, $\nu_{\mu}(\bar{\nu}_{\mu}) + N \rightarrow \mu^{-}(\mu^{+}) + $ hadrons, provided a rough determination of the masses of the two weak bosons. The W mass was predicted to be in the range 60 to 80 GeV/c$^2$ , while the estimate for the $Z$ was a value between 75 and 92 GeV/c$^2$. Both values were too large to make possible the production of such heavy states with the accelerators available at that time.

The idea suggested by Rubbia, Cline and McIntyre [1] in 1976 changed completely the scenario. They proposed to transform an existing hadron accelerator in a proton-antiproton collider. The aggressive approach would have allowed to achieve collision energies above the threshold for W and $Z$ production with a quick and relatively modest investment. The new scheme was proposed both to Fermilab and CERN, but it was approved only by this latter laboratory. The SPS, Super Proton Synchrotron accelerator, was transformed in a proton-antiproton collider reaching $\sqrt{s} = 540$ GeV in July 1981. At the end of 1982 it was possible to observe the first leptonic decays of the W, a few months later, in spring 1983 also a handful of leptonic decays of the $Z$ were observed.

In this article, the 1973 discovery of NC neutrino interactions is briefly described (Section 2). Section 3 presents the conception, construction and operation of the CERN $p\bar{p}$ collider. It is followed by the description of the two experiments, UA1 and UA2, of the *W* and *Z* discovery, and of the measurement of their properties (Sections 4 and 5).

## 2. The Discovery of Neutral-Current Neutrino Interactions

In the 1960s, high-energy neutrino beam experiments frequently observed events in which only hadronic activity was detected in the final state. These events were interpreted as resulting from interactions of neutrons, which were produced in $\nu_{\mu}$ or $\bar{\nu}_{\mu}$ charged-current interactions occurring near the downstream end of the shielding. In such cases, the associated muon escaped detection—typically due to its trajectory or energy—while the secondary neutron entered the detector and initiated a hadronic interaction [2].

In 1964 André Lagarrigue proposed to build a large-volume bubble chamber filled with heavy liquid, to be installed on the neutrino beam from the CERN 26 GeV Proton Synchrotron (PS). This chamber, named Gargamelle, was built in Saclay and installed at CERN in 1970. It had a cylindrical volume 4.8 m long, with a diameter of 1.8 m, located between two magnetic coils which provided a horizontal magnetic field of $2\,\mathrm{T}$ orthogonal to the beam axis (Figure 1). It was filled with liquid Freon-13, whose density is about $1.5\,\mathrm{g/cm^3}$ (in its boiling point at $-81.4$ C or $191.75$ K and atmospheric pressure), a radiation length of $11\,\mathrm{cm}$, and a mean nuclear interaction length of $78\,\mathrm{cm}$. Data-taking started in 1971, using $\nu_\mu$ and $(\bar{\nu}_\mu)$ beams with energy distributions between 1 and $10\,\mathrm{GeV}$.

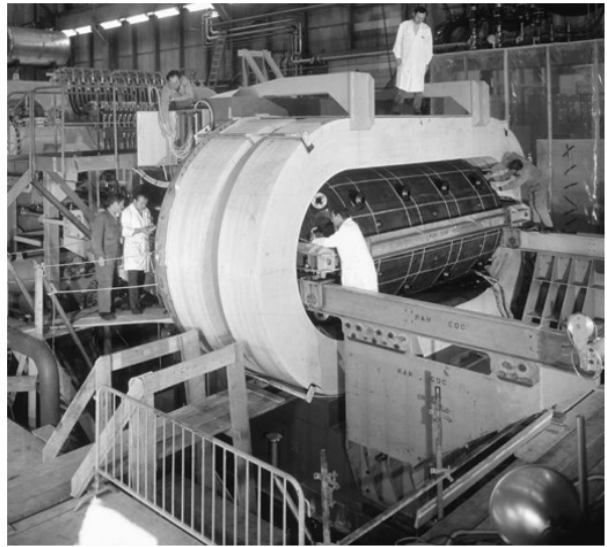

**Figure 1.** The Gargamelle body installed inside the magnetic coils.

The first hint for the existence of NC interactions was obtained in 1971 by the observation of an event consisting of a single electron only during a run with a $\bar{\nu}_\mu$ beam [3]. In this event, shown in Figure 2, the electron had an energy of $385 \pm 100\,\mathrm{MeV}$ and was emitted at an angle of $1.4° \pm 1.4°$ with respect to the beam axis. The most probable explanation of this event was $\bar{\nu}_\mu$ elastic scattering on an atomic electron, $\bar{\nu}_\mu + e^- \to \bar{\nu}_\mu + e^-$, which, to first order, could only be explained by the exchange of a neutral vector boson, as predicted by the electroweak theory. The background from CC processes involving incident $\nu_e$ or $\bar{\nu}_e$, a beam contamination of less than $1\%$, was estimated to amount to $0.03 \pm 0.02$ events.

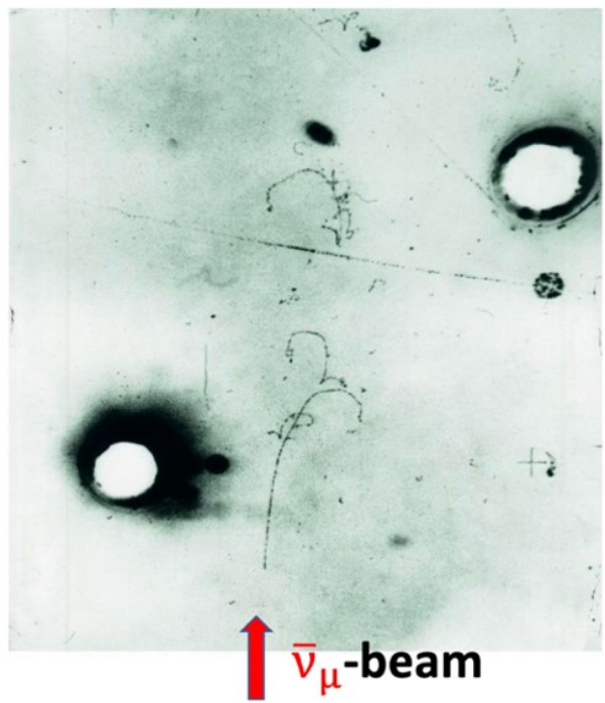

**Figure 2.** A single-electron event observed in Gargamelle during a $\bar{\nu}_\mu$ run in 1971.

The background from neutron interactions in events with final states consisting of hadrons only was studied by using CC $\nu_\mu$ $(\bar{\nu}_\mu)$ interactions occurring near the chamber entrance, with the outgoing muon clearly identified. The

                    

total energy measured in the neutron interactions was typically less than 500 MeV, and the longitudinal distribution along the chamber was consistent with the expected neutron mean free path. However, the $\nu_\mu$ and $\bar{\nu}_\mu$ interactions with only visible hadrons having a total energy of more than 1 GeV had a uniform longitudinal and radial distribution in the chamber [4], similar to the CC events with a visible muon, as expected for NC interactions (see Figure 3). The ratio NC/CC was measured to be $0.21 \pm 0.03$ for $\nu_\mu$, and $0.45 \pm 0.09$ for $\bar{\nu}_\mu$, both compatible with the same value of the weak mixing angle, $\sin^2 \theta_W$, in the range 0.3 to 0.4. A typical neutrino interaction with only hadrons in the final state is shown in Figure 4. In this event all three tracks are identified as hadrons from their interactions in the chamber.

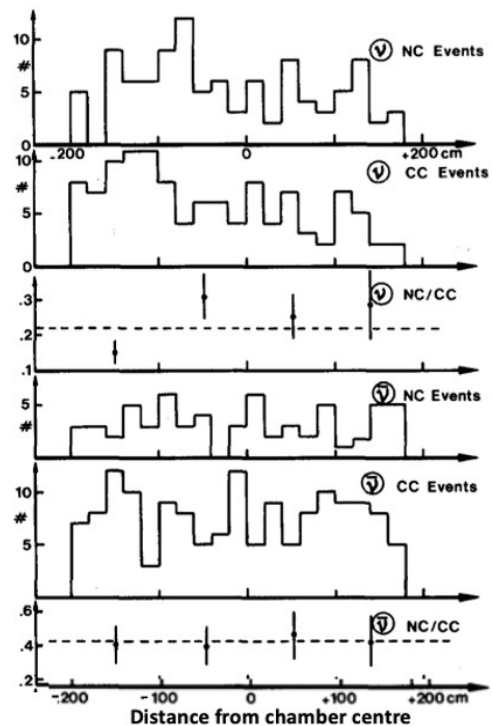

**Figure 3.** Event distributions and NC/CC ratios vs. distance from chamber centre [4].

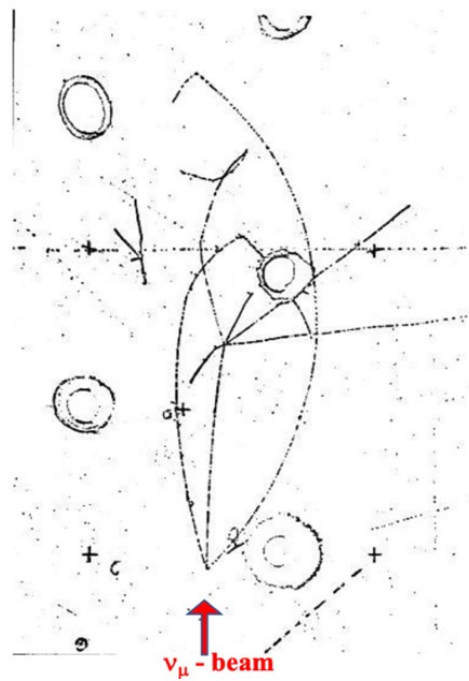

**Figure 4.** A neutrino interaction with three final-state hadrons and no muon.

### 3. The CERN $p\bar{p}$ Collider

The conception, construction and operation of the CERN proton-antiproton collider was a great achievement in itself. It is useful, therefore, to give a short description of this facility.

The production of $W$ and $Z$ bosons at a $p\bar{p}$ collider is expected to occur mainly as the results of quark-antiquark annihilation: $\bar{d}u \to W^+$; $\bar{u}d \to W^-$; $u\bar{u} \to Z$ ; $d\bar{d} \to Z$. Because 50% of the momentum of a high-energy proton is carried by three valence quarks, and the remainder by gluons, a valence quark carries, on average, about $1/6$ of the proton momentum. As a consequence, $W$ and $Z$ production should require a $p\bar{p}$ collider with a total center-of-mass energy equal to about six times the boson masses, or 500–600 GeV.

The detection of $Z \to e^+e^-$ decays determines the minimal collider luminosity: the cross-section for inclusive $Z$ production at ~600 GeV is ~1.6 nb, and the fraction of $Z \to e^+e^-$ decays is ~3%, hence a luminosity $L = 2.5 \times 10^{29}\,\mathrm{cm^{-2}s^{-1}}$ would give an event rate of ~1 per day. To achieve such luminosities one would need an antiproton source capable of delivering daily ~$3 \times 10^{10}\,\bar{p}$ distributed in few (3–6) tightly collimated bunches within the angular and momentum acceptance of the CERN SPS.

The 26 GeV CERN Proton Synchrotron (PS) was capable of producing antiprotons at the required rate. Each PS pulse accelerated approximately $10^{13}$ protons, delivered every 2.4 s to the antiproton production target. About $7 \times 10^6$ antiprotons with a momentum of 3.5 GeV/c were produced at $0°$ within a solid angle of $8 \times 10^{-3}$ sr and a momentum spread of $\Delta p/p = 1.5\%$. While this yield was sufficient in number, the resulting antiprotons occupied a phase-space volume more than $\geq 10^8$ times too large to be accepted by the SPS, even after acceleration to its injection energy of 26 GeV. To overcome this, the phase-space density of the antiproton beam had to be increased by at least a factor of $10^8$ before injection into the SPS. This process, known as cooling, reduces the spread of particle velocities—akin to compressing a hot gas—when viewed in the center-of-mass frame of the antiproton bunch.

The CERN collider project used the technique of stochastic cooling, invented by S. van der Meer in 1972 [5,6]. The developed principle is known as stochastic cooling and is illustrated in Figure 5. Let us consider, for simplicity, the horizontal "betatron" oscillations. If a transverse pick-up detector is located at the maximum amplitude of the oscillation, using fast processing electronics and transmission cables, it is possible to generate a signal triggering a transverse kicker exactly when the particle, whose orbit must be corrected, is passing by.

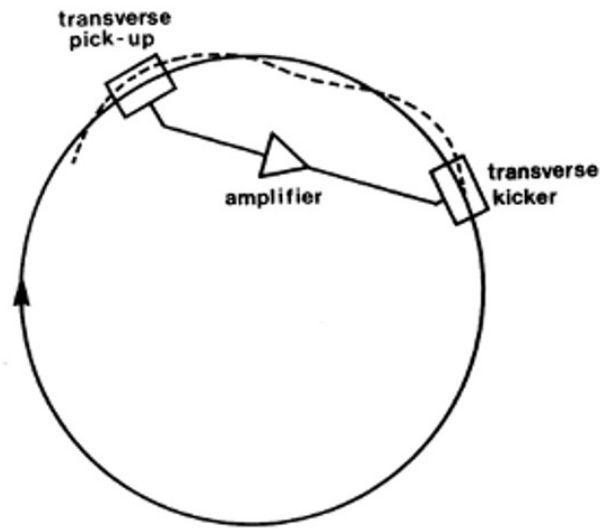

**Figure 5.** Cooling of a single particle (dotted line) horizontal oscillation.

Following the success of the so-called Initial Cooling Experiment (ICE) [7], which provided the experimental demonstration that stochastic cooling could indeed achieve the required increase of $\bar{p}$ phase-space density, the CERN proton-antiproton collider project was approved on May 28, 1978. For the CERN collider, stochastic cooling was achieved in a purpose-built machine called Antiproton Accumulator (AA), which included several independent cooling systems to cool both horizontal and vertical oscillations, and also to decrease the beam momentum spread (cooling of longitudinal motion), by using pick-up electrodes which provided signals proportional to $\Delta p$. The AA is a large aperture magnetic ring. A picture of the AA during construction is shown in Figure 6. Figure 7 illustrates the operation of cooling and accumulation of a $\bar{p}$ stack in the AA.

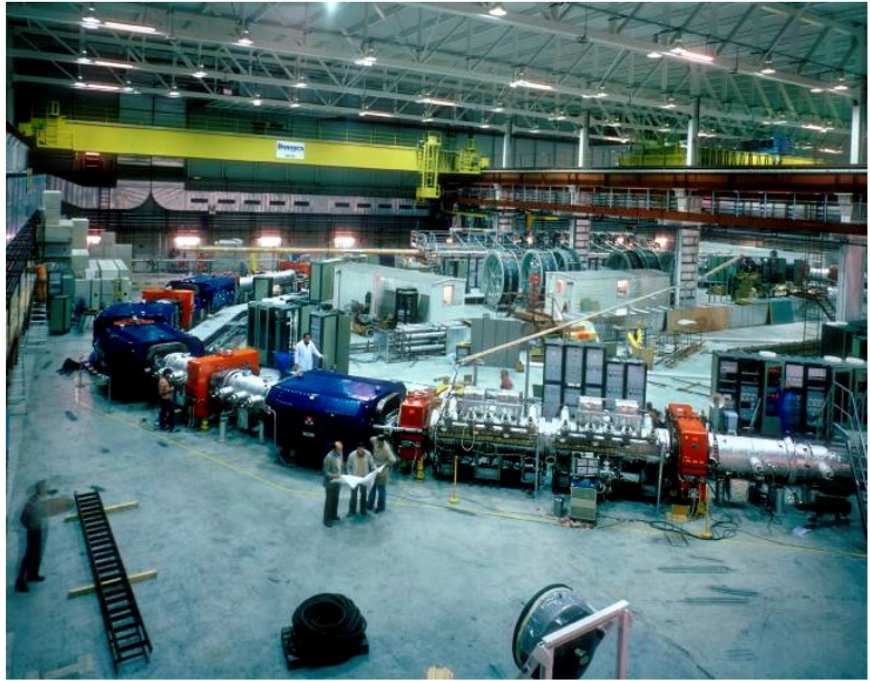

**Figure 6.** View of the Antiproton Accumulator during construction.

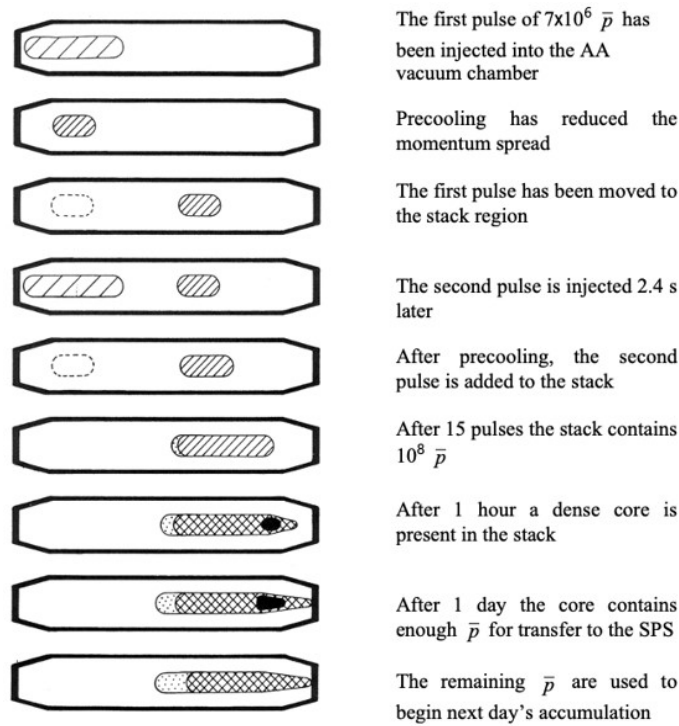

**Figure 7.** Schematic sequence illustrating antiproton cooling and accumulation in the AA [1].

Consecutive PS cycles achieved beam injection into the SPS, when AA accumulated a sufficiently dense $\bar{p}$ stack. Firstly, three proton bunches (six after 1986), each containing $\sim 10^{11}$ protons, are accelerated to 26 GeV in the PS and injected into the SPS (see Figure 8). Then three $\bar{p}$ bunches (six after 1986), of typically $\sim 10^{10}$ each, are extracted from the AA and injected into the PS. Here they are accelerated to 26 GeV in a direction opposite to that of the protons, and then injected into the SPS. The relative injection timing of the bunches is controlled with a precision of $\sim 1$ ns to ensure that bunch crossing in the SPS occurs in the centre of the detectors.

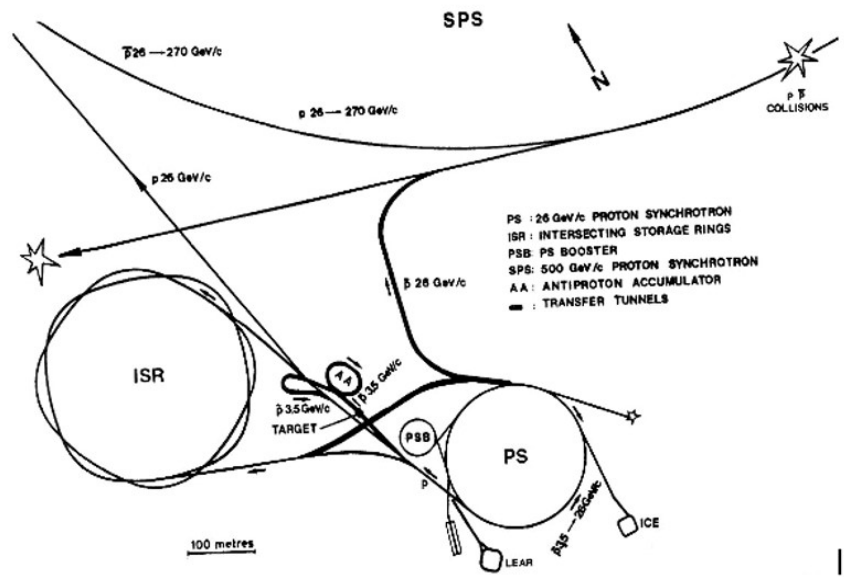

**Figure 8.** Layout of the three machines initially involved in the operation of the CERN $p\bar{p}$ collider: the PS, the AA, the SPS and the interconnecting transfer lines.

In 1987, to increase the luminosity of the machine, the source of antiprotons was upgraded. A second ring, the Antiproton Collector (AC) which had a much larger acceptance was built around the AA. As a result the stacking rate for antiprotons was increased by a factor of $\sim 10$. The main collider parameters and their evolution with time are shown in Table 1. The CERN proton-antiproton collider completed its operations in 1990 and was shut-down at the end of the year.

**Table 1.** CERN proton-antiproton collider operation, 1981–1990.

| Year | Collision Energy (GeV) | Peak luminosity (cm$^{-2}$ s$^{-1}$) | Integrated luminosity (cm$^{-2}$) |
|------|------------------------|--------------------------------------|-----------------------------------|
| 1981 | 546 | $\sim 10^{27}$ | $2 \times 10^{32}$ |
| 1983 | 546 | $5 \times 10^{28}$ | $1.5 \times 10^{35}$ |
| 1984–1985 | 630 | $3.9 \times 10^{29}$ | $1.0 \times 10^{36}$ |
| 1987–1990 | 630 | $3 \times 10^{30}$ | $1.6 \times 10^{37}$ |

## 4. The Discovery of the *W* and *Z* Bosons

Since the SPS is built in an underground tunnel at an average depth of $\sim 100$ m, the project also required the excavation of underground experimental areas to house the detectors. The first experiment, named UA1 for "Underground Area 1", was soon approved on 29 June 1978. UA2, the second experiment, followed a few months later and was approved at the end of the year.

### 4.1. The UA1 Experiment

The UA1 experiment is a general-purpose magnetic detector with an almost complete $4\pi$ coverage [8]. Figure 9 shows a schematic view of the detector. The magnet is a dipole with a horizontal field of $0.7$ T over a volume of $7 \times 3.5 \times 3.5$ m$^3$, perpendicular to the beam axis and produced by a warm aluminium coil to minimize absorption.

The magnet contained the central track detector, which was a system of drift chambers filling a cylindrical volume $5.8$ m long with a $2.5$ m diameter, reconstructing charged particle trajectories down to polar angles of $\sim 6°$ with respect to the beams. Tracks were sampled approximately every centimetre and could have up to 180 hits. This detector, at the cutting edge of technology in those days, was surrounded by electromagnetic and hadronic calorimeters down to $0.2°$ to the beam line. This "hermeticity", as it was called later, turned out to be very effective to reconstruct undetected neutrinos from $W \to e\nu$ decay, and also to search for possible new, as yet undiscovered neutral particles escaping direct detection. It became one of the basic features of all general-purpose detectors at the next-generation $e^+e^-$ and hadron colliders (LEP, the Fermilab $p\bar{p}$ collider and the LHC).

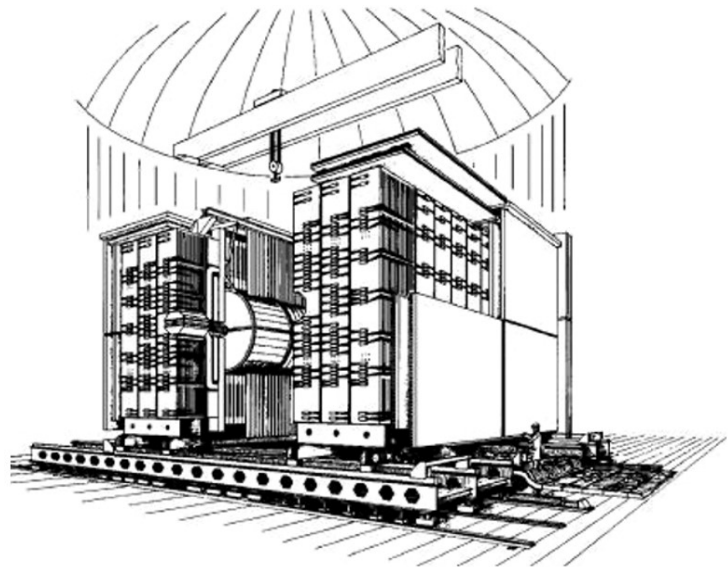

**Figure 9.** View of the UA1 detector with the two magnet halves opened up.

Electromagnetic calorimeters are installed inside the magnet and they consist of scintillators interleaved with sheets of lead to form a multi-layered sandwich. The central region is organised in two cylindrical half-shells surrounding the tracker, each subdivided into 24 elements ("gondolas"). The geometrical coverage is $180°$ in $\phi$ and $24$ cm along the beam line. The total thickness of the calorimeter is 26.4 radiation lengths ($X_0$). Two similar structures were present at smaller angles to the beam line, each consisting of 32 radial sectors. The energy resolution for electrons was measured to be $\sigma(E)/E = 0.15/\sqrt{E}$ ($E$ in GeV). A 40 GeV electron could therefore be measured with a typical ±2.5% energy resolution.

The hadronic calorimeter was built by inserting scintillator planes into the laminated structure of the magnet return yoke, and two iron walls located symmetrically at the two ends of the magnet. The hadron calorimeter was segmented into 450 independent cells.

A complex system of drift tubes surrounding the magnet joke, provided information to detect muons and measure their momentum. Muon detectors, consisting of systems of drift tubes, surrounded the magnet yoke. The typical momentum resolution for a 40 GeV/c muon track was ±20%.

A picture of UA1 during the assembly phase is shown in Figure 10.

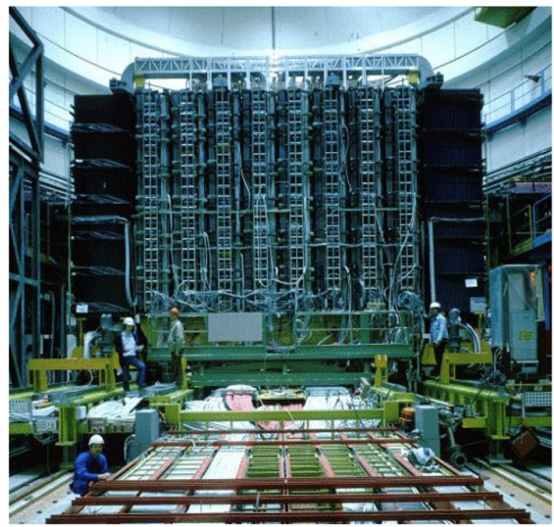

**Figure 10.** The UA1 detector during assembly.

### 4.2. The UA2 Experiment

UA2 was not designed as a general-purpose detector, but rather optimized for the detection of electrons from $W$ and $Z$ decays. The emphasis was on highly granular calorimetry with spherical projective geometry, which was well adapted also to the detection of hadronic jets.

Figure 11 shows the layout of the UA2 detector for the collider runs between 1981 and 1985. The central region contained a "vertex detector", which consisted of various types of cylindrical tracking chambers. A "preshower" counter, located just behind the last chamber and consisting of a tungsten cylinder followed by a multi-wire proportional chamber, was crucial for electron identification. The vertex detector was surrounded by the central calorimeter, which covered the full azimuth and was subdivided into 240 independent cells, each subtending the angular interval $\Delta\theta \times \Delta\phi = 10° \times 15°$ and consisting of an electromagnetic (Pb-scintillator) and a hadronic (Fe-scintillator) section. The calorimeter energy resolution for electrons was $\sigma(E)/E = 0.14/\sqrt{E}$ ($E$ in GeV), and was ∼10% for an 80 GeV hadron in the central calorimeters. There was no magnetic field in this region.

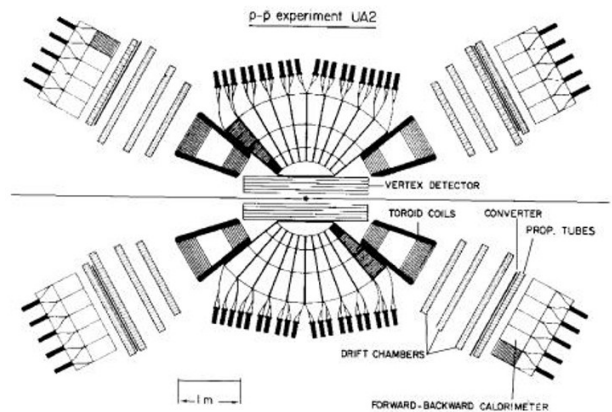

**Figure 11.** Sketch of the UA2 detector in the 1981–1985 configuration.

Twelve azimuthal sectors composed the two forward detectors. Each sector consisted of tracking chambers and an electromagnetic calorimeter equipped with a "preshower" detector. Individual coils generated a toroidal magnetic field in the forward/backward regions. The UA2 detector had no coverage for muons. The collaboration gathered, initially, about 60 scientists coming from three research institutes, CERN, Orsay and Saclay, and three universities, Bern, Copenhagen, Pavia. A picture of UA2 during assembly is shown in Figure 12.

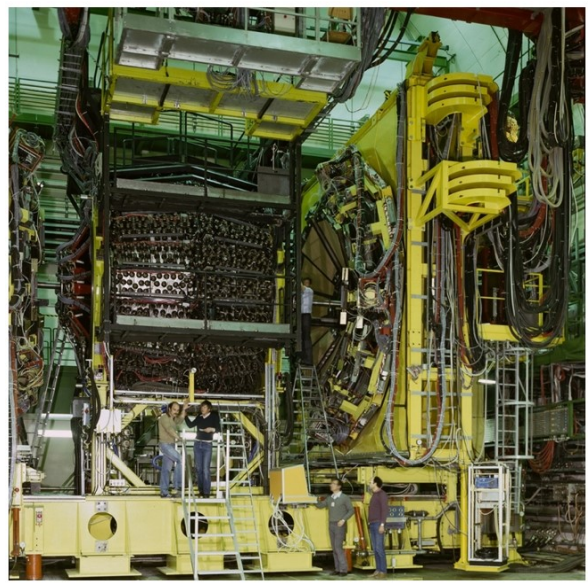

**Figure 12.** The UA2 detector in its 1981–1985 configuration.

### 4.3. Discovery of the W Boson

The dominant decay mode of the $W$ boson is to quark-antiquark pairs, i.e. hadronic jets. Unfortunately, this decay mode, that accounts for about 70% of the branching fraction, is overwhelmed by QCD processes yielding an irreducible two-jet background. This is the reason why both UA1 and UA2 focused on the leptonic decay of the $W$. Both experiments were optimised to detect high $p_T$ isolated electrons produced in the decay $W^\pm \to e^\pm \nu_e(\bar{\nu}_e)$ and high missing transverse due to the neutrino. UA1 was able to identify also muons produced in the decay

$W^\pm \to \mu^\pm \nu_\mu (\bar{\nu}_\mu)$.

The signal from $W \to e\nu_e$ decay is expected to have the following features:

- the presence of a high transverse momentum ($p_T$) isolated electron;
- peak in the electron $p_T$ distribution at $m_W/2$, where $m_W$ is the $W$ mass;
- the presence of high missing transverse momentum from the undetected neutrino.

It is important to note that the longitudinal component of missing momentum (along the beam axis) cannot be reliably measured at hadron colliders. This is because a significant number of high-energy secondary particles from the hadrons are produced at very small polar angles relative to the beam direction. These particles travel down the beam pipe and are not intercepted by the detector, as they remain inside the vacuum system of the accelerator. As a result, any momentum they carry in the longitudinal direction is not accounted for, making it impossible to reconstruct the total momentum along the beam axis. Only the transverse component of missing momentum can be inferred from the measured imbalance in the detector's transverse plane.

The missing transverse momentum vector ($\vec{p}_T^{\,miss}$) is defined as

$$\vec{p}_T^{\,miss} = -\sum_{cells} \vec{p}_T \tag{1}$$

where $\vec{p}_T$ is the transverse component of a vector associated to each calorimeter cell, with direction from the event vertex to the cell center and length equal to the energy deposition in that cell, and the sum is extended to all cells with an energy deposition larger than zero. For UA1 events with final-state muons, the muon transverse momenta are also added. In an ideal detector with perfect measurement, and assuming a final-state neutrino is the only undetected particle, momentum conservation implies that the $\vec{p}_T^{\,miss}$ corresponds exactly to the transverse momentum of the neutrino.

Figure 13 shows the $|\vec{p}_T^{\,miss}|$ distribution, as measured by UA1 from the 1982 data[9]. There is a component decreasing approximately as $|\vec{p}_T^{\,miss}|^2$, due to the effect of calorimeter resolution in events without significant $\vec{p}_T^{\,miss}$, followed by a flat component due to events with genuine $\vec{p}_T^{\,miss}$. Six events with high $p_T^{\,miss}$ in the distribution of Figure 13 contain a high $p_T$ electron. The $\vec{p}_T^{\,miss}$ vector in these events is almost back-to-back with the electron transverse momentum vector, as shown in Figure 14. These events are interpreted as due to $W \to e\nu_e$ decay. This result was first announced at a CERN seminar on January 20, 1983. Figure 15 shows the graphics display of one of these events. A fit to the distribution of these events using $m_W$ as a free parameter gives $m_W = 81 \pm 5\,\mathrm{GeV}/c^2$.

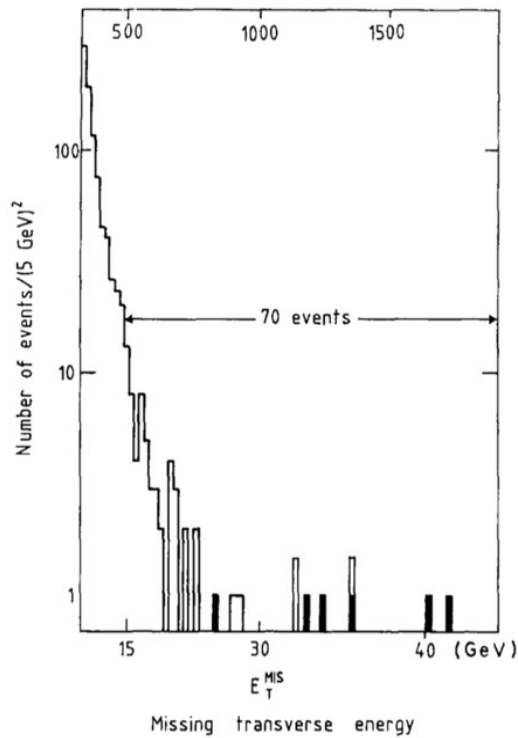

**Figure 13.** UA1 distribution of the missing transverse momentum (called $E_T^{MIS}$ in this plot) for equal bins of $(E_T^{MIS})^2$. The events shown as dark areas in this plot contain a high $p_T$ electron.

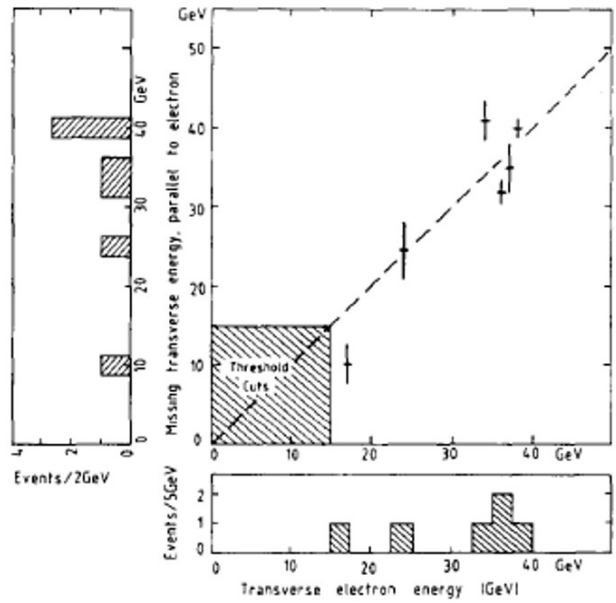

**Figure 14.** UA1 scatter plot of all the events from the 1982 data which contain a high $-p_T$ electron and large $\vec{p}_T^{miss}$. The abscissa is the electron $p_T$ and the ordinate is the $\vec{p}_T^{miss}$ component antiparallel to the electron $\vec{p}_T$ .

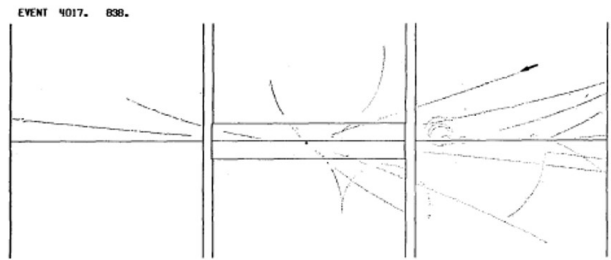

**Figure 15.** Display of a UA1 $W \to e\nu_e$ event. The arrow points to the electron track.

The results from the UA2 search for $W \to e\nu_e$ events[10] were presented at a CERN seminar on the day after the UA1 presentation. Six events containing an electron with $p_T > 15$ GeV/c were identified among the 1982 data. Figure 16a shows the distribution of the ratio between $|\vec{p}_T^{miss}|$ and the electron $p_T$ for these events, while Figure 16b displays the electron $p_T$ distribution for the events with $|\vec{p}_T^{miss}|$ larger than 80% of the electron $p_T$ (four events). These events have the properties expected from $W \to e\nu_e$ decay. A fit to the distribution of these events using $m_W$ as a free parameter gives $m_W = 80^{+10}_{-6}$ GeV/c$^2$.

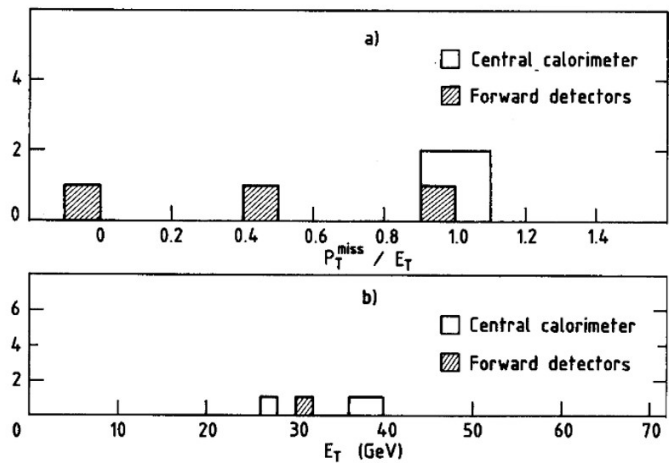

**Figure 16.** (**a**) Distribution of the ratio between the missing transverse momentum in the event and the electron transverse momentum (called $E_T$ in this plot) for six UA2 events containing an electron with $E_T > 15$ GeV; (**b**) Electron transverse momentum distribution for the four events with the highest $|\vec{p}_T^{miss}|/E_T$ ratio.

### 4.4. Discovery of the Z Boson

Figure 17 illustrates the search for the decay $Z \to e^+ e^-$ in UA1[11]. The first step of the analysis requires the presence of two calorimeter clusters consistent with electrons and having a transverse energy $E_T > 25$ GeV. Among the data recorded during the 1982–1983 collider run, 152 events are found to satisfy these conditions. The next step requires the presence of an isolated track with $p_T > 7$ GeV/c pointing to at least one of the two clusters. Six events satisfy this requirement, showing already a clustering at high invariant mass values, as expected from $Z \to e^+ e^-$ decay. Of these events, four are found to have an isolated track with $p_T > 7$ GeV/c pointing to each clusters. They are consistent with a unique value of the $e^+ e^-$ invariant mass within the calorimeter resolution. One of these events is displayed in Figure 18.

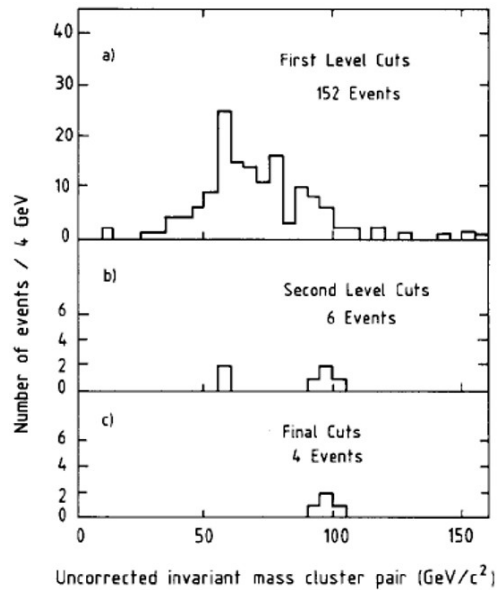

**Figure 17.** Search for the decay $Z \to e^+ e^-$ in UA1 (see text).

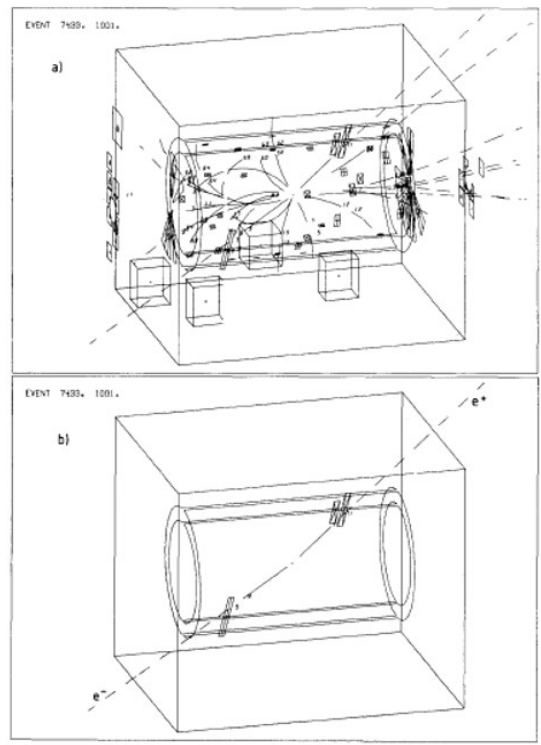

**Figure 18.** One of the $Z \to e^+ e^-$ events in UA1: (**a**) display of all reconstructed tracks and calorimeter hit cells; (**b**) only tracks with $p_T > 2$ GeV/c and calorimeter cells with $E_T > 2$ GeV are shown.

An event consistent with the decay $Z \to \mu^+ \mu^-$ was also found by UA1 among the data collected in 1983 [11]. Figure 19 shows the mass distribution of all lepton pairs found by UA1 from the analysis of the 1982–1983 data. The mean of these values is

$$m_Z = 95.2 \pm 2.5 \pm 3.0 \, \text{GeV/c}^2 \tag{2}$$

The first uncertainty arises from statistical, while the second originates from the systematic uncertainty associated with the calorimeter energy scale calibration.

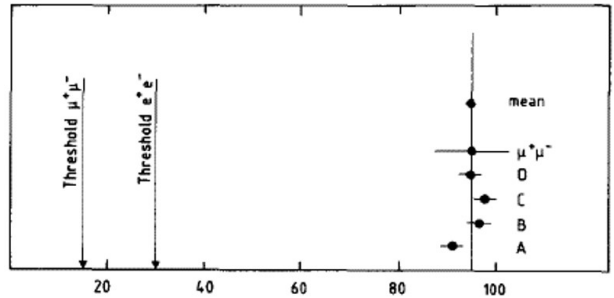

**Figure 19.** Invariant mass distribution of all lepton pairs detected by UA1 in the 1982–1983 data.

The UA2 search for the decay $Z \to e^+ e^-$ among the 1982–1983 data [12] is illustrated in Figure 20. First, pairs of energy depositions in the calorimeter consistent with two isolated electrons and with $p_T > 25 \, \text{GeV}$ are selected. Then, an isolated track consistent with an electron (from preshower information) is required to point to at least one of the clusters. Eight events satisfy these requirements: of these, three events have isolated tracks consistent with electrons pointing to both clusters. The weighted average of the invariant mass values for the eight events is

$$m_Z = 91.9 \pm 1.3 \pm 1.4 \, \text{GeV/c}^2 \tag{3}$$

where the first error is statistical and the second one originates from the systematic uncertainty on the calorimeter energy scale. The latter is smaller than the corresponding UA1 value because the smaller size of the UA2 calorimeter, and its modularity, allow frequent recalibrations on electron beams of known energies from the CERN SPS.

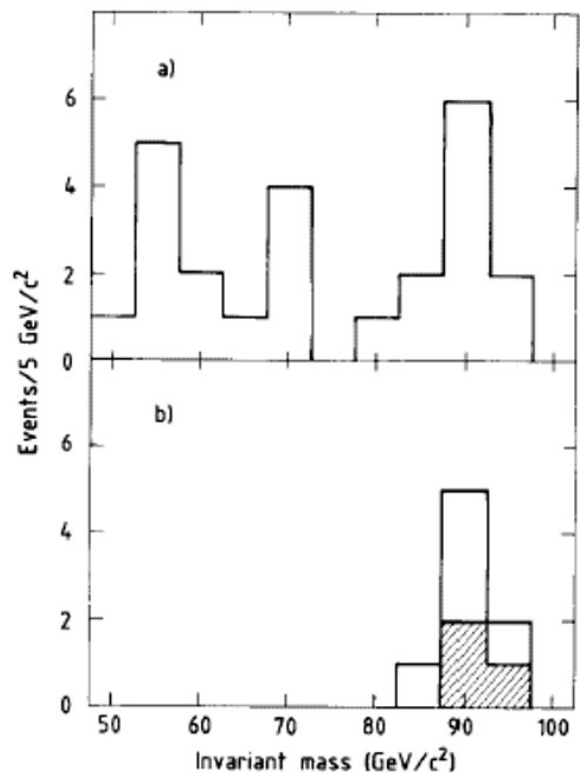

**Figure 20.** Search for the decay $Z \to e^+ e^-$ in UA2 (see text). The shaded area represents the three events with isolated electron tracks pointing to both energy clusters in the calorimeter.

Figure 21 shows the energy deposited in the UA2 calorimeter by a $W \to e\nu$ and by a $Z \to e^+e^-$ event. These distributions exemplify the distinctive topologies of such events, characterized by a substantial deposition of energy within a limited number of calorimeter cells, with minimal or no energy detected in the remaining cells.

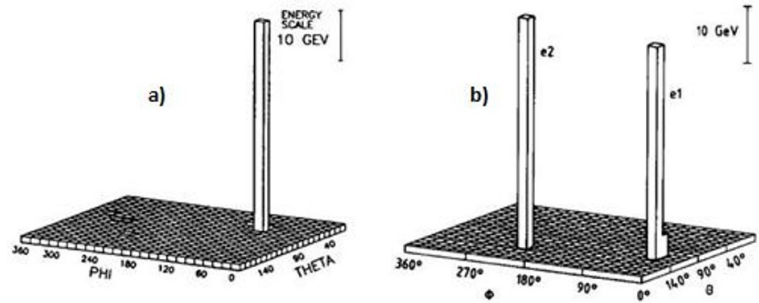

**Figure 21.** The energy deposited in the UA2 calorimeter for a $W \to e\nu$ (**a**) and a $Z \to e^+e^-$ event (**b**).

## 5. Physics Results from Subsequent Collider Runs

Following the historical runs in 1982–1983 which led to the discovery of the $W$ and $Z$ bosons, additional runs took place in the following years. In a first phase, up to the end of 1985, with the two detectors basically unchanged, the collider energy was raised from $\sqrt{s} = 540\,\text{GeV}$ to $630\,\text{GeV}$ and the peak luminosity doubled. Then, between 1987 and the collider shut-down at the end of 1990, more physics runs took place at higher luminosity (see Section 3, Table 1). The most important results on $W$ and $Z$ physics achieved during these runs are described in the next subsections.

### 5.1. W and Z Masses and Production Cross-Sections

At the end of 1985, UA1 had recorded 290 $W \to e\nu$, 33 $Z \to e^+e^-$, 57 $W \to \mu\nu$ and 21 $Z \to \mu^+\mu^-$ [13]. As an example, Figure 22 shows the $W \to e\nu$ transverse mass ($M_T$) distribution, where

$$M_T = \sqrt{2p_T^e p_T^\nu (1 - \cos\phi_{e\nu})} \tag{4}$$

and $\phi_{e\nu}$ is the azimuthal separation between electron and neutrino (the electron transverse momentum is instead by the transverse mass, due to its reduced sensitivity to the transverse momentum of the W boson).

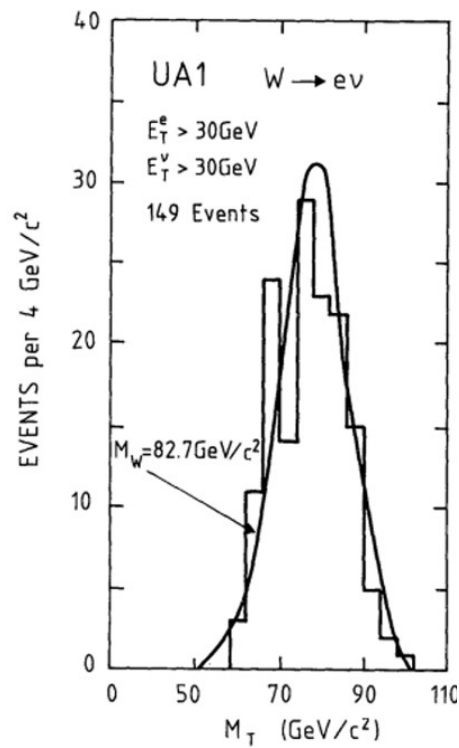

**Figure 22.** Transverse mass distribution for all 1982–1985 UA1 $W \to e\nu$ events.

Figure 23 shows the invariant mass distribution of all $e^+e^-$ pairs recorded by UA1 during the same period. The $W$ and $Z$ mass values obtained from fits to the distributions of Figures 22 and 23 were

$$m_W = 82.7 \pm 1.0 \pm 2.7 \,\text{GeV/c}^2 \tag{5}$$

$$m_Z = 93.1 \pm 1.0 \pm 3.1 \,\text{GeV/c}^2 \tag{6}$$

respectively, where the first error is statistical and the second one reflects the uncertainty on the calorimeter energy scale. The $W$ and $Z$ production cross-sections values, multiplied with the corresponding decay branching ratios (BR), as measured by UA1, were

$$\sigma_W \,\text{Br}\,(W \to e\nu) = 630 \pm 50 \pm 100 \,\text{pb} \tag{7}$$

$$\sigma_Z \,\text{Br}\,(Z \to e^+e^-) = 74 \pm 14 \pm 11 \,\text{pb} \tag{8}$$

UA1 has also observed 32 $W \to \tau\nu_\tau$ decays followed by $\tau$ hadronic decay [14]. These events appeared in the detector as a highly collimated, low multiplicity hadronic jet approximately back-to-back in azimuth to the $\vec{p}_T^{miss}$ vector.

In the same physics runs, from 1982 to 1985, the UA2 experiment had recorded samples of 251 $W \to e\nu$ and 39 $Z \to e^+e^-$ events[15]. The measured properties of these events were in good agreement with the UA1 results. The $W$ and $Z$ mass values, as measured by UA2, were

$$m_W = 80.2 \pm 0.8 \pm 1.3 \,\text{GeV/c}^2 \tag{9}$$

$$m_Z = 91.5 \pm 1.2 \pm 1.7 \,\text{GeV/c}^2 \tag{10}$$

respectively, where, as usual, the first error is statistical and the second one reflects the uncertainty on the calorimeter energy scale.

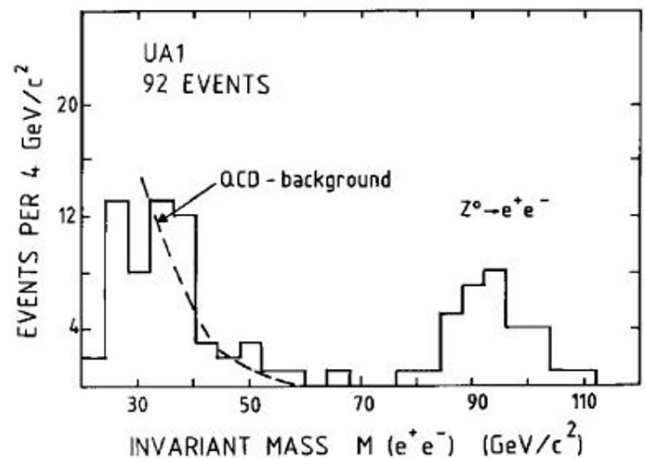

**Figure 23.** Invariant mass distribution of all 1982–1985 UA1 $e^+e^-$ pairs.

### 5.2. Charge Asymmetry in the Decay $W \to e\nu$

The angular distribution of the charged lepton in the $W$ rest frame can be written as

$$\frac{dn}{d\cos\theta^*} \propto (1 + q\cos\theta^*)^2 \tag{11}$$

where $\theta^*$ is the angle of the charged lepton measured with respect to the $W$ polarization, and $q = -1\,(+1)$ for electrons (positrons). This axis is practically collinear with the incident $\bar{p}$ direction if the $W$ transverse momentum is small.

A complication arises from the fact that the neutrino longitudinal momentum is not measured, and the requirement that the invariant mass of the $e\nu$ pair be equal to the $W$ mass gives two solutions for $\theta^*$. The UA1 analysis[13] retains only those events for which one solution is unphysical ($W$ longitudinal momentum inconsistent with kinematics), and the lepton charge sign is unambiguously determined. Figure 24 shows the distribution of the variable $q\cos\theta^*$ for 149 unambiguous events. The distribution agrees with the expected $(1 + q\cos\theta^*)^2$ form. It must be noted that this result cannot distinguish between V-A and V+A because in the latter case all helicities change sign and the angular distribution remains the same.

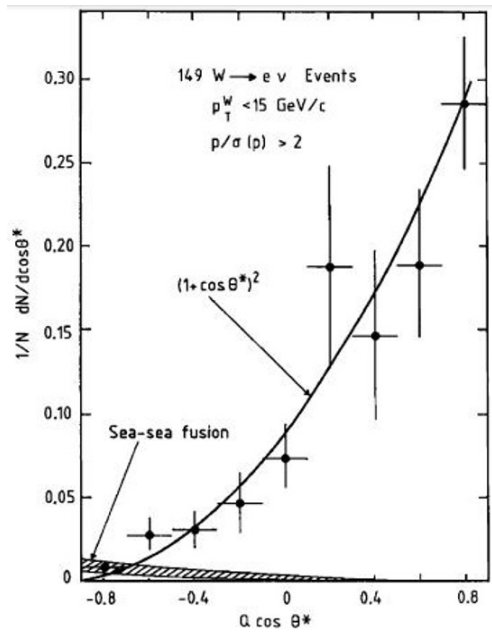

**Figure 24.** Decay angular distribution for the final UA1 $W \to e\nu$ event sample (see text). The shaded band shows the expected contribution of wrong polarization from the annihilation of a sea quark with a sea antiquark.

### 5.3. A Test of QCD: The W Boson Transverse Momentum

To lowest order, the $W$ and $Z$ bosons produced by $q\bar{q}$ annihilation are emitted with very low transverse momentum. However, gluon radiation from the initial quarks (or antiquarks) may result in $W$ and $Z$ production with a sizeable transverse momentum, which is equal and opposite to the total transverse momentum of all hadrons produced in association with the intermediate bosons.

Figure 25 shows the distribution of the $W$ transverse momentum, $p_T^W$, as measured by UA1 [13] using the $W \to e\nu$ event sample. A QCD prediction [16], also shown in Figure 25, agrees with the data over the full $p_T^W$ range. The $W$ bosons produced with high $p_T^W$ are expected to recoil against one or more jets, and such jets are indeed observed experimentally.

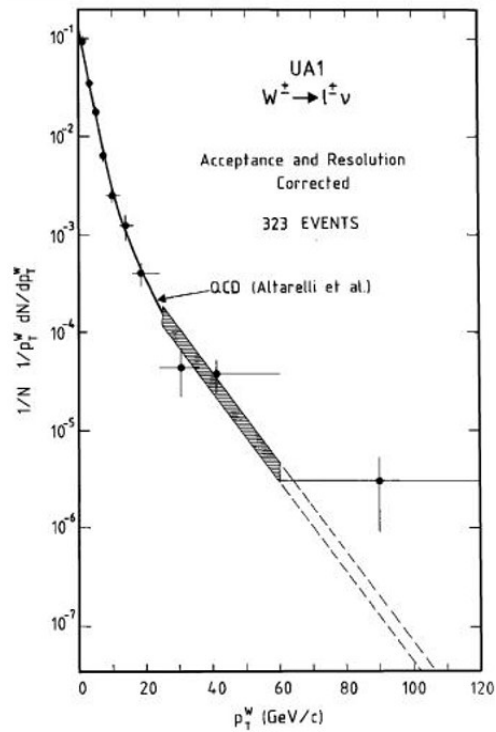

**Figure 25.** Distribution of $p_T^W$, as measured by UA1. The curve is a QCD prediction [16]. The shaded band shows the theoretical uncertainty in the region of high $p_T^W$.

*5.4. Precision Measurement of the W to Z Mass Ratio*

In view of the last collider runs at higher luminosity (1987–1990, see Section 3, Table 1), the UA2 detector underwent a major upgrade: new segmented calorimeters were installed in the forward regions to improve the hermeticity. The central tracking system was upgraded using silicon pad detectors, scintillating fibres and transition radiation detectors. Figure 26 displays the UA2 layout for the 1987–1988 collider run. University groups from Cambridge, Heidelberg, Milano, Perugia and Pisa joined the collaboration bringing the total number of physicists to about 100.

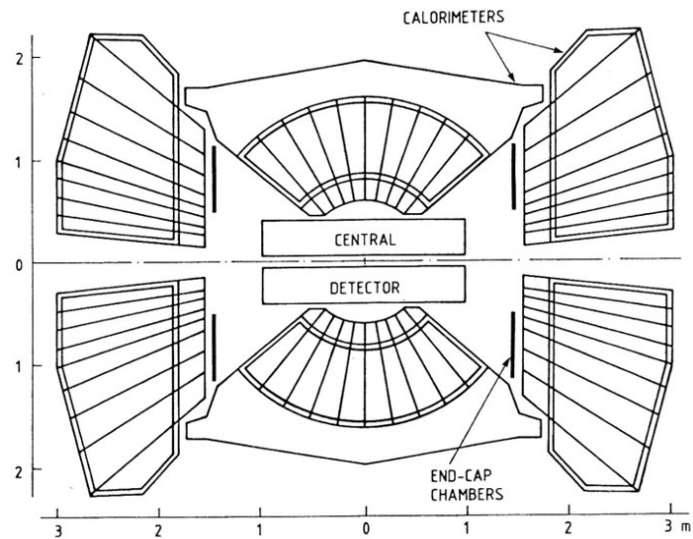

**Figure 26.** The UA2 detector layout for the 1987–1988 collider run.

From 1988 to 1990, UA2 collected large samples of $W \to e\nu$ and $Z \to e^+e^-$ events.

Figure 27 shows the transverse mass distribution for 2065 $W \to e\nu$ decays with the electron measured in the UA2 central calorimeter [17]. A best fit to this distribution using $m_W$ as a free parameter gives $m_W = 80.84 \pm 0.22 \mathrm{GeV/c}^2$ (statistical error only).

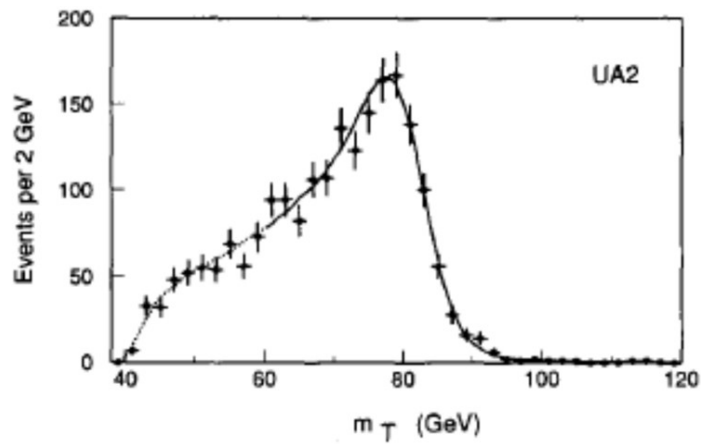

**Figure 27.** Transverse mass distribution for 2065 $W \to e\nu$ decays (see text). The curve is the best fit to the experimental distribution using $m_W$ as a free parameter.

The measured $e^+e^-$ invariant mass distribution [17] is shown in Figure 28, which displays two spectra: one containing 95 events in which both electrons fall in a fiducial region of the central calorimeter and their energies are accurately measured (Figure 28a); and the other spectrum containing 156 events in which one of the two electrons falls outside the fiducial region of the central calorimeter, resulting in a broader mass resolution (Figure 28b). Best fits to the two spectra give $m_Z = 91.65 \pm 0.34 \mathrm{GeV/c}^2$ and $m_Z = 92.10 \pm 0.48 \mathrm{GeV/c}^2$, respectively. The weighted mean of these two values is $m_Z = 91.74 \pm 0.28 \mathrm{GeV/c}^2$ (statistical error only).

The two independent measurements of $m_W$ and $m_Z$ give

$$\frac{m_W}{m_Z} = 0.8813 \pm 0.0036 \pm 0.0019 \tag{12}$$

where the first error is statistical and the second one is a small systematic uncertainty which takes into account a possible calorimeter non-linearity.

By 1991, a precise measurement of $m_Z$ from LEP experiments had become available [18]

$$m_Z = 91.175 \pm 0.021 \, \text{GeV/c}^2 \tag{13}$$

Multiplying this value with the ratio $m_W/m_Z$ measured by UA2 provided a determination of $m_W$ with a precision of 0.46%:

$$m_W = 80.35 \pm 0.33 \pm 0.17 \, \text{GeV/c}^2 \tag{14}$$

in agreement with a direct measurement, $m_W = 79.91 \pm 0.39 \, \text{GeV/c}^2$, by the CDF experiment at the Fermilab $p\bar{p}$ collider [19].

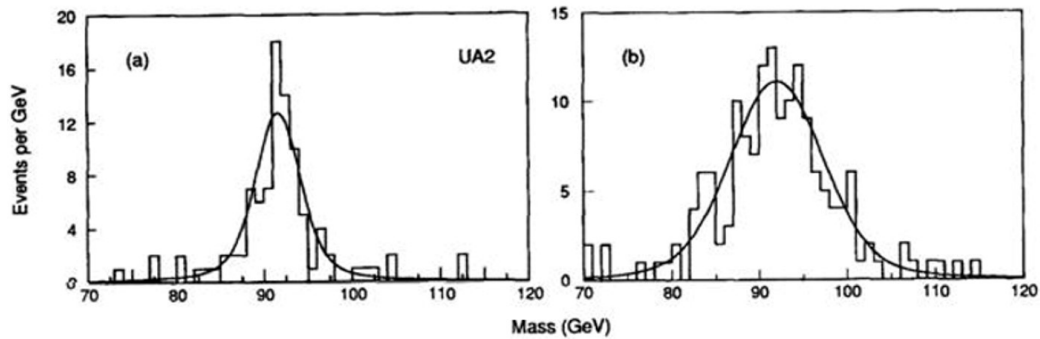

**Figure 28.** Invariant mass distributions for two $Z \to e^+e^-$ event samples, as measured by UA2 (see text). The curves are best fits to the data using $m_Z$ as a free parameter.

The precise determination of $m_W$ was used to obtain bounds on the top quark mass, for which early direct searches at the CERN and Fermilab $p\bar{p}$ colliders had only provided the lower bound $m_{top} > 89 \, \text{GeV/c}^2$ [20–22]. As shown by Veltman [23], within the frame of the Standard Model the value of $m_W$ for fixed $m_Z$ depends quadratically on the mass of the top quark through electroweak radiative corrections from virtual fermion loops (and also, to a much smaller extent, on the mass of the Higgs boson). As illustrated in Figure 29, the UA2 result gave

$$m_{top} = 160^{+50}_{-60} \, \text{GeV/c}^2 \tag{15}$$

suggesting a heavy top quark well before its discovery at the Fermilab $p\bar{p}$ collider with a measured mass $m_{top} = 174 \pm 10 \pm 13 \, \text{GeV/c}^2$ [24] (the present world average of all measurements from the experiments at the Fermilab $p\bar{p}$ collider and at the LHC, is $m_{top} = 172.57 \pm 0.29 \, \text{GeV/c}^2$ [25]).

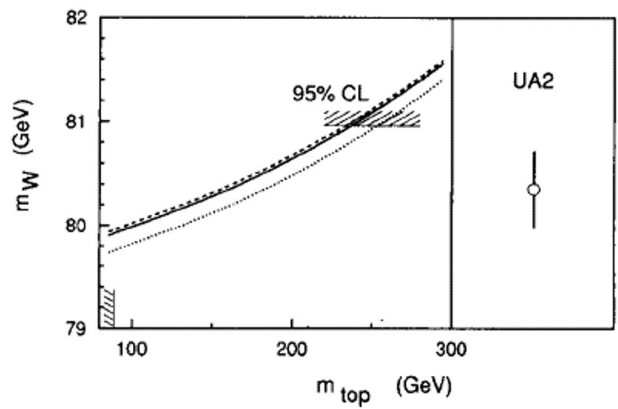

**Figure 29.** $m_W$ versus $m_{top}$ (curves on the left) and the determination of $m_W$ obtained by combining the precise UA2 measurement of $m_W/m_Z$ with an early precise measurement of $m_Z$ at LEP (the open point with error bar on the right). The curves are the Standard Model predictions for fixed $m_Z$ (as measured at LEP), and for different values of the Higgs boson mass [23]: 50 GeV (dashed curve); 100 GeV (solid curve); and 1000 GeV (dotted curve). Also shown are the 95% confidence level (CL) upper bound $m_{top} < 250$ GeV obtained from the error on $m_W$, and the lower bound $m_{top} > 89$ GeV from early direct searches at the CERN and Fermilab colliders.

## 6. Conclusions

The CERN proton-antiproton collider was initially conceived as an experiment to detect the $W$ and $Z$ bosons. Not only it beautifully fulfilled this task, but it also tested the electroweak theory to a level of few percent and provided important verifications of QCD predictions. In the end, it turned out to be a first-class general-purpose accelerator facility with a very rich physics programme, opening the way to the approval of LHC construction.

**Conflicts of Interest**

The author declares no conflict of interest.

**References**

1. Rubbia, C.; McIntyre, P.; Cline, D. In Proceedings of the International Neutrino Conference, Aachen, Germany, 8–12 June 1977; p. 683.
2. Haidt, D. The discovery of neutral currents. *Eur. Phys. J. C* **2004**, *34*, 25.
3. Hasert, F.J.; Faissner, H.; Krenz, W.; et al. Search for elastic muon-neutrino electron scattering. *Phys. Lett. B* **1973**, *46*, 121.
4. Hasert, F.J.; Kabe, S.; Krenz, W.; et al. Observation of neutrino-like interactions without muon or electron in the gargamelle neutrino experiment. *Phys. Lett. B* **1973**, *46*, 138.
5. van der Meer, S. *Stochastic Damping of Betatron Oscillations in the ISR*; CERN-ISR-PO 72-31; CERN: Geneva, Switzerland, 1972.
6. Van der Meer, S. Stochastic cooling and the accumulation of antiprotons. *Rev. Mod. Phys.* **1985**, *57*, 689.
7. Carron, G.; Herr, H.; Koziol, H.; et al. Stochastic cooling tests in ice. *Phys. Lett. B* **1978**, *77*, 353.
8. Astbury, A. Calorimeter facilities. *Phys. Scr.* **1981**, *23*, 397.
9. Arnison, G.; Astbury, A.; Aubert, B.; et al. Experimental observation of isolated large transverse energy electrons with associated missing energy at s = 540 GeV. *Phys. Lett. B* **1983**, *122*, 103.
10. Banner, M.; Battiston, R.; Bloch, P.; et al. Observation of single isolated electrons of high transverse momentum in events with missing transverse energy at the CERN. *Phys. Lett. B* **1983**, *122*, 476.
11. Arnison, G.T.; Astbury, A.; Aubert, B.; et al. Experimental observation of lepton pairs of invariant mass around 95 GeV/c$^2$ at the CERN SPS collider. *Phys. Lett. B* **1983**, *126*, 398.
12. Bagnaia, P.; Banner, M.; Battiston, R.; et al. Evidence for $Z^0 \to e^+e^-$ at the CERN pp collider. *Phys. Lett. B* **1983**, *129*, 130.
13. Albajar, C.; Albrow, M.G.; Allkofer, O.C.; et al. Studies of intermediate vector boson production and decay in UA1 at the CERN proton-antiproton collider. *Z. Phys. C* **1989**, *44*, 15.
14. Albajar, C.; Albrow, M.G.; Allkofer, O.C.; et al. Events with large missing transverse energy at the cern collider: I.W$\to \tau \upsilon$ decay and test of $\tau$ $\mu$ e universality at $Q^2 = mw^2$. *Phys. Lett. B* **1987**, *185*, 233.
15. Ansari, R.; Bagnaia, P.; Banner, M.; et al. Measurement of the standard model parameters from a study of W and Z bosons. *Phys. Lett. B* **1987**, *186*, 440.
16. Altarelli, G.; Ellis, R.K.; Greco, M.; et al. Vector boson production at colliders: A theoretical reappraisal. *Nucl. Phys. B* **1984**, *246*, 12.
17. Alitti, J.; Ambrosini, G.; Ansari, R.; et al. An improved determination of the ratio of W and Z masses at the CERN $pp$ collider. *Phys. Lett. B* **1992**, *276*, 354.
18. Carter, J. In Proceedings of the Joint International Lepton-Photon Symposium and Europhysics Conference on High Energy Physics, Geneva, Switzerland 25 July–1 August 1991; Volume 2, p. 3.
19. Abe, F.; Amidei, D.; Apollinari, G.; et al. Measurement of the $W$-boson mass in 1.8-TeV $\bar{p}p$ collisions. *Phys. Rev. D* **1991**, *43*, 2070.
20. Åkesson, T.; Alitti, J.; Ansari, R.; et al. Search for top quark production at the CERN $\bar{p}p$ collider. *Z. Phys. C* **1990**, *46*, 179.
21. Albajar, C.; Albrow, M.G.; Allkofer, O.C.; et al. Search for new heavy quarks in proton-antiproton collisions at $\sqrt{s} = 0.63$ TeV. *Z. Phys. C* **1990**, *48*, 1;
22. Abe, F.; Amidei, D.; Apollinari, G.; et al. Search for the top quark in the reaction $\bar{p}p \to$electron+jets at $\sqrt{s} = 1.8$ TeV. *Phys. Rev. Lett.* **1990**, *64*, 142.
23. Veltman, M. Limit on Mass Differences in the Weinberg Model. *Nucl. Phys. B* **1977**, *123*, 89.
24. Abe, F.; Albrow, M.G.; Amendolia, S.R.; et al. Evidence for top quark production in $\bar{p}p$ collisions at $\sqrt{s} = 1.8$ TeV. *Phys. Rev. D* **1994**, *50*, 2966.
25. Navas, S.; Amsler, C.; Gutsche, T.; et al. Review of Particle Physics. *Phys. Rev. D* **2024**, *110*, 030001.

*Perspective*

# A Personal History of CERN Particle Colliders (1972–2022)

Stephen Myers

European Organization for Nuclear Research (CERN), 1217 Geneva, Switzerland

**Abstract:** In the following, I recall my personal memories of the conception, design, construction and operation of particle accelerators and particle colliders over the 50 years from 1972 to 2022. This is not a technical report, but more a review to give an insight to the historical beginnings and endings of some of the world's most technically complex and expensive scientific instruments.

**Keywords**: ISR; SP$\overline{\text{P}}$S; LEP; LHC

## 1. Some Fundamentals

### 1.1. Colliders vs Stationary Target

Advances in accelerator technology have driven higher beam energies, often spurring new tech and cross-disciplinary innovations. A key development was the use of colliders**,** where two beams collide head-on, enabling much higher collision energies than single beams impinging on stationary targets.

However, collider technology is more complex. The two counter-circulating beams bring to collision packets of particles. This implies that high-intensity, tightly focused beams are needed to produce sizable interaction rates. To preserve the lifetime of circulating beams, ultra-high vacuum ($<10^{-9}$ Pa) is also essential to reduce beam-gas interactions.

### 1.2. Two Ring vs. Single Ring Colliders

If the collider has two separate rings, then the beams can be electromagnetically steered into collisions at a predefined number of locations thereby avoiding the problem, in a single-ring collider with beam lifetime due to the large number of collisions points (normally twice the value of the number of bunches). The disadvantages of a two-ring collider are technical complexity and cost which may be increased by up to a factor of two. However, with two separate rings the collision rate can be significantly increased.

### 1.3. The Non-Linear Beam-Beam Effect

In storage ring colliders, beams collide to produce high-energy interactions. Each charged beam generates electromagnetic fields that affect the opposing beam. Close to the beam center, this effect is approximately linear, acting like a focusing (or defocusing) magnet—this is known as the beam-beam tune shift. However, at greater distances, the effect becomes highly non-linear, degrading collider performance. There is a critical upper limit to the tune shift, beyond which performance declines, making the beam-beam limit a fundamental constraint in collider design.

### 1.4. High Energy Superconducting Hadron Colliders

A hadron collider's maximum energy depends on the ring's size and the strength of its magnetic bending fields. While ring size is limited by cost and geography, magnetic field strength is constrained by technology. Superconducting magnets, which enable higher fields and lower power use, have been key to increasing beam energy—though they come with added complexity and cost.

## 2. ISR the First pp and p$\overline{\text{p}}$ Collider

The CERN Intersecting Storage Rings (ISR) was the first proton–proton (and p$\overline{\text{p}}$) collider ever constructed and operated. The ISR collided first beams in 1971 and operation for physics was from 1971 to 1983. Table 1 gives the basic parameters of the ISR.

**Table 1.** Basic parameters of the ISR.

| | |
|---|---|
| Colliding particles | pp, dd, pd, αα, αp, ppbar |
| Particle momentum | 3.5 to 31.4 GeV/c |
| Circumference (m) | 942.5m (300 π) |
| Number of main magnets | 132/ring |
| Magnetic dipole field | 1.33T (max) |
| Length of main magnets | 4.88/2.44 m |
| Betatron oscillations/turn | 8.9 (h), 8.88 (v) |
| β* (h/v) | 21 m /12 m |
| β* (h/v) | 2.5 m / 0.28 m in sc low-beta section |
| RF system per ring | 7 cavities, 9.5 MHz, 16 kV RF peak voltage |

The ISR consisted of two independent storage rings intersecting at eight points with a crossing angle of 14.8 degrees (see Figure 1, the control room, and Figure 2, interaction points 1 and 8). The circumference of the rings was 943 metres (1.5 times that of the CERN Proton Synchrotron (PS) which supplied particles to the ISR). The larger circumference was needed to allow space for the long straight sections in the interaction regions and the injection sections. The first proton–proton collisions took place in 1971 with beam momenta up to 26.5 GeV/c, which is the maximum momentum available from the PS injector.

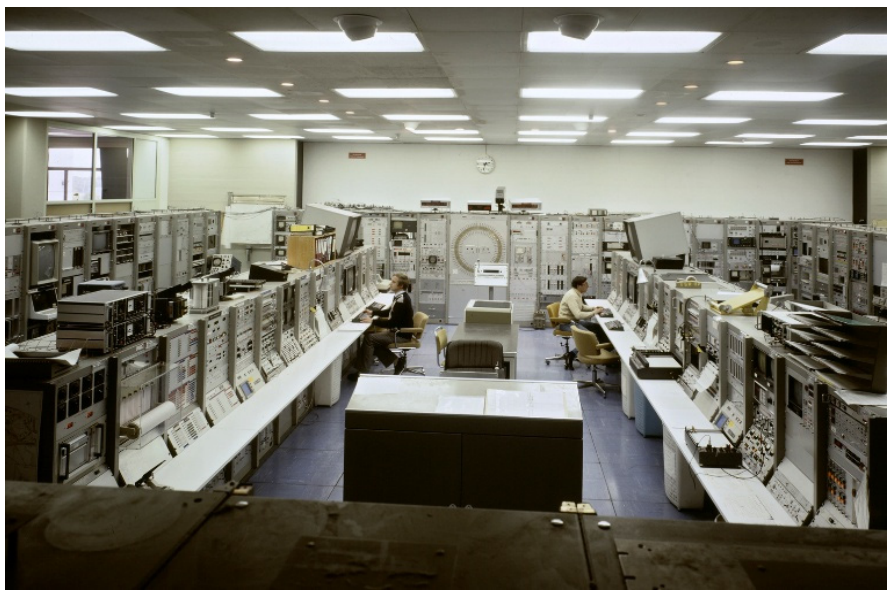

**Figure 1.** ISR Control Room.

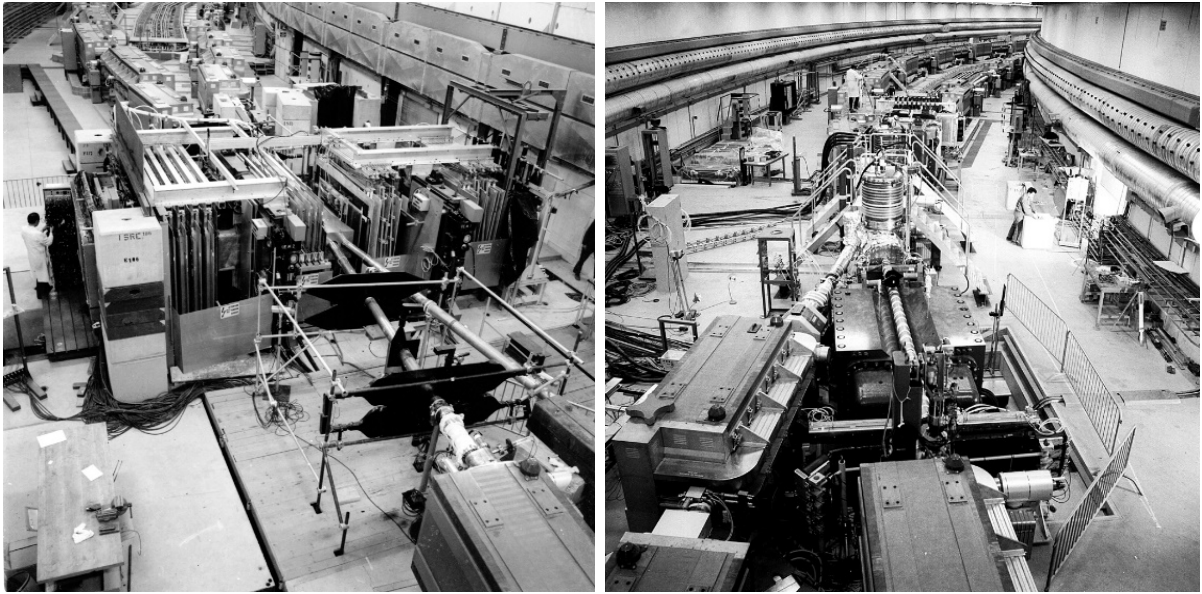

**Figure 2.** Physics Interaction points 1 (**left**) and 8 (**right**).

### 2.1. Lack of Diagnostics in ISR Due to Coasting Beam

High-precision beam diagnostics are crucial for particle colliders, enabling measurement of particle count, beam size, and position. Beams are grouped into highly charged packets called bunches, which interact with electromagnetic fields. Diagnostics detect the electromagnetic signals induced by these charged bunches as they pass, allowing evaluation of beam properties such as proton count, bunch position, and length.

The ISR was a "storage" ring. This meant it took a batch of (20) bunches from the injector PS, accelerated them to the storage aperture in the ISR and then "let them go". Each batch of bunches were "let go" by switching off the RF acceleration system to allow the RF system to capture and accelerate the next batch. This meant that the stored beam in the ISR was a "coasting" rather than a bunched beam. Hence none of the normal e-m beam diagnostics devices could be used to measure the parameters of the coasting beam. This was a huge disadvantage for efficient operation for physics.

At the start of ISR operation, operators could "see" the coasting beam using only two devices: a real-time transverse profile monitor and a DC current monitor [1]. The profile monitor employed a thin sodium gas curtain at one azimuthal position, where circulating protons ionized the gas, enabling a camera to capture the beam's cross-section. Meanwhile, the precise DC current monitor measured the total beam current, including both bunched and unbunched components.

### 2.2. Stacking

High currents in the ISR were built through momentum "stacking," accumulating about 200 PS batches across its large momentum aperture. Each cycle captured 20 PS bunches at −2% injection momentum, accelerated them to +2% by changing RF frequency, then debunched the beam by switching off the RF. The maximum single beam current reached an impressive 57 amperes.

### 2.3. Phase Displacement Acceleration

Phase displacement occurs when an RF "bucket" (phase stability area) traverses a debunched beam [2]. The particles in the debunched beam travel around the unstable trajectories associated with the bucket. Traversing a debunched beam from high momentum to low momentum produces an increase in the average momentum of the debunched beam by an amount equal to the phase space area of the phase displacing buckets. A useful analogy is the release of droplets of mercury (RF buckets) into a cylindrical container containing some water (coasting beam). The mercury droplets go from high energy to low energy and the water energy is increased by displacement.

Since the ISR circumference was larger than the PS, the maximum possible momentum was also higher (31.4 compared to 26.6 GeV/c). In the quest for higher collision energies, it was decided to attempt to increase the momentum of the accumulated beam in the ISR from 26.6 to 31.4 GeV/c. However, the small ISR RF (16kV maximum) system could only capture a tiny amount of the coasting which had a 3% momentum spread. Hence phase-displacement was the only option. The phase-displacement technique required several hundred traversals of the coasting beam by the RF system.

So, in our relative ignorance of the problems (space charge, changing tunes, chromaticity, orbits, RF noise effects, absence of diagnostics...) we decided to attempt to phase-displace high intensity stacks of protons. Initially the progress was very slow and frustrating, but after some better understanding and a few breakthroughs, 31.4 GeV/c became the preferred high luminosity operational momentum of the ISR during the latter years of operation. In the last years of ISR operation, beams of more than 30 amperes were accelerated by phase displacement to 31.4 GeV/c.

### 2.4. Schottky Scans

The Schottky noise spectra [3] result from the discrete nature of the particles in the beam. A sensitive high frequency longitudinal pick-up with long-term signal processing, could produce a signal proportional to the longitudinal phase space density of the debunched beam. Figure 3 shows the first Schottky scans taken operationally in the ISR. The three scans shown were taken at beam currents of 10, 15 and 19.2 amperes. The horizontal axis is the longitudinal frequency and allows evaluation of the beam $\Delta p/p$.

Soon after discovering longitudinal Schottky scans, transverse pickups were used to measure the transverse Schottky scans which gave information about the "tune" values in the stacked beams.

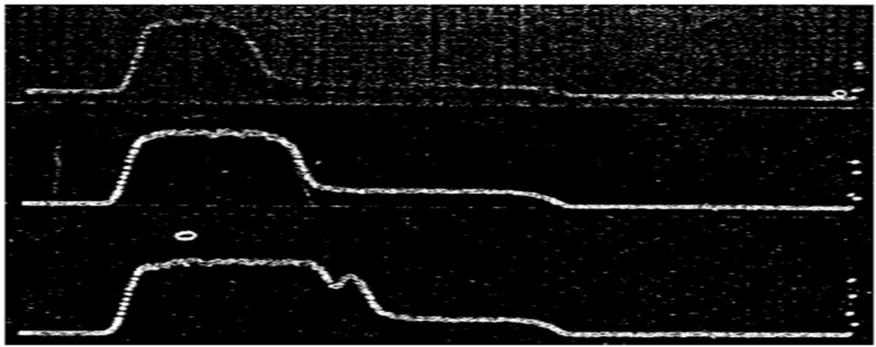

**Figure 3.** Longitudinal Schottky scans.

Schottky scans revolutionized ISR operation by providing the only quantitative beam diagnostics during long stable-beam fills, sometimes lasting up to five days. While the sodium curtain offered visual but non-quantitative cross-section views, Schottky scans enabled evaluation of longitudinal density versus $\Delta p/p$ using current meter data. Identifiable "markers" in the beam, initially at stack edges and later placed via phase displacement, allowed precise tune measurements at discrete momentum offsets, enabling continuous tracking of the tune "working line" throughout physics runs.

### 2.5. Stochastic Cooling

The first ISR Schottky scans sparked interest in damping particle oscillations via stochastic cooling. The technique was tested by Wofgang Schnell [4] that followed the initial idea proposed by Simon Van Der Meer [5]. Schnell's team built a test system demonstrating this concept. At the ISR the most sensitive measurement of transverse beam size was obtained through the normalized luminosity, inversely related to the beam "height".

Figure 4 shows the results of the first ever conclusive observation of stochastic cooling (in an ISR machine physics experiment [4]).

The inverse normalized luminosity (effective beam height) is shown over a 13-h period with stochastic cooling turned on and off every few hours. The effect is small but very significant: stochastic cooling worked! Very soon afterwards a similar system was designed for the Initial Cooling Experiment (ICE) with spectacular results as shown in Figure 5.

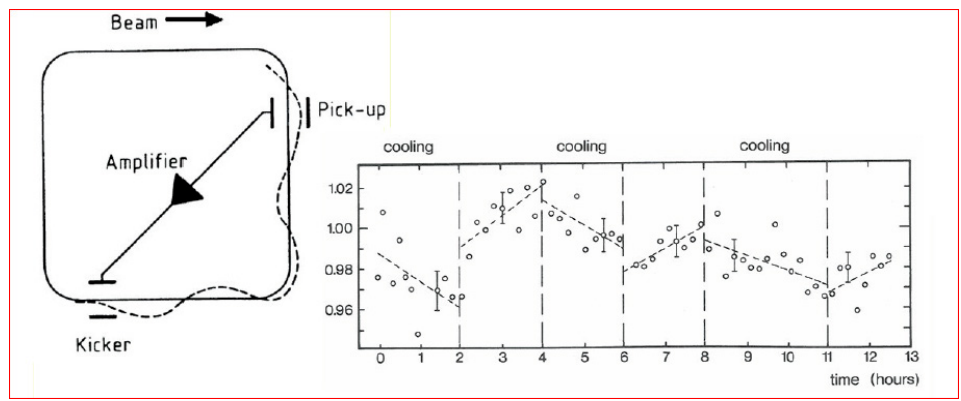

**Figure 4.** First observation of stochastic cooling.

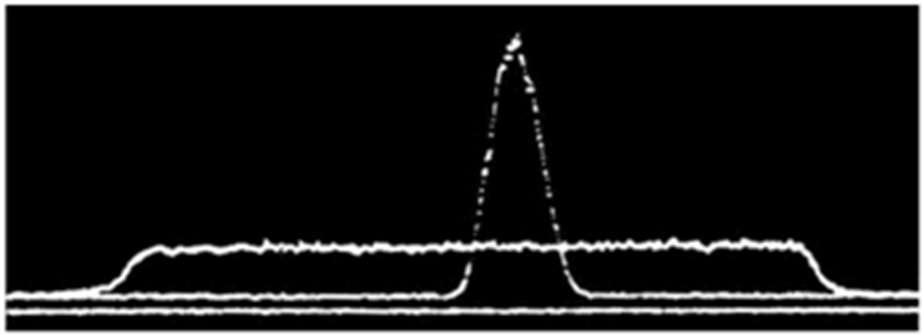

**Figure 5.** Fast momentum cooling in ICE.

## 2.6. The Legacy of the ISR

Although not known for discoveries, the ISR was a pioneering project as the first proton-(anti)proton collider, answering key questions and inspiring future collider designs. It also served as an excellent training ground for accelerator physicists. Below is a list of its most important lessons and legacy.

- Type of collider detectors

Before the ISR particle detectors were highly specialized, some looking at small angle events, others focused on high transverse momentum events etc. After the ISR all detectors were general-purpose "4π" systems aiming at collecting the maximum number of simultaneous events produced in the collisions. An immediate consequence of this was the UA1 and UA2 detectors for the SP$\overline{\text{P}}$S.

- Experimental proof of stochastic cooling.

Although the two previously mentioned legacies had, in my opinion, the greatest impact, the complete list of contributions to accelerator physics, operations, and engineering is long and impressive. Many of these have been successfully implemented by more modern accelerators. In an arbitrary order of importance or priority the legacies from the ISR are:

- A fantastic training ground for accelerator physicists and engineers.
- Safe operation of beams with high destructive stored energy.
- Background control by collimation.
- Absolute luminosity calibration using "Van der Meer scans".
- Longitudinal phase space stacking of protons.
- Production and maintenance of ultra-high vacuum in the presence of high intensity. beams (clearing electrodes, electron cloud, etc.)
- Space charge tune compensation.
- Impact of non-linear resonances driven by machine imperfections.
- Pulsed beam-beam effects (Overlap knock-out resonances).
- Instabilities and impedance.
- Phase displacement acceleration and deceleration.
- Proton-antiproton collisions: demonstrated for the first time.
- High precision, low noise, power converters.

- Low beta insertions.
- Computer control of acceleration in colliders.
- Optics measurements and corrections.
- Stacking and phase displacement of different particle species, protons, anti-protons, alfas, deuterons etc.

## 3. The CERN SPS Proton—Antiproton Collider

### 3.1. The Importance of Stochastic Cooling

Luminosity in a collider requires small beam sizes, but the antiproton beam for the SP$\overline{\text{P}}$S was naturally much larger than needed. Stochastic cooling, proven in the ISR for one plane, had to be extended to all three dimensions for success. The 1977-approved Initial Cooling Experiment (ICE) demonstrated 3D cooling by 1978, leading to the green light for the SPS proton–antiproton project. Following ICE's success, the Antiproton Accumulator (AA) was built rapidly (1979–1980) with strong stochastic cooling to prepare antiprotons for collisions.

### 3.2. Conversion of the SPS to the SP$\overline{\text{P}}$S

The SPS had been built as a proton synchrotron; not as a collider. The following upgrades were required if the SP$\overline{\text{P}}$S project was to be successfully operated as a p$\overline{\text{p}}$ collider.

- A new beam line was needed, to transfer the antiprotons from the PS to the SPS, and a new injection system for counterclockwise injection was added in the SPS.
- The SPS had been built for an injection energy of 14 GeV/c. The proton transfer line, TT10, and the injection system had to be upgraded to 26 GeV/c.
- To provide sufficient beam lifetime for stored beams, the vacuum system had to be improved by two orders of magnitude, from the design vacuum of the SPS (200 nTorr) to better than 2 nTorr,
- To increase the luminosity, tight focussing low-beta insertions were needed in straight sections 4 and 5 for the UA2 and UA1 experiments
- Beam diagnostics had to be improved to measure the beam parameters with the very low beam intensities, and new devices added, such as directional couplers for independent observation of protons and antiprotons.
- To reduce the beam-beam effect, high voltage electrostatic deflectors were required to separate the beams, in 9 of the 12 crossing points,
- The RF system had to be upgraded with reduced "RF noise".
- New transfer lines to and from the AA and from the PS to the SPS (TT70) were required.

The first proton–antiproton collisions were recorded in summer 1981, with the first physics run producing 0.2 nb$^{-1}$ of integrated luminosity by the end of the year. Although initially low, the luminosity was sufficient to discover the W and Z bosons in 1982–1983, earning Carlo Rubbia and Simon van der Meer the 1984 Nobel Prize. Key to this success were the large 4π detectors absent at the ISR. The performance of the accelerator was improved with the Anti-proton Collector (AC), boosting antiproton density and shortening cooling times. Beam energy rose from 273 GeV (1982–1985) to 315 GeV (1987–1991), when the program ended. The SP$\overline{\text{P}}$S was a highly ambitious and successful CERN project.

### 3.3. The Legacy of the SP$\overline{\text{P}}$S

Before its commissioning, doubts existed about operating a hadron collider with bunched beams due to beam–beam effects and RF noise. Although low-intensity bunched beams were tested in the ISR, the SP$\overline{\text{P}}$S proved such concerns unfounded for future colliders. Proton and antiproton orbits were separated using a "Pretzel" shape with electrostatic separators—a method later used at LEP and LEP2. Additionally, civil engineering experience from building large experimental caverns at SP$\overline{\text{P}}$S aided construction for LEP and LHC.

## 4. The Large Electron Positron Collider (LEP)

Over the past decades electron positron colliders have been ideal tools for studying mesons ($J/\psi$, Y) and leptons (τ). Although the actual discoveries have often occurred at proton machines, the precise and easily tunable beam energy, as well as the well defined initial state, are big assets of electron positron colliders.

Following the prediction of the existence of two massive vector bosons, the neutral Z$^0$ and the charged W$^\pm$, LEP was designed with the aim of discovering and studying those bosons, which were, however, first observed at the SPS proton-antiproton collider in 1982. With LEP it was possible to measure the properties of these bosons with excellent precision. A very important early result was that there are three types of light neutrinos and thus three fundamental fermion families. The precise determination of the standard model parameters from the LEP data allowed a prediction of the top mass and limits on the expected mass range for the standard Higgs boson.

Design studies of the LEP machine started at CERN in 1976 and the first practical design was published in 1978. The proposed machine had a cost-optimized energy of 70 GeV per beam and measured 22 km in circumference. After extensive discussions during the autumn of 1978 it was decided to embark on the design of a somewhat larger machine, 30 km in circumference, with a cost-optimized energy of about 90 GeV per beam. The energy of both these machines could be extended by using super-conducting RF cavities, were these to become available, to 100 GeV and above.

Studies of the 30 km machine were completed during 1979 and a design report was issued in August of that year. These studies covered not only machine design but also the design and development of the components of LEP. A much cheaper design for the main magnet system was developed, as well as a more economical system for the RF accelerating system using a storage cavity scheme. At the same time, it was decided to increase the effort on the development of super-conducting RF cavities for LEP by setting up a small team at CERN and establishing a collaboration with other European laboratories. The basic feature of the final design was a machine with a large circumference which could be installed in stages to match the particle physics requirements and new technological developments.

### 4.1. Beam-Beam Effect

Before LEP, electron-positron colliders struggled with beam-beam effects. Drawing on ISR experience, I developed a Monte Carlo simulation to track a large number of particles over many revolutions to study this impact on LEP. Thanks to CERN's upgraded computing power, I overcame past limitations. After extensive development and debugging, the simulation successfully matched experimental data from smaller colliders.
Insertion: a Proton-Proton Collider in the LEP Tunnel.

*(LEP Note 440 April 1983)*

*By mid-1981, the first $p\bar{p}$ collisions were observed in the SP$\bar{P}$S at CERN. It was now very clear that ISABELLE (BNL) was obsolete, and that the USA was pursuing a much higher energy collider (SSC).*

*Following an invitation to represent CERN at an SSC meeting in the US, I began to think about the 27 km tunnel we would have in Geneva and what sort of proton collider we could imagine installing there.*

*After some calculations and discussions, Wolfgang Schnell and I co-authored "LEP" note 440 entitled "Preliminary performance estimates for a LEP proton collider", which was published in April 1983 [7]. This was a short, 16-page report that provided performance estimates and limitations for the design of a proton collider in the LEP tunnel. As far as I am aware, this was the first document to address performance issues of the LHC. It raised many of the points that were subsequently part of the LHC design: 8 TeV per beam, beam–beam limitation (arguing the case for a twin-ring accelerator), twin-bore magnets and the need for magnet development, problems with pile-up (multiple collisions per bunch-crossing) and impedance limitations.*

*We continued, and wrote additional "LEP" notes (450, 460, and 470) on the parameters for other sub-systems for the LHC and then abandoned this subject to return to the design of LEP.*

*Nevertheless, these reports were never cited in any of the reports from the CERN LHC study group.*

*However, Burt Richter (Nobel Laureate 1976) referred to the LEP notes in 2014 [8]*

*"The Myers/Schnell paper started informal discussions at CERN that became more serious when the SSC was initially approved by the U.S. Congress, and turned into a major design effort when the SSC was cancelled by the U.S. Congress in 1993…."*

### 4.2. 1985 Start of LEP Construction

The construction of the LEP tunnel started in 1985 following a standard public enquiry in France (Figure 6a) shows the layout of LEP).

LEP's tunnel, then the longest built, faced disaster just 2 km in when unsuitable rock required blasting, causing a high-pressure underground river that delayed work for six months (Figure 6b). After many failed attempts, a solution was found by June 1987, allowing tunnel completion and accelerator installation. This incident complicated the original smooth construction schedule and served as a tough lesson for future collider tunnels.

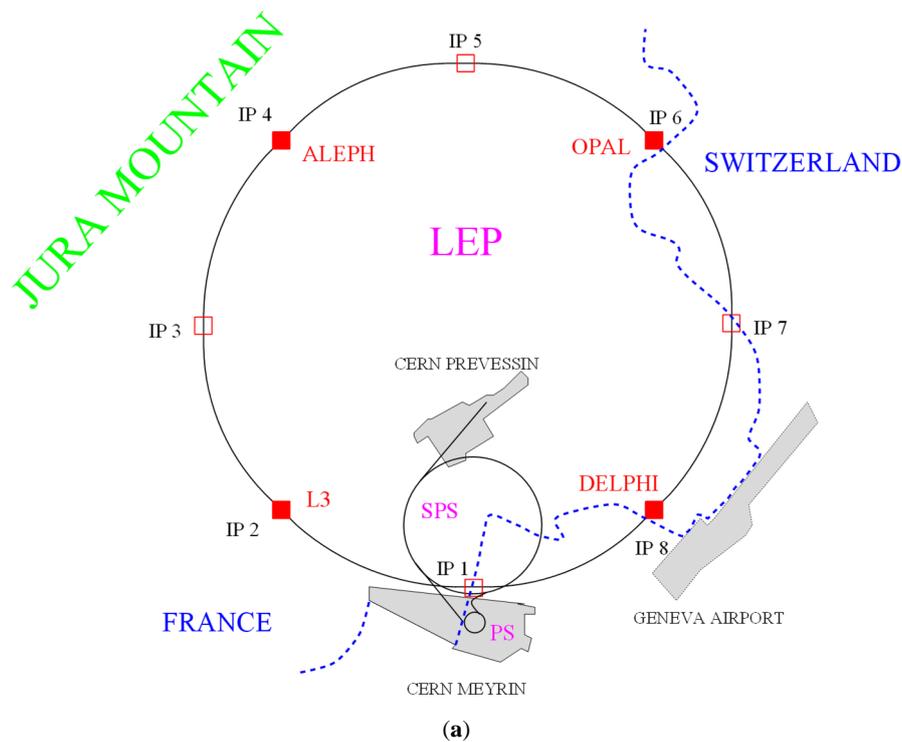

(**a**)

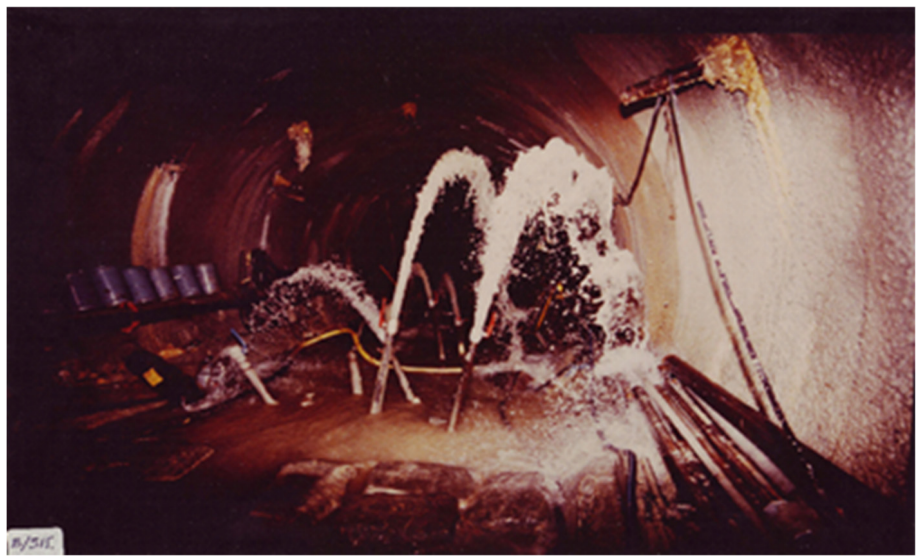

(**b**)

**Figure 6.** (**a**) Layout of the LEP ring, the 4 experiments and the LEP injectors. (**b**) Photo of water input to LEP tunnel.

### *4.3. 1988 LEP Octant Test*

The first major task was the controversial octant test—passing a positron beam through the first eighth of the accelerator. Despite skepticism from CERN leadership, octant 8 was completed by July 1988, shortly after finishing difficult Jura excavation. On 12 July, four positron bunches traveled 2.5 km successfully. Early beam measurements revealed harmful betatron coupling caused by a nickel magnetic layer inside the vacuum chamber—a crucial defect discovered a year before the first full beam circulation.

### *4.4. 1989 First Circulating Beam in LEP*

At 23:45, on 14 July 1989, the beam made its first turn on the first attempt. Soon afterwards the beam was circulating many turns, and we were ready to fine-tune the multitude of parameters needed to prepare beams for physics. The beams were brought into first collisions on August 14. After ten agonising minutes, we heard the long-awaited comment: "We have the first Z⁰!" A period of machine studies followed, allowing big improvements to be made in the collider's performance (on 14 July 1989—the 200th anniversary of the seizure of the Bastille).

*4.5. LEP Operations*

LEP remains the highest energy electron-positron collider ever built. It was commissioned in 1989 and finished operation in November 2000. During this period, it was operated in different modes, with different optics, at different energies, all with excellent performance. In the end, LEP surpassed all design parameters. It provided a large amount of data for the precision study of the standard model, first on the Z resonance, and then above the W pair threshold. Finally, with beam energies above 100 GeV, a tantalizing glimpse of what might have been the Higgs boson was observed.

### 4.5.1. LEP Performance

Performance at LEP was divided naturally into two regimes: 45.6 GeV per beam running around the Z boson resonance and high energy running above the threshold for W pair production. A summary of the performance through the years is shown in Table 2 below [9–12].

**Table 2.** Overview of LEP performance from 1989 to 2000. Note $E_b$ is the energy per beam, $k_b$ the number of bunches per beam and $\zeta$ is the peak luminosity. $\int \zeta dt$ is the luminosity integrated per experiment over each year and $I_{tot}$ is the total beam cur- rent 2 $k_b I_b$. The luminosity C is given in units of $10^{30} \mathrm{cm}^{-2} \mathrm{s}^{-1}$.

| Year | $\int \zeta dt$ (pb$^{-1}$) | $E_b$ (GeV/c2) | $k_b$ | $I_{tot}$ (mA) | $\zeta$ |
|------|------|------|------|------|------|
| 1989 | 1.74 | 45.6 | 4 | 2.6 | 4.3 |
| 1990 | 8.6 | 45.6 | 4 | 3.6 | 7 |
| 1991 | 18.9 | 45.6 | 4 | 3.7 | 10 |
| 1992 | 28.6 | 45.6 | 4/8 | 5.0 | 11.5 |
| 1993 | 40.0 | 45.6 | 8 | 5.5 | 19 |
| 1994 | 64.5 | 45.6 | 8 | 5.5 | 23.1 |
| 1995 | 46.1 | 45.6 | 8/12 | 8.4 | 34.1 |
| 1996 | 24.7 | 80.5–86 | 4 | 4.2 | 35.6 |
| 1997 | 73.4 | 90–92 | 4 | 5.2 | 47.0 |
| 1998 | 199.7 | 94.5 | 4 | 6.1 | 100 |
| 1999 | 253 | 98–101 | 4 | 6.2 | 100 |
| 2000 | 233.4 | 102–104 | 4 | 5.2 | 60 |

In the regime on or around the Z resonance, performance was constrained by the beam-beam effect which limited the bunch currents that could be collided. The beam-beam effect blew up beam sizes and the beam-beam tune shift saturated at around 0.04. Optimization of the transverse beam sizes was limited by beam-beam driven effects such as "flip-flop". The main breakthroughs in performance at this energy was an increase in the number of bunches. First with the Pretzel scheme (8 bunches per beam) commissioned in 1992, and then with the bunch train scheme (up to 12 bunches per beam) used in 1995. The optics (phase advance and tunes values) were also changed in attempts to optimize the emittance and the beam-beam behaviour.

With the increase in energy to above the W pair threshold, the beam-beam limit increased, and the challenge was to develop a low emittance optics with sufficient dynamic aperture to go to the 100 GeV regime. Luminosity production was maximized by increasing the bunch current to the limit while operating with four bunches per beam and rigorous optimization of vertical and horizontal beam sizes.

Between 1996 and 2000, LEP's beam energy increased from 80.5 to 104.4 GeV. At these energies, strong damping and photon emission reduced beam blow-up, allowing higher bunch currents and record beam-beam tune shifts above 0.08 at all four collision points. Superconducting cavities were pushed beyond design to provide over 3.6 GV accelerating voltage per turn. The progression in accelerating voltage can be seen in Figure 7. LEP surpassed all design goals, with peak luminosity nearly four times higher than expected. The design and achieved values for a number of crucial LEP performance parameters are summarized in Table 3.

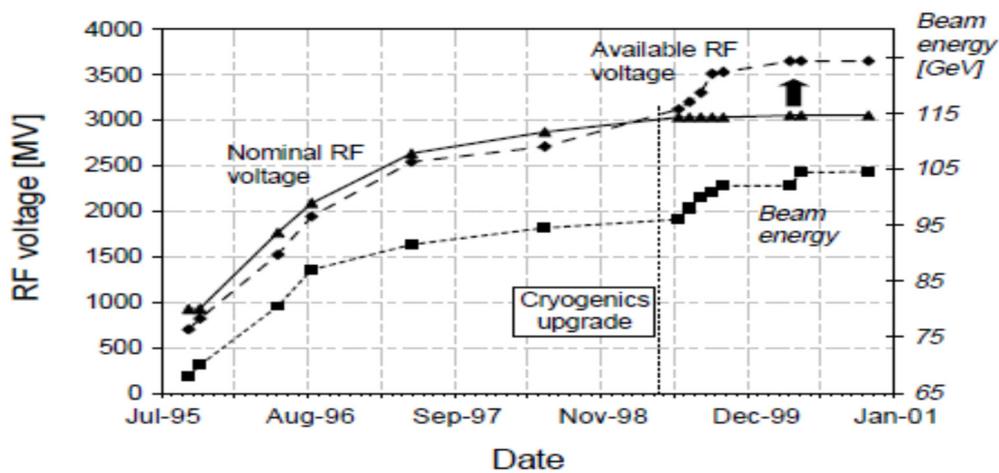

**Figure 7.** RF voltage per turn over the years.

**Table 3.** LEP Performance parameters.

| Parameter | Design (55/95 GeV) | Achieved (40/98 Gev) | | |
|---|---|---|---|---|
| Bunch current | 0.75 mA | 1.00 mA | | |
| Total beam current | 6.0 mA | 8.4/6.2 mA | | |
| Vertical beam-beam parameter | 0.03 | 0.045/ 0.083 | $\times 10$ | |
| Emittance ratio | 4.0% | 0.4 % | $\times 1.4/3.7$ | |
| Maximum luminosity | 16/27 $10^{30}$ cm$^3$s$^{-1}$ | 23/ 100 $10^{30}$ cm$^3$s$^{-1}$ | | |
| IP beta function b$_x$ | 1.75 m | 1.25 m | | |
| IP beta function b$_y$ | 7.0 cm | 4.0 cm | | |

### 4.5.2. Competition from the USA

There was also fierce competition from the more innovative Stanford Linear Collider (SLC) in California. But LEP got off to a fantastic start and its luminosity increase was much faster than at its relatively untested linear counterpart.

### 4.5.3. Performance Summary

LEP's performance was fundamentally limited by the beam-beam interaction, which directly affects luminosity. While the highest tune shift before LEP was 0.045, LEP achieved a record 0.083 at high energies, thanks to very fast synchrotron damping time.

### *4.6. LEP2*

In 1995, the LEP2 upgrade began, allowing beam energies above the WW threshold (161 GeV). As LEP2 Project Leader from 1996, I oversaw the construction of 288 superconducting cavities. LEP2 exceeded design goals in luminosity and beam energy, reaching 104.4 GeV per beam with over 3.5 GV RF voltage to offset synchrotron losses. Continuous improvements raised luminosity by increasing bunch intensity and tightening beam focus, though performance was ultimately limited by beam-beam nonlinear forces. The luminosity performance of LEP over the eight-year period 1993–2000 [13] is shown in the Figure 8 below. This performance was impressive given that the operational mode was changed every single year.

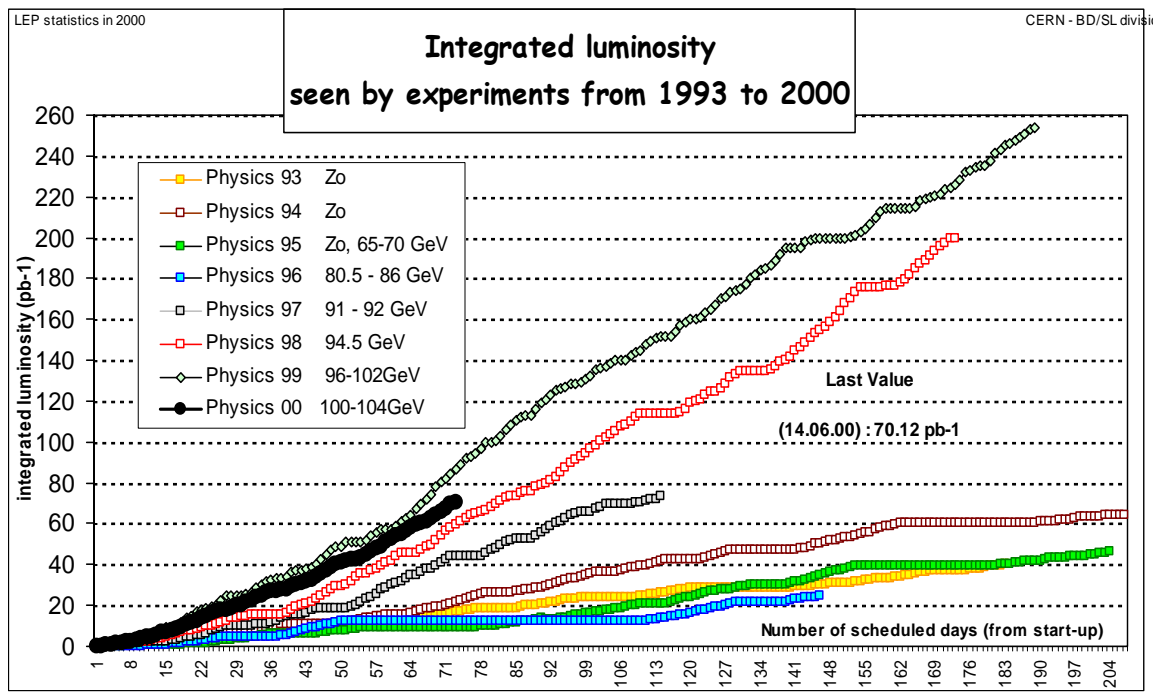

**Figure 8.** LEP daily integrated luminosity from 1993 to 2000.

Insertion: 2000, The Last Year of LEP2 Operation: On the verge of a great discovery?

*LEP's days were never fated to dwindle. Early on, CERN had a plan to install the Large Hadron Collider in the same tunnel, in a bid to scan ever higher energies and be the first to discover the Higgs boson. However, on 14 June 2000, LEP's final year of scheduled running, the ALEPH experiment reported a possible "Higgs" event during operations at a centre-of-mass energy of 206.7 GeV. On 31 July and 21 August ALEPH reported second and third events corresponding to a supposed reconstructed Higgs mass in the range 114–115 GeV/c².*

LEP was set to stop in mid-September, with two reserve weeks for new Higgs-like events. ALEPH requested a two-month extension but was granted one month, boosting data by 50% and increasing their signal excess to 2.6σ by October. L3 soon reported a missing energy candidate. LEP's collision energy was pushed to the limit, and by November ALEPH's excess grew to 2.9σ. A one-year extension was requested but met gridlock due to concerns over delaying the LHC by up to three years, potential Higgs discovery at Fermilab's Tevatron, and practical issues like resource shifts and costs.

*The impending closure of LEP, when many of us were sure that we were about to discover the Higgs, was perceived likethe death of a dear friend by most of the LEP-ers.*

*The CERN Research board met again on 7 November and again there was deadlock, with no unanimous recommendation, the vote being split 8-8. The next day CERN Director-General Luciano Maiani announced that LEP had closed "for the last time". It was a deeply unpopular decision, but history has shown it to be correct, the Higgs being discovered at the LHC 12 years later, with a mass of not 115 but 125 GeV/c².*

When LEP was finally laid to rest most of the LEP protagonists went into deep mourning, and we met one last time for an official wake (see Figure 9).

LEP was the highest-energy e⁺e⁻ collider ever built, leaving a vital legacy for current and future colliders. Its unmatched data quality, luminosity, and energy calibration set the standard and serve as the benchmark for all future e⁺e⁻ ring collider designs. The legacy of LEP can be listed below:

- The physics data on the Z and W (luminosity, energy, energy calibration).
- The ultra-precise beam energy determination.

- Operation at record beam-beam tune shifts
- The experience in running a very large collider.
    - Technical requirements to control a large-scale facility.
    - Operational procedures for high efficiency.
    - Orbit optimization in long machines
    - Alignment, ground motion and emittance stability in deep tunnels.
- Designing and efficiently operating a large SC RF system
- Impedance and Transverse Mode Coupling Instability (TMCI) evaluation in large colliders.
- Flexibility in beam optics designs with changed in the betatron phase advance per cell ranging from 60o/60o to 102o/90o and 102o/45o.
- Strong reminder of the need for quality beam instrumentation and controls for efficient commissioning.
- The use of the personnel experience and expertise gained in LEP/LEP2 to prepare beam commissioning and operation of the LHC collider
- Avoid tunnelling in non-solid rock terrain

LEP's experience running a large collider was invaluable for LHC preparation, highlighting the importance of shutdown planning, maintenance, and remote repairs. Its real-time beam monitoring was essential for LHC's safe, efficient operation, and expertise with superconducting and cryogenic systems greatly benefited LHC performance.

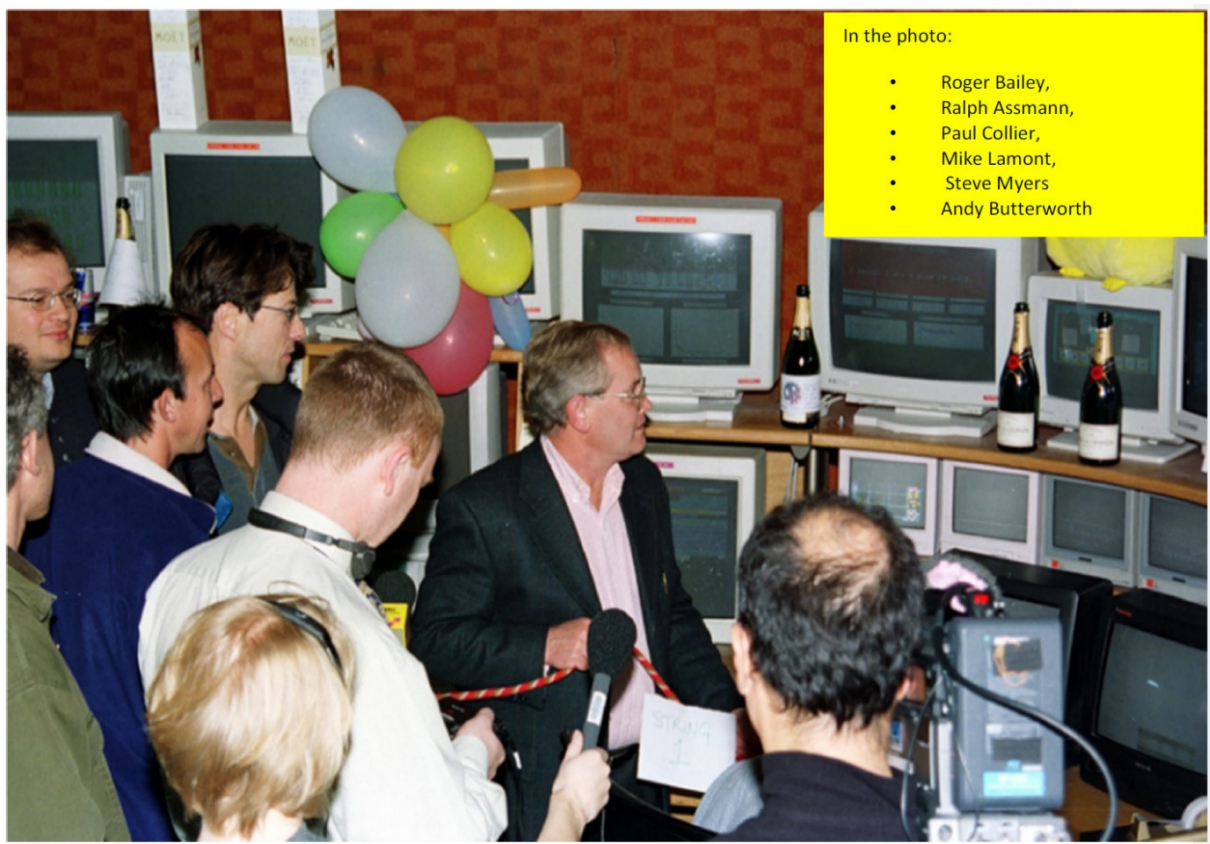

In the photo:
- Roger Bailey,
- Ralph Assmann,
- Paul Collier,
- Mike Lamont,
- Steve Myers
- Andy Butterworth

**Figure 9.** Last beam dump in LEP.

*4.7. Moving Focus and Personnel to the LHC*

The closure of LEP allowed massive redeployment of skilled and experienced CERN staff from LEP2 to the LHC design.

With the new focus from the closure of LEP, the design of the LHC gathered real momentum.

## 5. The Large Hadron Collider

### 5.1. 2008: the LHC Start Up on 10 September and the Terrible Accident of 19 September

September 10 was a fantastic day for CERN and LHC. Protons circulated in the ring very smoothly and everybody was happy. The LHC's start-up attracted massive global media attention, making it difficult to operate amid crowds on the first beam day—a momentous occasion for CERN. However, on 19 September 2008, just nine days after the initial success, during the test of sector 3–4, the last octants of magnets intended to be brought at 9.3 kA a treal disaster did happen. At 8.7 kA a resistive zone developed in the dipole bus-bar magnet interconnects. This caused a thermal runaway in one of the magnet interconnects, followed by electrical arc and sudden helium release. Several magnets and the whole vacuum system were damaged over 400 m.

The accident brought intense media scrutiny after the initial euphoria. As it happened, however, another even more important disaster took place only days before the LHC accident: the bankruptcy of Lehman Brothers (announced 15th September, just three-four days before the LHC accident!).

This was the largest bankruptcy filing in the history of the US with Lehman holding over $600 billion in assets, and of course took most of the journalists' attention. From a media point of view, the LHC accident went by unnoticed, luckily for CERN!

Post-Mortem of the Accident

An inquiry by CERN specialists [14] indicated several causes of the substantial damage to the machine:

*   There was an absence of solder on the offending magnet interconnect giving a contact resistance of 220 n$\Omega$ (design ~1 n$\Omega$).
*   There was poor electrical contact between the superconducting (SC) cable and the copper stabilizing busbar.
*   The fault detection of the interconnect was not sensitive enough. If the fault detection had been more sensitive, the accident would have been prevented.
*   The pressure relief ports were under-dimensioned for an accident of this magnitude.
*   The anchorage of the magnets to the tunnel floor was inadequate.

One of the sources of heat can be the movement of the coils, which can create friction. Another potential source of heat is the proton beam itself, which can deposit a huge amount of energy

The LHC accident occurred because the magnet protection system didn't work. There was a resistance runaway, and the energy wasn't transferred out of the magnets fast enough.

The damage inside the tunnel was impressive but horrific: many large heavy components had been blasted out of their places by the force of the "explosion". We were very concerned about inspecting the magnets due to their weight and their precarious positioning after the accident. For the repair, replacing the magnets was of course much more complicated than installing them for the first time, because they had to be extracted from the full tunnel, in the restricted space with little room for manoeuvre.

### 5.2. Chamonix 2009: The Repair

Following the initial investigation of the resulting damage, a crash programme was set up to repair and consolidate the LHC. The teams included many CERN partners, collaborators, detector people as well as the accelerator sector.

Two external panels were created, the first on Technical Risk was headed by Don Hartill from Cornell, and the second on Quench Protection headed by Jay Theilacker from Fermi National Accelerator Laboratory. Most of the members of these panels participated in the Chamonix meeting.

### 5.3. Chamonix January 2010: Restart

After the Christmas break it was time for the Chamonix retreat once again. This time we had, amongst others, two very hot topics; the beam energy scenario for the LHC and a project proposal to replace the injector chain of the LHC by a new superconducting linac and an upgraded Proton Synchrotron.

For the LHC, two energy scenarios were compared. The first was to run at 3.5 TeV/beam to accumulate as much data as possible at this energy, and to delay the consolidation of the whole machine for 7 TeV/beam in the foreseen Long Shutdown 1 (LS1).

The second option was to run until the second half of 2010 then do the minimum repair on splices to allow 5 TeV/beam in 2011 (7 TeV/beam comes much later).

The discussion on the choice of these two options was heated and emotional. Indeed, some people openly disagreed with either of the two options and insisted that LHC had been tested and proven up to 5 TeV per beam and operation in 2010 should be at this energy. They dismissed and ignored the measurements and simulations which clearly showed that operating at 5 TeV was highly risky. Like most of my colleagues, I was totally convinced that this 5 TeV proposal was not only risky but foolhardy. I was also convinced that if we ever had a second accident like that of 2008, CERN's long-term future would be jeopardised. Fortunately, common sense prevailed, and the experiments agreed to follow the first of the official proposals which was the lower risk for data taking.

*As a result of measurements done, some years later, during LS1, it was clearly shown that if we had decided to run LHC at 5 TeV in 2010 we would almost certainly have provoked another serious accident (Insertion).*

The second major topic was the replacement of the injector system for the LHC. This was the second phase of a larger project under the previous DG (R. Aymar) which had also included replacement of the existing LINAC2 by a new LINAC4. I was not in agreement with these projects as I could not justify them from the LHC performance point of view. I had asked the injector specialists for a clear objective comparison of the performance limits for the LHC, between an upgrade of the injector systems and their replacement. The result was very clear: the proposed replacement was highly more resource intensive, much riskier and would not give any clear performance improvement in the LHC when compared with the much simpler less risky upgrade of the existing and well tested present injectors. The LHC Injector Upgrade (LIU) was born as a result. As always, Chamonix had been a tough but very useful and productive retreat.

First LHC 7 TeV Collisions after the Repair 2010:

(from CERN Bulletin 06-07/2010)

The first collisions at 7 TeV centre-of-mass energy were recorded on 30 March 2010. During 2010, operation was continuously divided between machine studies to increase the luminosity and physics data taking.

We had already suffered from the stored energy in the magnet system which produced the accident in 2008, however many of us were more concerned about the stored energy in the beams. Although the amount of stored energy in the beams was much lower than that in the magnets, the type of energy in the beams was potentially much more risky and destructive.

The LHC beam protection system is the most intricate accelerator protection system ever developed. It has literally thousands of beam abort triggers and relies on very stringent control of the optics of the machine, both locally and globally. The collimation system is part of the machine protection system and must intercept almost the totality of any beam losses if they are to protect the rest of the machine. When the collimators are well set up and the hierarchy of losses is correct, the vast majority of all losses are localized at the collimator.

The rate of progress was impressive, nevertheless before each step was taken to increase the intensity and hence the stored beam energy, all machine protection systems were validated up to the new higher level.

### 5.4. 2011–2012 LHC Operation

We continued to operate the collider in 2011 and were now increasing the performance much more rapidly (see Figure 10 comparing 2010 performance with 2011: note the different vertical scaling).

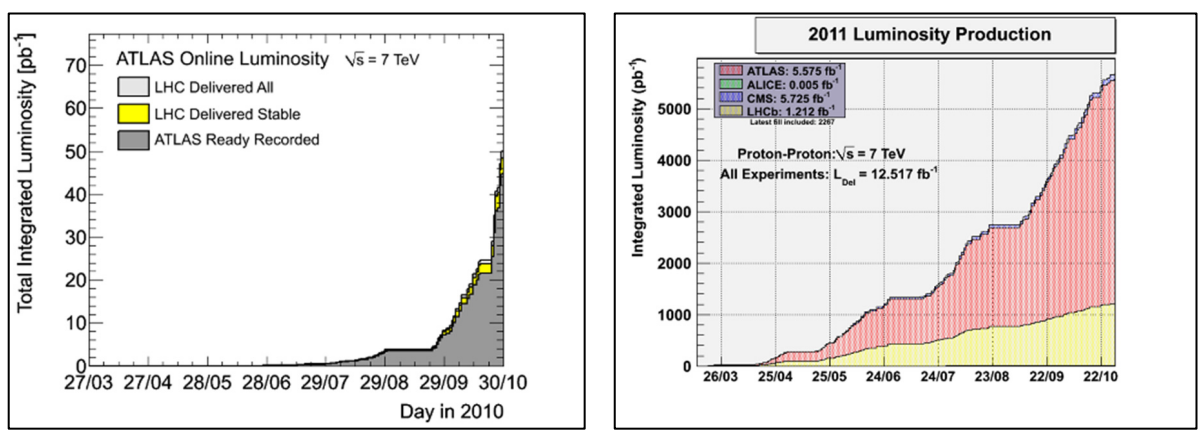

**Figure 10.** Plots show the daily progress in performance of the LHC in 2010 and 2011, reaching a maximum of 44 in 2010 and 6000 in 2011.

During the second half of 2011, the physics world was waiting for 2012; the last year of LHC operation before the long shutdown. The looming question was: will the LHC discover the Higgs boson or prove that it doesn't exist? I ran my performance simulation code with the parameters foreseen for running in 2012. The results shown in Figure 11a indicated that the luminosity needed to discover the Higgs boson would be reached during 2012. I presented these results to the CMS and ATLAS collaborations on 21/22 November 2011 respectively.

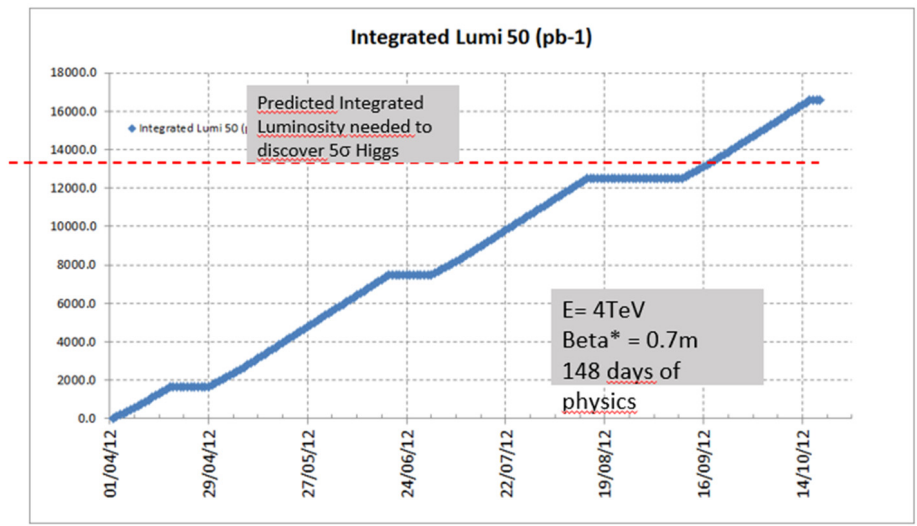

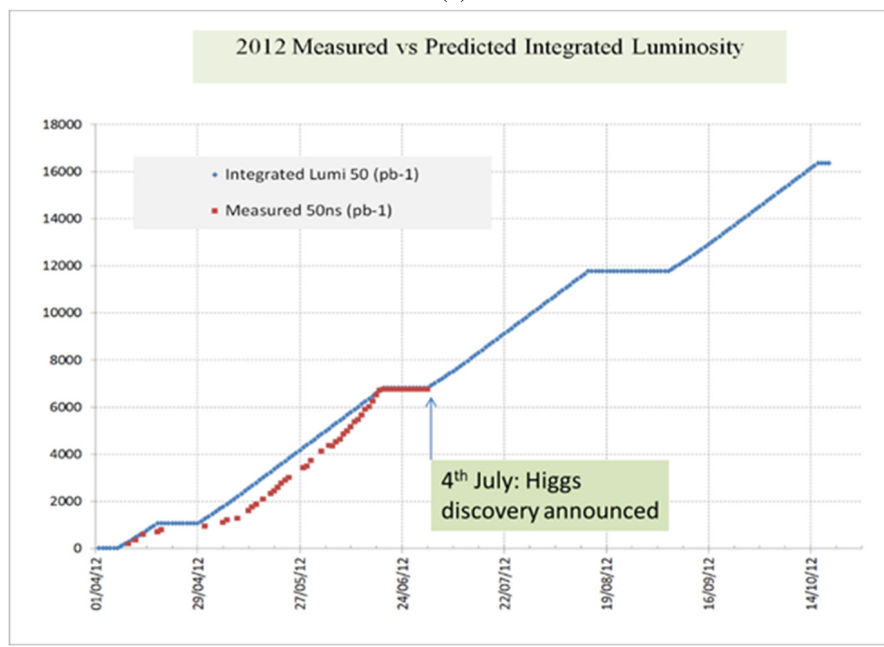

**Figure 11.** (**a**) Prediction of Integrated Luminosity for 2012. (**b**) Plot of predicted integrated luminosity in blue and measured in red.

Operation of the LHC restarted in April 2012 and I was carefully watching the performance every day. Initially the preformance was lagging my predictions, but then during May the slope changed and by the start of the 2nd technical stop on 17 June, the actual performance was exactly as predicted (see Figure 11b).

It was now fairly clear that we would reach the threshold for Higgs' discovery before the end of the 2012 run. It normally takes some weeks or months before the experiments can analyse all of the events they have recorded, so I was more than surprised to hear the rumours in late June that the data had been analysed and there were signs of a Higgs' discovery. The annual high energy physics conference was scheduled for 4th July in Melbourne and a video link had been set up from the main CERN auditorium to allow the LHC results to be transmitted to the physics world. Everyone in the high energy physics' world was waiting with bated breath. The main auditorium had to be locked to stop the summer students camping out for days beforehand in the conference room and blocking it for anyone else. I arrived at CERN at 5am on the morning of the 4th July and the queue for a place in the main auditorium was already long. Fortunately I had a reserved seat so I did not need to join the queue.

The main auditorium was packed to capacity long before the start of the presentations, with many legendary physicists including Peter Higgs as well as most of the living previous CERN DGs.

CMS had won, by the toss of a coin, to be the first to present. Joe Incandela nervously started his presentation and then, after some explanations on how they analysed the data, showed the small bump which was the five sigma signal for the discovery of the Higgs'. Everyone present stood up and applauded for minutes. Then Joe came over to me sitting in the front row and shook my hand and thanked the accelerator team. After Joe, Fabiola Gianotti made the presentation on behalf of the ATLAS experiment: again a five sigma signal and the auditorium erupted.

CERN was now triumphant having produced perhaps the most important physics result of the past century.

After the euphoria of the Higgs' discovery the collider resumed operation with a new approval for a run extension of nearly two months at the end of the year. The final integrated performance at the previously planned end of run was almost exactly at the level of the predictions. (see left plot in Figure 12).

During the first three years of operating LHC, (see right plot in Figure 12) the performance increased (in inverse pico-barns, pb$^{-1}$) from 44 in 2010, to 6000 in 2011 and then to 23000 in 2012 [15,16]. An incredible collider!

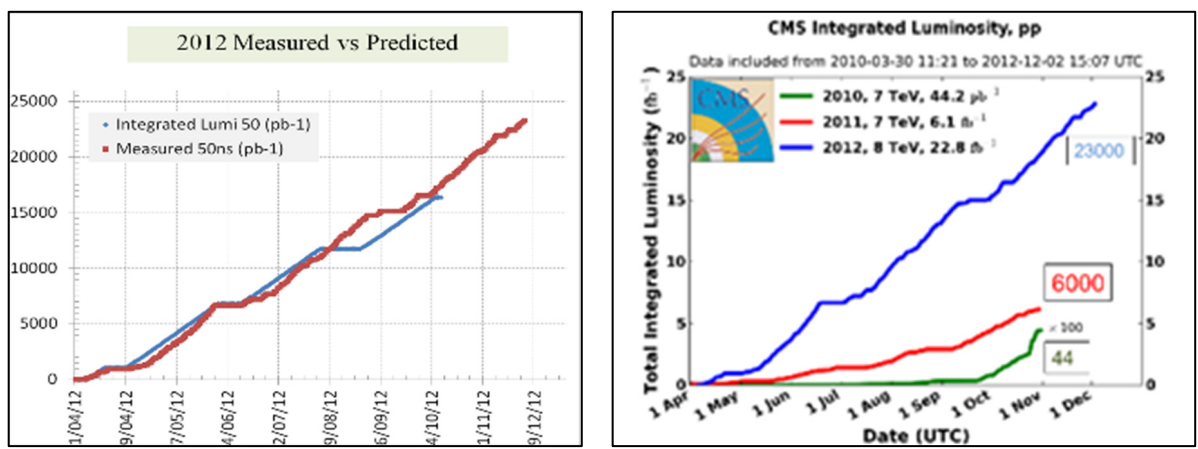

**Figure 12.** LHC performance in 2012 (predicted and achieved) and over 2010 to 2012.

## Conflicts of Interest

The author declares no conflict of interest.

## References

1. Unser, K. A Toroidal DC Beam Transformer with High Resolution. *IEEE Trans. Nucl. Sci.* **1981**, *28*, 2344–2346.
2. Ciapala, E.; Myers, S.; Wyss, C. Phase Displacement of High Intensity Stacks in the CERN ISR. *IEEE Trans. Nucl. Sci.* **1977**, *24*, 1431–1433.
3. Borer, J.; Bramham, P.; Schnell, W.; et al. 1974. Non-Destructive Diagnostics of Coasting Beams with Schottky Noise. In Proceedings of the IXth Internation Conference on High-Energy Accelerators SLAC, Stanford, CA, USA 2–7 May 1974.
4. Bramham, P.G.; Carron, H.G.; Hereward, K.; et al. Stochastic cooling of a stored proton beam. *Nucl. Instrum. Methods* **1975**, *125*, 201–202.
5. Van der Meer, S. *Stochastic Damping of Betatron Oscillations in the ISR*; Internal Report CERNISR-PO-72-31; CERN: Geneva, Switzerland, 1972.
6. Myers, S. Simulation of the Beam-Beam Effect for Electron-Positron Storage Rings. *Nucl. Instrum. Methods Phys. Res.* **1983**, *211*, 263–282.
7. Myers, S.; Schnell, W. *Preliminary Performance Estimates of a LEP Proton Collider*; Schnell LEP Note 440; CERN: Geneva, Switzerland, 1983.
8. Chao, A.W.; Chou, W. (Eds.) *Reviews of Accelerator Science and Technology: Vol. 7: Colliders*; World Scientific Pub.: Singapore, 2014.
9. Myers, S.; Picasso, E. *The CERN LEP Collider*; Scientific American: New York, NY, USA, 1990; pp. 54–61.

10. Myers, S.; Picasso, E. The Design, Construction and Commissioning of the CERN Large-Electron-Positron Collider. *Contemp. Phys.* **1990**, *31*, 387–403.

11. Myers, S. *The LEP Collider, from Design to Approval and Commissioning*; CERN: Geneva, Switzerland, 1990; pp. 91–108.

12. Myers, S. The LEP Machine—Present and Future. *Philos. Trans. R. Soc. London. Ser. A Phys. Eng. Sci.* **1991**, *336*, 191–200.

13. Assmann, R.; Lamont, M.; Myers, S. A Brief History of the LEP Collider. *Nucl. Phys.* B **2002**, *109*, 17–31.

14. Bajko, M.; Rossi, L.; Schmidt, R.; et al. *Report of the Task Force on the Incident of 19 September 2008 at the LHC*; LHC Project Report 1168; CERN: Geneva, Switzerland, 2009.

15. Myers, S. Large Hadron Collider commissioning and first operation. *Philos. Trans. R. Soc. A Math. Phys. Eng. Sci.* **2012**, *370*, 859–875.

16. Brüning, O.; Burkhardt, H.; Myers, S. The Large Hadron Collider. *Prog. Part. Nucl. Phys.* **2012**, *67*, 705–734.

17. Myers, S.; Wyss, C. Prospects for Energy and Luminosity at LEP2; CERN 96-01; CERN: Geneva, Switzerland, Volume 1, p. 26.

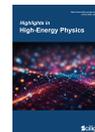

*Review*

# The Age of Gravitational Wave Astronomy

Frédérique Marion [1,2]

[1] Laboratoire d'Annecy de Physique des, Univ. Savoie Mont Blanc, CNRS, 74941 Annecy, France; marionf@lapp.in2p3.fr
[2] Particules-IN2P3, 74940 Annecy, France

**Abstract:** Detecting gravitational waves has been a very long experimental quest, which came to fruition with the second generation of the LIGO and Virgo ground-based interferometric detectors. These groundbreaking instruments have opened a new window on the Universe, revealing cataclysmic astrophysical events. They are spearheading a large-scale endeavor to explore the gravitational-wave spectrum at all frequencies, with rich science cases linked to fundamental physics, astrophysics, cosmology and the physics of the early Universe.

**Keywords:** gravitational waves; LIGO; Virgo; gravitational-wave astronomy; multi-messenger astronomy

## 1. Introduction

It feels very appropriate that a conference on the rise of particle physics, taking place at La Sapienza University in Rome, also covered the topics of gravitational waves, as La Sapienza was the home university of Edoardo Amaldi, one of the pioneers of the field. High-energy physics studies the fundamental particles and forces at play in Nature. Gravitation is one of the latter and holds a special place, being weaker than the other interactions by dozens of orders of magnitude and being described by a theoretical framework—Einstein's theory of general relativity—that has no overlap with the quantum field theory modelling the other forces. Although gravitation plays a negligible role when studying particles, due to its extreme weakness, it does play a major role in shaping the distribution of matter in the Universe. It is also intertwined with two major mysteries in modern physics, dark matter and dark energy. This explains why particle physicists make a strong part of the gravitational-wave community, along with astrophysicists, cosmologists and instrumentalists.

General relativity describes gravitation as a geometrical property of a dynamical space-time, summarized in the words of American physicist John Wheeler as "space tells matter how to move and matter tells space how to curve". The dynamics of space-time allows for the possibility of gravitational waves, i.e. ripples in space-time curvature that propagate through space at the speed of light, with their amplitude decreasing with distance. They are transverse waves, their physical effects manifesting in a plane perpendicular to their direction of propagation, where space is elongated and contracted along two orthogonal axes. They can have two independent polarization states, differing by an angle of 45° between their reference axes. Physically their effect is a strain, i.e. their amplitude $h$ corresponds to a rate of elongation or contraction. The longer the distance $L$ between free-floating objects, the larger the distance variation $\delta L$ induced by gravitational waves ($h = 2 \, \delta L / L$), which leads to the need for long-baseline detectors.

Gravitational waves are generated through mass acceleration. High-luminosity emission requires compact and relativistic sources with some asymmetry, so that the mass quadrupole of the system varies strongly with time. These criteria are met in some astrophysical sources and a canonical gravitational-wave source is a system of two compact objects orbiting each other. Although gravitational waves can carry enormous energy, space is so rigid that the resulting strain is very small. The maximum strain experienced on Earth, from the most powerful astrophysical sources, is typically at the level of $10^{-21}$, which has made experimental detection an extremely difficult task.

The sources of signals detected so far are indeed compact object binaries, involving stellar-mass black holes or neutron stars, believed to be the remnants of massive stars in both cases. As the compact objects orbit each other, the system emits gravitational waves, which carries energy away and drives the system to smaller and smaller orbits until the objects eventually merge. The associated gravitational-wave signal is very specific and is often referred to as a chirp, because both the frequency and amplitude increase with time until merger. This signature shows up as a typical pattern in a time-frequency plot of the measured strain, as illustrated in Figure 1.

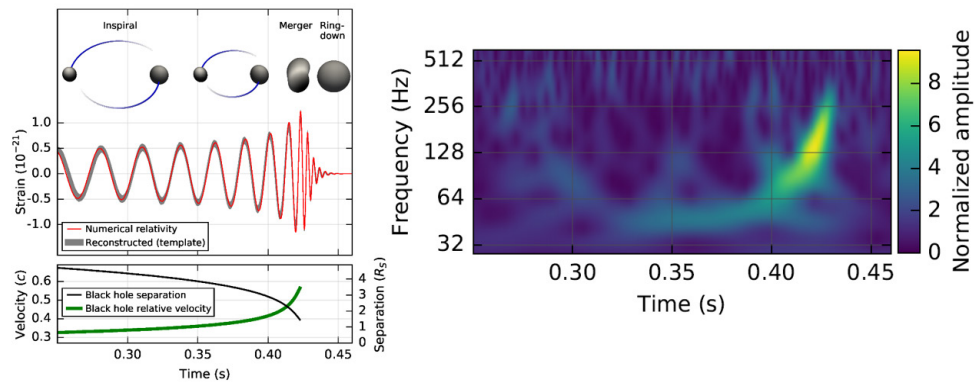

**Figure 1. Left**: Estimated gravitational-wave strain signal from GW150914 in LIGO Hanford and illustration of the system dynamics. **Right**: Time-frequency representation of the LIGO Livingston data at the time of the GW150914 event [1].

A key feature of the chirp is that the characteristic frequency at merger decreases as the total mass of the system increases and also as the cosmological distance to the source increases (the expansion of the Universe shifting the signal to lower frequencies, similarly to light being redshifted). Ground-based detectors operate at frequencies that make them sensitive to the mergers of neutron stars or stellar-mass black holes (weighing up to a few tens of the mass of the Sun). Sensitivity at lower frequencies is required to capture similar systems at an earlier stage of their evolution (long before merger), to capture systems at much larger distances or to capture much heavier systems involving super-massive black holes. The latter, weighing millions or billions of solar masses, are believed to be present at the center of most galaxies and can emit low-frequency gravitational waves if they are in close binaries.

Overall, gravitational waves are expected to manifest over a broad spectrum covering many orders of magnitude in wavelength and frequency (see Figure 2). Different frequency ranges have different detection techniques. Ground-based interferometers cover the band around 100 Hz. Space interferometry will cover the band around 1 mHz, while pulsar timing arrays cover the low-frequency range at the nHz level, and gravitational waves with periods comparable to the age of the Universe are expected to leave a subtle imprint on the polarization of the cosmic microwave background. The various frequency bands include specific astrophysical sources, with several cases of interest beyond compact binaries. Another speculative but fascinating prospect is the possibility to detect gravitational waves emitted in the very early Universe. There is a discovery potential at all frequencies, which is of great interest to high-energy physicists, the physics of the primordial Universe being largely unknown.

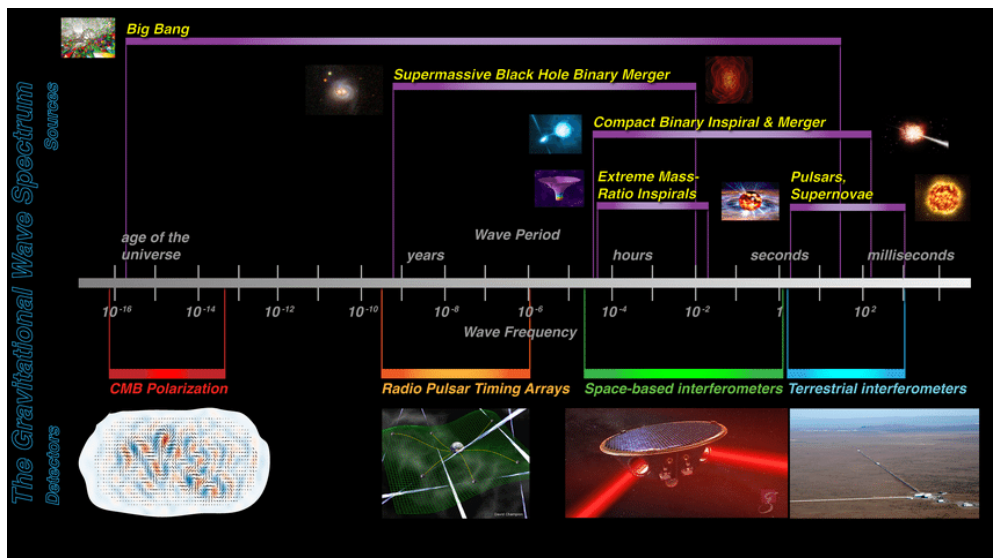

**Figure 2.** The gravitational-wave spectrum, highlighting detection techniques and various sources of interest. Figure courtesy of NASA/J. I. Thorpe.

## 2. Detecting Gravitational Waves

Einstein's theoretical prediction of gravitational waves dates back to 1916, but the experimental efforts toward detection only started in the 60', when Joe Weber built the first resonant-mass detectors in the US. Sophisticated versions of such detectors were developed and operated for several decades around the world. In the 70', interferometers were proposed as an alternative detection method, and Ray Weiss produced the first realistic study of an interferometric detector. The 80' brought observational confirmation of gravitational waves, through the orbital decay of the PSR 1913+16 binary, where a pulsar orbits another neutron star in a system losing energy in perfect agreement to general relativity's prediction based on gravitational-wave emission [2]. This confirmation prompted the design of large-scale interferometers—this was the beginning of the LIGO project in the US and of Virgo in Europe (the latter driven by Adalberto Giazotto and Alain Brillet). The 90' were decisive in that the funding agencies decided to build the first generation of LIGO and Virgo, initial detectors meant to evolve.

The first-generation detectors became operational in the noughties and demonstrated that it was possible to reach a level of sensitivity making gravitational-wave detection a possibility, even though they were not quite sensitive enough for a detection. They paved the way with the first science observations and the agreement of the various detectors to form a network, share their data and analyze them together. Success came in the following decade with the second-generation detectors, which had been upgraded to their so-called 'advanced' configurations. The discovery was made in 2015 with the two LIGO detectors and was celebrated by the 2017 Nobel Prize awarded to Barry Barish, Kip Thorne and Ray Weiss.

The past decade has been very successful—with the fields of gravitational-wave astronomy and multi-messenger astronomy on the rise—but in many ways, this is only the beginning. A new generation of interferometers is already being designed and other wavelengths are targeted as well, so that in the future it will be possible to observe the Universe in depth with gravitational waves, across a broad spectrum. There are currently four long-baseline, ground-based interferometric detectors around the world: the two LIGO detectors [3] in the US (at the Hanford and Livingston sites), Virgo [4] in Italy, and KAGRA [5] in Japan gearing up to join observations. The basic principle is that of a Michelson interferometer with suspended mirrors serving as test masses. As a gravitational wave passes through the plane of the interferometer, one arm gets longer as the other arm gets shorter and vice versa, which changes the interference pattern at the output port and modifies the output power in a measurable way. The principle is simple, but the target is to measure strains at the level of $10^{-21}$ at best. With arms that are 3 (in Virgo and KAGRA) or 4 (in LIGO) kilometers long, this translates into variations in the arm length at the level of $10^{-18}$ m, which is 12 orders of magnitude smaller than the laser wavelength of 1 µm. The change in the interference pattern is therefore tiny and extreme care is needed so that it is not buried in noise.

The optical configuration is actually more complex than that of a Michelson interferometer, with the addition of mirrors to form Fabry-Perot cavities in the arms, as well as power and signal recycling cavities (see Figure 3). The purpose of these cavities is to amplify the gravitational-wave signal and minimize sensing noise. On top of this, the detectors need a whole range of advanced technologies to reach good sensitivity. They need powerful and extremely stable laser sources, as well as high levels of vacuum to propagate the laser beams and host the mirrors. The mirrors themselves need to be near perfect, so that they can reflect the light beams thousands of times without disturbing them. They need to be seismically isolated, as typical ground motion is orders of magnitude larger than the $10^{-18}$ m target. They also need to be suspended in a way that minimizes thermal noise.

The most recent versions of LIGO and Virgo also implement sophisticated techniques to reduce quantum noise, which comes in two ways. Photons arrive in a random way on the photodetector, which results in power fluctuations—this is shot noise, which decreases with laser power. In addition, fluctuations of photons reflecting from a suspended mirror cause random mirror motion—this is radiation pressure noise, which increases with laser power. Both therefore cannot be reduced at the same time, unless a technique called frequency-dependent squeezing is employed.

LIGO, Virgo and KAGRA operate as a network, for several reasons. One of them is that when searching for rare and weak signals, redundancy helps discriminate signals from noise. The main reason however is that more than one detector is needed to infer where a signal is coming from. A given detector is not very directional but has instead a broad antenna pattern showing a modest dependence of the wave-detector coupling on the source direction (except for blind spots along the bisector of the interferometer arms). Therefore, locating a source on the sky requires comparing the signals received in different detectors, primarily in terms of their arrival times, and performing triangulation. This is crucial to get a chance to know the full astrophysical context of an event.

The performance of detectors is characterized by their sensitivity curves, which give the level of noise as a function of frequency, over a frequency band that goes roughly from 10 Hz to 10 kHz for ground-based interferometers. They result from a mixture of fundamental and environmental noise sources, as well as many technical noise sources. They are improved on iteratively, through commissioning work and upgrades to the detectors. The sensitivity curve can be compared to the expected spectrum of a signal to estimate the signal-to-noise (SNR) ratio. Using by convention the signal of a 1.4 + 1.4 $M_\odot$ binary neutron star (BNS) merger, the sensitivity curve can be summarized into a figure of merit called the BNS range, giving the typical distance where such a merger can be detected with an SNR of 8. This figure of merit has doubled over the past decade. The Advanced LIGO detectors started with a BNS range around 80 Mpc back in their first observing run (O1) in 2015. It is now around 160 Mpc. A factor 2 on the reach of the detectors means a factor 8 on the volume probed, and consequently on the rate of detections. The latter has indeed increased with sensitivity improvement, up to a level of a couple per week of observing time in the on-going fourth observing run (O4). The current tally of confirmed detections and detection candidates is approaching 300.

The detectors are currently in the third part of the O4 run and are preparing the upgrades coming next, in view of the O5 run. Beyond O5, there are plans for an "ultimate" set of upgrades, to improve the sensitivities to the limits imposed by the current infrastructures; these are the A♯ concept for LIGO and the Virgo_nEXT concept for Virgo. Pushing further will require a new generation of instruments, built in new infrastructures, which is the purpose of the Cosmic Explorer project in the US and the Einstein Telescope in Europe.

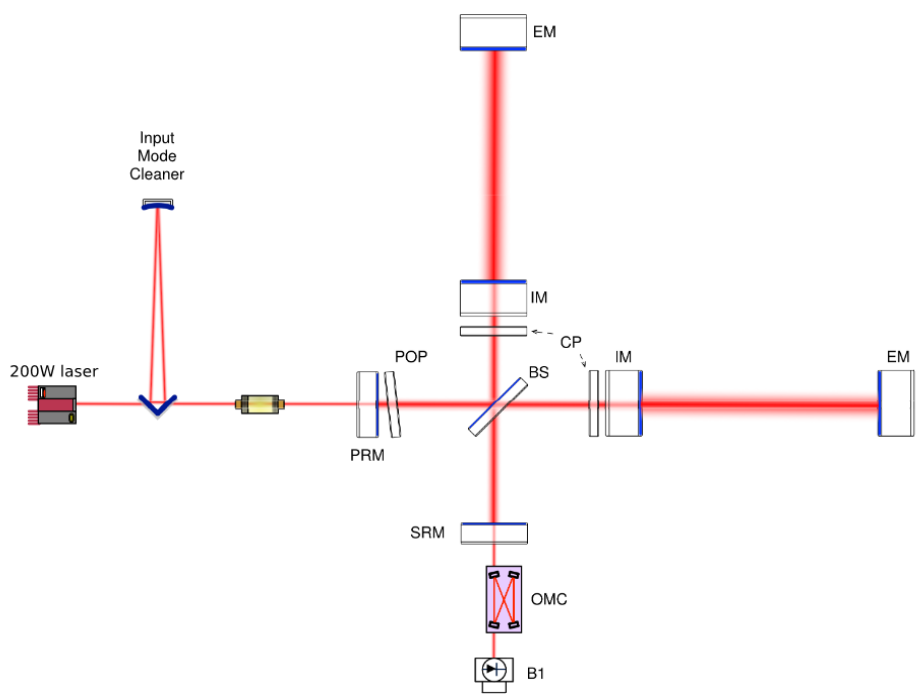

**Figure 3.** Simplified optical layout of the Advanced Virgo interferometer. Each of the long cavities in the arms is formed by an input mirror (IM) and an end mirror (EM). The recycling cavities are formed by the power-recycling mirror (PRM) or the signal-recycling mirror (SRM) and the two input mirrors [4].

## 3. Searching for Gravitational-Wave Signals

It is common practice in the LIGO-Virgo-KAGRA (LVK) collaboration to classify gravitational-wave sources both according to whether they produce transient or persistent signals and according to whether the expected waveform is known or not. Compact binary coalescences are not the only possible sources of transient signals, a category that also includes the generic class of "bursts", which are mostly unmodeled transients. On the persistent signal side, LVK data are searched for continuous waves from spinning neutron stars, and for gravitational-wave stochastic backgrounds produced by unresolved sources.

Compact binary coalescences are the only class of events detected to date. It is a prime case where the expected signal is known, as waveforms can be computed accurately, solving general relativity's equations both analytically and numerically. This allows using matched filtering, namely cross-correlating the data with the expected signal – called a template – giving different weights to different frequencies to take the sensitivity curve into account. The output of the matched filter will peak when the signal is indeed present in the data. The binary system parameters are not known a priori, though, which requires trying many templates to cover the parameter space of interest.

The case of generic transients, what we call bursts, is more difficult. They include many possible astrophysical events. A primary target is the gravitational-wave signal emitted during a core-collapse supernova, as it would be a powerful tool to understand the dynamics of the collapse. Other targets include the post-merger signal after a binary coalescence, instabilities in neutron stars, or a signal associated with long Gamma-ray bursts. The signals are typically poorly modeled, which requires a robust search strategy. This involves looking—in the time-frequency space—for excess power that is coherent in multiple detectors, meaning that the signal amplitudes and phases are consistent with a single sky location.

A primary target for persistent signals is spinning neutron stars, which are expected to emit gravitational waves if they are not symmetric around their rotation axis. There are many neutron stars in our own Galaxy, and detecting such a signal would be a way to probe their structure. The strength of the signal depends on how elliptical the neutron star can be, which is unknown. The signal is likely to be very small, but it can be integrated over a long time to increase the SNR. It is a search that is computationally limited, though, as the signal needs to be tracked coherently, and although the signal itself is quasi-monochromatic, the time-varying Doppler shift arising from the motion of the Earth-based detector relative to the source needs to be taken into account. This amounts to scanning an enormous parameter space, which cannot be done in practice and requires trade-offs. A fully coherent search is possible if the target is a known pulsar where both the sky position and the signal frequency are known a priori, but a semi-coherent search is the way to go if conducting an all-sky search where no parameters are known a priori.

Stochastic gravitational-wave backgrounds can be of two kinds. Astrophysical backgrounds are expected to arise from the superposition of many unresolved sources; examples could be pulsars in our Galaxy or binary mergers too far away to be resolved. Their detection would complement individual source detection. Even more fascinating is the second kind, backgrounds arising from gravitational waves produced in the primordial Universe. The physics of the latter is largely unknown and there are various speculative models for gravitational waves generated during inflation, during phase transitions, by topological defects, etc. The mechanisms typically involve energy scales way beyond what can be reached in particle colliders, therefore offering a unique discovery potential. Stochastic backgrounds would basically show up in the data as extra noise, but noise that is correlated across detectors and shows a signature spectrum. The search method is similar to the matched filtering we use for compact binary coalescences, but using the data stream of one detector as a template for another detector.

## 4. Highlights of Gravitational-Wave Science

The hundreds of events detected so far have led to a rich variety of results. The science relies on measuring the source properties by analyzing the detailed features of the signal, which depends on the parameters of the system, both the intrinsic parameters that drive the dynamics, like the component masses and spins, and the extrinsic parameters, like distance and sky location. This parameter estimation is done through Bayesian analysis. Results are then derived in terms of astrophysics, fundamental physics or cosmology, based either on some individual events or on a statistical analysis of the sample of events.

With event names based on the date they were detected, GW150914 was the discovery event, detected on 14 September 2015 [1]. It was the merger of two black holes weighing about 30 solar masses ($M_\odot$) each, which was surprisingly heavier than the black holes previously known in our Galaxy. Though unique is that it was the first ever detected, the event later turned out to be quite typical. Our sample of detected events is indeed dominated by binary black hole mergers, most of them with roughly equal masses—asymmetric systems are rarer (see Figure 4).

GW170817 [6] was the first binary neutron star merger, a strong signal with a well-localized source, as it was one of the first events observed at a time when three detectors were online (LIGO Hanford, LIGO Livingston and Virgo). What made this event particularly remarkable is that it was immediately followed by a short Gamma-ray burst—a flash emitted by an outflow of relativistic particles—and was later followed by a kilonova, the light emitted by the matter ejected during the collision, which is a site of rapid nucleosynthesis, heated by radioactivity. For astrophysics, GW170817 therefore confirmed that BNS mergers are linked to short Gamma-ray bursts and to kilonovae, and that kilonovae are a site of heavy elements production. BNS mergers are also crucial for fundamental physics, as they are a laboratory to study the structure of matter at the extreme densities found in neutron stars. GW170817 and the almost simultaneous Gamma-ray burst confirmed to an excellent precision that gravitational waves and light propagate at the same speed. The event also illustrated the potential that BNS mergers have for cosmology, in measuring the current rate of expansion of the Universe.

GW190521 [7] is another remarkable event for several reasons; one being that its remnant—the black hole formed by the merger—has a mass close to 150 $M_\odot$, which falls in the intermediate-mass range between stellar-mass black holes and super-massive black holes. This is a range where hardly any black holes are known yet some are expected, if the lighter black holes are seeds to form the heavier ones.

Beyond the features of individual events, looking at the sample of detections [8] as a whole allows addressing a broad range of questions. It provides measurements of how often mergers occur and of the merger rate per binary

type—mergers have now been observed from three types of systems: pairs of black holes, pairs of neutron stars, and mixed systems with a black hole and a neutron star. Moreover, as the sample keeps growing and includes sources at further distance, it becomes possible to study how the merger rate evolves with redshift and therefore with cosmic time.

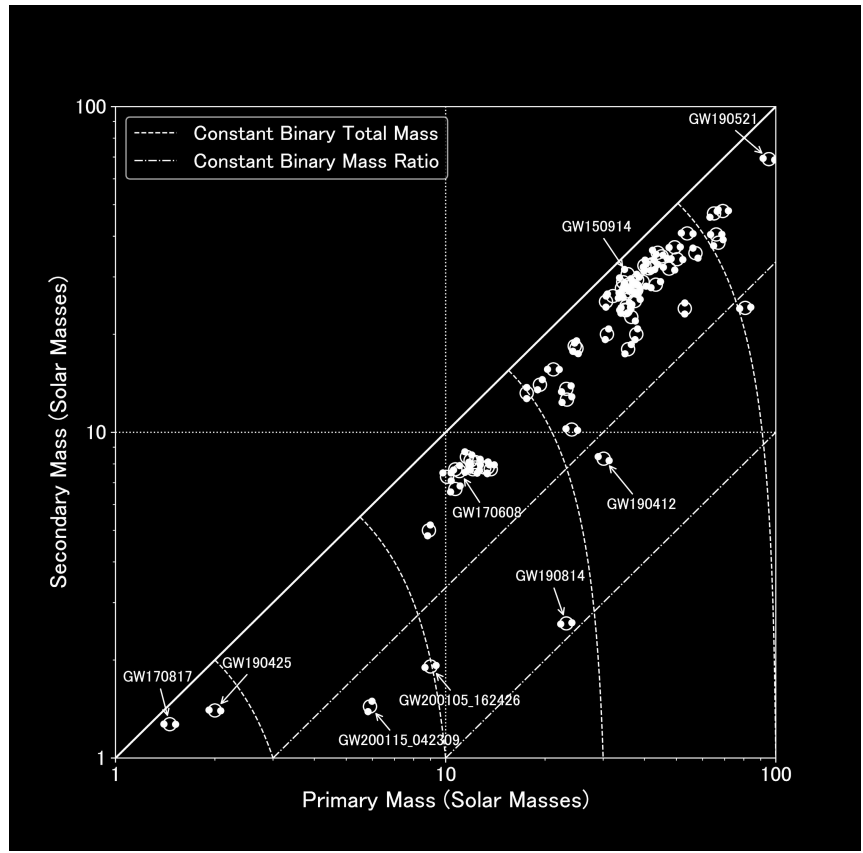

**Figure 4.** The sample of binary systems observed in the LIGO-Virgo O1, O2 and O3 observing runs and used to characterize the population of sources, shown in the component mass space. Figure courtesy of LIGO-Virgo-KAGRA Collaboration/IGFAE/Thomas Dent.

Inferring the mass distribution of the compact objects involved in merging binaries helps shed light on a number of astrophysical open questions: What is the maximum mass for neutron stars? What is the minimum mass for black holes born from stars? Is there a gap between the two? Although difficult, it is also interesting to measure how fast those compact objects are spinning, as this might be a key to understand not only their formation but also how the binaries themselves were formed, with two main scenarios: either the stars co-evolved in a pre-existing binary, or the binary was formed dynamically after the stars had become compact remnants.

An aspect that is of major interest to particle physicists is to understand how the strong nuclear force behaves in neutron stars, where matter reaches extreme densities, higher than in atomic nuclei, and the structure is unknown. Depending on their structure, neutron stars will be more or less deformable, leading to stronger or weaker tidal effects during the late inspiral of the binary. Tidal effects are expected to leave a subtle imprint on the gravitational-wave signal, which can only be constrained today but will be measurable in hundreds of events with the next generation of detectors. One can also explore if the compact objects involved in mergers are actually neutron stars or ordinary black holes, or if they could be exotic objects, and look for signatures of dark matter.

Another major application is cosmology, with the prospect of measuring the expansion rate of the Universe using compact binary coalescences as standard candles, i.e. the source distance can be inferred from the gravitational wave signal. The redshift cannot, but GW170817 gave a proof of concept that an electromagnetic counterpart can point to the host galaxy, whose redshift can be used to measure the Hubble constant. The measurement was only at the $\sim$15% level but the prospect of reaching a precise measurement is exciting, given that the main methods to measure the Hubble constant disagree with each other, creating tension in the standard model of cosmology.

Finally, gravitational waves are a unique tool to test general relativity. GW170817 has already provided tight constraints that gravitational waves and light not only travel at the same speed but also are affected by gravitational potentials in the same way. Still regarding the propagation of gravitational waves, one can look for signs of dispersion, i.e. different frequencies travelling at slightly different speeds. Thinking of gravitation in a framework similar to the other interactions, it would be mediated by a particle, called the graviton. Even though our detectors sense waves, not gravitons, the graviton mass can be constrained as a non-zero mass would result in dispersion in the signal. With no observational evidence for the latter, the mass of the hypothetical graviton is constrained to be less than $10^{-23}$ eV. One can test if gravitational-wave signals are consistent with two independent polarizations and, more generally, if the waveforms are consistent with general relativity predictions, in an extreme regime of space-time dynamics. An important test is to check if black holes behave as predicted by general relativity. Binary mergers leave behind a remnant black hole that reach the quiescent state by emitting a set of damped sinusoids, whose frequencies and damping times depend only on the black hole mass and spin, according to general relativity. Doing precise spectroscopy of those black hole ringdowns will allow testing this 'no-hair' prediction.

## 5. Outlook

In the future, gravitational-wave astronomy is going to evolve towards a multi-wavelength and multi-messenger landscape. As parts of the gravitational spectrum are opened or explored with improved sensitivity, new astrophysical sources will be revealed, and possibly gravitational-wave backgrounds of cosmological origin, which would be a major breakthrough.

On Earth, there are plans for a new generation of detectors, in new infrastructures that would provide baselines up to an order of magnitude longer. Europe has the Einstein Telescope project, designed as an underground triangle with 10 km sides, while the US have the Cosmic Explorer project, with one or two 40 km L-shaped detectors on the ground. The long baseline will make a major difference in the volume probed. Current detectors are only skimming the surface whereas future detectors will essentially probe the whole population of merging binaries, far into the past of the Universe, and will likely reveal novel types of gravitational-wave sources.

Current detectors are very sensitive and can measure very weak signals, but they do not measure them with high precision, as the recorded signal-to-noise ratio remains modest. SNRs will reach hundreds and thousands for nearby sources in Cosmic Explorer and Einstein Telescope, which will also provide a much larger sample of events and, with a bandwidth extended at low frequencies, more precise parameter estimation. The next generation will therefore bring us into the high-statistics, high-precision regime that has made the success of high-energy physics. Moreover, a sizeable subset of the sample will be multi-messenger events, which enable so much science.

Other observational windows are already opening at lower frequencies. Pulsar-timing arrays are like Galactic-scale detectors. Pulsars are excellent clocks, with their radio emission periodically received on Earth like a rotating lighthouse, with a stability at the level of ∼ 100 ns for the most stable of them. With the Earth and pulsars in free fall, monitoring the arrival times of radio pulses is sensitive to perturbations in the metric due to gravitational waves passing through the Galaxy, which affect the arrival times from different pulsars in a correlated way. This is sensitive to frequencies around the nHz and allows probing a stochastic background of gravitational waves emitted by close binaries of super-massive black holes. The various pulsar-timing arrays in the US, Australia and Europe are indeed seeing growing evidence of such a signal in their data [9–12]. If confirmed, this will provide a new handle on the growth of massive black holes and the evolution of galaxies.

The frequency band around the mHz is expected to open in about a decade from now, when the LISA mission [13] of the European Space Agency (with contributions from NASA) will be launched. LISA will perform interferometry in space between three satellites separated by 2.5 million kilometers. The mission will probe a diversity of compact-object binaries: light systems long before they merge (revealing the population of binary stars in our Galaxy); massive black hole binaries as they merge, at all cosmic times; extreme-mass ratio inspirals, where a stellar-mass black-hole orbits a massive black hole (the complex gravitational-wave signal offering a map of space-time around the massive object).

With the detection of gravitational waves, a new field of physics is born, which relies on incredibly sensitive detectors. This age of gravitational-wave astronomy brings a wealth of science within reach, but fulfilling this potential will need a new generation of instruments. Plans exist but will also need a new generation of physicists to make them a reality. It is a must, as gravitational waves are a goldmine for a very broad community sitting at the crossroads of fundamental physics, cosmology and astrophysics.

**Conflicts of Interest**

## References

1. Abbott, B.P.; Abbott, R.; Abbott, T.D.; et al. Observation of Gravitational Waves from a Binary Black Hole Merger. *Phys. Rev. Lett.* **2016**, *116*, 061102. https://doi.org/10.1103/PhysRevLett.116.061102.

2. Weisberg, J.M.; Huang, Y. Relativistic Measurements from Timing the Binary Pulsar PSR B1913+16. *Astrophys. J.* **2016**, *829*, 55. https://doi.org/10.3847/0004-637X/829/1/55.

3. Aasi, J.; Abbott, B.P.; Abbott, R.; et al. Advanced LIGO. *Class. Quantum Gravity* **2015**, *32*, 074001. https://doi.org/10.1088/0264-9381/32/7/074001.

4. Acernese, F.; Agathos, M.; Agatsuma, K.; et al. Advanced Virgo: A Second-Generation Interferometric Gravitational Wave Detector. *Class. Quantum Gravity* **2015**, *32*, 024001. https://doi.org/10.1088/0264-9381/32/2/024001.

5. KAGRA Collaboration. KAGRA: 2.5 Generation Interferometric Gravitational Wave Detector. *Nat. Astron.* **2019**, *3*, 35–40. https://doi.org/10.1038/s41550-018-0658-y.

6. Abbott, B.P.; Abbott, R.; Abbott, T.; et al. GW170817: Observation of Gravitational Waves from a Binary Neutron Star Inspiral. *Phys. Rev. Lett.* **2017**, *119*, 161101. https://doi.org/10.1103/PhysRevLett.119.161101.

7. Abbott, R.; Abbott, T.D.; Abraham, S.; et al. GW190521: A Binary Black Hole Merger with a Total Mass of 150 $M_\odot$. *Phys. Rev. Lett.* **2020**, *125*, 101102. https://doi.org/10.1103/PhysRevLett.125.101102.

8. Abbott, R.; Abbott, T.D.; Acernese, F.; et al. GWTC-3: Compact Binary Coalescences Observed by LIGO and Virgo during the Second Part of the Third Observing Run. *Phys. Rev. X* **2023**, *13*, 041039. https://doi.org/10.1103/PhysRevX.13.041039.

9. Arzoumanian, Z.; Baker, P.T.; Blumer, H.; et al. The NANOGrav 12.5 yr Data Set: Search for an Isotropic Stochastic Gravitational-wave Background. *Astrophys. J. Lett.* **2020**, *905*, L34. https://doi.org/10.3847/2041-8213/abd401.

10. Goncharov, B.; Shannon, R.M.; Reardon, D.J.; et al. On the Evidence for a Common-spectrum Process in the Search for the Nanohertz Gravitationalwave Background with the Parkes Pulsar Timing Array. *Astrophys. J. Lett.* **2021**, *917*, L19. https://doi.org/10.3847/2041-8213/ac17f4.

11. Chen, S.; Caballero, R.N.; Guo, Y.J.; et al. Common-red-signal Analysis with 24-yr High-precision Timing of the European Pulsar Timing Array: Inferences in the Stochastic Gravitational-wave Background Search. *Mon. Not. R. Astron. Soc.* **2021**, *508*, 4970–4993. https://doi.org/10.1093/mnras/stab2833.

12. Agazie, G.; Anumarlapudi, A.; Archibald, A.M.; et al. The NANOGrav 15 yr Data Set: Evidence for a Gravitational-wave Background. *Astrophys. J. Lett.* **2023**, *951*, L8. https://doi.org/10.3847/2041-8213/acdac6.

13. Colpi, M.; Danzmann, K.; Hewitson, M.; et al. LISA Definition Study Report. *arXiv* **2024**, arXiv:2402.07571.

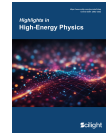

*Review*

# Precision Physics in the Era of (HL)LHC

Maurizio Pierini

The European Organization for Nuclear Research (CERN), 1217 Geneva, Switzerland; maurizio.pierini@cern.ch

**Abstract:** The LHC was conceived as the ultimate discovery machine, designed mainly to find the Higgs boson and extensively search for SUSY and its alternatives at the electroweak (EW) energy scale. After a decade of operation, thanks to improved detector performance and advanced algorithms, the LHC experiments entered a precision era, surpassing LEP and Tevatron in many areas. Innovations such as deep learning and new data-taking methods have further extended their reach, beyond assumed limitations such as the trigger bandwidth. The luminosity upgrade of LHC, HL-LHC, promises even greater precision with upgraded detectors. The LHC experiments will remain leading in high-energy physics for the foreseeable future, leaving a lasting impact on various fronts.

**Keywords:** hadron collider physics; new physics searches; higgs boson

## 1. The Initial Mission

Our modern understanding of high energy physics before the Large Hadron Collider (LHC) has long rested on two central theoretical pillars: the Higgs mechanism for EW symmetry breaking and the expectation of EW-scale supersymmetry (SUSY) as a natural solution to stabilize the Higgs vacuum expectation value. On the experimental side, progress has been driven along two major fronts: precision measurements at $e^+e^-$ colliders to indirectly probe signs of new physics, and direct searches for new phenomena at hadron colliders.

There have been notable exceptions that blurred the line between these approaches, such as EW SUSY searches at LEP and the precise measurement of the $W$ boson mass at the Tevatron. However, it was the construction of the Large Hadron Collider (LHC) that marked a transformative step forward. Designed as the ultimate discovery machine, the LHC exceeded expectations early in its run. ATLAS and CMS discovered the Higgs boson [1,2] sooner than anticipated—with only about half the design energy and a fraction of the projected dataset. Furthermore, large portions of the EW-scale natural SUSY parameter space have been excluded. In particular, gluino searches have ruled out all models in which the gluino is accessible at LHC energies, and even when the gluino is assumed to decouple, most of the viable SUSY parameter space remains tightly constrained.

A key strength of the LHC lay in its unprecedented collision energy and luminosity, enabling the collection of enormous datasets. This, however, introduced significant computing and data handling challenges, addressed by novel High-Level Trigger systems and the globally distributed LHC Computing Grid. The LHC environment also posed harsh experimental conditions, characterized by high particle multiplicity and event pileup. Remarkably, many of these initially daunting challenges are now considered routine, thanks to substantial progress in detector performance and analysis techniques. The LHC experiments are then capable to still deliver outstanding results, having transitioned into a new era of precision physics, something that was far from obvious when the LHC started.

## 2. The Rise in Precision

The transition of the LHC from a discovery machine to a precision instrument has been driven by two complementary developments: improved data quality and access to increasingly larger datasets in addition to high performance detectors. Together, these advancements have enabled a new class of precision measurements that were previously beyond reach.

The effort to achieve the ambitious goals of the LHC program, along with the extensive exploitation of Run 2 data, led to numerous experimental breakthroughs. Among these are advanced event processing techniques such as sophisticated pileup subtraction schemes, and enhanced reconstruction algorithms including state-of-the-art jet tagging methods. Since around 2015, the adoption of deep learning algorithms has significantly accelerated progress in many of these areas. These methods have been deployed across a wide range of tasks, from object identification to regression and classification problems central to event reconstruction.

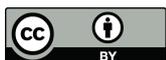

These developments, shown in Figure 1, allowed physicists to extract precise measurements from increasingly complex data. This has paved the way for the LHC's growing role in the domain of precision physics.

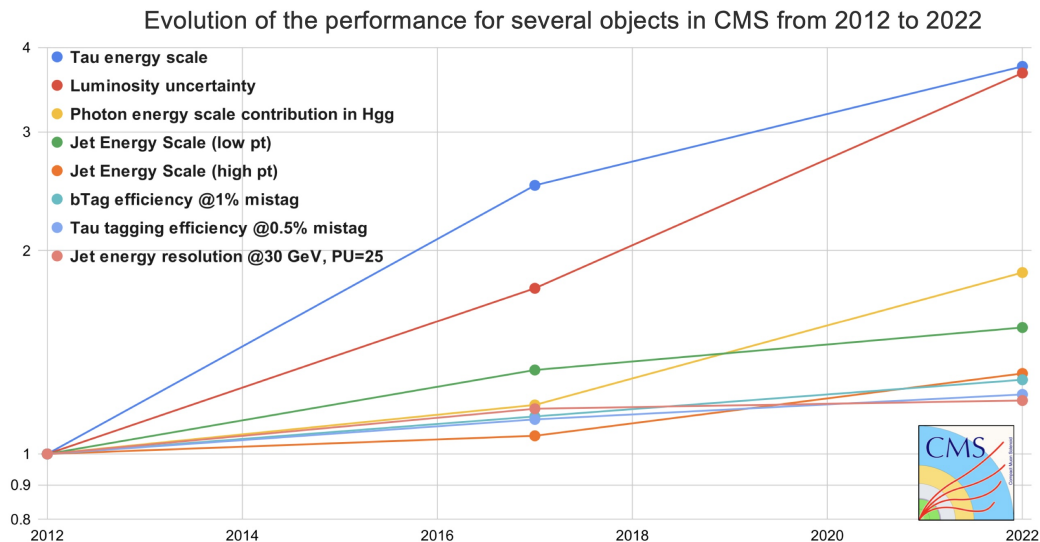

**Figure 1.** Evolution of experimental performance vs. time for the CMS experiment, quantified quoting the ratio of the efficiency or of the inverse of the uncertainty over the corresponding 2012 value, for various inputs to physics analyses. Plot extracted from Ref. [3] .

The LHC has consistently delivered more collisions than the experiments were originally designed to process in real-time. By design, this should have been a problem: traditionally, detectors were configured to record only a subset of "interesting" events, constrained by bandwidth and computing limits. To overcome this bottleneck, LHC experiments began breaking this paradigm as early as Run 1, turning the problem into an opportunity. One notable strategy is the use of parking [4] (knows in ATLAS as delayed reconstruction), wherein more events than can be promptly processed are recorded to tape and analyzed later, typically during accelerator shutdowns when computing resources are more readily available. Another complementary approach is the use of trigger-level reconstruction for physics analysis. The strategy consists in maximizing the number of triggered events by reducing the event data format to the minimum required to perform physics studies (as opposed to store the whole event in so-called RAW format). This approach opens new frontiers for both discovery and precision measurements without exceeding real-time processing limits. This approach was pioneered by CMS already in Run 1 with the introduction of the so-called data scouting stream, and since Run 2 it is exploited also in LHCb's turbo stream [5] and ATLAS' trigger-level analyses [6].

Machine learning (ML), particularly deep learning (DL) techniques, has played a transformative role in enhancing the sensitivity of the LHC experiments beyond initial expectations. One notable advancement has been the development of novel DL-based taggers, which have obtained improved tagging capabilities on traditional tasks such as $b$-jet tagging (see Figure 2) and extended the reach of experiments to probe previously unexplored topologies, such as $c$, $bb$, and $cc$ jets. These advances have enabled incredible progress in the study of Higgs physics, as discussed in Section 3. In particular, the application of DL to boosted topologies has been instrumental in analyzing events with high transverse momentum, a challenging regime for standard reconstruction techniques [7,8]. The study of resolved topologies, which involve lower momentum objects and more complex event structures, have also seen improvements due to better identification of $b$-jets [9,10] and hadronic tau decays [11,12]. These developments have collectively led to an unprecedented level of precision in measurements.

In addition, the use of ML algorithms created unexpected opportunities. For instance, CMS managed to compensate for the absence of a dedicated particle identification (PID) for flavor physics system with cutting-edge DL techniques, developing the most accurate $b$-flavor tagger at hadron colliders, exploiting both same-side and opposite-side topologies in dedicated binary classifiers based on DeepSets, an architecture designed to handle sets of particles and their relationships. The $b$-flavor tagging power was reached an impressive 5.6%, increased by a factor of four over previous methods.

These are only a few examples to highlight the indispensable role of ML in pushing the boundaries of precision physics at the LHC, opening new avenues for exploring both Standard Model (SM) processes and potential new physics.

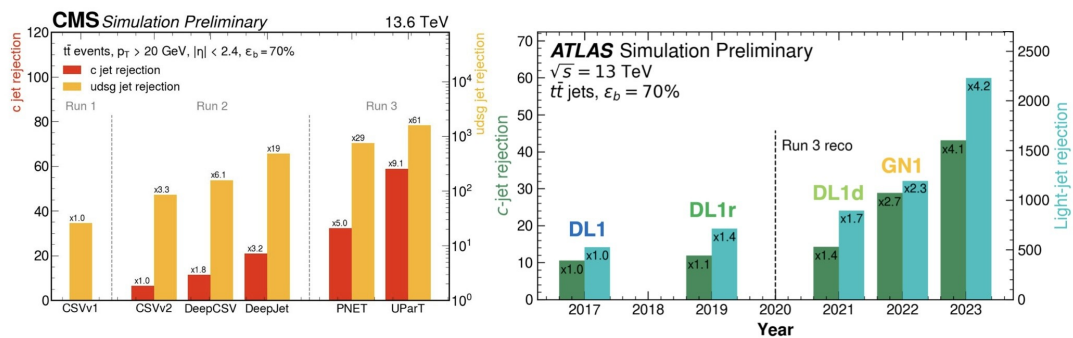

**Figure 2.** Evolution of *b*-jet tagging efficiency for CMS (**left**) and ATLAS (**right**), from the Run-1 algorithms to the DL architectures.

## 3. Physics Highlights from LHC Run 1 and Run 2

The precision in Higgs boson coupling measurements has seen remarkable progress in Run 2, with current uncertainties already below 10% for most couplings—despite exploiting only about 5% of the total dataset expected from the HL-LHC program. This rapid advancement reflects not only the quantity of data collected but also significant improvements in analysis techniques. These innovations have extended the sensitivity of the LHC experiments well beyond initial expectations, enabling, for instance, the probing of second-generation fermions through decays such as $H \to \mu\mu$ and $H \to c\bar{c}$ [13,14] (see the top row of Figure 3).

The search for double Higgs boson (HH) production has made remarkable progress since the first round of analyses. Initially a highly challenging measurement, recent developments have significantly improved sensitivity (see section 2). By the end of Run 2, the precision achieved in HH searches matched the expectations originally set for an integrated luminosity of 1000 fb$^{-1}$ at the HL-LHC [15], highlighting the power of experimental innovation and analysis improvements in pushing the limits of what is experimentally achievable (see the bottom row of Figure 3).

The status of the global EW fit as of 2023 reflects a landscape shaped by EW precision observables (EWPO) measured at lepton colliders, complemented increasingly by contributions from hadron colliders. While hadron colliders have historically played a secondary role in measuring EWPO, the high precision reached by the LHC experiments has inverted this tendency. For instance, CMS recently measured the branching ratios of hadronic *W* decays [16], outperforming LEP-era results and marking a clear step forward in the precision EW program at hadron colliders.

A recent full Run 2 (2015-2018) measurement of the effective EW mixing angle ($\sin^2 \theta_{\text{eff}}$) was performed by the CMS Collaboration using the forward-backward asymmetry ($A_{FB}$) in Drell–Yan events [17]. This result achieves a precision greater than that of the LEP combination for the same quantity and is now comparable to the most precise determinations from LEP $A_{FB}^b$ and SLD $A_{LR}$. The extracted value, $\sin^2 \theta_{\text{eff}} = 0.23157 \pm 0.00010$ (stat) $\pm$ 0.00015 (syst) $\pm$ 0.00009 (theory) $\pm$ 0.00027 (PDF), sits between the LEP and SLD measurements and is in excellent agreement with the SM prediction. This result contributes valuable insight into a long-standing tension among EW precision measurements and showcases the growing impact of hadron collider data in this domain.

Last summer, LHCb released its own precise determination of $\sin^2 \theta_{\text{eff}}$, also obtained through a measurement of $A_{FB}$ in Drell–Yan events [18]. The result, $\sin^2 \theta_{\text{eff}} = 0.23152 \pm 0.00044$ (stat) $\pm$ 0.00005 (syst) $\pm$ 0.00022 (theory), is remarkably competitive despite a significantly larger statistical uncertainty (due to the lower integrated luminosity at LHCb, owing to the beam separation necessary to keep pileup under control). This result stands out because of its notably smaller theoretical uncertainty—especially with respect to parton distribution functions (PDFs)—thanks to the unique forward phase space probed by the LHCb detector. This complementary approach enhances the global picture of EW precision and underscores the value of diverse experimental environments at the LHC.

In addition, hadron colliders traditionally provide unique inputs to the EW fit, such as the *W* boson mass ($m_W$), the top quark mass ($m_t$), and the Higgs boson mass ($m_H$). However, these contributions have not come without controversy. Notable tensions include discrepancies between $m_t$ measurements from the Tevatron and the LHC, and the striking difference between the CDF measurement of $m_W$ and the global average from other experiments.

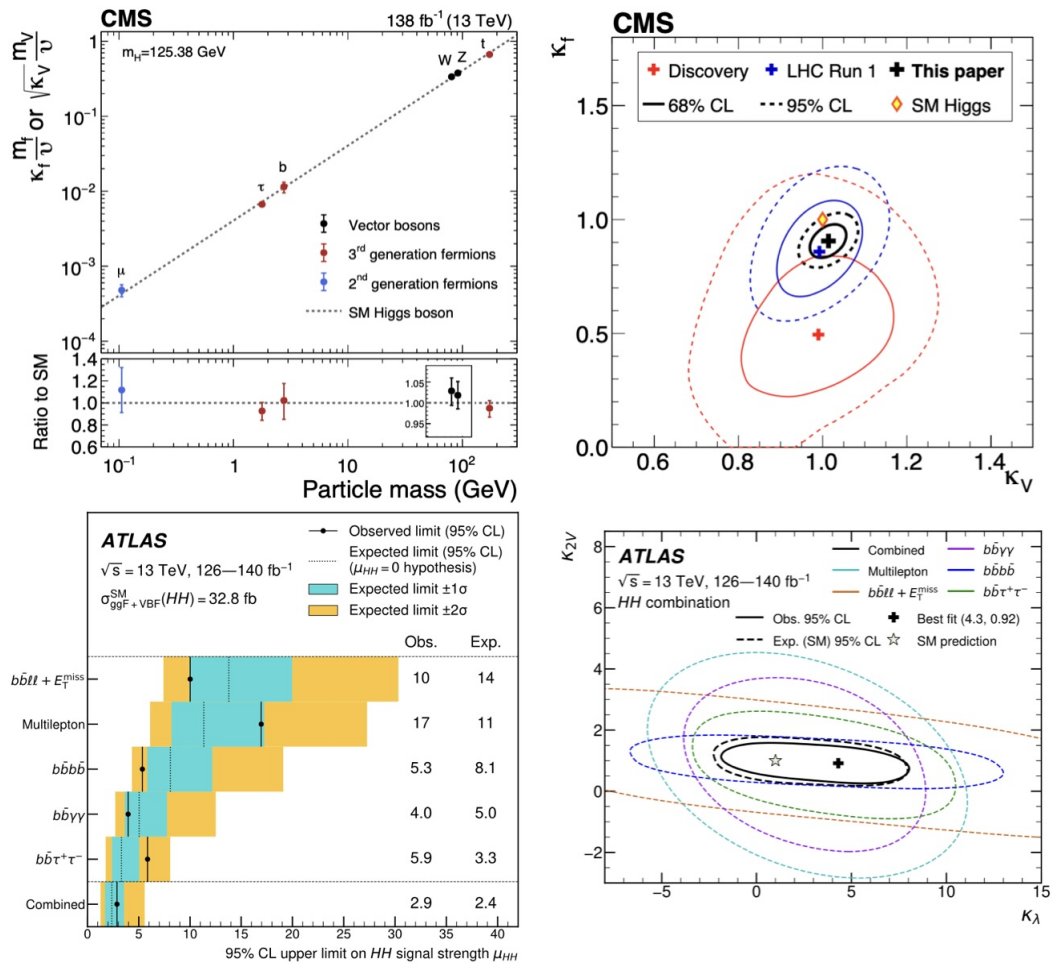

**Figure 3.** (**Top-left**): CMS measurements of the Higgs couplings to bosons and fermions as a quadratic and linear function of the mass, respectively. (**Top-right**): extraction of the universal fermion and boson coupling modifiers from a combined fit to the CMS coupling measurements. (**Bottom-left**): 95% CL upper limit on the $HH$ signal strength (normalized to the SM prediction) from ATLAS searches to various final states. (**Bottom-right**): Two-dimensional bounds on $\kappa_{2V}$ and $\kappa_{\lambda}$ coupling modifiers, obtained from ATLAS $HH$ searches.

The precision reached by the LHC experiments has been crucial to solve these issues. ATLAS utilized the low-pileup dataset collected in Run 1 (2010–2012) to perform a high-precision measurement of both $m_W$ and the $W$ width [19]. The analysis leveraged decays to both muons and electrons, extracting information from the transverse mass ($m_T$) and lepton transverse momentum ($p_T^\ell$) distributions to maximize sensitivity. More recently, CMS released the most precise determination of $m_W$ at the LHC to date [20]. This measurement was based on approximately half of the 2016 dataset and focused exclusively on the muon decay channel, using the $p_T^\mu$ distribution to minimize sensitivity to pileup effects. In addition to the main result, CMS provided a complementary analysis using relaxed theoretical assumptions, which yielded a consistent outcome. Overall, the CMS measurement is in agreement with previous LHC results and aligns well with SM predictions, reinforcing the robustness of the LHC's EW precision program.

In parallel, a comprehensive program exists to determine $m_t$, employing multiple techniques across different final states (leptonic and hadronic), production processes (cross-section based vs. kinematic reconstructions), and event topologies (resolved vs. boosted) [21]. The most precise determination to date comes from a combined Run 1 ATLAS and CMS analysis [22]. Run 2 measurements, enhanced by modern statistical techniques such as full profiling of systematics (similar to the Higgs discovery strategy), have independently achieved comparable precision.

At the LHC, $m_H$ has been precisely measured using the two golden decay channels: $H \to \gamma\gamma$, which exploits the excellent electromagnetic calorimeter resolutions, and $H \to ZZ^* \to 4\ell$, which leverages the high-precision tracking system. ATLAS has achieved the most precise determination on $m_H$, combining the two channels ($m_H = 125.11 \pm 0.11$ GeV) [23]. CMS has delivered the best single-channel measurement so far

($m_H = 125.04 \pm 0.12$ GeV) [24].

Besides providing valuable inputs to the EW fit, these measurements give a unique hint on the history of our Universe [25]. Assuming the validity of the SM up to the Planck scale, the values of $m_H$ and $m_t$ are crucial inputs for determining the stability of the EW vacuum. Current best-fit values for $m_H$ and $m_t$ place the vacuum at the boundary between stability and metastability [26]. However, a definitive statement requires a further improvement in precision, both on the input masses and on the strong coupling constant, $\alpha_s$.

The knowledge of $\alpha_s$ has significantly improved at the LHC. CMS recently performed a combined analysis of inclusive jet production across four center-of-mass energies—2.76, 7, 8, and 13 TeV—yielding the most precise determination of $\alpha_s$ from jet events to date. The analysis was carried out at next-to-next-to-leading order (NNLO) in perturbative QCD and simultaneously constrained the parton distribution functions (PDFs) in situ, further reducing systematic uncertainties [27]. Despite this achievement, the ultimate precision still falls short of that obtained from measurements based on the $Z$ boson transverse momentum ($p_T^Z$), which are less sensitive to jet energy scale uncertainties. In this context, ATLAS has released the most precise determination of $\alpha_s$ at the LHC, based on the $p_T^Z$ spectrum [28]. While discussions are ongoing regarding whether this measurement qualifies as N³LO, the improvement in experimental precision is unquestionable and marks a significant leap forward in the determination of one of the fundamental parameters of the SM.

One of the most remarkable achievements in precision flavor physics has come from LHCb's improved determination of the CKM unitarity triangle (UT) angle $\gamma$ [29]. Historically, $\gamma$ was the least precisely known angle of the UT, but it is now measured with an uncertainty of only a few degrees. This represents a significant step forward in our understanding of the CKM matrix, as $\gamma$ is determined from tree-level processes and is therefore largely insensitive to potential new physics contributions. As such, it provides a clean SM reference point for testing consistency with indirect measurements affected by possible beyond-the-SM (BSM) effects. With this improvement, the precision of tree-level determinations of the CKM parameters has reached a level comparable to that of the full pre-LHCb global fits. For instance, the tree-level analysis now yields $|\bar{\rho}| = 0.158 \pm 0.026$ and $|\bar{\eta}| = 0.358 \pm 0.012$ [30], illustrating how the CKM matrix can now be reliably determined from purely SM-safe observables. In addition, further precision measurements such as CP violation in neutral meson mixing provide stringent bounds on potential new physics amplitudes in $|\Delta F| = 2$ processes, further constraining the BSM landscape.

CP violation measurements in neutral meson mixing have now reached an astonishing level of precision, with the three main LHC experiments—ATLAS, CMS, and LHCb—contributing results at comparable precision. The field has now entered the regime of $\mathcal{O}(10^{-3})$ sensitivity, enabling clear evidence of CP violation in the $B_s$ system [31].

The first phase of the LHC physics program is nearing completion. By the end of 2025, both ATLAS and CMS aim to have collected over 300 fb$^{-1}$ of integrated luminosity EACH, combining Run 2 and Run 3 (2022–2026) data. This milestone is expected to be achieved within the current year, marking a successful conclusion to the initial exploration and precision program of the LHC.

## 4. A Look into the Future

Preparations are already well underway for the HL-LHC phase. The upgraded machine, together with substantially enhanced detectors, is targeted to deliver an integrated luminosity of 3000 fb$^{-1}$ by 2041. In the meantime, both experiments are already pushing their current detectors far beyond their original design. For instance, CMS has operated stably while recording events with up to 63 simultaneous proton-proton collisions—2.5 times its original design tolerance and 45% of what is expected during HL-LHC conditions. Data-taking rates have also ramped up significantly, with CMS reaching 7 kHz, which corresponds to 70% of HL-LHC targets and nearly 7 times the standard operational rate during Run 2. In practice, many HL-LHC challenges—particularly related to pileup—are already being addressed during Run 3 operations.

Looking ahead to Run 4 and beyond, further increases in data throughput will come at the cost of even more intense pileup environments, potentially reaching an average of 140 simultaneous collisions per bunch crossing. This is substantially larger than the original LHC design, which anticipated tolerances around 20 pileup events. However, both ATLAS and CMS have already demonstrated their ability to handle pileup levels around 60 in routine operations, thanks to rapid algorithmic developments and advanced reconstruction techniques. In addition, the upcoming detector upgrades are designed precisely with these challenges in mind. Key improvements include:

- Larger angular coverage, especially in tracking detectors, enabling robust reconstruction in the forward regions.
- Higher granularity in calorimeters and trackers, facilitating a better tracking and vertexing accuracy, an improved pileup suppression, and more precise jet and lepton identification.

- Precision timing layers, providing time-of-flight information for vertex separation and PID, crucial in mitigating pileup effects and for flavor and heavy ion physics.
- Particle-flow algorithms extended to the forward region, enhancing object reconstruction and energy resolution.
- Hardware-based *Track Triggers*, allowing tracking information to be used directly at the earliest stages of the trigger system.

Thanks to these improvements, the HL-LHC era will not only deliver more data, but also higher quality data, enabling continued exploration of the SM and extending our reach into potential new physics territories.

Currently, ATLAS and CMS work along two independent but complementary fronts to search for new physics at the LHC. The first, *direct searches for new physics*, involves assuming a specific new physics model and searching for it through hypothesis testing. This approach relies heavily on data-driven models of background processes, using techniques such as template fits for bump hunting, where an excess in the invariant mass spectrum would point to new particles or resonances. These searches are highly model-dependent, assuming particular signatures of new physics such as new particles or interactions, and they provide direct evidence or exclusion of these models.

The second front, *indirect bounds from measurements*, focuses on precise measurements of SM processes. By measuring absolute and differential cross-sections for SM processes and comparing them to theoretical predictions, experiments can constrain new physics models by ensuring that no significant deviations from the SM predictions are observed. This is where the growing interest in Effective Field Theory (EFT) plays a crucial role. EFT allows for a systematic framework to encode potential new physics effects as higher-dimensional operators in the Lagrangian, offering a way to probe deviations from the SM in a model-independent manner. In this context, EFT analyses effectively combine both direct searches and precision measurements, as they search for discrepancies between the data and the SM predictions that might signal new physics beyond the SM. A key challenge in this area is minimizing systematic uncertainties, especially those arising from theoretical uncertainties in the calculation of SM processes.

The increasing interest in EFT and its application to high-energy processes has blurred the lines between direct searches and indirect measurements. For example, searches for large extra dimensions or broad resonances can be interpreted within the EFT framework. Similarly, high-precision measurements of differential cross-sections can provide sensitivity to new physics in a similar way to direct searches for new particles. On the timescale of the HL-LHC, the direct and indirect search programs will likely merge, with EFT analyses becoming a central tool for interpreting deviations from the SM across a wide range of processes. Recasting studies [32]—where the same experimental data is analyzed under different new physics scenarios—will be crucial in understanding the full implications of the data.

Looking ahead, the Higgs factory will improve precision by a factor of 2–3 for coupling measurements involving the $W$, $Z$, and $g$ bosons, with a particular focus on third-generation quarks. However, it is important to note that the Higgs factory will not improve LHC measurements for rare decays, as these are primarily loop-mediated processes. The LHC will remain the primary probe for rare Higgs decays and serve as an indirect tool for new physics searches. Assuming no other hadron collider before 2070, the LHC will remain the only machine capable of probing this and other crucial aspects of SM physics at the highest energies. Additionally, the study of Higgs boson pair (HH) production and its implication on the shape of the Higgs potential will be even more central, providing a unique opportunity to probe new physics scenarios related to baryogenesis and other aspects of the physics of the early universe [33]. The top quark Yukawa coupling, which can be probed in several processes (such as $t\bar{t}H$ production), is another important target for precision measurements. The LHC will also continue to offer crucial insights into multi-top production, vector boson scattering (VBS), and other precision measurements of SM couplings. Before the advent of a dedicated Higgs factory, the LHC is poised to deliver precise measurements of these couplings [33].

Historically, new physics searches were focused on processes with small SM amplitudes. With the HL-LHC, however, we expect to see an unprecedented improvement in precision, particularly in the determination of the Unitarity Triangle (UT) parameters, with LHCb's precision step-up pushing the UT analysis below 1% precision [34]. This will set a new milestone for the SM and provide stringent constraints for beyond-the-SM model building. Further efforts from CMS to contribute to this area are expected, though the long-term implications of new parking strategies are still under active assessment.

## 5. Conclusions

The LHC began its journey as the ultimate discovery machine, tasked with probing the deepest open questions of particle physics. Its early successes, such as the landmark discovery of the Higgs boson, marked a monumental achievement for the SM. The LHC also provided a unique platform for exploring the potential existence of new physics, including the search for supersymmetry (SUSY) and its various alternatives. These endeavors demonstrated the LHC's unparalleled ability to push the boundaries of our knowledge, both through direct searches and precision measurements.

As the LHC entered its precision era, significant improvements in detector technology and novel computational algorithms, including advances in Deep Learning, have enabled unprecedented levels of precision in data analysis. With the advent of new data-taking strategies, such as scouting and parking, the experiments could maximize their reach and overcome challenges related to the immense data volumes. The combination of these innovations has allowed LHC experiments to surpass the precision of previous machines like LEP and Tevatron on many fronts, setting the stage for even more refined tests of the SM.

Looking toward the future, the upcoming HL-LHC will further enhance precision through new detector capabilities, allowing the experiments to continue to explore fundamental questions such as the Higgs potential and vacuum stability. With the HL-LHC, the ATLAS and CMS detectors will be equipped with new capabilities, enabling them to remain the leader in high-energy physics experiments for the foreseeable future, maintaining an unchallenged position until the next major collider. The legacy of the LHC, both in terms of its direct contributions to understanding particle physics and its critical role in shaping future collider experiments, will leave an enduring mark on the field.

**Acknowledgments**

I would like to thank the organizers of the "Rise of Particle Physics" workshop for the opportunity to speak at such an engaging event. It was an honor to share the stage with some of the pioneers of the SM, whose remarkable contributions to particle physics I studied in the very same place where this workshop was held.

**Conflicts of Interest**

**References**

1. ATLAS Collaboration. Observation of a new particle in the search for the Standard Model Higgs boson with the ATLAS detector at the LHC. *Phys. Lett. B* **2012**, *716*, 1–29.
2. CMS Collaboration. Observation of a new boson at a mass of 125 GeV with the CMS experiment at the LHC. *Phys. Lett. B* **2012**, *716*, 30–61.
3. Gianotti, F. Talk at ICHEP 2022, Bologna, Italy. Available online: https://agenda.infn.it/event/28874/contributions/171915/attachments/95072/130540/ICHEP-Higgs-2022-Fabiola.pdf (accessed on 30 March 2025).
4. CMS Collaboration. Enriching the physics program of the CMS experiment via data scouting and data parking. *Phys. Rep.* **2025**, *1115*, 678–772.
5. Aaij, R.; Benson, S.; Cian, M.D.; et al. A comprehensive real-time analysis model at the LHCb experiment. *J. Instrum.* **2019**, *14*, P04006.
6. Boveia, A.; ATLAS Collaboration. Trigger Level Analysis Technique in ATLAS for Run 2 and Beyond. 2019. Available online: https://cds.cern.ch/record/2703715 (accessed on 20 November 2024).
7. CMS Collaboration. *Performance of Heavy-Flavour Jet Identification in Boosted Topologies in Proton-Proton Collisions at $\sqrt{s} = 13$ TeV*; CMS-PAS-BTV-22-001; CERN: Geneva, Switzerland, 2023.
8. ATLAS Collaboration; CERN. ATL-PHYS-PUB-2023-021. 2023. Available online: https://cds.cern.ch/record/2866601 (accessed on 20 November 2024).
9. CMS Collaboration. Identification of heavy-flavour jets with the CMS detector in pp collisions at 13 TeV. *J. Instrum.* **2018**, *13*, P05011.
10. ATLAS Collaboration. ATLAS flavour-tagging algorithms for the LHC Run 2 $pp$ collision dataset. *Eur. Phys. J. C* **2023**, *83*, 681.
11. CMS Collaboration. Identification of hadronic tau lepton decays using a deep neural network. *J. Instrum.* **2022**, *17*, P07023.
12. ATLAS Collaboration. *Identification of Hadronic Tau Lepton Decays Using Neural Networks in the ATLAS Experiment*; ATL-PHYS-PUB-2019-033; CERN: Geneva, Switzerland, 2019.
13. CMS Collaboration. A portrait of the Higgs boson by the CMS experiment ten years after the discovery. *Nature* **2022**, *607*, 60–68.
14. ATLAS Collaboration. A detailed map of Higgs boson interactions by the ATLAS experiment ten years after the discovery. *Nature* **2022**, *607*, 52–59.
15. ATLAS; CMS Collaborations; CERN. CMS-PAS-FTR-22-001. 2022. Available online: https://cds.cern.ch/record/2806962 (accessed on 20 November 2024).

16. CMS Collaboration. Measurement of the W boson decay branching fraction ratio $\mathcal{B}(\mathrm{W} \to cq)/\mathcal{B}(\mathrm{W} \to \mathrm{q}\bar{\mathrm{q}}')$ in proton-proton collisions at $\sqrt{s} = 13$ TeV. *arXiv* **2024**, arXiv:2412.16296.

17. CMS Collaboration. Measurement of the Drell-Yan forward-backward asymmetry and of the effective leptonic weak mixing angle in proton-proton collisions at $\sqrt{s} = 13$ TeV. *arXiv* **2024**, arXiv:2408.07622.

18. LHCb Collaboration. Measurement of the effective leptonic weak mixing angle. *J. High Energy Phys.* **2024**, *12*, 026.

19. ATLAS Collaboration. Measurement of the W-boson mass and width with the ATLAS detector using proton–proton collisions at $\sqrt{s} = 7$ TeV. *Eur. Phys. J. C* **2024**, *84*, 1309.

20. CMS Collaboration. High-precision measurement of the W boson mass with the CMS experiment at the LHC. *arXiv* **2024**, arXiv:2412.13872.

21. CMS Collaboration. Review of top quark mass measurements in CMS. *Phys. Rep.* **2025**, *1115*, 116–218.

22. ATLAS; CMS Collaborations. Combination of inclusive top-quark pair production cross-section measurements using ATLAS and CMS data at $\sqrt{s} = 7$ and 8 TeV. *J. High Energy Phys.* **2023**, *7*, 213.

23. ATLAS Collaboration. Combined Measurement of the Higgs Boson Mass from the and Decay Channels with the ATLAS Detector Using , 8, and 13 TeV Collision Data. *Phys. Rev. Lett.* **2023**, *131*, 251802.

24. CMS Collaboration. Measurement of the Higgs boson mass and width using the four-lepton final state in proton-proton collisions at $\sqrt{s} = 13$ TeV. *arXiv* **2024**, arXiv:2409.13663.

25. Espinosa, J.R.; Giudice, G.F.; Morgante, E.; et al. The cosmological Higgstory of the vacuum instability. *J. High Energy Phys.* **2015**, *9*, 174.

26. Hiller, G.; Höhne, T.; Litim, D.F.; et al. Vacuum stability in the Standard Model and beyond. *Phys. Rev. D* **2024**, *110*, 115017.

27. CMS Collaboration. Determination of the strong coupling and its running from measurements of inclusive jet production. *arXiv* **2024**, arXiv:2412.16665.

28. ATLAS Collaboration. A precise determination of the strong-coupling constant from the recoil of Z bosons with the ATLAS experiment at $\sqrt{s} = 8$ TeV. *arXiv* **2023**, arXiv:2309.12986.

29. LHCb Collaboration. Simultaneous determination of CKM angle $\gamma$ and charm mixing parameters. *J. High Energy Phys.* **2021**, *12*, 141,

30. UTfit Collaboration. New UTfit analysis of the unitarity triangle in the Cabibbo–Kobayashi–Maskawa scheme. *Rend. Lincei. Sci. Fis. Nat.* **2023**, *34*, 37–57.

31. CMS Collaboration. Evidence for $CP$ violation and measurement of $CP$-violating parameters in $\mathrm{B}_s^0 \to \mathrm{J}/\psi, \phi(1020)$ decays in pp collisions at $\sqrt{s} = 13$ TeV. *arXiv* **2024**, arXiv:2412.19952.

32. Cranmer, K.; Yavin, I. RECAST—extending the impact of existing analyses. *J. High Energy Phys.* **2011**, *4*, 38.

33. ATLAS; CMS Collaborations. Highlights of the HL-LHC physics projections by ATLAS and CMS. *arXiv* **2025**, arXiv:2504.00672.

34. ATLAS; Belle II; CMS; et al. Projections for Key Measurements in Heavy Flavour Physics. *arXiv* **2025**, arXiv:2503.24346.

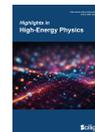

*Review*

# Recent Developments in Flavor Physics, the Unitary Triangle Fit, Anomalies and All That

Guido Martinelli [1,2]

[1] INFN, Sezione di Roma, P.le A. Moro 2, 00185 Roma, Italy; guido.martinelli@roma1.infn.it
[2] Dipartimento di Fisica, Università di Roma La Sapienza, P.le A. Moro 2, 00185 Roma, Italy

**Abstract:** Flavor physics represents one of the main fields of investigation to answer some fundamental questions like the baryon anti-baryon asymmetry or the presence of dark matter in our universe. Precision studies of the Cabibbo-Kobayashi-Maskawa matrix offer a very important testing ground of the Standard Model and, in the light of recent theoretical progress and experimental measurements, the status of the Unitarity Triangle is a fundamental tool to uncover physics beyond the Standard Model. The results of the most recent global fits performed by the **UT**fit Collaboration, including all the up-to-date experimental and theoretical inputs are reported and stringent constraints on New Physics from the generalized $|\Delta F| = 2$ effective Hamiltonian are presented.

**Keywords:** flavor physics; lattice gauge theory

## 1. Introduction

Flavor physics is a fundamental part of the Standard Model (SM) and one of the most powerful tools that we have at disposal to discover New Physics (NP) effects. In this talk I will discuss our present understanding of flavor physics and some of the processes that we are studying in order to discover signals of physics beyond the SM.

The plan of the talk is the following: after a brief review of lessons from the past, I will discuss the present situation of flavor physics within the SM by reviewing many different processes and then I will conclude with the searches of signals of physics beyond the SM, expecially in radiative or rare decays, such those involving flavor changing neutral currents.

### 1.1. Lessons from the Past

The first example of the importance of flavor physics, although at the time it was not called in this way, is the Fermi theory of $\beta$ decays. This is an example of a non-renormalisable, effective field theory that becomes non unitary at a scale of $O(100)$ GeV, needs to be unitarised by new degrees of freedom at this energy scale and thus it represents an example of indirect evidence for the existence of NP, in this case the Standard Model itself with the $W$ and $Z$ vector bosons. After the Fermi theory many fundamental experimental discoveries and theoretical advances opened the way to new concepts and understanding of our universe from sub-atomic to cosmological distances. Just to mention some of them

- 1963 The Cabibbo angle;
- 1964 The discovery of CP violation in neutral $K$ decays;
- 1970 The GIM Mechanism;
- 1973 Kobayashi-Maskawa showed that CP violation needs at least three quark families;
- 1975 The discovery of the $\tau$ lepton, the third lepton family;
- 1977 The discovery of the $b$ quark, the third quark family;
- 2003/2004 The discovery of CP violation in $B$ meson decays.

On the basis of the above progress the present SM has been built as a magnificent construction able to explain and describe an enormous variety of phenomena. In a more recent past theoretical calculations allowed also to predict, prior to their measurements, several quantities like for example the value of $\sin 2\beta$ [1] and the value of the $B_s - \bar{B}_s$ mixing amplitude $\Delta M_s$ [2], see below. These predictions, subsequently confirmed by their experimental values, remain a very important test of the model at the quantum level.

### 1.2. The Standard Model

The modern approach is to consider the Standard Model, a renormalisable gauge theory, as an effective theory valid at low energies, namely at energy scales well below the TeV, above which particles corresponding to new, still undiscovered, degrees of freedom will become manifest. The model can be written in terms of a Lagrangian, based on a $SU(3)_c \times SU(2)_W \times U(1)_Y$ (color $\times$ weak $\times$ hyper-charge) gauge symmetry, spontaneously broken to $SU(3)_c \times U(1)_{em}$, where only the color and the electromagnetic symmetries survive. The symmetry breaking is obtained through the classical Higgs mechanism, with the Higgs field belonging to a weak isospin doublet. The $W^\pm$ and $Z^0$ vector mesons and the quarks (three families) acquire a mass proportional to their coupling to the Higgs particle. Schematically we may write

$$\mathcal{L} = \Lambda^4 + \Lambda^2 H^2 + \lambda H^4 + (D_\mu H)^2 +$$
$$\bar{\psi} \slashed{D} \psi + + Y H \bar{\psi} \psi + F_{\mu\nu}^2 + F_{\mu\nu} \tilde{F}_{\mu\nu} +$$
$$\frac{1}{\Lambda} (\bar{L} H)^2 + \frac{1}{\Lambda^2} \sum_i C_i O_i + \dots , \tag{1}$$

where in the first line we have the vacuum energy and the interactions of the Higgs particle with itself and with the gauge fields; in the second line the interactions of the Fermions with the gauge fields and with the Higgs particle are shown togheter with the gauge fields Lagrangian (including the source of strong CP violation); the third line represents the expansion of the (non-renormalisable) effective Lagrangian in terms of operators of higher and higher dimensions ($D \geq 4$) which are the low energy manifestation of physics Beyond the Standard Model (BSM). Thus, for example, the first term in the third line of Equation (1) will give a mass to the neutrinos.

If we only impose the gauge symmetries to our model, this theory is very elegant relying only in three parameters namely the three gauge couplings corresponding to the $SU(3)_c \times SU(2)_W \times U(1)_Y$ symmetry. One can imagine that these couplings will be unified in a single parameter within a Grand Unification scheme (GUT). The introduction of the Higgs field, with the accompanying symmetry breaking, a necessary step in order to give a mass to the vector mesons as dictated by the experiments, introduce 19 arbitrary, measurable parameters

$$3 g_i + (\lambda, M_H) + 6 m_q + 3 m_\ell + 3\theta_{CKM} + \delta + \theta_{QCD} = 19 , \tag{2}$$

which correspond to the three gauge couplings, $g_i$, the Higgs self coupling and mass, $\lambda$, $M_H$, the quark and lepton masses, $m_q$, $m_\ell$, the Cabibbo-Kobayashi-Maskawa (CKM) angles [3,4], $\theta_{CKM}$, and phase, $\delta$, and the coupling of the strong CP violating term $\theta_{QCD}$. The introduction of the neutrino masses and mixing increases by nine the number of arbitrary parameters $3 m_\nu + (2 + 1) \delta_{PMNS} + 3 \theta_{PMNS} = 9$.

Thus with the Higgs field, the model becomes chaotic with many unanswered questions: why we have three families and not one or five or any other number (Rabi); why the model includes the fundamental breaking of parity (Landau); the model has too many arbitrary features for its predictions to be taken too seriously (Weinberg). There are other puzzling aspects in the model: the Higgs is the only particle the couplings of which are not gauge couplings and this is at the basis of the instability of the SM at the quantum level and of the hierarchy problem; the masses of the Fermions do not follow any simple rational rule, their values look like the result of a lottery drawing, see Table 1, spanning about five orders of magnitude just for the quark sector; unlike the electro-magnetic, neutral currents and strong interactions couplings, the couplings of weak charged currents are hierarchical with a strong correlation to the quark masses. Flavor physics is indeed the study of the weak couplings and CP violation with the aim of understanding the above puzzles.

**Table 1.** Full lattice inputs. The values of the quark masses have been obtained by taking the weighted averages of the $N_f = 2 + 1$ and $N_f = 2 + 1 + 1$.

| Input | Lattice/Exp |
|---|---|
| $m_u^{\overline{\text{MS}}}(2\,\text{GeV})$ (GeV) | 2.20(9) MeV |
| $m_d^{\overline{\text{MS}}}(2\,\text{GeV})$ (GeV) | 4.69(2) MeV |
| $m_s^{\overline{\text{MS}}}(2\,\text{GeV})$ (GeV) | 93.14(58) MeV |
| $m_c^{\overline{\text{MS}}}(2\,\text{GeV})$ (GeV) | 993(4) MeV |
| $m_c^{\overline{\text{MS}}}(m_c^{\overline{\text{MS}}})$ (GeV) | 1277(5) MeV |
| $m_b^{\overline{\text{MS}}}(m_b^{\overline{\text{MS}}})$ (GeV) | 4196(19) MeV |
| $m_t^{\overline{\text{MS}}}(m_t^{\overline{\text{MS}}})$ (GeV) | 163.44(43) GeV |

## 2. Precision Flavor Physics and the Search for New Physics Signals

In addition to gauge symmetry and renormalizability, the Standard Model is characterised by the so called accidental symmetries, which are present in the model although not required in its construction. Before the introduction of the Higgs and of the symmetry breaking, the flavor symmetries are $U(3)^5$, corresponding to all the possible internal rotations of the Fermion fields. With the symmetry breaking, the only remaining symmetries are those associated to baryon, $B$, and lepton, $L$, number conservation, although, beyond perturbation theory, only $B - L$ is conserved. Since at zero temperature $B$ violation is negligible, we may still consider baryon number as a realised accidental symmetry of the SM. Similarly, by neglecting neutrino masses and mixing, we have the separate conservation of the different lepton flavours, $L = L_e \times L_\mu \times L_\tau$.

The most interesting processes to identity signals of physics BSM are those which are forbidden in the Standard Model. For instance, because of an accidental symmetry, proton decay is not allowed if all the symmetries of the Standard Model are preserved, together with its renormalizability: in order to trigger proton decay, maintaining gauge invariance, you would need a new operator with at least dimension 6. In the SM, even if $\mu \to e\,\gamma$ is not forbidden since neutrinos have a mass, the branching ratio $BR(\mu \to e\,\gamma) \sim \alpha_{em} m_\nu^4/m_W^4 \sim 10^{-52}$ is, however, so small that even a single event would be an indication of new physics. This is precisely what the experiment MEG is looking for.

After the forbidden or almost forbidden processes the most interesting ones to study are those which are heavily suppressed, in particular due to the GIM mechanism, see below, like processes induced by flavour-changing neutral currents, i.e., weak currents where the charge does not change, as for instance

$$q_i \to q_k\,\nu\,\overline{\nu} \qquad q_i \to q_k\,\ell^+\,\ell^- \qquad q_i \to q_k\,\gamma \tag{3}$$

where $i$ and $k$ are flavor indices related to equal charge quarks. These processes are forbidden at three level and, in general, Cabibbo suppressed, see below. For this reason they are particular sensitive to new physics, so that an accurate calculation of the relevant matrix elements from lattice simulations is of particular interest.

Coming back to the Standard Model, one of the salient and most interesting objects of investigation is the CKM matrix, $(\mathbf{V}_{\text{CKM}})_{ij} = V_{ij}$, which controls the strength of weak charged currents

$$L^{cc} = \frac{g_W}{\sqrt{2}} \left( \bar{u}_L^i V_{ij} \gamma^\mu d_L^j W_\mu^+ + h.c. \right), \tag{4}$$

where $i$ and $j$ are the flavor indices, $u^i \equiv (u, c, t)$ and $d^i \equiv (d, s, b)$. The amount of CP violation entailed by a certain process is all contained, after the diagonalization of the mass matrix, in the CKM matrix. With $N$ generations of quarks the CKM matrix is characterised by $N(N-1)/2$ Euler angles and $(N-1)(N-2)/2$ phases which generate CP violation. With $N = 3$ we have three angles and one phase. The standard representation of the CKM matrix is

$$\mathbf{V}_{\text{CKM}} = \begin{pmatrix} c_{12}c_{13} & s_{12}c_{13} & s_{13}e^{-i\delta_{13}} \\ -s_{12}c_{23} - c_{12}s_{23}s_{13}e^{i\delta_{13}} & c_{12}c_{23} - s_{12}s_{23}s_{13}e^{i\delta_{13}} & s_{23}c_{13} \\ s_{12}s_{23} - c_{12}c_{23}s_{13}e^{i\delta_{13}} & -c_{12}s_{23} - s_{12}c_{23}s_{13}e^{i\delta_{13}} & c_{23}c_{13} \end{pmatrix}. \tag{5}$$

If you look at the strength of weak couplings, you discover that the CKM matrix is very close to the identity matrix since, as you go out of the main diagonal, i.e., as you proceed from the lightest to heaviest quarks, the matrix elements become smaller and smaller. This suggested to Wolfenstein that the CKM matrix $\mathbf{V}_{\text{CKM}}$ can be expanded in a small parameter given by $\lambda = s_{12} = \sin\theta_c$ ($s_{23} = \sin\theta_{23} = A\lambda^2$, $s_{13}e^{-\delta_{13}} = \sin\theta_{13}e^{-\delta_{13}} = A\lambda^3(\rho - i\eta)$), where $\theta_c$ is the Cabibbo angle. $\mathbf{V}_{\text{CKM}}$ can be written therefore as

$$\mathbf{V}_{\text{CKM}} = \begin{pmatrix} 1 - \frac{1}{2}\lambda^2 & \lambda & A\lambda^3(\rho - i\eta) \\ -\lambda & 1 - \frac{1}{2}\lambda^2 & A\lambda^2 \\ A\lambda^3(1 - \rho - i\eta) & -A\lambda^2 & 1, \end{pmatrix} + \mathcal{O}(\lambda^4), \tag{6}$$

where we have $\lambda \approx 0.2$, $\eta \approx 0.2$, $A \approx 0.8$ and $\rho \approx 0.3$. Indeed one would expect $\rho$ and $\eta$ of order one whereas their value is small and of the order of the Cabibbo angle.

Everything I will discuss in the following will be expressed in term of $\rho$ and $\eta$. In particular $\eta$ corresponds to the phase that is necessary to provide CP violation in the CKM matrix. Note that the condition that in the SM at least three generations are needed to have CP violation [4] is necessary but not sufficient because the value of $\delta_{13}$ ($\eta$) is not a priori determined and could be zero. In this case the observed CP violation could arise from some other mechanism from physics beyond the SM.

As a last remark on the CKM matrix let me recall the properties related to its unitarity. This implies that the scalar product of any two columns or rows is zero, e.g., $V_{11}V_{12}^* + V_{21}V_{22}^* + V_{31}V_{32}^* = 0$. Each of the terms summed in the scalar product is in general a complex number, namely a vector in the complex plane, that we may define as $a_1 = V_{11}V_{12}^*$, $a_2 = V_{21}V_{22}^*$ and $a_3 = V_{31}V_{32}*$. The condition $a_1 + a_2 + a_3 = 0$ defines a *unitary triangle* and you may define as many triangles as scalar products obtained by multiplying any two rows or two columns. If you change the convention on quark phases a triangle will rotate in space and the only physical quantities are those which remain invariant under these rotations, as for instance the length of the sides, the angles between sides and the area. The measure of the area in particular is related to the amount of CP violation and it is zero if $\eta = 0$. The main investigation effort is in the attempt to extract from data such unitary triangles. Among these Unitary Triangles (UT), the most studied, since is the easiest to be measured (none of its sides is too small) and it has been overdetermined by several experimental measurements, is the one defined by the product $V_{ud}V_{ub}^* + V_{cd}V_{cb}^* + V_{td}V_{tb}^* = 0$, shown in Figure 1 together with its three related angles.

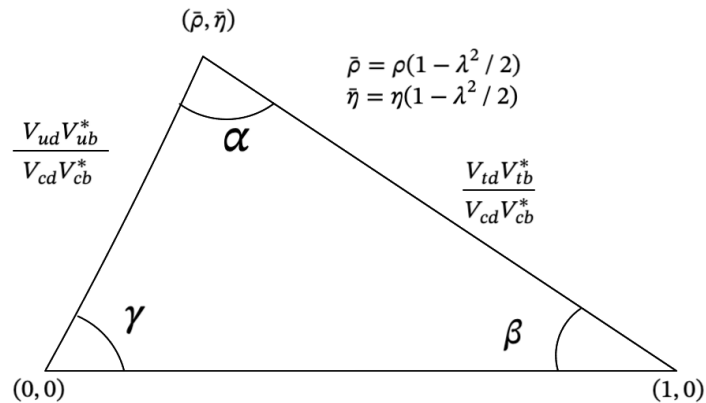

**Figure 1.** The "standard" unitarity triangle of the Standard Model from the three generations CKM matrix.

In the SM, another possible source of CP violation is the $\theta_{QCD}$-term of strong interactions which was introduced before, i.e., $\mathcal{L}_\theta(x) = \theta_{QCD}\,\text{Tr}[\tilde{G}^{\mu\nu}G_{\mu\nu}] \sim \theta_{QCD}\vec{E}^a \cdot \vec{B}^a$, where the dual strength tensor $\tilde{G}^{\mu\nu}$ is defined as $\tilde{G}^{\mu\nu\,a} = \varepsilon^{\mu\nu\rho\sigma}G_{\rho\sigma}^a$. A non-zero $\mathcal{L}_\theta(x)$ induces an electric dipole moment in the neutron, $e_n$. Given the present experimental upper limit $e_n < 3 \cdot 10^{-26}$ e cm, see for example [5], and based on lattice calculations which are still rather uncertain, one can put on the value of the $\theta_{QCD}$ parameter an upper limit $\theta_{QCD} < 10^{-10}$.

One among the most important results in flavor physics is the determination of the allowed regions in the plane of the two parameters $\rho$ and $\eta$ of the Wolfenstein parametrisation, in particular the allowed region with respect to $\eta$, which controls the amount of CP violation. Accurate theoretical estimates and measurements of CP-even and CP-odd observables from neutral meson oscillations are of particular interest for the analysis of the unitarity triangle characterised by the determination of $V_{ud}V_{ub}^* + V_{cd}V_{cb}^* + V_{td}V_{tb}^* = 0$. Being the $\lambda$ and $A$ parameters well-constrained by leptonic and semi-leptonic meson decays, the UT analysis boils down to the investigation of all possible constraints in the $(\bar{\rho}, \bar{\eta})$ plane [6] ($\bar{\rho} = \rho\,(1 - \lambda^2/2)$ and $\bar{\eta} = \eta\,(1 - \lambda^2/2)$, see Figure 1). The sensitivity of the CKM metrology is then driven by $|V_{ub}/V_{cb}|$ from semi-leptonic $B$ decays, $\Delta M_d$ and $\Delta M_s$ from $B^0_{d,s}$-$\bar{B}^0_{d,s}$ mixing, $\varepsilon$ from neutral $K$ mixing, the angle $\alpha$, see Figure 1, from charmless ($\pi\pi$, $\pi\rho$, $\rho\rho$) non leptonic decays, $\gamma$ from $B$ decays with an open charm in the final state and $\sin 2\beta$, where $\beta$ is defined in Figure 1, from the asymmetries in decays like $B^0 \to J/\psi K^0$. The UT*fit* Collaboration has recently published a comprehensive study on the SM UT [6], and presented a related one beyond the SM in [7]. Most of the non-perturbative quantities necessary for this analysis are provided by lattice calculations which became more and more accurate over the years [8].

The structure of Yukawa couplings of the SM implies a rich phenomenology, characterized in the quark sector by the appearance of flavor Changing Neutral Currents (FCNC) only at the loop level, and further suppressed due to the Glashow-Iliopoulos-Maiani (GIM) mechanism [9], rooted in the approximate $U(2)^3$ symmetry of the first two generations. Transitions with units of flavor violation $|\Delta F| \neq 0$ as well as CP-violating observables can be studied within a scheme with six quark masses, $m_{u,d,s,c,b,t}$, five of which determined by lattice calculations, and four mixing parameters [10], $\lambda, A, \bar{\rho}, \bar{\eta}$, required to describe the unitary CKM matrix [3,4], $V_{ij}$, with $i = u, c, t$ and $j = d, s, b$.

The hierarchical structure of the CKM and the fact that the $\bar{\eta}$ parameter is the only source for CP violation in weak interactions, make processes like $|\Delta F| = 2$ transitions very sensitive probes of NP. Indeed, an active interplay of all three generations is required in order to be sensitive to CP-violating effects in the SM, strengthening the important role of loop-induced processes like FCNCs in the phenomenology of weak interactions.

## 3. Updated Inputs and Measurements

A detailed description of the experimental and theoretical inputs entering in the UT analysis can be found in [2,6]. Here we limit ourselves in highlighting the novelties for the global fits presented in the next sections. The most important theoretical updates for the analyses presented in this work comprise:

- New averages for quark masses accounting for the latest progress from lattice QCD [8]; See the online results from FLAG 2023.
- Form factors for semileptonic $B$ decays related to the exclusive determination of $|V_{cb}|$ and $|V_{ub}|$ in line with the updates from the dispersive matrix method of [11]; See Table 2.
- A novel estimate of radiative corrections to neutron decay as recently obtained by the authors of [12] in relation to the extraction of $V_{ud}$.

**Table 2.** Results for the SM global fits. In the first column we report all key observables for the determination of the UT, with corresponding experimental UT*fit* averages provided in the next column. The third and fourth column reports the outcome for each observable with or without its statistical weight in the likelihood of the global fit. In the last column we show the pull of the SM predictions with respect to the measurements.

| Observable | Measurement | Full Fit | Prediction | Pull (#$\sigma$) |
|---|---|---|---|---|
| $|V_{ud}|$ | $0.97433 \pm 0.00017$ | $0.97431 \pm 0.00017$ | $0.9737 \pm 0.0011$ | 0.6 |
| $|V_{ub}|$ | $0.00375 \pm 0.00026$ | $0.003702 \pm 0.000081$ | $0.003696 \pm 0.000087$ | 0.2 |
| $|V_{cb}|$ | $0.04132 \pm 0.00073$ | $0.04194 \pm 0.00041$ | $0.04221 \pm 0.00051$ | 1.0 |
| $\alpha\,[^\circ]$ | $93.8 \pm 4.5$ | $92.4 \pm 1.4$ | $92.3 \pm 1.5$ | 0.9 |
| $\sin 2\beta$ | $0.689 \pm 0.019$ | $0.705 \pm 0.014$ | $0.739 \pm 0.027$ | 1.5 |
| $\gamma\,[^\circ]$ | $65.4 \pm 3.3$ | $65.1 \pm 1.3$ | $65.2 \pm 1.5$ | 0.1 |
| $\Delta M_d\,[\text{ps}^{-1}]$ | $0.5065 \pm 0.0019$ | $0.5067 \pm 0.0020$ | $0.519 \pm 0.022$ | 0.6 |
| $\Delta M_s\,[\text{ps}^{-1}]$ | $17.741 \pm 0.020$ | $17.741 \pm 0.021$ | $17.89 \pm 0.65$ | 0.2 |
| $\varepsilon$ | $0.002228 \pm 0.000011$ | $0.002227 \pm 0.000014$ | $0.00200 \pm 0.00014$ | 1.6 |
| $\text{Re}\,(\varepsilon'/\varepsilon)$ | $0.00166 \pm 0.00033$ | $0.00160 \pm 0.00028$ | $0.00146 \pm 0.00045$ | 0.3 |
| $\overline{\text{BR}}(B_s \to \mu\mu) \times 10^9$ | $3.41 \pm 0.29$ | $3.44 \pm 0.12$ | $3.45 \pm 0.13$ | 0.1 |
| $\text{BR}(B \to \tau\nu) \times 10^4$ | $1.06 \pm 0.19$ | $0.872 \pm 0.041$ | $0.865 \pm 0.041$ | 1.0 |

Notice that in the UT analysis we employ unitarity in order to determine $|V_{us}|$ from $|V_{ud}|$; the latter is obtained via a skeptical average à la D'Agostini [13] from the study of neutron decay and super allowed $0^+ \to 0^+$ nuclear $\beta$ processes as well as from a joint analysis of $K_{\mu 2}$, $K_{\ell 3}$ and $\pi_{\mu 2}$ decays. Regarding other key measurements adopted in our study, we update:

- The constraint on $\alpha$, using the most recent outcome from the isospin study of hadronic $B$ decays into $\pi\pi$, $\rho\rho$

and $\pi\rho$ channels from PDG and HFLAV; after Bayesian marginalization, this yields: $\alpha = (93.8 \pm 4.5)°$;

- The constraint on $\beta$ including a new measurement from LHCb on time-dependent CP violation from $B$ decays into charmonium-kaon final states[14], weighting it with Cabibbo-suppressed penguin corrections [15]; we obtain: $\sin 2\beta = 0.689 \pm 0.019$;

- The constraint on $\gamma$ from a preliminary combined analysis of $B \to D^{(*)}K^{(*)}$ modes with $D$ meson oscillations [16], along the lines of what done by LHCb in [17]; For more details, see the dedicated EPS-HEP2023 contribution. we report $\gamma = (65.4 \pm 3.3)°$ and negligible correlation with $D$ mixing parameters (relevant for NP fits).

## 4. Standard Model Global Fits

The main message of the present UT analysis in the SM is that there is a general consistency, at the percent level, between theory predictions and the experimental measurements. This fact is exemplified in Figure 2. Using all the most informative constraints in order to determine the apex of the UT in the $(\bar\rho, \bar\eta)$ plane as precise as possible, we actually reach 3% precision in the inference of CP violation, namely:

$$( \bar\rho = 0.160 \pm 0.009 \ , \ \bar\eta = 0.346 \pm 0.009 ) \ \textbf{SM fit} , \qquad (7)$$

with the other Wolfenstein parameters determined to be: $\lambda = 0.2251 \pm 0.0008$, $A = 0.828 \pm 0.010$. It is remarkable that the determination of the UT angles $\alpha$, $\beta$ and $\gamma$ allows for the same level of precision in constraining CP violation from weak interactions in the SM:

$$( \bar\rho = 0.159 \pm 0.016 \ , \ \bar\eta = 0.339 \pm 0.010 ) \ \textbf{angles} . \qquad (8)$$

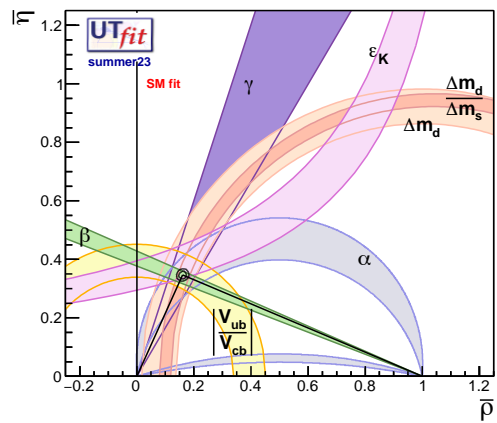

**Figure 2.** State-of-the-art UT analysis in the SM implementing all the most relevant constraints in the $(\bar\rho, \bar\eta)$ plane. Contour regions are shown at the 95% probability. Further details on the fit are reported in Table 1.

We observe that such a bound on CP violation still holds at the 6% level when one restricts the UT fit only to CP-conserving observables, and marginally improves with the addition in the fit of the observable $\epsilon$, parametrizing CP violation from the mixing in the neutral kaon system, see Figure 3.

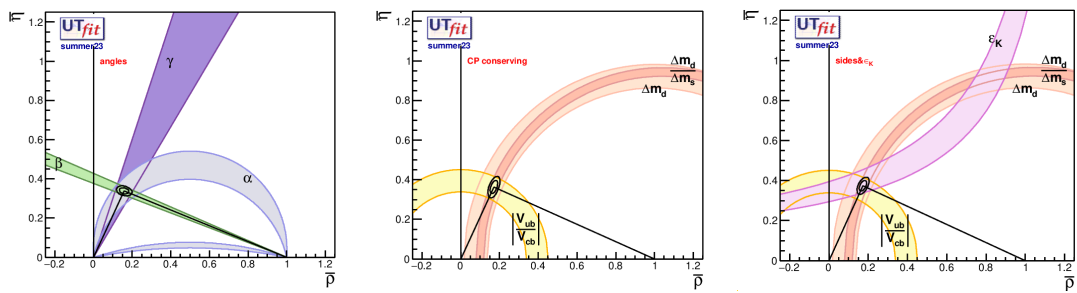

**Figure 3.** Determinations of the SM UT using partial information from the constraints available.

In Table 1 we report all the key observables for the SM global fits, with the measurements adopted in the analysis, the mean and standard deviation of the posterior from the full fit, and the corresponding predictions obtained removing the statistical weight of the observable under scrutiny from the likelihood. Comparing in absolute value the SM prediction against the corresponding measurement over the theoretical and experimental standard deviations summed in quadrature, we can define a pull for each observable as reported in the last column of Table 1, and perform compatibility tests as the ones pictured in Figure 4.

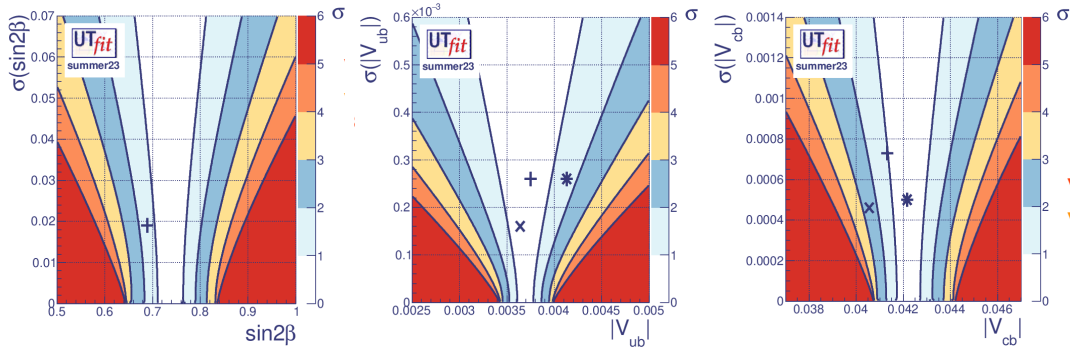

**Figure 4.** Highlight on the compatibility plots for the observables predicted in the SM UT analysis. For the case of $|V_{ub}|$ and $|V_{cb}|$ we also report the adopted exclusive and inclusive measurements with "x" and "*".

We observe that the tension between exclusive and inclusive determination of $|V_{ub}|$ and $|V_{cb}|$, related to the tree-level partonic processes $b \to u\ell\nu$ and $b \to c\ell\nu$, is no longer as severe as in the past. In particular, we report the following pulls from the fit:

$$\text{pull}(\#\sigma) = 2.4\ (0.1) \text{ for } |V_{cb}^{\text{excl}}| \times 10^3 = 40.55 \pm 0.46 \text{ (for } |V_{cb}^{\text{incl}}| \times 10^3 = 42.16 \pm 0.50) \,,$$

$$\text{pull}(\#\sigma) = 1.6\ (0.3) \text{ for } |V_{ub}^{\text{incl}}| \times 10^3 = 4.13 \pm 0.26 \text{ (for } |V_{ub}^{\text{excl}}| \times 10^3 = 3.64 \pm 0.16) \,,$$

underlying an agreement of the SM with data always within the $3\sigma$ level. This improved situation with respect to the past might be partly ascribed to an overall better understanding of the systematics in the measurement of the moments of some differential distributions for the semileptonic $B$ decays under the spotlight; most importantly, in this regard a better handle on the theoretical uncertainties stemming from lattice QCD and unitarization techniques adopted for the computation of the relevant form factors has been playing a crucial role [18]. According to Table 1, the largest discrepancies from the outcome of the UT analysis actually shows up in the observables $\sin 2\beta$ and $\varepsilon$, both pointing to a mild $\sim 1.5\sigma$ tension of the SM against the respective measurements.

Given the high precision of the experimental measurements and of the theoretical calculations, we may ask the following question: is the UT analysis in agreement within the SM theoretical expectations? Until one year ago I would have said no, mainly because of the so-called *anomalies*, all of them seemingly connected to leptonic final states. These anomalies not only concerned the difference of the value of $|V_{cb}|$ as determined from exclusive and inclusive decays but also the tension between the measured values of the semi-leptonic ratios of branching ratios

$$R(D) = \frac{\text{BR}(B \to D\tau\nu_\tau)}{\text{BR}(B \to D\ell\nu_\ell)} \,, \qquad R(D^*) = \frac{\text{BR}(B \to D^*\tau\nu_\tau)}{\text{BR}(B \to D^*\ell\nu_\ell)} \,, \tag{9}$$

together with the corresponding SM predictions based on the lattice calculation of the relevant form factors and also the violation of Lepton Flavour Universality (LFU) in $b \to s\ell^+\ell^-$, where $\ell$ are light leptons (either $e^\pm$ or $\mu^\pm$). The apparent violation of LFU was given by a coherent pattern of tensions based on the ratios of BRs

$$R_X = \frac{\text{BR}(b \to s\mu^+\mu^-)}{\text{BR}(b \to se^+e^-)} \,, \tag{10}$$

which were expected to be very close to one, contrary with the experimental findings. For example, if we define the ratio of branching fractions

$$R_{K^{(*)}}^{[q_1^2, q_2^2]} = \frac{BR(B \to K^{(*)}\mu\mu)}{BR(B \to K^{(*)}ee)} \,, \tag{11}$$

corresponding to a chosen lepton invariant mass interval $q_1^2 \leq q^2 \leq q_2^2$, the values from LHCb were [19,20]

$$R_K^{[1.,1.6]} = 0.847 \pm 0.042 \,; \quad R_{K^*}^{[0.045,1.0]} = 0.68 \pm 0.10 \,; \quad R_{K^*}^{[1.0,1.6]} = 0.71 \pm 0.10 \,. \tag{12}$$

Order thousands papers and many new physics models were written in the last few years to explain such anomalies.

"Unfortunately" for us, unfortunately because anomalies might have indicated new physics, all such anomalies did disappear and the new experimental values of the $B \to K\ell^+\ell^-$ ratios are very close to one, in full agreement with the SM expectations (no explanation was given for the discrepancy with previous measurements of the same quantities) [21]. Moreover a re-examination of the values and uncertainties of the predictions for $|V_{cb}|$ and $R(D^{(*)})$ lowered the discrepancies below the $2.5\sigma$ level [11]. The new uncertainties took into account the still large differences in the calculation of the relevant form factors and also to the differences between the experimental data.

On the side of the successful predictions of the SM, it is worth noticing that the branching ratio of the FCNC process $B_s \to \mu^+\mu^-$ shows now remarkable agreement between theory and data, an impactful result for the phenomenology of weak interactions in light of the recent discussion on rare $B$ decay anomalies [22,23]. Eventually, it is also important to stress the excellent agreement of the current measurement of direct CP violation in the kaon system against the SM prediction via the implementation of $\varepsilon'/\varepsilon$ as a novel observable in the global fit of the UT, see [6] for more details.

## 5. New Physics Global Fits

The UT analysis can be generalized to the case of NP under the key assumption that tree-level flavor violating processes used to constrain the $(\bar\rho, \bar\eta)$ plane should not be significantly affected by physics beyond the SM. On the one hand, one can enlarge the number of fitted parameters and deal with additional $\mathcal{O}(10)$ ones capturing the effects of heavy new dynamics on the phase and the absolute value of $|\Delta F| = 2$ amplitudes. At the same time, one can include semileptonic charge and same-side dilepton asymmetries measured for the $B_{(s)}$ system, which are helpful in disentangling possible degeneracies in the NP UT fit, as well as $D$-$\bar{D}$ mixing observables, which provide the only genuine probe of flavor violation coming from the up-quark sector. Finally, one needs to tame long-distance contributions plaguing the estimate of the amplitudes of $K$-$\bar{K}$ and $D$-$\bar{D}$ mixing, treating them in a conservative fashion.

Following Ref. [24] and implementing the latest theoretical updates and measurements listed in the previous section, the NP UT analysis provides us today a constraint on the SM CP-violating parameter at the level of 8% of precision:

$$( \bar\rho = 0.167 \pm 0.025 \;\; , \;\; \bar\eta = 0.361 \pm 0.027 ) \;\; \mathbf{NP\ fit} \;, \tag{13}$$

which stems from the determination of the allowed $\bar\rho$-$\bar\eta$ region using $|V_{ub}/V_{cb}|$ and $\gamma$ only, together with the information provided in particular by the charge asymmetries in semileptonic $B$ decays, see Figure 5.

The presence of NP in meson mixing amplitudes can be simply parametrized as:

$$\mathcal{A}_{\Delta F=2} = \left( 1 + |\mathcal{A}^{\mathrm{NP}}|/|\mathcal{A}^{\mathrm{SM}}|e^{i2(\phi^{\mathrm{NP}}-\phi^{\mathrm{SM}})} \right) |\mathcal{A}^{\mathrm{SM}}|e^{i2\phi^{\mathrm{SM}}} \;, \tag{14}$$

and from the NP UT analysis it follows that at present the relative size of NP effects with respect to the SM, $|\mathcal{A}^{\mathrm{NP}}|/|\mathcal{A}^{\mathrm{SM}}|$, in $B_{d(s)}$ mixing amplitudes–characterized in the SM by the short-distance contribution of the top-quark in the loop–is constrained to be at most 30(25)% at 95% probability.

Barring accidental cancellations, the constraints on the NP phase and amplitudes in $|\Delta F| = 2$ processes can be then translated into a bound on the Wilson coefficient of dimension-six effective operators parametrizing in a model-independent fashion the effect of NP in neutral meson mixing:

$$
\begin{aligned}
\mathcal{O}_1 &= \left( \bar{q}_i^\alpha \gamma_\mu P_L q_j^\alpha \right) \left( \bar{q}_i^\beta \gamma^\mu P_L q_j^\beta \right) \;, \\
\mathcal{O}_2 &= \left( \bar{q}_i^\alpha P_L q_j^\alpha \right) \left( \bar{q}_i^\beta P_L q_j^\beta \right) \;, \\
\mathcal{O}_3 &= \left( \bar{q}_i^\alpha P_L q_j^\beta \right) \left( \bar{q}_i^\beta P_L q_j^\alpha \right) \;, \\
\mathcal{O}_4 &= \left( \bar{q}_i^\alpha P_L q_j^\alpha \right) \left( \bar{q}_i^\beta P_R q_j^\beta \right) \;, \\
\mathcal{O}_5 &= \left( \bar{q}_i^\alpha P_L q_j^\beta \right) \left( \bar{q}_i^\beta P_R q_j^\alpha \right) \;,
\end{aligned}
\tag{15}
$$

where $P_{L,R} = (1 \pm \gamma_5)/2$; the pair $i, j$ and $\alpha, \beta$ runs over flavor and color indices, and the independent set of operators obtained via the substitution $P_L \to P_R$ in $O_{1,2,3}$ is not reported for brevity.

In Figure 5 we show the state-of-the-art bounds on the real and imaginary part of the Wilson coefficient of each of the NP operators entering in the $|\Delta F| = 2$ effective Hamiltonian of $K$-$\bar{K}$ and $D$-$\bar{D}$ mixing, and the constraint directly on the absolute value of the Wilson coefficient for the set of NP operators related to $B_{d,s}$-$\bar{B}_{d,s}$ mixing (whose SM amplitude is not plagued by long-distance effects). We show in Figure 6 with empty histograms the scenario where the UV theory does not enjoy any particular protection against novel sources of flavor and CP violation: in such a case, CP violation from the mixing in the neutral kaon system yields the strongest constraint on the scale of NP, $\Lambda \gtrsim 5 \times 10^5$ TeV, assuming $\mathcal{O}(1)$ couplings between the SM fields and the heavy new degrees of freedom. While the constraints in Figure 5 can be dramatically relaxed within the ansatz of Minimal Flavor Violation [25], a similar protection in the UV where however new $\mathcal{O}(1)$ phases in the flavor violating coupling are allowed is a tightly constrained possibility, probing scales as high as $\Lambda \gtrsim 110$ TeV, still way beyond the reach of present and next-generation colliders.

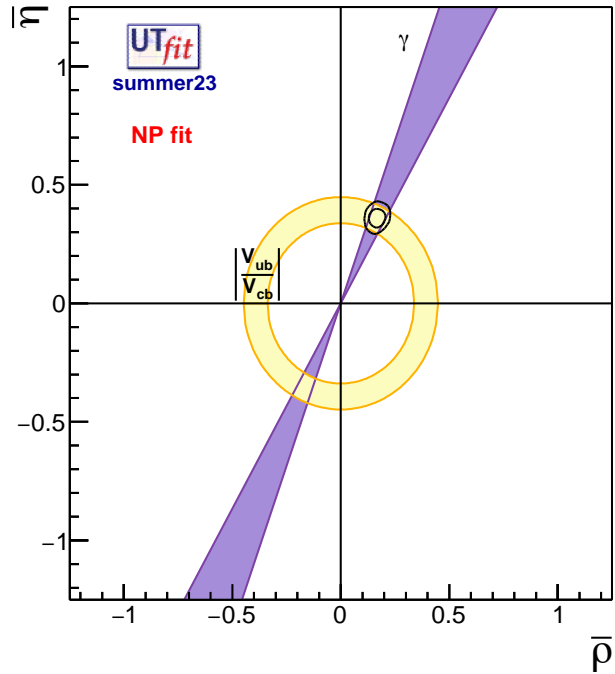

**Figure 5.** Constraints in the $(\bar{\rho}, \bar{\eta})$ plane at the 95% probability using tree-level determinations for the UT and generalizing SM loop-induced amplitudes as the ones for meson mixing to account for NP effects.

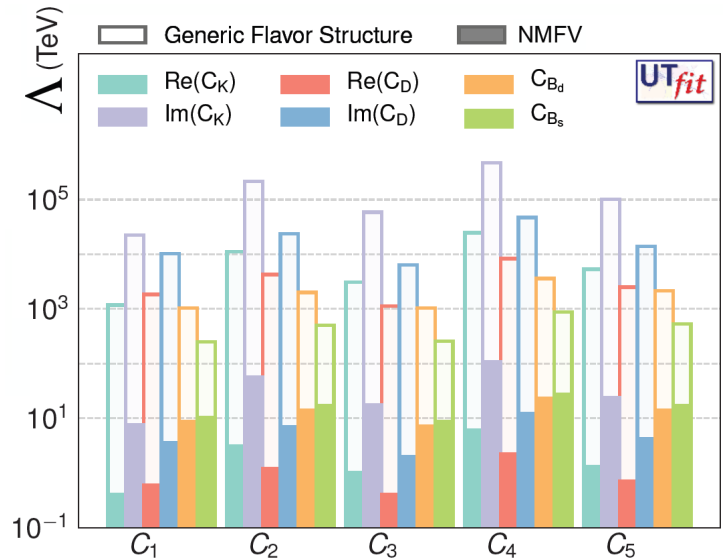

**Figure 6.** Constraints from the NP UT analysis on the set of dimension-six operators that generalizes the effective Hamiltonian for $|\Delta F| = 2$ transitions beyond the SM. Filled histograms correspond to bounds on local operators affecting the short-distant physics of neutral meson oscillation amplitudes in the scenario of Next-To-Minimal Flavor Violation, while empty ones apply to a generic flavor structure in the UV.

## 6. Future Prospects

The accuracy of the experimental measurements and the precision reached by theoretical calculations, especially in lattice QCD, allows to put stringent constraints on new physics even for radiative and rare decays. A satisfactory

NP model should produce the strong hierarchy of the Fermion coupling and masses while simultaneously explain the low scale suppression of FCNC and CP violation at an acceptable level. In this regard, the analysis of the unitarity triangle, and its extension/generalisation beyond the Standard Model, remains one of the best phenomenological tools to explore, via virtual effects, very large energy scales that cannot be reached by present and future colliders. Now that the *anomalies* practically disappeared, the quest for going beyond percent precision in the determination of the SM UT is dictated by the absence of any signal of new physics with present data. The hope to constrain NP amplitudes in FCNC processes like meson anti-meson oscillation, including long distance effects, at the level of few percent is a foreseeable achievement for the next decade. Rare processes like $K \to \pi \nu \bar{\nu}$ will provide further information about the triangle, while theoretical progress from lattice QCD will be necessary in order to bring the NP UT at the percent level of precision [26].

From the theoretical point of view, it is important to note that one of the main messages from flavor physics is perhaps the existence of a large gap between the electroweak energy scale and the scale of physics beyond the SM. If this is the case, an Effective Field Theory approach remains the most suitable framework to study NP constraints from precision measurements, including the ones from the UT analysis [27]. While some specific quantitative work along this direction has already been carried out, see e.g., [28,29], a more general study of the Standard Model Effective Field Theory including flavor constraints is still probably only at its infancy [30,31].

**Funding**

**Conflicts of Interest**

The author declares no conflict of interest.

**References**

1. Ciuchini, M.; Franco, E.; Martinelli, G.; et al. Estimates of $\epsilon'/\epsilon$. *arXiv* **1995**, arXiv:hep-ph/9503277.
2. Ciuchini, M.; D'Agostini, G.; Franco, E.; et al. 2000 CKM-triangle analysis a critical review with updated experimental inputs and theoretical parameters. *JHEP* **2001**, 7, *13*. https://doi.org/10.1088/1126-6708/2001/07/013.
3. Cabibbo, N. Unitary Symmetry and Leptonic Decays. *Phys. Rev. Lett.* **1963**, *10*, 531–533, https://doi.org/10.1103/PhysRevLett.10.531.
4. Kobayashi, M.; Maskawa, T. *CP*-Violation in the Renormalizable Theory of Weak Interaction. *Prog. Theor. Phys.* **1973**, *49*, 652–657, https://doi.org/10.1143/PTP.49.652.
5. Dar, S. The neutron EDM in the SM: A review. *arXiv* **2000**, arXiv:hep-ph/0008248.
6. Bona, M.; Ciuchini, M.; Derkach, D.; et al. New **UT***fit* analysis of the unitarity triangle in the Cabibbo–Kobayashi–Maskawa scheme. *Rend. Lincei Sci. Fis. Nat.* **2023**, *34*, 37–57. https://doi.org/10.1007/s12210-023-01137-5.
7. Bona, M.; Ciuchini, M.; Derkach, D.; et al. Unitarity Triangle global fits beyond the Standard Model: **UT***fit* 2021 new physics update. *PoS* **2022**, *EPS-HEP2021*, 500. https://doi.org/10.22323/1.398.0500.
8. Aoki, Y.; Blum, T.; Colangelo, G.; et al. FLAG review 2021. *Eur. Phys. J. C* **2022**, *82*, 869. https://doi.org/10.1140/epjc/s10052-022-10536-1.
9. Glashow, S.L.; Iliopoulos, J.; Maiani, L. Weak Interactions with Lepton-Hadron Symmetry. *Phys. Rev. D* **1970**, *2*, 1285–1292. https://doi.org/10.1103/PhysRevD.2.1285.
10. Wolfenstein, L. Parametrization of the Kobayashi-Maskawa Matrix. *Phys. Rev. Lett.* **1983**, *51*, 1945, https://doi.org/10.1103/PhysRevLett.51.1945.
11. Martinelli, G.; Simula, S.; Vittorio, L. Updates on the determination of $|V_{cb}|$, $R(D^*)$ and $|V_{ub}|/|V_{cb}|$. *arXiv* **2023**, arXiv:2310.03680.
12. Cirigliano, V.; Dekens, W.; Mereghetti, E.; et al. Effective field theory for radiative corrections to charged-current processes: Vector coupling. *Phys. Rev. D* **2023**, *108*, 53003. https://doi.org/10.1103/PhysRevD.108.053003.
13. D'Agostini, G. Skeptical combination of experimental results using JAGS/rjags with application to the $K^{\pm}$ mass determination. *arXiv* **2020**, arXiv:2001.03466.
14. Aaij, R.; Abdelmotteleb, A.S.W.; Abellan, Beteta, C.; et al. Measurement of Violation in Decays. *Phys. Rev. Lett.* **2024**, *132*, 021801. https://doi.org/10.1103/PhysRevLett.132.021801.
15. Ciuchini, M.; Pierini, M.; Silvestrini, L. Effect of Penguin Operators in the Asymmetry *Phys. Rev. Lett.* **2005**, *95*, 221804. https://doi.org/10.1103/PhysRevLett.95.221804.
16. Kagan, A.L.; Silvestrini, L. Dispersive and absorptive *CP* violation in $D^0 - \bar{D}^0$ mixing. *Phys. Rev. D* **2021**, *103*, 053008. https://doi.org/10.1103/PhysRevD.103.053008.

17. Aaij, R.; Abdelmotteleb, A.S.W.; Abellan, Beteta, C.; et al. Measurement of the ratio $R_{D^{(*)}}$ in semileptonic $B^0 \rightarrow D^{(*)} \tau^- \bar{\nu}_\tau$ decays. *JHEP* **2021**, *12*, 141. https://doi.org/10.1007/JHEP12(2021)141.

18. Martinelli, G.; Simula, S.; Vittorio, L. Simultaneous determination of CKM angle $\gamma$ and charm mixing parameters. *Phys. Rev. D* **2022**, *105*, 034503. https://doi.org/10.1103/PhysRevD.105.034503.

19. Aaij, R.; Adeva, B.; Adinolfi, M.; et al. Addendum: Test of lepton universality in beauty-quark decays *Nat. Phys.* **2023**, *19*, 1517. https://doi.org/10.1038/s41567-023-02095-3.

20. Aaij, R.; Adeva, B.; Adinolfi, M.; et al. Test of lepton universality with $B^0 \rightarrow K^{*0} \ell + \ell$-decays. *JHEP* **2017**, *8*, 55. https://doi.org/10.1007/JHEP08(2017)055.

21. Aaij, R.; Abdelmotteleb, A.S.W.; Abellan, Beteta, C.; et al. Test of Lepton Universality in Decays. *Phys. Rev. Lett.* **2023**, *131*, 51803. https://doi.org/10.1103/PhysRevLett.131.051803.

22. Ciuchini, M.; Fedele, M.; Franco, E.; et al. Constraints on lepton universality violation from rare decays. *Phys. Rev. D* **2023**, *107*, 055036. https://doi.org/10.1103/PhysRevD.107.055036.

23. Greljo, A.; Salko, J.; Smolkovič, A.; et al. Rare $b$ decays meet high-mass Drell-Yan. *JHEP* **2023**, *5*, 87. https://doi.org/10.1007/JHEP05(2023)087.

24. Bona, M.; Ciuchini, M.; Franco, E.; et al. Model-independent constraints on $\Delta F = 2$ operators and the scale of new physics. *JHEP* **2008**, *3*, 49. https://doi.org/10.1088/1126-6708/2008/03/049.

25. Buras, A.J.; Gambino, P.; Gorbahn, M.; et al. Universal unitarity triangle and physics beyond the standard model. *Phys. Lett. B* **2001**, *500*, 161–167. https://doi.org/10.1016/S0370-2693(01)00061-2.

26. Kou, E.; Urquijo, P.; Altmannshofer, W.; et al. The Belle II Physics Book. *PTEP* **2019**, *2019*, 123C01. https://doi.org/10.1093/ptep/ptz106.

27. Descotes-Genon, S.; Falkowski, A.; Fedele, M.; et al. The CKM parameters in the SMEFT. *JHEP* **2019**, *5*, 172. https://doi.org/10.1007/JHEP05(2019)172.

28. Silvestrini, L.; Valli, M. Model-independent bounds on the standard model effective theory from flavour physics. *Phys. Lett. B* **2019**, *799*, 135062. https://doi.org/10.1016/j.physletb.2019.135062.

29. Aebischer, J.; Bobeth, C.; Buras, A.J.; et al SMEFT atlas of $\Delta F = 2$ transitions. *JHEP* **2020**, *12*, 187. https://doi.org/10.1007/JHEP12(2020)187.

30. Garosi, F.; Marzocca, D.; Rodriguez-Sanchez, A.; et al. Indirect constraints on top quark operators from a global SMEFT analysis. *JHEP* **2023**, *12*, 129. https://doi.org/10.1007/JHEP12(2023)129.

31. Allwicher, L.; Cornella, C.; Isidori, G.; et al. New Physics in the Third Generation: A Comprehensive SMEFT Analysis and Future Prospects. *arXiv* **2023**, arXiv:2311.00020.

*Review*

# About BSM Physics, with Emphasis on Flavour

## Riccardo Barbieri

Scuola Normale Superiore , Piazza dei Cavalieri 7, 56126 Pisa, Italy; riccardo.barbieri@sns.it

**Abstract:** This article is an account of a talk given at the Conference *The rise of Particle Physics* celebrating the 50th anniversary of the $J/\Psi$ discovery. It contains some reflections on BSM physics and flavour in particular.

**Keywords:** BSM physics; flavour physics

## 1. A View of BSM Physics

The 50th anniversary of the November Revolution, marked by the discovery of the $J/\Psi$ particle [1–3], represents a turning point in the history of physics. Since then, the rapid emergence of the Standard Model (SM) has established it as the reference theory for an entire quadrant of nature: Particle Physics. This is particularly evident when comparing how concisely the SM can be defined—see Figure 1—with the vast catalogue of independent observables it explains, often with remarkable numerical precision.

**1. Symmetry group $L \times \mathcal{G}$**

$L$ = Lorentz (space-time)
$\mathcal{G} = SU(3) \times SU(2) \times U(1)$ **(local)**

**2. Particle content (rep.s of $L \times \mathcal{G}$)**

|  | $h$ | $q_i$ | $l_i$ | $u_i$ | $d_i$ | $e_i$ |
|---|---|---|---|---|---|---|
| Lorentz | 0 | $1/2_L$ | $1/2_L$ | $1/2_R$ | $1/2_R$ | $1/2_R$ |
| $SU(3)$ | **1** | **3** | **1** | **3** | **3** | **1** |
| $SU(2)$ | **2** | **2** | **2** | **1** | **1** | **1** |
| $U(1)$ | $-1/2$ | $1/6$ | $-1/2$ | $2/3$ | $-1/3$ | $-1$ |

**3. All local operators of dimension d≤4 in $\mathcal{L}$**

**Figure 1.** The SM unambiguously defined in the context of field theory. Each fermion field occurs in 3 replicas.

Not surprisingly however, as it happens for all great theories of nature, the SM leaves open a number of important questions, both of observational and of structural nature, as summarised in Figure 2. Since the first are well known, I briefly comment on the structural questions:

- **Which is the rationale for matter quantum numbers?** In particular the presence of an Abelian $U(1)$ factor in the gauge symmetries of the SM leaves in some way unexplained the quantization of electric charge, $Q = T_{3L} + Y$. The absence of gauge anomalies is enough to guarantee charge quantization in the case of a single fermion family [4] but not in the full SM, as defined in Figure 1 , due to the presence of additional anomaly-free global symmetries [5]. This is in sharp contrast with the bounds on the neutrality of matter, at the level of $10^{-21}$ relative to the electron charge, or on the neutrino charge, of about $10^{-14}$, from plasmon decays into neutrinos in stars.

- **A single lacking operator of dimension** $d \leq 4$. The dimensionless coefficient of the operator $G_{\mu\nu}\tilde{G}^{\mu\nu}$, odd under CP, is bound to be less than $10^{-10}$ by the absence of any signal, so far, of an electric dipole moment of the neutron, equally odd under CP.

**Publisher's Note:** Scilight stays neutral with regard to jurisdictional claims in published maps and institutional affiliations.

- **What about operators of dimension** $d > 4$? The stability of the Higgs potential, the convergence of the perturbative series and the Landau-pole problem [6] indicate that the SM cannot be valid at all energies. This leads to the expectation that higher dimensional operators could be present, weighted by dimensionful coefficients, which may, in turn, be related to some striking new phenomena. The neutrinos masses, which in the SM are predicted to be massless at $d = 4$, can be considered the first evidence of higher-dimensional operators.

- **A matter of calculability**. Of the seventeen particles of the SM two are massless, the photon and the gluon, due to gauge invariance. None of the remaining particle masses is predicted by the SM. The mass of the Higgs boson suffers from its sensitivity to any higher mass scale coupled to the Higgs, the so-called *naturalness problem*. All the fermion masses, aswell as the four physical parameters of the CKM matrix, constitute the *flavour puzzle*, itself strongly intertwined with the Higgs boson via the Yukawa couplings.

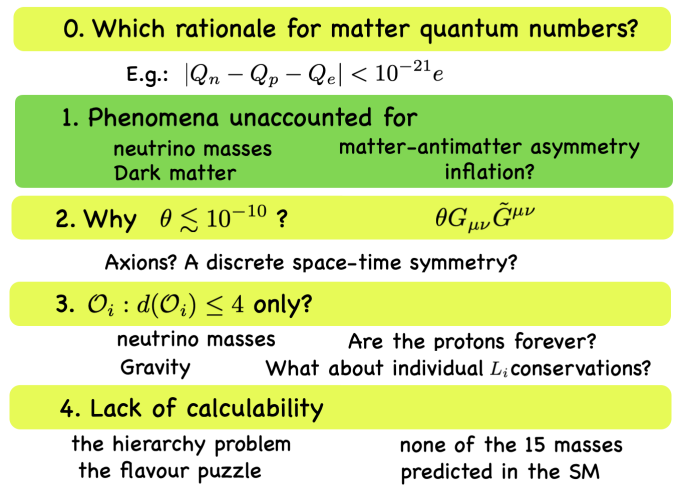

**Figure 2.** Questions raised by the SM, of observational (green) or structural (yellow) origin.

The significance of these questions cannot be overstated. Unsurprisingly, they have long been, and continue to be, the driving force behind numerous ideas and inquiries in BSM physics—too many to list—though, as of yet, without any direct or unambiguous experimental evidence. Here I focus on the last item in Figure 2 with particular emphasis on the *flavour puzzle*, though not before expressing a personal view about the SM Lagrangian $\mathcal{L}_{SM}$. If one separates $\mathcal{L}_{SM}$ into two sectors, as in Figure 3, one notices that:

- (i) the "lack of calculability" alluded to in Figure 2 mostly resides in the Higgs sector, which includes as well the Cosmological Constant problem ("Λ"), of similar origin, from an EFT point of view, as the naturalness problem of the Higgs mass;

- (ii) the precision, experimental and theoretical, to which the two sectors have been tested so far is unequal: many observables in the gauge sector are correctly predicted at 1 ppm level or better [7], whereas flavour [8] or Higgs couplings [7] tests are more at about 10% level.

$$\mathcal{L}_{SM} = -\frac{1}{4} F_{\mu\nu}^a F^{a\mu\nu} + i\bar{\Psi} \slashed{D} \Psi \qquad \boxed{\text{The "gauge sector"}}$$

$$+ |D_\mu \phi|^2 + M^2 |\phi|^2 - \lambda |\phi|^4 + \Lambda + \lambda_{ij} \phi \bar{\Psi}_i \Psi_j \qquad \boxed{\text{The "Higgs sector"}}$$

**Figure 3.** The SM Lagrangian with its two sectors defined: the "gauge" and the "Higgs" one.

Both sectors are each an unavoidable pillar of the SM. Nevertheless, jointly with the fact that the Higgs sector is where the Fermi scale originates, these considerations represent, in my view, a strong motivation for the next high energy collider. The LHC is currently exploring the Fermi scale, but a next step in precision and energy appears mandatory.

## 2. Flavour and BSM

From a generic EFT point of view, the strongest lower bounds on the BSM scale come from flavour physics. Figure 4, from the UTfit Collaboration [8], shows these bounds from actually observed processes related to $\Delta F = 2$ transitions. Bounds from null observations of $\mu \to e$ transitions are comparable or slightly stronger. Taken at face value, as well known, these bounds set the possible scale of BSM physics very far from the Fermi scale or even the MultiTeV scale. Is this contradicting the view expressed in the last paragraph of the previous Section?

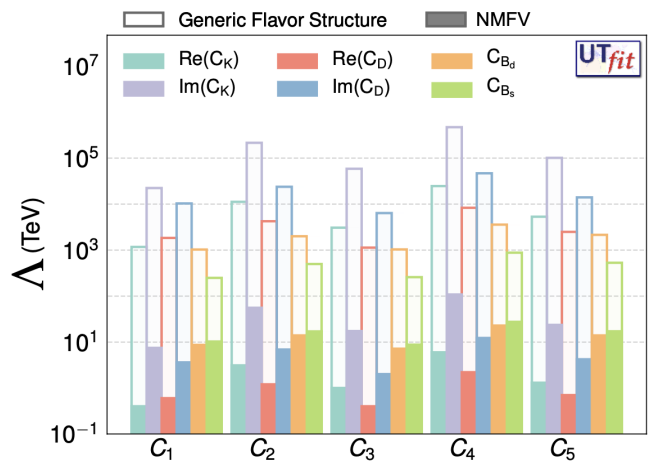

**Figure 4.** Constraints on the scale $\Lambda$ weighting the $d = 6$ operators that mediate $\Delta F = 2$ transitions, taken from Ref. [8], for a generic EFT (empty) or by including in each operator an *ad hoc* CKM factor (coloured).

To try to address this question, let us consider what we know of the Yukawa couplings $Y^f$, $f = u, d, e$ in the SM, as defined in Figure 1. Diagonalising $Y^f$ as $Y_f = U_L^{f\dagger} Y_f^{diag} U_R^f$, the entries of $Y_f^{diag}$ are strongly hierarchical and the CKM matrix $V_{CKM} = U_L^u (U_L^d)^+$ is close to the unit matrix. Figure 5 shows the structure of the quark Yukawa couplings if one takes $[U_L^{u,d}]_{i \neq j} \lesssim [V_{CKM}]_{i \neq j}$ but no special structure in $U_R^{u,d}$ (Figure 5 left) and if also $[U_R^{u,d}]_{i \neq j} \lesssim [U_L^{u,d}]_{i \neq j}$ (Figure 5 right).

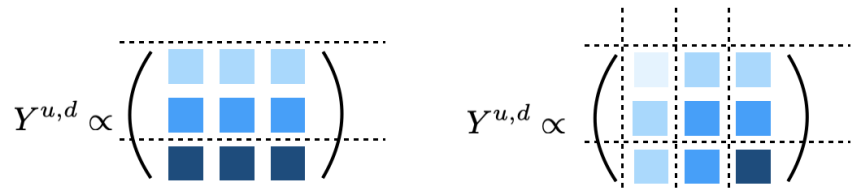

**Figure 5.** Representation of the Yukawa couplings with the colour intensity reflecting the typical size of the corresponding matrix elements, assuming $[U_L^{u,d}]_{i \neq j} \lesssim [V_{CKM}]_{i \neq j}$ (**left**) and also $[U_R^{u,d}]_{i \neq j} \lesssim [U_L^{u,d}]_{i \neq j}$ (**right**). The dotted lines indicate the emergence of approximate global symmetries, see text.

To the extent that the relatively smaller matrix elements can be neglected, Figure 5 left shows the emergence of an (approximate) $U(2)_q$ symmetry acting on the first two generations of left handed quarks as a doublet, whereas Figure 5 right an (equally approximate) $U(2)_q \times U(2)_u \times U(2)_d$ symmetry. In fact a further suppression of the elements of the first column in Figure 5 right can be associated with a $U(1)_u \times U(1)_d$ subgroup acting on the first generation of right-handed $u, d$ quarks.

The potential relevance of these symmetries, suitably broken, in reducing the scale associated with BSM flavour changing interactions has been pointed out in Ref. [9] and recently confirmed in general EFT analyses [10,11]. In particular their possible role in the case of Higgs compositeness, with the *flavour* and the *naturalness problem* strongly tied to each other, has also been emphasised [12] and keeps being examined in general [13] and in specific constructions [14,15]. All this leads to wonder about the origin, if any, of these symmetries, about the source of their breaking and, last but not least, about specific experimental manifestations at scales not too far from the Fermi scale.

## 3. Flavour Deconstruction

Recent models [16–19] that try to understand the origin of these approximate symmetries share the following features:

- The $SU(3) \times SU(2) \times U(1)$ gauge interactions at high energies are (fully or in part) flavour non-universal. Note that this is unlike the case of an additional flavour non-universal gauge group that commutes with $SU(3) \times SU(2) \times U(1)$ (See, e.g., [20] and references therein).
- In the unbroken gauge limit the Higgs field couples to one single chiral generation only: $y_3^f f_{L3} H f_{R3}$.
- The flavour universal gauge interactions observed so far are a low energy manifestation of a stepwise breaking of the gauge group at different scales. It is this stepwise breaking that is responsible for the hierarchical structure of the Yukawa couplings.

### 3.1. An Example

For concreteness an explicit example of this picture, fully based on $d = 4$ and specifically aimed at generating the pattern of Figure 5 right with its approximate flavour symmetries, is as follows [19]:

- The gauge group is

$$G = SU(3) \times SU(2) \times U(1)_Y^{[3]} \times U(1)_{B-L}^{[12]} \times U(1)_{T_{3R}}^{[2]} \times U(1)_{T_{3R}}^{[1]}, \qquad (1)$$

  where $SU(3)$ and $SU(2)$ act universally on the three fermion families, as in the SM, whereas the $U(1)$ groups act non-universally only on one or two families, as indicated by the corresponding superscripts. E.g. $q_3 = (3, 2)_{(1/6,0,0,0)}$ and similarly for all other chiral fermions.

- The full particle content, scalars and vector-like (VL) fermions, other than the usual chiral fermions (which include two right-handed neutrinos $\nu_{1,2}$ needed to cancel the gauge anomalies associated with $U(1)_{T_{3R}}^{[2]} \times U(1)_{T_{3R}}^{[1]}$) is shown in the following Tables (The abundance of U(1) charges in these Tables as well as in the implicit table for the standard chiral fermions raises the question of electric charge quantization. At variance with the SM with three families, however, the model under consideration, with the inclusion of the most general Yukawa couplings, does not have any non-anomalous global symmetry, hence electric charge is quantised. In particular one can show that $Q = T_{3L} + Y^{[3]} + (B - L)^{[12]}/2 + T_{3R}^{[2]} + T_{3R}^{[1]}$ is not a convention (as $Q = T_{3L} + Y$ in the SM with a single family)).

#### Scalars

| Field | $U(1)_Y^{[3]}$ | $U(1)_{B-L}^{[12]}$ | $U(1)_{T_{3R}}^{[2]}$ | $U(1)_{T_{3R}}^{[1]}$ | $SU(3) \times SU(2)$ |
|---|---|---|---|---|---|
| $H_{u,d}$ | $-1/2$ | 0 | 0 | 0 | $(\mathbf{1}, \mathbf{2})$ |
| $\chi^q$ | $-1/6$ | $1/3$ | 0 | 0 | $(\mathbf{1}, \mathbf{1})$ |
| $\chi^l$ | $1/2$ | $-1$ | 0 | 0 | $(\mathbf{1}, \mathbf{1})$ |
| $\phi$ | $1/2$ | 0 | $-1/2$ | 0 | $(\mathbf{1}, \mathbf{1})$ |
| $\sigma$ | 0 | 0 | $1/2$ | $-1/2$ | $(\mathbf{1}, \mathbf{1})$ |

#### VectorLike fermions

| | | $U(1)_Y^{[3]}$ | $U(1)_{B-L}^{[12]}$ | $U(1)_{T_{3R}}^{[2]}$ | $U(1)_{T_{3R}}^{[1]}$ | $SU(3) \times SU(2)$ |
|---|---|---|---|---|---|---|
| light VL ($\alpha = 1, 2$) | $U_\alpha$ | $1/2$ | $1/3$ | 0 | 0 | $(\mathbf{3}, \mathbf{1})$ |
| | $D_\alpha$ | $-1/2$ | $1/3$ | 0 | 0 | $(\mathbf{3}, \mathbf{1})$ |
| | $E_\alpha$ | $-1/2$ | $-1$ | 0 | 0 | $(\mathbf{1}, \mathbf{1})$ |
| heavy VL | $U_3$ | 0 | $1/3$ | $1/2$ | 0 | $(\mathbf{3}, \mathbf{1})$ |
| | $D_3$ | 0 | $1/3$ | $-1/2$ | 0 | $(\mathbf{3}, \mathbf{1})$ |
| | $E_3$ | 0 | $-1$ | $-1/2$ | 0 | $(\mathbf{1}, \mathbf{1})$ |

The scalar fields are responsible for the breaking of the $U(1)$ factors of the gauge group in two steps, by $\langle \sigma \rangle >> \langle \phi, \chi \rangle$, as well as for EW symmetry breaking by the two doublets $H_{u,d}$, distinguished by a softly broken $Z_2$ symmetry which makes them couple to the up-type quarks/neutrinos and to the down-type quarks/charged leptons respectively.

The full set of Yukawa-like couplings and fermion mass terms is determined by the transformation properties of these scalars and of the fermions, chiral or VL. For example in the up-quark sector ($i = 1, 2, \alpha = 1, 2$)

$$\begin{aligned} \mathcal{L}_Y^u &= (y_3^u \, \bar{q}_3 u_3 H_u + y_{i\alpha}^u \, \bar{q}_i U_\alpha H_u + y_\alpha^{\chi_u} \, \bar{U}_\alpha u_3 \chi^q + y_{\alpha 2}^{\phi_u} \, \bar{U}_\alpha u_2 \phi + y_{\alpha 3}^{\phi_u} \, \bar{U}_{R\alpha} U_{L3} \phi \\ &\quad + \bar{y}_{\alpha 3}^{\phi_u} \, \bar{U}_{L\alpha} U_{R3} \phi + y_1^{\sigma_u} \, \bar{U}_3 u_1 \sigma + \text{h.c.}) + M_{U_3} \bar{U}_3 U_3 + M_{U_\alpha} \bar{U}_\alpha U_\alpha \end{aligned} \qquad (2)$$

where, unless specified, the chirality component is left understood since non ambiguous ($q \equiv q_L, u \equiv u_R$), and similarly in the down and charged lepton sector.

The emerging overall picture is summarised in Figure 6 with the following noteworthy points:

- As manifest from Equation (2) in the limit of infinitely heavy VL fermions, $M_{\alpha,3} \to \infty$, the Yukawa couplings exhibit a $U(2)^5 \equiv U(2)_q \times U(2)_u \times U(2)_d \times U(2)_l \times U(2)_e$ global symmetry, reduced to $U(1)^3 \equiv U(1)_u \times U(1)_d \times U(1)_e$ if only $M_3 \to \infty$.

- For finite $M_{\alpha,3}$, after integrating out the heavy VL fermions, the breaking of these symmetries is controlled by three parameters,

$$\epsilon_\phi = \frac{\langle \phi \rangle}{M_\alpha}, \quad \epsilon_\chi = \frac{\langle \chi \rangle}{M_\alpha}, \quad \epsilon_\sigma = \frac{\langle \sigma \rangle}{M_3}, \qquad (3)$$

  represented in Figure 6 by the vertical lines with two arrows.

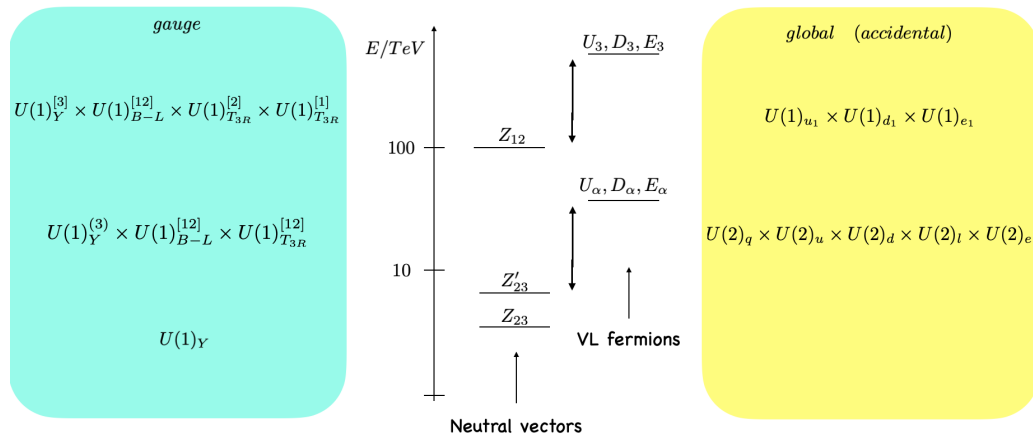

**Figure 6.** Overall representation of the model. On the left and on the right are shown the different gauge $(SU(3) \times SU(2))$ and global symmetries (Universal $U(1)_B \times U(1)_L$), which appear as (almost) unbroken at a given energy. In the centre are the masses of the new particles: neutral vectors, $Z_{23}, Z'_{23}, Z_{12}$, and VL $SU(2)$-singlet fermions, $U_i, D_i, E_i, i = 1, 2, 3$. The vertical lines with two arrows denote the separation of scales which controls the breaking of the global symmetries.

By integrating out the heavy VL fermions one gets the true Yukawa couplings of the chiral fermions. In the up-type case one obtains

$$Y_u \approx \begin{pmatrix} y_{1\alpha}^u \tilde{y}_{\alpha 3}^{\phi_u} y_1^{\sigma_u} \epsilon_\sigma \epsilon_\phi & -y_{1\alpha}^u y_{\alpha 2}^{\phi_u} \epsilon_\phi & -y_{12}^u y_2^{\chi_u} \epsilon_\chi \\ y_{2\alpha}^u \tilde{y}_{\alpha 3}^{\phi_u} y_1^{\sigma_u} \epsilon_\sigma \epsilon_\phi & -y_{2\alpha}^u y_{\alpha 2}^{\phi_u} \epsilon_\phi & -y_{22}^u y_2^{\chi_u} \epsilon_\chi \\ \approx 0 & \approx 0 & y_3^u \end{pmatrix} \tag{4}$$

and similarly for $Y_{d,e}$. For $v_u/v_d \approx 10$ and $\epsilon_\phi \approx \epsilon_\chi \approx \epsilon_\sigma \approx 0.05 - 0.1$ the charged fermion masses and quark mixings are described by all the Yukawa couplings $y$'s in Equation (2) in the 0.1–1 range. In particular the matrix elements of $U_L^{u,d,e}$ have similar size to the matrix elements of $V_{CKM}$ (with $U_L^u U_L^{d\dagger} = V_{CKM}$) and $[U_R^{u,d,e}]_{i \neq j} << [U_L^{u,d,e}]_{i \neq j}$.

### 3.2. Phenomenology

Phenomenological effects of Flavour Deconstruction at the TeV scale are due to the exchanges of the lightest new gauge bosons ($Z_{23}^{(\prime)}$ in the example above) and to their mixing with the $Z$-boson. They occur in:

- ElectroWeak Precision Tests. In the above example, a particularly important effect is the correction to the $Z$-mass proportional to $(m_Z/m_{Z_{23}})^2$ [21];
- High $p_T$ effects in $pp \to ll, l = e, \mu$ via Drell-Yan $q\bar{q} \to Z_{23}^{(\prime)} \to ll$. In the example under consideration the negative searches by ATLAS [22] and CMS [23] set the nominally stronger lower bound on $m_{Z_{23}}$ at about 5 TeV [21];
- Flavour changing effects in $\Delta F = 2, b \to sll(\nu\nu), K \to \pi\nu\nu, \tau \to 3\mu, \mu \to 3e$, controlled by the matrix elements $[U_L^f]_{i3}[U_L^f]_{j3}^*$ with $i, j; f$ depending on the process under consideration. A peculiarity of the model above is the cancellation in the $b \to sll$ transitions of the operator $O_{10} = \bar{b}_L \gamma^\mu s_L \bar{\mu} \gamma_\mu \gamma_5 \mu$ that contributes, e.g., to $B_s \to \mu\mu$ [21].

The exchange of the heavier gauge boson, $Z_{12}$, produces effects in $\Delta S = 2$ transitions, controlled by the matrix elements $[U_R^d]_{12}[U_R^d]_{22}^*$, which require the $Z_{12}$-mass to be heavier than about 100 TeV [19].

Loop effect are potentially important both in dipole moments and in $\Delta S = 2$ effective operators due to the exchange of heavy VL fermions. In the special case of the example described above, however, they do not set bounds more significant than the tree level ones. In the case of the dipoles this is because of a strong alignment of the dipole operators with the corresponding effective Yukawa couplings [21].

## 4. Summary

During the half-century since the November Revolution, the catalogue of observables correctly accounted for by the Standard Model (SM)—in many cases with great numerical precision—has increased enormously. Alongside the lack of any observed deviation from SM expectations, this establishes the SM as one of the most successful theories of a quadrant of nature ever formulated. At the same time, and entirely consistent with this success, the SM leaves us with a number of unanswered questions, both of observational and, at least equally important, of structural nature.

The variety of these questions, as summarized in Section 1 and Figure 2, suggests a wide front of attack. In this talk, I have drawn attention to the distinction between the two pillars of the SM—the gauge sector and the Higgs sector (see Figure 3)—emphasizing two points: the accumulation of questions emerging from the Higgs sector and the differing numerical precision to which the two sectors have been tested so far.

In general terms, these considerations, together with the fact that the Higgs sector is where the Fermi scale originates, strongly motivate the development of the next high-energy collider. On a shorter timescale, but based on similar considerations, increased precision in flavour tests appears highly motivated as well. An extended Table of Flavour Precision Tests at a typical 1% level in many observables is eagerly awaited and may be within reach of the so-called *mid-term* flavour program. The hope is that the emergence of clear deviations from the SM in such a table will lend credence and give substance to daring hypotheses like flavour non-universal $SU(3) \times SU(2) \times U(1)$ gauge interactions, as described in Section 3.

**Acknowledgments**

I am indebted to Gino Isidori for his collaboration and for many useful discussions on the general subject touched upon in this paper.

**Conflicts of Interest**

The author declares no conflict of interest.

**Appendix A. Raul Gatto and the November Revolution**

It so happened that Raul Gatto and I ended up together at CERN around the time of the November revolution. Raul, visiting from Roma La Sapienza, was already well known as a strong theorist and an effective mentor of students and young collaborators. I was a fellow of the CERN Theory Division, after having worked on QED and QED bound states in particular.

The discovery of the $J/\Psi$ and the subsequent works of Appelquist, De Rujula, Glashow, and Politzer [24–26] naturally led to our collaboration, focusing on the non-relativistic (NR) $c\bar{c}$ bound-state interpretation of the new resonances. First we worked on the spectrum of the $c\bar{c}$ resonances, with only the first two $J = 1^{--}$ states, $\Psi, \Psi'$, observed at that time [27,28]. As other groups [29,30] we were using a NR potential that included a Coulomb-like one-gluon exchange term and a long-distance linear term, complemented with suitable relativistic corrections in part also due to the one-gluon exchange.

Most of all, however, our attention was attracted to the idea that the total width of a charmonium state into light hadrons is due to the annihilation of a pair of $c\bar{c}$ into gluons, capable of explaining the narrowness of the $J/\Psi$. With my experience on positronium, it was then straightforward to apply this idea to the $0^{++}, 2^{++}$ P-waves decaying into two gluons, obtaining the result [31]

$$\frac{\Gamma(2^{++})}{\Gamma(0^{++})} = \frac{4}{15},$$ (5)

independent from the charmonium wave function, at leading order in $\alpha_S$. The case of the remaining P-wave was more tricky, since the $1^{++}$ state does not decay into two gluons due to a generalisation of the Landau-Yang theorem which forbids the decay of a $J = 1$ state into two massless vectors. The decay into three gluons is readily computed but, at the same order in $\alpha_S$, the decay into one gluon and a pair of light $q\bar{q}$ has a logarithmic divergence when the momentum of the emitted gluon goes to zero and the external $c\bar{c}$ quarks are assumed free. In the charmonium case, however, the $c$-quark propagator is not free due to its localisation inside the charmonium radius, which is what provides the cutoff of the logarithmic divergence. This allowed a rough estimate of $\Gamma(1^{++})$ as well, which is summarised into the overall leading-order prediction [32]

$$\Gamma(2^{++}) : \Gamma(0^{++}) : \Gamma(1^{++}) = 15 : 4 : 1$$ (6)

for the widths of the annihilation into light hadrons. The relatively precise determination of these widths had to wait for the production of these states in $p\bar{p}$ collisions at Fermilab [33,34]. The current values give

$$\Gamma(2^{++}) : \Gamma(0^{++}) : \Gamma(1^{++}) = 12 : 2.4 : 1$$ (7)

with a typical 10% experimental uncertainty. Corrections of different origins to Equation (6) have been introduced in the following decades. For a review see Ref. [35].

**References**

1. Aubert, J.J.; Becker, U.; Biggs, P.J.; et al. Experimental Observation of a Heavy Particle *J*. *Phys. Rev. Lett.* **1974**, *33*, 1404–1406. https://doi.org/10.1103/PhysRevLett.33.1404.

2. Augustin, J.E.; Boyarski, A.M.; Breidenbach, M.; et al. Discovery of a Narrow Resonance in $e^+e^-$ Annihilation *Phys. Rev. Lett.* **1974**, *33*, 1406–1408. https://doi.org/10.1103/PhysRevLett.33.1406.

3. Bacci, C.; Celio, R.B.; Bernardini, M.; et al. Preliminary Result of Frascati (ADONE) on the Nature of a New 3.1-GeV Particle Produced in $e^+e^-$ Annihilation *Phys. Rev. Lett.* **1974**, *33*, 1408. https://doi.org/10.1103/PhysRevLett.33.1408.

4. Bouchiat, C.; Iliopoulos, J.; Meyer, P. An Anomaly-Free Version of Weinberg's Model. *Phys. Lett. B* **1972**, *38*, 519–523. https://doi.org/10.1016/0370-2693(72)90532-1.

5. Foot, R.; Lew, H.; Volkas, R.R. A Model with Fundamental Improper Space-Time Symmetries *J. Phys. G* **1993**, 19, 361–372. https://doi.org/10.1088/0954-3899/19/3/005.

6. Abrikosov, A.A.; Landau, L.D.; Khalatnikov, I.M. On the elimination of infinities in quantum electrodynamics. *Dokl. Akad. Nauk SSSR* **1954**, *95*, 497.

7. de Blas, J.; Ciuchini, M.; Franco, E.; et al. Electroweak precision observables and Higgs-boson signal strengths in the Standard Model and beyond: present and future. *J. High Energy Phys.* **2016**, *12*, 135. https://doi.org/10.1007/JHEP12(2016)135.

8. Bona, M.; Ciuchini, M.; Derkach, D.; et al. Overview and theoretical prospects for CKM matrix and CP violation from the UTfit Collaboration. *PoS* **2024**, *457*, 7. https://doi.org/10.22323/1.457.0007.

9. Barbieri, R.; Isidori, G.; Jones-Perez, J.; et al. $U(2)$ and minimal flavour violation in supersymmetry. *Eur. Phys. J. C* **2011**, *71*, 1725. https://doi.org/10.1140/epjc/s10052-011-1725-z.

10. Greljo, A.; Palavrić, A.; Thomsen, A.E. Adding Flavor to the SMEFT. *JHEP* **2022**, *10*, 005. https://doi.org/10.1007/JHEP10(2022)005.

11. Allwicher, L.; Cornella, C.; Stefanek, B.A.; et al. New Physics in the Third Generation: A Comprehensive SMEFT Analysis and Future Prospects. *arXiv* **2023**, arXiv:2311.00020.

12. Barbieri, R.; Buttazzo, D.; Sala, F.; et al. A 125 GeV composite Higgs boson versus flavour and electroweak precision tests. *JHEP* **2013**, *5*, 069. https://doi.org/10.1007/JHEP05(2013)069.

13. Glioti, A.; Rattazzi, R.; Ricci, L.; et al. Exploring the Flavor Symmetry Landscape. *arXiv* **2024**, arXiv:2402.09503.

14. Davighi, J.; Isidori, G. Non-universal gauge interactions addressing the inescapable link between Higgs and flavour. *JHEP* **2023**, *7*, 147. https://doi.org/10.1007/JHEP07(2023)147.

15. Covone, S.; Davighi, J.; Isidori, G.; et al. Flavour deconstructing the composite Higgs. *JHEP* **2025**, *1*, 41. https://doi.org/10.1007/JHEP01(2025)041.

16. Davighi, J.; Isidori, G.; Pesut, M. Electroweak-flavour and quark-lepton unification: a family non-universal path. *JHEP* **2023**, *4*, 30. https://doi.org/10.1007/JHEP04(2023)030.

17. Navarro, M.F.; King, S.F. Tri-hypercharge: a separate gauged weak hypercharge for each fermion family as the origin of flavour. *JHEP* **2023**, *8*, 20. https://doi.org/10.1007/JHEP08(2023)020.

18. Davighi, J.; Stefanek, B.A. Deconstructed Hypercharge: A Natural Model of Flavour. *arXiv* **2023**, arXiv:2305.16280.

19. Barbieri, R.; Isidori, G. Minimal flavour deconstruction. *JHEP* **2024**, *5*, 33. https://doi.org/10.1007/JHEP05(2024)033.

20. Belfatto, B.; Berezhiani, Z. How light the lepton flavor changing gauge bosons can be. *Eur. Phys. J. C* **2019**, *79*, 202. https://doi.org/10.1140/epjc/s10052-019-6724-5.

21. Barbieri, R. Phenomenology of Minimal Flavour Deconstruction at the lowest new scale. *arXiv* **2024**, arXiv:2409.08657.

22. Aad, G.; Abbott, B.; Abbott, D.C.; et al. Search for high-mass dilepton resonances using 139 fb-1 of pp collision data collected at $\sqrt{s}$ = 13 TeV with the ATLAS detector *Phys. Lett. B* **2019**, *796*, 68–87. https://doi.org/10.1016/j.physletb.2019.07.016.

23. Sirunyan, A.M.; Tumasyan, A.; Adam, W.; et al. Search for resonant and nonresonant new phenomena in high-mass dilepton final states at $\sqrt{s}$ = 13 TeV. *JHEP* **2021**, *7*, 208. https://doi.org/10.1007/JHEP07(2021)208.

24. Appelquist, T.; Politzer, H.D. Heavy Quarks and $e^+e^-$ Annihilation. *Phys. Rev. Lett.* **1975**, *34*, 43. https://doi.org/10.1103/PhysRevLett.34.43.

25. Rujula, A.D.; Glashow, S.L. Is Bound Charm Found? *Phys. Rev. Lett.* **1975**, *34*, 46–49. https://doi.org/10.1103/PhysRevLett.34.46.

26. Appelquist, T.; Rujula, A.D.; Politzer, H.D.; et al. Spectroscopy of the New Mesons. *Phys. Rev. Lett.* **1975**, *34*, 365. https://doi.org/10.1103/PhysRevLett.34.365.

27. Barbieri, R.; Gatto, R.; Kogerler, R.; et al. Meson hyperfine splittings and leptonic decays. *Phys. Lett. B* **1975**, *57*, 455–459. https://doi.org/10.1016/0370-2693(75)90267-1.

28. Barbieri, R.; Kogerler, R.; Kunszt, Z.; et al. Meson masses and widths in a gauge theory with linear binding potential. *Nucl. Phys. B* **1976**, *105*, 125–138. https://doi.org/10.1016/0550-3213(76)90064-X.

29. Eichten, E.; Gottfried, K.; Kinoshita, T.; et al. Spectrum of Charmed Quark-Antiquark Bound States. *Phys. Rev. Lett.* **1975**, *34*, 369–372. https://doi.org/10.1103/PhysRevLett.34.369.

30. Kang, J.S.; Schnitzer, H.J. Dynamics of light and heavy bound quarks. *Phys. Rev. D* **1975**, *12*, 841. https://doi.org/10.1103/PhysRevD.12.841.

31. Barbieri, R.; Gatto, R.; Kogerler, R. Calculation of the annihilation rate of P wave quark-antiquark bound states. *Phys. Lett. B* **1976**, *60*, 183–188. https://doi.org/10.1016/0370-2693(76)90419-6.

32. Barbieri, R.; Gatto, R.; Remiddi, E. Singular binding dependence in the hadronic widths of $1^{++}$ and $1^{+-}$ heavy quark antiquark bound states *Phys. Lett. B* **1976**, *61*, 465–468. https://doi.org/10.1016/0370-2693(76)90729-2.

33. Bagnasco, S.; Baldini, W.; Bettoni, D.; et al. New measurements of the resonance parameters of the $\chi_{c0}(1^3P_0)$ state of charmonium. *Phys. Lett. B* **2002**, *533*, 237–242. https://doi.org/10.1016/S0370-2693(02)01657-X.

34. Andreotti, M.; Bagnasco, S.; Baldini, W.; et al. Measurement of the resonance parameters of the $\chi_1(1^3P_1)$ and $\chi_2(1^3P_2)$ states of charmonium formed in antiproton–proton annihilations. *Nucl. Phys. B* **2005**, *717*, 34–47. https://doi.org/10.1016/j.nuclphysb.2005.03.042.

35. Brambilla, N.; Eidelman, S.; Heltsley, B.K.; et al. Heavy quarkonium: progress, puzzles, and opportunities. *Eur. Phys. J. C* **2011**, *71*, 1534. https://doi.org/10.1140/epjc/s10052-010-1534-9.

*Perspective*

# The Discovery of the Antiproton between Rome and Berkeley

## Gianni Battimelli

Dipartimento di Fisica, Sapienza Università di Roma, 00184 Rome, Italy; gianni.battimelli@gmail.com

**Abstract:** Solid confirmation for the discovery of the antiproton came shortly after its first detection in September 1955, through the visual evidence offered by the observation of annihilation stars in nuclear emulsions exposed to the Bevatron beam. The emulsion work was a result of a cooperative effort between Emilio Segrè's team in Berkeley and the group of physicists working under the guidance of Edoardo Amaldi in Rome, who had already observed a possible antiproton annihilation star in emulsions exposed to cosmic rays. The origin and development of the Rome-Berkeley collaboration are presented, in the wider context of the changing balance between cosmic ray investigation and accelerator research in the mid-fifties.

**Keywords:** antiproton; emulsions; cosmic rays

It is fair to state that basically every fundamental discovery in experimental particle physics, up to the early fifties, had been the result of cosmic-ray investigation; each new particle that had enriched the growing zoo of the "elementary" constituents of matter had been found by means of cloud chambers and emulsion plates, the standard tools created to catch the signals coming from outer space. By the mid-fifties, the development of the big particle accelerators gradually dictated a changed hierarchy in the tools of the trade, providing more efficient, more controllable (and much more expensive) instruments to research in particle physics.

I would say that the Sixth Rochester Conference (3–7 April 1956) marked the transition from "little science" to "big science" in particle physics. Until the sixth conference, the decision to give equal treatment to accelerator physics, cosmic ray physics and particle theory had served its purpose. Indeed, during the first half- dozen Rochester conferences, it was a common experience for the cosmic ray experimentalists to describe qualitative features of some new discoveries at high energies, for the theorists to articulate these results into a set of model options and, finally, for the accelerator physicists to present at the same, or the very next conference, the quantitative data that enabled one to select the most likely theoretical model. But, at Rochester VI, it was clear that the stream of results from the Berkeley bevatron and the Brookhaven cosmotron would monopolize strange particle physics and Bob Leighton was led to remark that "next year those people still studying strange particles using cosmic rays had better hold a rump session of the Rochester Conference somewhere else—that the machine work had been pretty hard on cosmic-ray people" . . . It should be noted that 1956 was the year when the production of the antiproton was achieved with the Berkeley bevatron—after years of frustration with cosmic ray experiments [1] (pp. 755–756).

And there is also no doubt that the first half of the fifties was regarded in this respect as a dramatic moment of transition by the protagonists themselves; such was, in particular, the perception of European physicists. Trained in the "poor" research carried out on cosmic rays, they saw themselves pursued ever more closely by the infinitely "richer" research conducted with particle accelerators on the other side of the Atlantic [2]. A significant example is offered by the recollections of Edoardo Amaldi and Charles Peyrou on the International Conference on Particle Physics held in Pisa in July, 1955:

> *A striking fact that emerged in Pisa was that the time for important contributions to subnuclear particle physics from the study of cosmic rays was very close to an end. A few papers presented by physicists from the U.S.A. showed clearly the advantage for the study of these particles presented by the Cosmotron of*

*Brookhaven National Laboratory (3 GeV) but even more by the Bevatron of the Lawrence Radiation Laboratory in Berkeley (6.3 GeV)* [3] (p. 117).

*. . . at the Pisa Conference in July 1955 . . . the cosmic ray physicists could be proud; they had found just in time all possible decays of the heavy mesons, and made it very plausible that there was one and only one K particle. But their triumph was a swan's song. At the same conference the Berkeley physicists brought better proofs of that idea* [4] (p. 631).

In the words of Peyrou, "better proofs" could be replaced by "better tools"; the Berkeley physicists were able to bring "better proofs" because they could avail themselves of the "better tool" that was the Bevatron, recently put into operation with a peak energy of 6.3 GeV, the most powerful accelerating machine in existence at that time in the world. Not only the Bevatron was a "better" tool; it was also fundamentally "new", in the sense that it allowed to perform in radically altered conditions the process from the collection of empirical indications on the world of new particles toward their transformation into conclusive evidence: from the observation of the phenomenon, reading the traces left in an unpredictable and uncontrolled way by cosmic radiation, one could move to its artificial production, in copious quantities and under controlled and repeatable conditions. With the construction of the new apparatus, not only the dimensions of science built around it changed, but the cognitive procedures themselves were redefined. In particular, the conditions that allowed what Peter Galison refers to as a "change in the status of evidence" were modified. This change made it possible to transform "evidence . . . from a hint to a demonstration" [5] (p. 1); In other words, it created the necessary framework to move, as Peyrou puts it, from a "very plausible" conjecture to actual "proof" or, to echo Galison once more, to construct convincing arguments about the world around us.

*In denying the old Reichenbachian division between capricious discovery and rule-governed justification, our task is neither to produce rational rules for discovery - a favorite philosophical pastime - nor to reduce the arguments of physics to surface waves over the ocean of professional interests. The task at hand is to capture the building up of a persuasive argument about the world around us, even in the absence of the logician's certainty* [5] (p. 277).

If discovery techniques and justification strategies walk together, the appearance of new tools (new mediators between "what nature tells us" and "what we say about it") helps to redraw the boundary and interaction between the two contexts. The story of the "discovery" of the antiproton, and of the "justification" of its existence, is an excellent example in this regard. Located exactly in the mid-fifties, it represents a decisive moment, perhaps the most significant, of the transition from cosmic ray research to particle accelerator experiments, allowing to grasp some of its most characteristic features. Furthermore, for its developments and its outcomes, this story illustrates some aspects of the way in which the new technology available intervenes in redefining the rules with which "persuasive arguments" are constructed.

An official recognition—the Nobel Prize awarded to Emilio Segrè and Owen Chamberlain in 1959—identifies in a date, a place and an experimental device the discovery of the antiproton: in Berkeley in September 1955 [6], thanks to the identification of particles of negative charge and protonic mass in a sophisticated detection apparatus mounted at the exit of the beam generated by the Bevatron, the only machine able at the time to reach the threshold energy necessary for the production of proton-antiproton pairs (The official version of the Radiation Laboratory has emphatically maintained, starting in 1955, that the energy of the Bevatron had been fixed at the value of 6 GeV from the earliest stages of design, in view of the production of antiprotons. On this aspect of the problem see [7–9]).

We will not dwell on the developments that led to the September experiment of the Segrè group (developments that have already been object of historical research, besides having acquired a certain resonance in public opinion for being at the origin of one of the first legal cases linked to a dispute on scientific priorities [8]) (Reports and impressions of the protagonists are found in [10–13]). In the present paper we are interested in following some threads of a parallel path, which winds throughout 1955 and part of the following year, starting from the probable detection of an antiproton trace in emulsions exposed to cosmic rays by the group of physicists in Rome led by Edoardo Amaldi up to the collaboration between this group and the group of Segrè in Berkeley aimed at searching similar events in emulsions irradiated by the Bevatron beam.

The antiproton did not appear on the scene as an unexpected guest, contrary to the puzzling array of new particles that had enriched the zoology of fundamental physics in previous years. The theoretical doubts about its existence, linked to the difficulty of extending Dirac's theory to objects other than the electron, had been largely overcome at the turn of the fifties; although, in the scientific literature of the time, the terms "negative proton" and "antiproton" still significantly coexisted to distinguish different features of this elusive particle [8]. When it was finally detected, "*it was certainly no surprise*" [11] (p. 283). By the mid-fifties, the hunt for the antiproton was open, even if not with the intensity and determination attributed a posteriori. Still at the end of 1954 E.J. Lofgren declared that at the Bevatron (the machine "built to find the antiproton") "*there are no defined plans for looking for anti-protons*" [8] (p. 185). Long before the new Berkeley accelerator came into play, the hunt was carried out using the cheap tools available to cosmic ray physicists: cloud chambers and nuclear emulsions. In the traps set up to catch the fleeting prey, someone had even believed to have caught something.

Between 1947 and 1955, a series of "strange" events recorded by the detectors of different experiments carried out on cosmic radiation had led to advance, with different levels of conviction and determination, the hypothesis of having observed a phenomenon interpretable as evidence for an antiproton annihilation process [14–18]. Everything was however still at the level of "hints" that did not reach the status of "demonstration", as it is exemplarily proved by the precautions with which the results were presented. "*Such an event does not seem unlikely*", "*other possible explanations are that it is a negative proton*" [17] (pp. 937–941), "*one should consider the possibility that the event represents the annihilation process . . . for example, the incident particle might be an antiproton*" [15] (p. 1103), "*one possibility is that it may be produced by an annihilation process*" [18] (p. 857). Claims were thus in the domain of "reasonable possibilities", but no one had yet "persuasive arguments". It is worth pointing out that the arguments which might appear persuasive to some physicists (as in the case of Bruno Rossi, leader of the MIT group, which fully defended the validity of his interpretation of the data) (For example, at the 1956 Rochester Conference Rossi claimed that "*. . . there is thus little doubt that the M.I.T. event was indeed the annihilation of an antiproton*" [19]; see [20] (pp. 95–96)), were not- or were no longer- persuasive arguments for all.

Fairly persuasive appeared the evidence shown by the last of the "strange events" mentioned above. In one of the emulsions exposed to cosmic rays during the campaign in Sardinia in 1953 (On the European collaborations of the early fifties, see [2, 21]), and examined by the group of physicists under the guidance of Edoardo Amaldi in Rome, a double star was found in February 1955, which could be interpreted in terms of the process of "*production, capture and annihilation of a negative proton*". The conclusions reached did not allow to exclude with certainty that the event (known as "Faustina" ("Fausta" is a female proper name, which corresponds to the adjective "fausto", meaning "auspicious"; "Faustina" is the diminutive of "Fausta")) might be an accidental coincidence, but the connected probability was so low that the Roman group felt entitled to "*look for an interpretation of the observed event in terms of physical process and not of an accidental coincidence*" [14] (p. 497).

> *This value* (the expected number of similar events due to casual spatial coincidences in the volume explored) *is sufficiently small to entitle us to look for an interpretation of the observed event in terms of a physical process and not of an accidental coincidence. We are left to consider the star B as produced by the track p. Then the corresponding particle either has a rest energy of the order of 1.5–2 GeV, or, being an antiproton, it has been annihilated by a nucleon, releasing* $2\,m_p c^2 = 1876\,MeV$. *We do not have any argument in favour of one or the other of these two possibilities apart from the fact that unstable particles of rest energy of the order of 1.5–2 Gev have never been observed; nor has the antiproton,but this, at least, is expected to exist as a consequence of very general arguments based on symmetry with respect to the sign of the electric charge . . .*
>
> *We are glad to express our thanks to Prof. B. Ferretti, Dr. B. Touschek, Dr. G. Morpurgo and dr. R. Gatto for various criticisms, and enlightening discussions.*

Caution, however, was essential: the title originally envisaged for the paper to be published in *Il Nuovo Cimento* was "Unusual Event Produced by Heavy Particle at Rest", but was soon changed in the more cautious "Unusual Event Produced by Cosmic Rays". In the final remarks it was stated that "the many questions raised by the discussion of this event will obviously find their final answer only if other similar events will be observed" [14] (p. 499).

To "observe similar events" and obtain the "final answer" the physicists of Rome decided to use the same observation technique (i.e., the exposure of nuclear emulsions), but in order to avoid the whims of cosmic radiation they aimed at employing a more reliable and controllable source. A few days after the publication of the note on Faustina, Amaldi wrote to Segrè at Berkeley proposing a collaboration to search for a definitive proof of the annihilation of antiprotons, exposing the nuclear emulsion plates to the proton beam of the Bevatron.

> *Now the meaning of our work is the following: we cannot rule out the possibility that Faustina be a casual coincidence, but in case it is due to a real antiproton one should conclude that the corresponding production cross section is large at an energy of about 10 GeV, which is likely the energy of the primary of Faustina's A star. One can then think of trying to produce them also with your machine. True, the energy is much lower, but there is still a good probability to observe them . . .*

*Now my proposal is as follows: we make an agreement that you set up the experience and make the irradiations, and we take care of development and scanning; if anything worth comes out of the work, we publish together. When I say "you", I mean you Emilio Segrè, or Gerson Goldhaber who works on emulsions and is with you, or both . . .*

*Here all the matter has been discussed extensively with our theoreticians (Ferretti and Touschek) and with the emulsions group* (Amaldi to Segrè, 29 March 1955 [22]).

Segrè accepted the proposal, saying he was "impressed" by Faustina:

*I have looked carefully to Faustina and I am also impressed by it. I would like to cooperate in the experiment you suggest; Goldhaber would also like to work on it, and Warren Chupp would almost certainly work on it . . . Coming to the practical program: there are at least two programs, of which I know, for hunting the negative protons. One is a photographic one initiated by Rosen of Los Alamos, who has already made an exposure practically identical to your proposal, without the magnet . . . The other method is based on a measurement of momentum and velocity, with a possible photographic check* (Segrè to Amaldi, 15 April 1955 [22]).

Segrè had his reasons for being "impressed". Faustina was a further indication that cosmic ray physicists had good chances of ending the hunt for the antiproton even before his experiment at Bevatron started the operational phase (the experience plan had just been approved by the management of the laboratory (Lawrence Berkeley Laboratory Report (UCRL 2920, November 1954, January 1955))). The question, of course, did not only involve the personal projects and scientific ambitions of Segrè and his group, but invested the entire Radiation Laboratory: a successful conclusion on the existence of the antiproton by the cosmic ray physicists would have nullified most of the scientific arguments advanced by the leaders of the laboratory, Lawrence in the first place, to obtain by the Atomic Energy Commission the provision of the amount of money necessary for the construction of the most expensive experimental apparatus ever made in a physics laboratory.

Starting in March, therefore, the timing of the experiment of the group of Segrè (which we will call the "counter experiment", and is the one that will provide in September the results that will lead to the recognition of the Nobel Prize) intertwined with the timing of the experiment carried out by the collaboration between Rome and Berkeley, which used the same machine through a different technique ("emulsion experiment"). Meanwhile, the composition of the American group in the collaboration was defined. While Amaldi had advanced, besides of course the name of Segrè, the only name of Gerson Goldhaber—the Berkeley expert in emulsions—Segrè extended the participation to Owen Chamberlain, Clyde Wiegand and Warren Chupp (Segrè to Amaldi, 28 june 1955 [22]). Thus, with the sole exception of Tom Ypsilantis, the whole group of the counter experiment was also present in the work with emulsions.

Towards the end of July the emulsions were exposed to the Bevatron beam, and at the beginning of August some of them were sent to Rome to be studied by the Italian team led by Amaldi, which consisted of Giustina Baroni, Carlo Castagnoli, Carlo Franzinetti and Augusta Manfredini. The scanning of the emulsions began in August, while in Berkeley the experiment with the counters started to run. Towards the end of September, the latter provided the first positive data. Amaldi was visiting Segrè in Berkeley at that time, and hastened to inform his team in Rome:

*There are 7 experiments to find the antiproton . . .* (among them) *one of the Segrè group based on a measurement of velocity from the time of flight between two scintillation counters and a measurement of momentum by deflection through a magnet. Yesterday this experiment started giving results that look positive: nothing is for sure yet, and therefore nothing should be circulated, but possibly a definitive answer will arrive in two or three days: should the thing be confirmed, there must be about one antiproton in 25.000-30-000 negative pions in the conditions of exposition A, that is in the conditions of the stacks 63 and 64 you are scanning . . . Therefore, keep your eyes open and go ahead full force . . .* (Amaldi to Baroni et al., Berkeley, 22 September 1955 [22]).

Only on November 18 the slow work of scanning the emulsions provided the desired result, producing the event called "Letizia" ("Letizia" is another female proper name in Italian language. The word stands for "joy"): a clear annihilation star with a visible energy release that left little doubt about its interpretation.

*Found Letizia similar Faustina particle protonic mass enters stack 62 left side leading edge comes to rest after $9.31\,cm$ and produces star consisting 6 black particles 1 grey proton 1 pion 80 MeV 1 minimum ionization particle stop lower limit energy release 800 MeV stop measurements not yet finished letter follows Amaldi* (Telegramme, Amaldi to Segrè, 18 November 1955 [22]).

Other similar events showed up in the following months, as the Roman and Berkeley groups proceeded in the work of examining the emulsion plates, while in the meantime other Berkeley physicists became involved (The final results of the collaboration are in [23, 24]). On 11 January 1956, an annihilation star was found in the plates scanned by the Berkeley team clearly showing an energy release that dispelled any trace of remaining doubts:

> *This event turned out to be particularly important because it gave the conclusive proof ("sufficient condition" for those who were still in doubt) of the annihilation process. The visible energy release in this star was $1300 \pm 50\,MeV$. Clearly greater than the mass of the incident negative particle! . . . Chamberlain gave an invited talk at the 1956 New York meeting of the American Physical Society. There he reported on both the counter experiment and our annihilation event. He told me afterward that the proof supplied by the annihilation event was an important ingredient in the minds of the audience* [25].

Goldhaber's remarks point to two key issues: the kind of evidence deemed necessary to claim the discovery, and the effectiveness of how the evidence was presented in creating a consensus about what actually had been observed. The second point is related to the wider issue of the complementary interplay of visual and logic means to provide information on the piece of nature under investigation. In the case here considered, "seeing" the annihilation, beside the intrinsic convincing power given by visualisation, had the advantage of showing the physical process that allowed to properly label the particles as antiprotons and not just negative protons, something that from the "logic" inference offered by the counter experiment could not be derived (The standard reference on the general issue is [26]. For the specific case of the antiproton discovery, a strong case is found in [27]).

When was the antiproton "discovered"? Excessive attention to priority disputes has produced historiographical practices of dubious reputation, but scientific priority is not the main historical issue to be addressed here. Indeed, behind the above question stands precisely the problem of the building of persuasive arguments, able to finally transform "hints" into "demonstrations" and to allow the formation of consensus around a new stabilized piece of knowledge. Is the identification of particles having protonic mass and negative charge (the evidence provided by the experiment with the counters) a "demonstration" of the existence of the antiproton? It is worthwhile to listen to some of the protagonists:

> (E. Segrè was able) *to establish the existence of a small but clearly observable number of negative protons, among the particles produced by the collision of protons of 6.3 GeV, accelerated with the Bevatron, against quiet nuclei . . . That they were antiprotons in Dirac's sense, i.e. corpuscles capable of annihilating themselves with as many protons, was demonstrated in an experiment carried out with the technique of nuclear emulsions by the same group extended with the addition of G. Goldhaber et al. in Berkeley and by E. Amaldi et al. in Rome* [28] (p. 121, original text in Italian).

> *By October 1955, the counter experiment had clearly demonstrated the following:*
> - *There were negative particles of protonic mass within an accuracy of 5 percent.*
> - *There was a threshold for the production of these particles at about 4 GeV of incident-proton-beam kinetic energy.*
>
> *These were necessary conditions for the identification of antiprotons.*

> *Then, in November 1955, our efforts in the emulsion experiment . . . yielded one event, found in Rome, that came to rest and produced a star with a visible energy release of about 826 MeV. Again a* <u>necessary condition</u> (Here and in the following quotations, underscores are made by the author of the present paper) *for antiprotons* [25] (p. 267).

To evaluate the margins of uncertainty within which the discussion was still moving at the end of 1955, it is useful to compare some crucial steps of the conclusions of the preliminary work in which Letizia was presented, in the different versions we have: the transcription of the report made by Amaldi at the monthly session of the Accademia dei Lincei in December, the Note published in the *Rendiconti* of the Accademia (both in Italian) and the translation of the latter, which appeared soon after in the *Physical Review*. In the Amaldi report we read:

> *It can therefore be* <u>concluded</u> *that this process is due to an antiproton . . . This observation is in a sense complementary and integrates the discovery of Chamberlain, Segrè, Wiegand and Ypsilantis announced in mid-October by the Radiation Laboratory.*

> *On the other hand the disintegration observed in the emulsions exposed to the Bevatron has the same characteristics as that observed at the beginning of 1955 by the group of Rome in emulsions exposed to* *cosmic radiation and therefore it can be concluded that the interpretation of that event, proposed at the time, in terms of an antiproton annihilation process, was correct. (Report by Amaldi at the meeting of the Accademia dei Lincei, December 10, 1955, p. 3; Amaldi Archive (section Archivio Amaldi Eredi, box 21) Original document in Italian).*

These statements sound less conclusive in the text published in the *Rendiconti*:

*This event confirms, even if not definitively, the interpretation . . . that the new particles observed at the Bevatron are antiprotons. It also confirms the hypothesis that the star described in (5)* (i.e., Faustina, authors' note) *was actually due to an antiproton* [29] (p. 386, original document in Italian).

Finally, the statements appearing in the *Physical Review* are clearly weaker:

*This event is corroborating evidence, but not final proof, for the interpretation . . . that the new particles observed at the Bevatron are antiprotons. It also gives support to the hypothesis that the star described in ref. 5 was indeed due to an antiproton* [30] (p. 910).

There is a certain difference between drawing a "conclusion" and having "corroborating evidence". A conclusion is definitive; it does not require further "final proof". Even leaving out the transcription, understandably stronger, of the report to the Italian Academy and concentrating on the two published works, it is not possible to ignore the subtle linguistic discrepancies (and this despite Segrè had explicitly insisted, arousing Amaldi's irritation, that the note for the *Rendiconti* should be the "literal translation" of the one prepared for the *Physical Review*) (Segrè to Amaldi, 29 November 1955; Amaldi to Segrè, 5 December 1955 [22]). The two "confirmations" become "corroborating evidence" and "support"; one may argue that "confirmation" applies to what is already firmly established, while "support" is needed for what cannot stand on its own.

The point is that it was not simply a question of subtle linguistic discrepancies. The apparent inconsistencies can be explained if one takes into account the fact that what was said emerged as the result of a mediation in which not everyone wanted to say the same things. Different expectations were at stake, which triggered different priorities and led the two groups to place different emphasis on the various aspects of the results achieved. For the physicists in Rome, in particular, it was central to highlight the "similarity" of Letizia with Faustina; precisely what the Berkeley physicists were not willing to concede. Physicists in Rome were looking for annihilation stars in order to confirm their interpretation of the dubious result they already had—Faustina, the uncertain annihilation star observed in the cosmic radiation—and regarded Letizia as sufficient final proof for their hunt; physicists in Berkeley were looking for annihilation stars in order to prove that the solid result they already had—the detection with counters of negative protons—was indeed the discovery of the antiproton, and judged Letizia not yet "final proof" for that purpose.

These tensions may easily be documented in more detail by resorting to the extensive exchange of correspondence that took place since November 1955 between Amaldi and Segrè, concerning the forms of publication of the scientific results of the collaboration (The exchange of letters is in [22, 31]). What we wish to emphasise here is that these difficulties arise naturally from the inherent tension between the desire to make definitive claims about nature and the inevitable ambiguity with which nature reveals itself; in Rome and Berkeley, attempts were being made to construct "persuasive arguments" in ways that do not coincide.

Be it as it may, in the end a consensus was built, or at least a mediation was reached, given that a text was published, in which all the authors shared with equal weight credits and responsibilities, and the disagreements on the interpretations of the results were canceled in the unanimous version that was exposed to the judgment of the scientific community. If for the group in Rome Letizia represented necessary and sufficient proof, while this was not the case for the team in Berkeley, then it remains to be understood which was the shared "persuasive argument", on which the groups based the agreement reached in the final version of the published work. Indeed they agreed to insert the controversial statement that the observed event was to be considered "not final proof". The authority of the facts was, by itself, insufficient to impose a decision; and then the intervention of some other kind of authority was needed to release the tension and resolve the issue. It is an authority that doesn't come out of the lines of the published paper; however, we can identify it with reasonable certainty from other sources:

*This laboratory accepts first change but not omission words but not final proof or equivalent please cable whether we should mail letter Physical Reviews we want to see Italian text nota Licei (sic) before publication Segrè* (Telegramme, Segrè to Amaldi, 14 December 1955 [22]).

The final agreement was no longer (not only, anymore) the result of a comparison between the arguments of Amaldi and those of Segrè. It was "this laboratory", the authority of the prestigious- and rich, and powerful—Lawrence Radiation Laboratory, which was putting all its weight on the plate of the discussion to close it; thus intervening, not as a spurious element introducing irrationality into a process regulated by pure reason, but as a concrete part of a framework in which the game was played according to rules that did not have the clear "certainty of logic".

That the "big bosses" in Berkeley were playing the game, and not just his old friend Segrè, Amaldi made clear in a letter sent to Gian Carlo Wick:

> We have actually found here in Rome a nice star due to a negative corpuscle with mass $(1830 \pm 55)\, m_e$ very similar to the one we found in January in the cosmic radiation. We are in the publication process but we have some small difficulties as to the final text. Judging from what is happening these days it seems that the big bosses in Berkeley are rather difficult to deal with. You might possibly tell me that you were already well aware of that! (Amaldi to Wick, 15 December 1955; Amaldi Archive (section Archivio Dipartimento Fisica, box 5). Following his emigration to the States after the war, in 1948 Wick had replaced Robert Oppenheimer in Berkeley as professor of theoretical physics, a post from which he was dismissed in 1951 for refusing to swear the anticommunist loyalty oath required by the State of California).

"When" was the antiproton discovered? Everything suggests that, abandoning the easy punctual attributions possible *a posteriori*, it makes sense to answer by transforming the discovery from an event into a long-term process: and to argue that the "change in the status of evidence" that led to accumulating arguments persuasive for everyone, changing a series of "hints" into a definitive "demonstration"—and into a new cognitive acquisition—was the terminal stage of a journey that began somewhere in 1954 and ended at some point in the first half of 1956. Through this path two novelties were established in parallel: the existence of a new element of nature, and the emergence of a new way of questioning nature and formulating convincing assertions about it. Perhaps involuntarily, but certainly in a strongly symbolic way, this passage was well represented by the titles of the papers which announced Letizia to the world. The preliminary version of the work, published in English and in Italian respectively on the *Physical Review* and on the *Rendiconti* of the Accademia dei Lincei, was entitled "Antiproton Star Observed in Emulsion"; the accent was on "emulsion", the revelation technique that in the previous years had established itself as the workhorse of cosmic ray physics. But the title of the most complete work that appeared in *Il Nuovo Cimento* [32] sounded "On the Observation of an Antiproton Star in Emulsion Exposed at the Bevatron"; the chief protagonist was no longer the emulsion, the technique that allowed to see the antiproton, but the Bevatron, the machine that made it possible to make it.

Awarding of the Nobel prize to Segrè and Chamberlain "for their discovery of the antiproton", and the ensuing focus on the "ingenious methods" employed in their "logic" counter experiment, has left in the shade the contribution given to the whole process of discovery by the "visual" work with the emulsions (In his Nobel lecture [10], entirely devoted to an accurate description of the techniques used in the counter experiment, Chamberlain devotes only a few closing lines to the emulsion work, highlighting the "final proof" provided by the annihilation star found in Berkeley in January 1956, and mentioning that "other important work closely related to the same subject has occupied Professor Amaldi with his colleagues at the University of Rome"); as a consequence, in turn, any reference to the previous evidence gained by the work done on cosmic rays has been wiped off in the collective memory of the community. In the process, the role played by the emulsion scanners in Rome, and their previous pioneering work on cosmic rays, has been, if not neglected at all, largely underestimated. A good example of this is given by Segrè in his autobiography:

> My group had for some time studied the problem and prepared for it. I decided to attack the problem in two ways. One was based on the determination of the charge and mass of the particle. The other concentrated on the observation of the phenomena attendant on the annihilation of a stopping antiproton . . .
>
> For the first attack, Chamberlain, Wiegand, Ypsilantis and I designed and built a mass spectrograph with several technically new features. For the second attack, Gerson Goldhaber, who was then in my group, exposed photographic emulsions in a beam enriched in antiprotons by our apparatus. Many other people were involved in the enterprise, and we had agreements on how to publish the results and give appropriate credit to everyone . . .
>
> The mass-spectrograph experiment concluded on 1 October 1955, having proved the existence of the antiproton, and soon thereafter the emulsion work confirmed it . . . At the time of the antiproton

> *experiment, Amaldi and his wife Ginestra were at our home in Lafayette as our guests. He and I established a collaboration for the study of photographic emulsions exposed at Berkeley, taking advantage of the numerous well-trained scanners available in Rome* [13] (pp. 256–258).

There is nothing wrong in these lines, but, in my opinion, the overall picture, although not incorrect, can be misleading. It is true that Segrè "decided to attack the problem in two ways", but the second decision (hunting the antiproton with the emulsions) was taken as a result of a suggestion by Amaldi, as their correspondence clearly indicates. In Segrè's reconstruction of the events, Amaldi only appears at the end, and it sounds like their collaboration started after his visit to Berkeley in September, while it had origin with Amaldi's proposal in March. It is well known that perceptions of the past become altered in the course of time, even in the memory of the historical actors. As historian of physics John Heilbron has effectively put it,

> *one can understand that most historians do not consider the unsupported recollections of former participants very good evidence about events in the distant past. The problem of partial observation is in this case compounded by failing and selective memory* [7].

And so it also happens that "partial observation" and "failing and selective memory" sometimes cooperate to lead former protagonists to present reconstructions of past events giving credit to narrations diminishing their own personal contributions, as is the case with Giulio Cortini, one of the members of the Roman team that found Faustina:

> *The antiproton was in the air... A group of leading experimental physicists in Berkeley designed and performed an experience aimed at the final demonstration of its existence. The experiment was successful and was rewarded with a Nobel prize. Nonetheless, they wanted a more sensational confirmation: producing in their nuclear plates phenomena analogous to "our" ... Amaldi was in touch with the Berkeley group, and thanks to his prestige our group was associated to their "second" experiment: they sent us plates that had been exposed to the beam of antiprotons produced by their 6.3 GeV machine, and we found there the "first" event similar to "Faustina": telegram, congratulations. But naturally the prestige of this new result, and of those who followed, fell largely on them ...* [33] (pp. 84–87).

Again, nothing wrong, but a picture that, in my opinion, could be misleading. The Berkeley physicists wanted indeed a more sensational confirmation, but the nuclear plates were not "their" plates. Cortini should have said "our" plates; it was Amaldi who ordered the emulsions from the Ilford Company in Britain, and had them shipped to Berkeley, and he did so not because he had been "associated to their second experiment", but because of that second experiment he had been the prime mover.

Yes, as Cortini puts it, "*the prestige of this new result, and of those who followed, fell largely on them*". So much so that, in the Feature article "Fifty years of antiprotons" published in the November 2005 issue of the CERN Courier, one finds a beautiful picture of "the first annihilation star imaged in the photographic-emulsion stack experiments, led by Gerson Goldhaber of the Segrè group, which confirmed the discovery of the antiproton"; it is a reproduction of Letizia, found in Rome in the photographic-emulsion stack experiments led by Edoardo Amaldi, the annihilation star which at the time was considered by the Berkeley physicists as being "not final proof" for the confirmation of the discovery of the antiproton.

A last remark is probably in order. Regardless of the scarce mark left in posthumous memories, the work on emulsions performed in the fifties at the physics institute in Rome was an exciting time for all the experimentalists involved. Most likely, not only for them. From the closing lines of the Faustina paper, and from the recollections of some of the protagonists, it is clear that their results, and their meaning and possible interpretations, were subject of discussions with the small group of theoretical physicists active in the institute and at the new born laboratory of INFN in Frascati: Bruno Ferretti, Bruno Touschek, Giuseppe Morpurgo and Raoul Gatto. Cortini in particular remembers the "very important contribution" to these "heated discussions" given by Touschek, who "took the thing very seriously". With hindsight, knowing the fundamental contributions given by the Roman school of theoretical physics in the following years, it is tempting to assume that the involvement of theoreticians in Rome in the discussion about the antiproton findings of their experimentalist colleagues most likely contributed to strengthen their confidence in symmetry arguments. And the actual making of the antiproton turned antimatter from a theoretical speculation into a manageable tool. It is possible to suggest that in this respect the discovery of the antiproton contributed to pave the way in Rome for theoretical and experimental developments that followed, from the consequences of CPT theorem to matter-antimatter physics.

## Acknowledgments

This work is a revised and updated version of a paper originally published (in Italian): G. Battimelli, D. Falciai, Dai raggi cosmici agli acceleratori; il caso dell'antiprotone, in A. Rossi (ed.), Atti del XIV e XV Congresso Nazionale di Storia della Fisica (Udine1993-Lecce 1994), Ed. Conte, Lecce 1995, pp. 375–386. I wish to thank Donatella Falciai, who co-authored the original paper and whose dissertation was the starting point of this research, and Adele La Rana for her translation of the original paper into English).

## Conflicts of Interest

## References

1. Marshak, R.E. Scientific Impact of the First Decade of the Rochester Conferences (1950–1960). In *Pions to Quarks: Particle Physics in the 1950s*; Brown, L.M., Dresden, M., Hoddeson, L., Eds.; Cambridge University Press: Cambridge, UK, 1989; pp. 645–667.
2. Belloni, L.; Dilworth, C. From Little to Big. A Story of a European Post-War Collaboration with Nuclear Emulsions. In *The Restructuring of Physical Sciences in Europe and the United States 1945–1960*; Maria, M.D., Grilli, M., Sebastiani, F., Eds.; World Scientific: Hackensack, NJ, USA,1989; pp. 732–744.
3. Amaldi, E. The Beginning of Particle Physics: From Cosmic Rays to CERN Accelerators. In *40 Years of Particle Physics*; Foster, B., Fowler, P.H., Eds.; Adam Hilger: Shildon, UK, 1988; pp. 109–119.
4. Peyrou, C. The Early Times of Strange Particles Physics. In *The Restructuring of Physical Sciences in Europe and the United States 1945–1960*; Maria, M.D., Grilli, M., Sebastiani, F., Eds.; World Scientific: Hackensack, NJ, 1989; pp. 604–651.
5. Galison, P. *How Experiments End*; University of Chicago Press: Chicago, IL, USA, 1987.
6. Chamberlain, O.; Segrè, E.; Wiegand, C.; et al. Observation of Antiprotons. *Phys. Rev.* **1955**, *100*, 947–950.
7. Heilbron, J.L. An Historian's Interest in Particle Physics. In *Pions to Quarks: Particle Physics in the 1950s*; Brown, L.M., Dresden, M., Hoddeson, L., Eds.; Cambridge University Press: Cambridge, UK, 1989; pp. 47–54.
8. Heilbron, J.L. *The Detection of the Antiproton*, In *The Restructuring of Physical Sciences in Europe and the United States 1945–1960*; Maria, M.D., Grilli, M., Sebastiani, F., Eds.; World Scientific: Hackensack, NJ, USA,1989; pp. 161–217.
9. Seidel, R. The Postwar Political Economy of High-Energy Physics. In *Pions to Quarks: Particle Physics in the 1950s*; Brown, L.M., Dresden, M., Hoddeson, L., Eds.; Cambridge University Press: Cambridge, UK, 1989; pp. 497–507.
10. Chamberlain, O. *The Early Antiproton Work*; Nobel Lecture; The Nobel Foundation: Stockholm, Sweden, 1959.
11. Chamberlain, O. The Discovery of the Antiproton. In *Pions to Quarks: Particle Physics in the 1950s*; Brown, L.M., Dresden, M., Hoddeson, L., Eds.; Cambridge University Press: Cambridge, UK, 1989; pp. 273–284.
12. Piccioni, O. On the Antiproton Discovery. In *Pions to Quarks: Particle Physics in the 1950s*; Brown, L.M., Dresden, M., Hoddeson, L., Eds.; Cambridge University Press: Cambridge, UK, 1989; pp. 285–298.
13. Segrè, E. *A Mind Always in Motion: The Autobiography of Emilio Segrè*; University of California Press: Berkeley, CA, USA, 1993.
14. Amaldi, E.; Castagnoli, C.; Cortini, G.; et al. Unusual Event Produced by Cosmic Rays. *Il Nuovo C.* **1955**, *1*, 492–500.
15. Bridge, H.S.; Courant, H.; DeStaebler, H.D., Jr.; et al. Possible Example of the Annihilation of a Heavy Particle. *Phys. Rev.* **1954**, *95*, 1101–1103.
16. Cowan, E.W. A V-Decay Event with a Heavy Negative Secondary, and Identification of the Secondary V-Decay Event in a Cascade. *Phys. Rev.* **1954**, *94*, 161–166.
17. Hayward, E. Ionization of High Energy Cosmic-Ray Electrons. *Phys. Rev.* **1947**, *72*, 937–942
18. Schein, M.; Haskin, D.M.; Glasser, R.G. Narrow Shower of Pure Photons at 100,000 Feet. *Phys. Rev.* **1954**, *95*, 855–857.
19. Rossi, B. M.I.T. Cloud-Chamber p- Event. In Proceedings of the 6th Annual Rochester Conference on High Energy Nuclear Physics, Rochester, NY, USA, 3–7 April 1956; Section VII, p. 10.
20. Rossi, B. *Momenti nella Vita di uno Scienziato*; Zanichelli: Bologna, Italy, 1987.
21. Grilli M.; Sebastiani F. Collaboration among Nuclear Emulsion Groups in Europe during the 1950's; Internal Note no. 973; Department of Physics, University of Rome "La Sapienza", 1991
22. Amaldi Archive (section Archivio Amaldi Dipartimento, box 156), Department of Physics, University "La Sapienza", Rome.
23. Barkas, W.H.; Birge, R.W.; Chupp, W.W.; et al. Antiproton-Nucleon Annihilation Process (Antiproton Collaboration Experiment). *Phys. Rev.* **1957**, *105*, 1037–1058.
24. Chamberlain, O.; Chupp, W.W.; Ekspong, A.G.; et al. Example of an Antiproton-Nucleon Annihilation, *Phys. Rev.* **1956**, *102*, 921–923.
25. Goldhaber, G. Early Work at the Bevatron: A Personal Account. In *Pions to Quarks: Particle Physics in the 1950s*; Brown, L.M., Dresden, M., Hoddeson, L., Eds.; Cambridge University Press: Cambridge, UK, 1989; pp. 260–272.
26. Galison, P. *Image and Logic*; University of Chicago Press: Chicago, IL, USA, 1997.

27. Orrman-Rossiter, K. Observation and Annihilation: The Discovery of the Antiproton. *Phys. Perspect.* **2021**, *23*, 3–24.

28. Amaldi, E. Il caso della fisica. In *Le Conseguenze Culturali Delle Leggi Razziali*; Atti dei convegni lincei; Roma, 1991; Volume 81; pp. 107–133. .

29. Chamberlain, O.; Chupp, W.W.; Goldhaber, G.; et al. Su di una stella provocata da un antiprotone osservata in emulsioni nucleari. *Atti Della Accad. Naz. Lincei* **1955**, *19*, 381–386.

30. Chamberlain, O.; Chupp, W.W.; Goldhaber, G.; Segrè, E.; et al. Antiproton Star Observed in Emulsion. *Phys. Rev.* **1956**, *101*, 909–910.

31. Amaldi Archive (section Archivio Amaldi Dipartimento, box 155), Department of Physics, University "La Sapienza", Rome.

32. Chamberlain, O.; Chupp, W.W.; Goldhaber, G.; et al. On the Observation of an Antiproton Star in Emulsion Exposed at the Bevatron. *Il Nuovo C.* **1956**, *3*, 447–467.

33. Cortini, G. Interview with L. Bonolis, September 21, 2005. In *Maestri e Allievi Nella Fisica Italiana del Novecento*; Bonolis, L., Ed.; La Goliardica Pavese: Pavia, Italy, 2008; pp. 69–94.

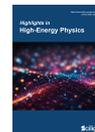

*Review*

# Raoul Gatto and Bruno Touschek: the Rise of $e^+e^-$ Physics

Luisa Bonolis [1], Franco Buccella [2] and Giulia Pancheri [3,*]

[1] Max Planck Institut für Wissenschaftsgeschichte, 14195 Berlin, Germany
[2] INFN, Sezione di Roma I, c/o Dipartimento di Fisica, Sapienza Universita' di Roma, 00184 Rome, Italy
[3] INFN, Laboratori Nazionali di Frascati, 00044 Frascati, Italy
* Correspondence: Giulia.Pancheri@lnf.infn.it

**Abstract:** The story of the collaboration between Raoul Gatto and Bruno Touschek, before during and after the construction of AdA, the first electron-positron collider built at the Frascati National Laboratories in 1960, is only partially known. A brief outline is presented here to show how electron-positron physics was influenced by Gatto and Touschek's common early interest in the CPT theorem. Their legacy is also illustrated with examples of their lasting impact as mentors to their students and collaborators. Starting with the early days after Fermi's departure, we describe the various physics scenarios behind the beginning of a deep relationship between Touschek and Gatto in Rome in 1953, the years of AdA between Rome, Frascati and Orsay, up to the construction of ADONE, the more beautiful and powerful collider, where multihadron production was first discovered in 1968/69 and the existence of the charm quark was confirmed in 1974.

**Keywords:** history of physics; elementary particles; electron-positron colliders

## 1. Introduction

AdA, the first electron-positron collider, was born in Frascati on 17 February 1960. On that day, the scientists of the Frascati Laboratory met to decide on the creation of a theoretical physics group in Frascati, where an electron synchrotron had been operating since the spring of the previous year. The discussion was initiated by Bruno Touschek. Having rejected both the proposal for a dedicated theory group and for a theoretical physics school, the former being insufficiently motivated and the latter unnecessary, he put forward a completely orthogonal idea: to do a new experiment, something that could attract theorists, not only from the University of Rome, but also from other Italian universities and beyond. His vision was to do something new that had never been done before: to study $e^+e^-$ collisions with two counter-rotating beams in the same vacuum chamber. Bruno was no newcomer to the novel science of particle accelerators, an expertise stemming from wartime and early postwar experience, first in Germany and later in the United Kingdom. However, to better understand the articulated motivations and scientific background behind Touschek's bold proposal, it is essential to place it within the broader and dynamic Italian scenario of the 1950s. The developments that followed, the construction and operation of AdA, the first electron-positron collider, and its main achievements have been reconstructed in [1,2]. A recent paper [3] has highlighted how the synergy between Touschek and Raul Gatto was instrumental in the rise of electron-positron physics and its impact on the construction and early operation of the larger ADONE collider (For full bibliographic references to the narrated events, see Bonolis et al. in Raoul Gatto and Bruno Touschek's joint legacy in the rise of electron positron physics, EPJH 2004 [3]).

This paper aims to shed light on previously unexplored aspects of the collaboration between the two physicists. It will explore the differences and convergences of the collaboration, as well as sketch out the backstage on which they appeared in the physics community and how their legacies were created and continued.

## 2. Fermi's Legacy and the Role of Bruno Ferretti

Bruno Touschek and Raoul Gatto owed their contemporary presence in Rome in the early 1950s to Bruno Ferretti, who had been Fermi's assistant in the last days of the "Via Panisperna group". Ferretti somehow represented the continuity with modern theoretical physics that had been initiated in Rome in the 1930s by Fermi and other younger theorists such as Giovanni Gentile Jr, Ettore Majorana, Ugo Fano and Gian Carlo Wick, who had rotated around Fermi at different times.

A graduate of the University of Bologna, Ferretti joined the cosmic ray group founded by Fermi in 1937, shortly before Fermi left Italy for the United States. Ferretti was influenced by Gian Carlo Wick and Gilberto Bernardini, with whom he worked on theoretical problems related to subnuclear physics with cosmic rays. These were attracting general attention after the discovery in 1936 of the "mesotron", a new-and unstable-particle whose nature and identity would remain unknown until its rebirth as the "muon" in 1947 [4]. During the war and in the early post-war years, the study of cosmic rays ensured the survival of the physics community in Rome and other centers in Italy.

Beginning in 1948, Ferretti began teaching theoretical physics in the chair left by Fermi and his successor Wick-when the latter took a position in the United States-and assisted Edoardo Amaldi in carrying on Fermi's legacy in Rome. They were joined by Gilberto Bernardini, an expert in nuclear physics and cosmic rays, whose postwar experience in particle physics at accelerators in the United States would later be instrumental in future plans to revive Fermi's frustrated prewar dreams of a modern competitive laboratory equipped with a high-energy accelerator [5]. As Amaldi recalled, Ferretti helped to strengthen the group of young theoretical physicists that included Bruno Zumino, Giacomo Morpurgo, Raul Gatto, Elio Fabri, Benedetto De Tollis and Carlo Bernardini [1] (p. 13), in addition to the young experimentalist Marcello Conversi. In various ways, the new generation and its students ensured the continuity of the new era inaugurated by the fathers of modern physics in Italy, and would in turn ensure the full revival of Italian physics in the post-war period.

Towards 1947, during a stay in the UK, Ferretti collaborated with Rudolf Peierls (Radiation Damping Theory and the Propagation of Light [6]), who was later to be external examiner for Bruno Touschek's doctoral thesis in Glasgow [7]. Ferretti was also Edoardo Amaldi's closest collaborator on the international scene, particularly in promoting the birth of CERN. We see him in Figure 1, in the last row behind Rudolf Peierls and Homi Bhabha, at the 1948 Solvay Conference, together with other eminent theoretical physicists whose work played an important role in Touschek's scientific development.

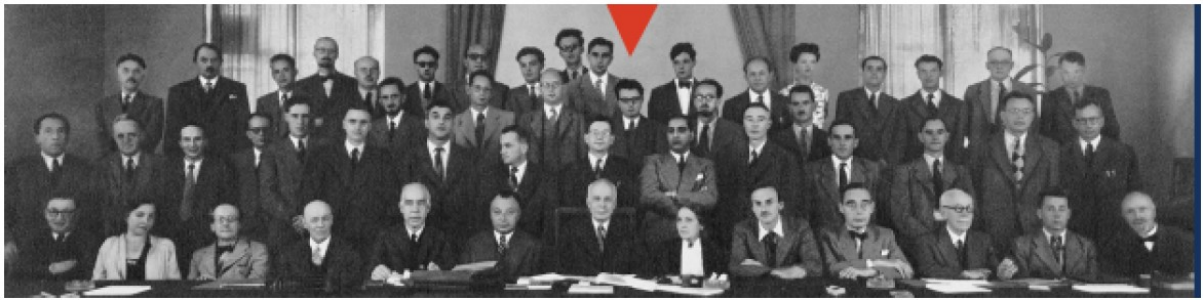

**Figure 1.** 8th Solvay Conference on Elementary Particles, 1948. Wikimedia.

After receiving his Doctorate from the University of Glasgow in November 1949, Bruno Touschek became Nuffield Lecturer, giving a contribution on weak interactions to Max Born's Atomic Physics book, and collaborating with Walter Thirring [8] on a problem related to Ferretti's radiation damping paper with Peierls. The mutual interest in the emerging field of Quantum Field Theory, which tied Touschek with Ferretti, gave probably rise to a plan for Touschek's sabbatical year in Italy. In September 1952 Touschek visited Ferretti in Rome. After a few hours spent discussing scientific issues of mutual interest, "they established such a marked professional respect and personal attachment for each other that Touschek decided to remain permanently in Rome" [1] (p. 13). Touschek, seen with Edoardo Amaldi in Figure 2, moved to Rome at the end of 1952, and this was made possible because he was given a contract as a researcher thanks to Amaldi, who was director of the Rome Section of the newly established National Institute of Nuclear Physics, INFN.

When Touschek arrived in Rome in December 1952, Raul Gatto, Figure 2, had just become Ferretti's assistant after graduating from the University of Pisa under the supervision of Marcello Conversi, then a professor at Pisa, and with Ferretti as external advisor in theoretical physics. The intellectual interaction established between the two is evidenced by Gatto's later writings about Touschek, by the letter reproduced in the Appendix A and by the fact that Gatto acknowledged discussions with Touschek in his early articles on nuclear and subnuclear physics based on the cosmic ray research of Amaldi's group in Rome. In this sense he was following Ferretti's example, but at the same time he was moving towards current problems in theoretical particle physics, while Bruno Touschek added his unique experience in subnuclear physics with accelerators to a deep mathematical and theoretical physics knowledge, developed from such mentors as A. Sommerfeld, W. Heisenberg, M. Von Laue, and M. Born, during his years in Germany and the United Kingdom. This unique expertise was highly valued in Italy, where INFN was being established, including ambitious plans to build a powerful accelerator and a national laboratory to house it. While waiting for this facility to become operational, the traditional work with cosmic rays continued in all the Italian centers, with the support of theoreticians.

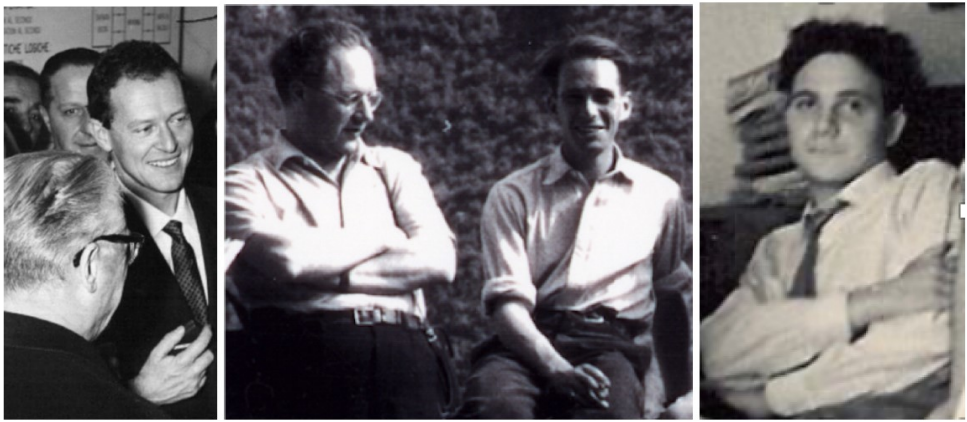

**Figure 2.** From left: Marcello Conversi in the 1960's, Edoardo Amaldi with Bruno Touschek soon after his arrival in Rome, and young Raoul Gatto, family photos.

## 3. New Challenges for Theoretical Physics

In the early 1950s, when Touschek and Gatto began their new scientific life in the exciting context of the reconstruction of Italian physics and its revival on a completely new basis, they became involved in the new, enigmatic physics of strange particles and the emergence of the $\theta$-$\tau$ puzzle, where data were still being derived from both cosmic ray and accelerator physics. But inexorably, the accelerators took over and particle physics moved beyond the study of high-energy cosmic rays, while the theorists began to face new, unprecedented challenges. Touschek found a very congenial atmosphere in the Physics Institute at Sapienza University of Rome. He immediately started collaborating with visiting scientists, such as Matthew Sands, or Giacomo Morpurgo and Luigi Radicati di Brozolo. The latter, considered Touschek one of the persons who had the greatest influence on his scientific life [9] (p. 67).

Radicati, who had graduated in 1943, with Enrico Persico in Turin, had discussed the time reversal problem with Peierls in UK, but they did not continue to work on that until, once in Italy, he resumed these topics with Touschek and Morpurgo [10–12]. Morpurgo had met Touschek at the end of August 1953, in the almost empty physics Institute, and, in less than one hour [9] (p. 80), a collaboration had started on a topic of common interest, concerning strong interactions through non-perturbative methods. By January 1954, their interests had shifted to the study of space-time symmetry properties, in particular time reversal. According to Morpurgo, the interest in time reversal had been sparked by Touschek, who had read a paper by Lüders [13] and may also have discussed the subject with Radicati, who was in Rome at the time. Meanwhile Touschek was also involved in work on $K$ mesons with several experimentalists. In April 1954 he attended the Padua Conference on Unstable Heavy Particles and High-Energy Events in Cosmic Rays [Particelle Instabili Pesanti e Sugli Eventi di Alta Energia nei Raggi Cosmici], Figure 3, where he contributed two papers [14,15].

Both out of genuine interest in a puzzling problem, and his commitment to give advice to the experimentalists, he followed ongoing experimental work on cosmic ray searches for anti-protons and strange particle decays, and his commitment culminated in August 1958 when he organized and directed the Fermi International Summer School on Pion Physics, held in Varenna.

In 1956 Gatto won a Fulbright fellowship and went to the United States, first to Columbia and then for a longer period to Berkeley, where the powerful Bevatron was operating. During this time, Gatto wrote several articles discussing the decays of K mesons and hyperons [16] within the strangeness scheme proposed independently by Gell-Mann and Nishijima [17,18]. Related experiments at Berkeley were followed by the first observation of the antiproton [19] and later of the antineutron [20], which consolidated the question of antimatter and strengthened the proof of its existence, almost twenty years after the discovery of the positron. In fact, an event interpreted as a proton-antiproton annihilation had been observed by the cosmic ray group in Rome [21], and it was the subject of an article written by Gatto before he left for Berkeley [22].

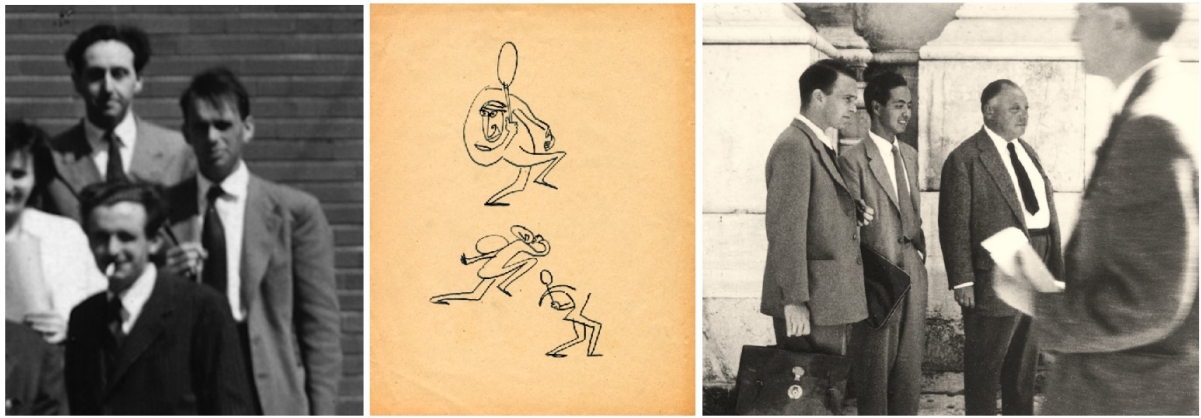

**Figure 3.** Bruno Touschek in Padua: at left in April 1954 at the Conference on Unstable Heavy Particles and High-Energy Events in Cosmic Rays in [23], and, at right, in September 1957 with T. D. Lee, W. Pauli and R. Marshak (at right) at the Padua-Venice Conference on Mesons and Recently Discovered Particles organized by the Italian Physical Society, courtesy of M. Baldo Ceolin. At the center, a contemporary drawing by Bruno Touschek [1] (p. 15), © Touschek's family.

In the same 1956, Touschek travelled with Amaldi to New York and attended the Rochester Conference, where anti-nucleons were discussed and the idea of parity non-conservation in weak processes was aired by Richard Feynman during the theoretical physics session chaired by C.N. Yang. Soon after, Lee and Yang, Figure 3, reviewed the experimental evidence in detail and suggested experiments that could settle the problem [24].

As high-energy nuclear physics evolved into particle physics, and accelerators-together with new kind of detectors - gradually replaced cosmic rays as the main source of high-energy particles, both Touschek and Gatto continued to work on the new puzzling phenomenology derived by experiments and on the classification of the new particles, which also raised the need to know more about symmetries and conservation laws.

Topics such as the annihilation process, time reversal, charge conjugation invariance, the $\theta$-$\tau$ puzzle and parity non conservation, and more generally the weak hyperon decay interactions under P, C, and T were studied by Gatto in articles between 1956–1958 [25,26], and a close look at Touschek's articles from the same period reveals a remarkable similarity between the two in terms of the underlying fundamental themes and issues addressed.

## 4. Setting the stage for AdA's Proposal: The Inspiring Role of the CPT Theorem

As mentioned above, Touschek's attention had long been drawn to various aspects related to fundamental symmetries, also stimulated by German theorists such as Pauli and Gerhardt Lüders. Beginning in 1953, Touschek discussed with Pauli questions related to time reversal, whose correct formulation in relativistic Quantum Field Theory was actually related to the roots of the CPT theorem (Blum et al. 2022). He wrote more than one paper on this subject [10,27] and discussed with Morpurgo the extension of the procedure to Parity and Charge Conjugation [11]. In 1957/1958, as evidenced by his personal papers, Touschek exchanged several letters with Lüders, Pauli and Zumino discussing topics related to symmetry properties of physical theories.(See the correspondence folders in Box 1 of Bruno Touschek Archives in Sapienza University of Rome, Archives of the Physics Department).

For his part, Gatto mentions how he became aware of the CPT theorem [28–31] through Bruno Zumino and Gerhard Lüders. Zumino had also graduated with Ferretti, but soon left for the United States, occasionally returning to Rome.

In a later paper by Zumino with Gerhart Lüders entitled "Some Consequences of TCP-Invariance", a direct reference is made to Zumino's having suggested in early 1953 the original formulation [32].

The CPT theorem, deeply connected with the physics of the weak interactions, acquired a central importance after the experimental discovery of parity violation in 1957 [33–35]. Here we would only like to emphasize how the theorem and its implications formed the backbone of Touschek's thought from which the idea of studying particle-antiparticle annihilations as a channel to new physics emerged. A clue to how things intermingled in his mind is provided by a letter that he wrote to Pauli on 31 January 1957. In the last lines, after discussing Lee and Yang's work, Abdus Salam's recent article on the neutrino, and the problem of $K$-decay, Touschek wrote: "*I have been trying for about a week to figure out whether invariance under CP (and not under P) means that one can distinguish between particles and antiparticles . . .*".

After the actual detection of the neutrino as a particle in 1956 and the explosion of interest surrounding the problem of parity violation, he focused on such relevant discoveries in the framework of the weak interactions. Alongside with a renewed interest in the symmetry properties of Fermi-Dirac fields [36,37], Touschek wrote several articles discussing a massless two-component neutrino. He was the first to introduce the concept of chiral symmetry as a consequence of parity violation [38] and in 1958 had begun a work with Pauli which was published only after Pauli's death [39]. Interestingly, in that same period, Gatto wrote an article with Lüders on "Invariants in Muon Decay" based on the assumption of a vanishing neutrino mass [40], which again shows how strong was their intellectual and scientific interaction during the 1950s. On the wave of parity violation, Touschek assigned three joint dissertations on weak interactions to Paolo Guidoni, Nicola Cabibbo, and Francesco Calogero.

At the end of the 1950s, after wandering between the puzzles in new elementary particles, and working in the "most abstract field of theoretical research [. . . ] the discussion of symmetries", Touschek wanted to get his feet "out of the clouds and onto the ground again" and get back to what he thought "[he] really understood: elementary physics".(B. Touschek , "Ada and Adone are storage rings", Bruno Touschek Papers, Box 11, Folder 3.92.4, p. 7.) Bruno Touschek was also coming to terms with Pauli's death in December 1958, an event which prompted him to write that "Without him [Pauli], physics is really only half as interesting for me" (*Ohne ihn ist die Physik für mich wirklich nur halb so interessant*, letter to Bruno's father on 24 December 1958, relating Wolfgang Pauli's recent passing on December 15th). In Touschek's mind CPT represented the central argument in his proposal of electron-positron annihilations as an alternative to the electron-electron collisions planned in the Princeton-Stanford project, presented by Wolfgang Panofsky at ICHEP 1959, in Kiev, and during a seminar in Frascati in October of that year. Gatto recalled how, after the seminar, "Bruno kept insisting on CPT invariance, which would grant the same orbit for electrons and positrons inside the ring". Nicola Cabibbo, who had recently graduated with Touschek with a thesis on "Pauli invariants in the decay of the $\mu$ meson" also testified: "Bruno Touschek came up with the remark that an $e^+e^-$ machine could be realized in a single ring, because of the CTP theorem".

These and similar remarks, by Gatto and Rubbia in 1987 [9] after so many years, highlight Touschek's firm belief in CPT as the tool that guaranteed the soundness of his proposal for a collider in which electron and positron beams would meet and annihilate. Something that was not taken for granted at the time as Carlo Bernardini always pointed out (Bernardini was a member of Enrico Persico's Theory group, which had contributed to the design of the Frascati electron synchrotron, and was in the AdA team led by Touschek from the beginning).

## 5. Experimental and Theoretical Challenges: AdA in Frascati and Orsay, 1960–1964

After the Frascati scientists gave the green light to Touschek's "experiment", Gatto and Touschek still had an uphill road ahead of them: AdA had to be built, but demonstrate the project's feasibility was far to have been established, both in the sense of making AdA as "proof of principle" for future colliders, and in making sure that physics could be extracted and be of interest for fundamental research. Touschek and Gatto's collaboration was the building stone. Establishing the dignity and richness of the physics came from the talks and papers they gave at international conferences, the thesis work they assigned to their students and their complementary insight in the processes to study. Crucial were the talks given by Gatto and Touschek at the Geneva International Conference in June 1961 [41], and the talk given by Gatto in September 1961 at the Aix-on-Provence Conference [42], when plans for moving the Frascati collider to the Laboratoire de l'Accélérateur Linéaire d'Orsay were laid out.

In the minutes of the Meeting of the Frascati Scientific Council held on 17 February 1960 [2] (ch. 10, p. 319), Touschek is recorded as having put forward the proposal for an experiment "that would be truly first order and that would be capable of attracting theorists to Frascati (not only him [Touschek] but also Gatto and certainly others) . . . [and] would be an experiment intended for the study of electron positron collisions." On February 18, 1960, the very day after the meeting, Touschek started a new notebook, and wrote SR for "Storage Ring" on the cover. On the first page, after the scheme of the experimental reactions to be studied, he wrote: "Ask Gatto. . . ", as shown in Figure 4. Together with the drawings on the last page, these two words indicate that Touschek considered Gatto his alter ego in the task of studying the physics governing the electron-positron "experiment".

And indeed, after Panofsky's seminar, discussions among theorists about electron-positron physics had taken place in Rome and, a few days before the Frascati meeting, Gatto and Cabibbo had sent an article to the *Physical Review Letters*. They were the first to systematically study $e^+e^-$ physics [43], and in the famous paper that became known as "The Bible" they discussed all possible experiments with high-energy colliding beams of electrons and positrons [44]. Their exploratory work not only confirmed Touschek's intuitions, but clarified that $e^+e^-$ machines would open up a whole world of physics to be explored.

Such was the enthusiasm and determination at Frascati Laboratory that by the end of 1960, while AdA's magnet was on its way from Terni and AdA was still to be assembled, Touschek had already prepared a preliminary report for a larger collider, ADONE, and plans for its design and construction began in early 1961. Two months later the memo had become an actual proposal which included both Gatto and Touschek's name [45].

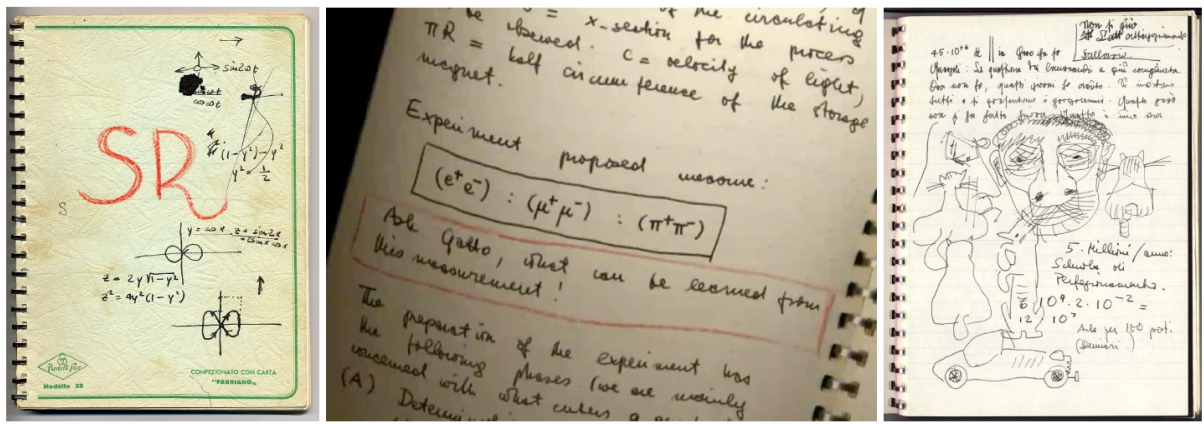

**Figure 4.** Cover and two pages from AdA's *Storage Ring* Notebook, started by Bruno Touschek on 18 February 1960. © Touschek Family, and Touschek Papers, Sapienza University of Rome, Archives of the Physics Department, all rights reserved.

## 5.1. Moving to Orsay

One of the reasons behind the final success of the AdA experiment is that Frascati was not alone in developing Italy's pathways to high energy accelerators, as crucial roles were played through CERN, at European level, and at the Laboratoire de l'Accélérateur Linéaire d'Orsay, in France.

CERN had been officially approved in 1954, after a rather long gestation period of at least five years [46]. Its planning during this period was complemented by various national initiatives, which the result that at the end of 1959 three high energy modern type accelerators in continental Europe were ready to take data, in as many large laboratories: the Proton Synchrotron (PS) at CERN in Geneva, the electron LINAC in Orsay [41] and the Frascati electron synchrotron [47,48]. This almost contemporary appearance offered an international stage to European scientists, who could exchange ideas and communicate their results to a wide audience.

This is precisely what happened with AdA. After the initial excitement of observing electrons (or positrons) circulate in AdA in February 1961, gloom descended on the Frascati group which had constructed AdA, because of the feeble luminosity obtained by using the synchrotron as an injector. Prospects changed after Touschek and Gatto attended a Conference in Geneva and both gave a talk in the session about Electrodynamics experiments. Following Burton Richter from SLAC, who described the Stanford electron electron project, Bruno's talk on electron positron storage rings in Frascati (AdA and ADONE), and the one by Gatto, on their physics prospects, fascinated two French physicists and started spreading some enthusiasm among others, planning experiments with proton beams. Thus, while at CERN a group of enthusiasts began gathering around Kjell Johnsen, Pierre Marin from the Laboratoire de l'Accélérateur Linéaire and Georges Charpak, decided to go to Frascati and see with their eyes the little jewel, un petit bijoux, as Marin later called AdA [41]. The photon beam from the Orsay linear accelerator (the LINAC) was soon recognized as the way forward to improve AdA's luminosity and, thus, the probability to observe collisions. One year later, in July 1962, AdA was moved to the Laboratoire de l'Accélérateur Linéaire, with all dowry and endowments, namely vacuum pumps, oscillographs, etc. The final leg of the journey towards high energy physics with electron positron colliders had begun.

The move to Orsay proved to be a winner, but, once more, the final success did not come without Gatto's playing a part, namely a dissertation at the University of Rome under Gatto's supervision on the cross-section for "Single photon emission in high-energy $e^+e^-$ collisions".

The French team in Orsay consisted of the two physicists Pierre Marin and François Lacoste, who had been enthused to join von Halban's linear collider team, and welcomed AdA in July 1962. They were supported by the new director André Blanc-Lapierre and highly skilled technicians, some of them having come from working in the UK during the war and who had built the LAL linear collider. When Lacoste left to pursue other interests, Jacques Haïssinski joined in. His doctoral thesis would become the main document describing in detail AdA's operation in Orsay and its success in proving the observation of electron positron collisions [49].

The Italian team consisted of Bruno Touschek, Giorgio Ghigo, Gianfranco Corazza, Carlo Bernardini, Ruggero Querzoli, his student Giuseppe Di Giugno and the technicians Giorgio Cocco, Bruno Ilio, Mario Fascetti and Angelino Vitale. At first they felt to have done the right move, but at the turn of the year, during the January–February 1963 run, the unexpected striked. They realised that the high-intensity photon beam from the Orsay LINAC (which was essential for creating high energy positrons) would not yield the anticipated evidence of annihilation into pion pairs, or even into two photons. When the team was ready for reaching the high current in the doughnut which could break the threshold of sufficient luminosity, the beam life-time started to decrease. Another run confirmed the presence of a collective effect, since then known as the Touschek effect, namely a decrease in the beam life-time even while increasing the beam current in the doughnut. The effect seemed to shatter the team's hopes, although it did not affect future colliders, since it was lessening as the beam energy increased. They could have stopped there and waited for the construction of higher energy colliders. But Bruno did not give up. Once more, what he knew and had thought came to his mind, and he understood that proof of feasibility of collision did not only come from annihilation into new particles. There was a process, which had not been listed in his note book Figure 4 and with a higher cross-section, for which experimental evidence could be gathered by the existing set up, namely single photon emission in elastic positron scattering. Emission of a photon in coincidence with the final $e^+e^-$ pair is in fact proof that the initial particles have disappeared and have been recreated with emission of one or more photons. The hitch was that while an approximate calculation showed the process could be measured with the LINAC photon beam, a precise calculation had never been done. What to do? Ask Gatto once more.

There were at the time many promising physics students at the University of Rome, looking for a thesis, among them Guido Altarelli and Franco Buccella. After some initial contacts with Gatto and Bruno Touschek [2] (Ch. 12, p. 377), they joined forces under Gatto's supervision and Touschek occasional crucial advice [50] and calculated the cross-section for the process, needed to confirm the experimental proof of collisions in AdA [51,52], through an ultra relativistic approximation for the final leptons, which made possible the computation of the differential cross-section in the energy and angle of the emitted photon. They graduated in November 1963 with a thesis on "Single photon emission in high-energy $e^+e^-$ collisions".

The approximation neglecting the annihilation diagrams has the consequence of predicting an equal cross section for electron-electron and electron-positron beams: indeed, the work is cited in the book by Landau and his collaborators for the emission of photons with electron-electron beams.

## 6. Formation of Young Theorists

Gatto and Touschek's influence on theoretical physics is heralded by the quality and number of the young people who graduated with them, their collaborators and the lectures in statistical mechanics, which Touschek started teaching at the University of Rome in November 1959 [53–55], while his former student Nicola Cabibbo was starting work with Raoul Gatto on the physics relevance of electron positron collisions.

After an initial period at the chair of theoretical physics in Cagliari, in the academic year $1962 - 63$ Gatto moved to the University of Florence, where young graduates and new students came together in a group which came to be know as those of the "Gattini", the kittens in English. In Florence Gatto gathered some of the most brilliant graduates from Rome, Cagliari and Florence, leaving a lasting legacy to theoretical physics. Memories of this period [56–59] highlight Gatto's legacy. As for Touschek, after AdA's confirmation of collisions in 1964, he turned his full attention to ADONE, the new collider, whose construction had been approved by INFN in 1963. Not wanting to change his citizenship status (he was Austrian by birth) he could not become professor in Italy until 1968, when the law changed [1]. This made him turn his full attention to develop the tools needed to explore the higher energy landscape where ADONE would operate. His contributions include participation to meetings to plan for future experiments, theoretical insight about the working of the new machine, and mostly formation of students and young graduates from the University of Rome. Thus, in 1966, while ADONE was being built across the street from the synchrotron, Figure 5, he started a theoretical physics group in Frascati, for which he asked for positions and physical space, as in the letter seen in Figure 6.

At this time Touschek's planning included Gian De Franceschi, who had graduated with Marcello Cini and was already in Frascati, Paolo Di Vecchia [60,61], Giancarlo Rossi [55,62], Francesco Drago, Etim Gabriel, Etim G.P., and Mario Greco [63], who had been supervised by Benedetto De Tollis for his 1964 thesis on new vector mesons photo production and had then been hired by Frascati in the accelerator division. Shortly after, and for a brief period, the group also included Maria Grazia (Pucci) De Stefano, who had graduated with Francesco Calogero -a former graduate of Touschek's [64]-with a thesis on the problem of scattering on singular potentials [65].

Among the young cohort he assembled, a remarkable expertise in QED calculation was present and with him or with his input a series of papers on the problem of radiative corrections to electron positron experiments emerged. Once more, Gatto's help came from the papers he kept writing to highlight the new field of electron positron physics, such as the one he presented at a meeting in Hamburg in 1964 [66] on "Theoretical aspects of colliding beam experiments". In 1966, when Touschek started to prepare his treatment of soft photon resummation, Gatto's paper was among those he suggested to his two young collaborators in the work on the infrared radiative corrections to electron and positron experiments [67]. As also discussed in these Proceeding by M. Greco, Touschek's input and insistence for the need to go beyond perturbative calculations for higher and higher energy collisions led the way to the development of resummation techniques in Quantum ElectroDynamics and later inspired analogous application to Quantum Chromodynamics.

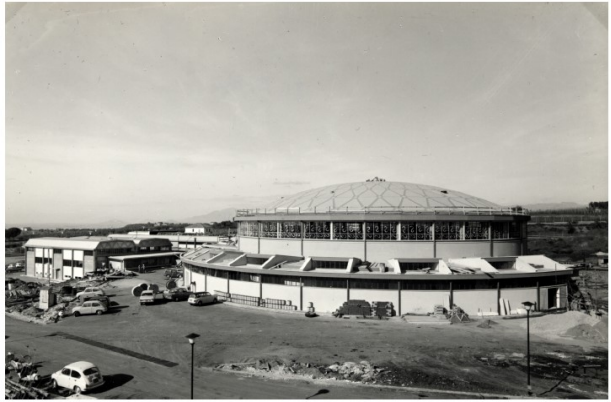

**Figure 5.** Exterior of the ADONE building in 1966, © INFN-LNF, all rights reserved.

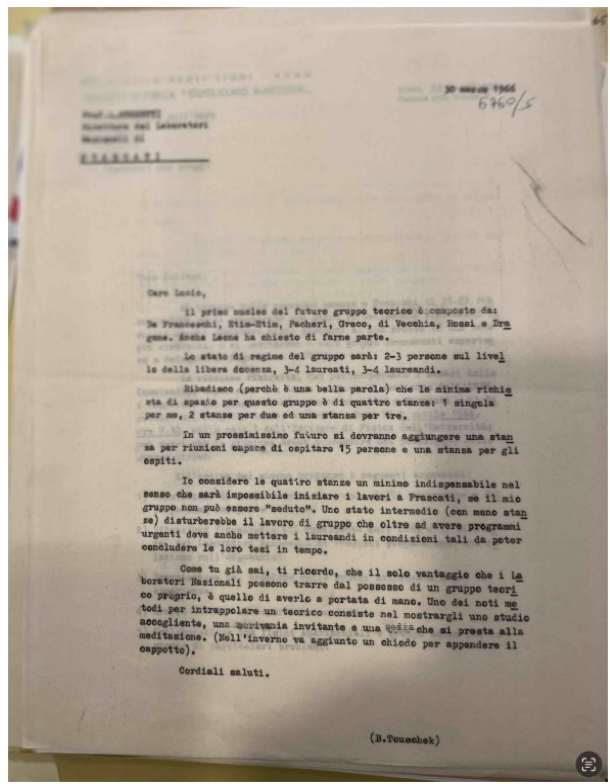

**Figure 6.** May 1966 letter by Touschek to Lucio Mezzetti, Frascati Laboratories director, Sapienza University of Rome, Archives of the Physics Department, all rights reserved.

In the fall of 1968, two beams, of electrons and positrons, circulated in ADONE. The long road Italian physicists had started in 1953, with the approval of the construction of the Frascati National Laboratories to host an electron synchrotron, opened the world stage to the Frascati Laboratories. At the time, and for few years to come, ADONE was the electron positron collider operating at the highest energy in the world. It would soon show that a new physics threshold had been reached, sparking the interest of theorists. ADONE set the experimental stage for further discoveries culminating in the detection of the $J/\psi$, which confirmed the existence of a fourth quark [68] and Touschek's vision that the quantum vacuum should be explored beyond the nucleon anti-nucleon threshold, as Heisenberg's had urged in his summary contribution to the 1953 Conference in Geneva [69].

ADONE gave green light to new physics arising from $e^+e^-$ collisions with the unexpected discovery of the multi hadron production which immediately sparked the interest of the new generation of theorists such as Giorgio Parisi and Massimo Testa, who had graduated with Nicola Cabibbbo [70]. In 1971 Gatto was called back to Rome and gathered a new group including Aurelio Grillo, Sergio Ferrara and Giorgio Parisi [71]. He also started to look for correlations between deep-inelastic scattering and (what he called) "deep-inelastic electron-positron annihilation", together with Giuliano Preparata [72].

His friendship and admiration for Bruno Touschek never wavered and he was deeply moved by Touschek's death in 1978. A few year later, Gatto moved to University of Geneva, where he continued his mentoring in theoretical physics as Editor of Physics Letters B, receiving wide recognisance for his scientific life, Figure 7.

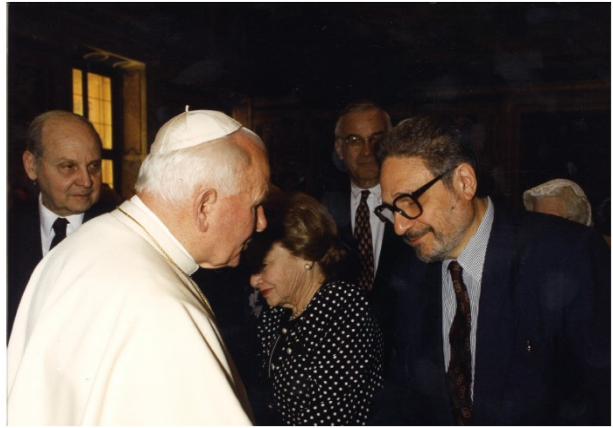

**Figure 7.** Raoul Gatto, receiving blessings from Pope John Paul II, INFN-LNF images.

## 7. Conclusions

We have presented a short overview of how Raul Gatto and Bruno Touschek together contributed to the rise of electron positron physics in the 1960s. Although they never wrote a paper together, their collaboration and mutual understading were deep and highly productive, resulting in a lasting legacy. We complete this paper with an Appendix A, where we reproduce a letter written by Raoul Gatto to Bruno Touschek, which testifies to the profound friendship between these two great scientists.

## Acknowledgments

We thank Yogendra Srivastava for contributions on Benedetto De Tollis. We are grateful to Raoul Gatto and Bruno Touschek's families, and to the Archives of the Sapienza University of Rome and INFN Frascati National Laboratories for permission to reproduce photographs and documents.

## Conflicts of Interest

The authors declare no conflicts of interest.

## Appendix A. Gatto's 1972 Letter to Touschek

A moving testimony of Gatto's feelings about Bruno Touschek appears in a December 1972 letter from Gatto to Touschek, written on the occasion of an incident occurring during the student unrest which took place in Italian universities, starting from 1968, lasting a few years. When a crude accusation of being a 'Nazi baron' was directed at Touschek by some students motivated by wanting to pass the exam for his course despite their ignorance, Touschek's decided to resign from his position at University of Rome (he was by that time "Professore aggregato"). In solidarity, and to deter him from leaving, Gatto wrote to him the letter shown in Figure A1.

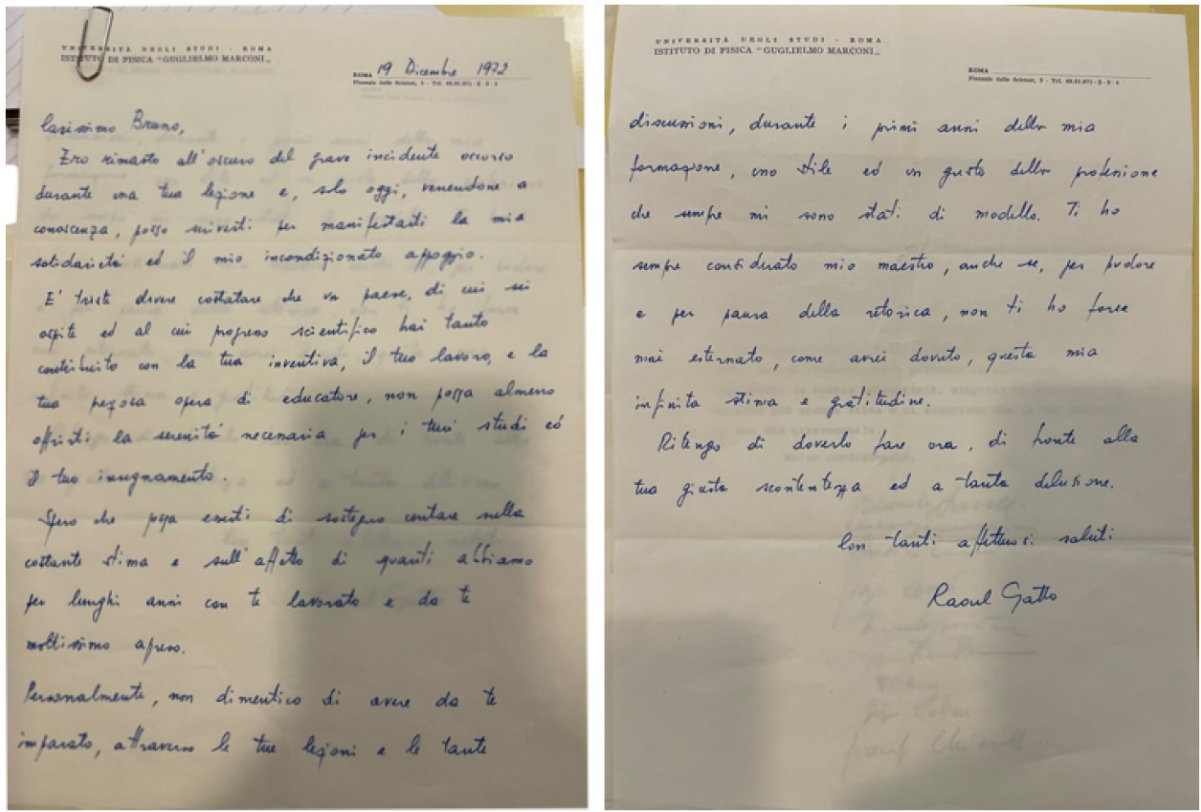

**Figure A1.** Raoul Gatto's letter to Bruno Touschek on 19th December 1972, ... *Personally, I do not forget to have learnt from you, during the early early years of my formation as a physicist, through your lectures and many conversations, a style and a sense of the our profession which have been my model to follow. I have always considered you my mentor and teacher, although my natural reserve and shyness have prevented me, in the past, to express my infinite admiration and gratefulness. I think it's right to do it now, facing your justly felt discontent and disillusionment ....;* Sapienza University of Rome, Archives of the Physics Department, all rights reserved.

## References

1. Amaldi, E. *The Bruno Touschek Legacy (Vienna 1921-Innsbruck 1978)*; No. 81-19 in CERN Yellow Reports: Monographs; CERN: Geneva, Switzerland, 1981.

2. Pancheri, G. *Bruno Touschek's Extraordinary Journey*; Springer Biographies: Cham, Switzerland, 2022.

3. Bonolis, L.; Buccella, F.; Pancheri, G. Bruno Touschek vs. machine politics: Science and technology in post-war Italy. *Eur. Phys. J. H* **2024**, *49*, 2311.01293.

4. Lattes, C.M.G.; Muirhead, H.; Occhialini, G.P.S.; et al. Processes Involving Charged Mesons. *Nature* **1947**, *159*, 694. https://doi.org/10.1038/159694a0.

5. Battimelli, G.; Gambaro, I. Il gruppo di Fermi e l'inizio dell'era nucleare in Italia. *Quad. Stor. Fis.* **1997**, *1*, 319.

6. Ferretti, B.; Peierls, R. The Properties of Slow Mesons. *Nature* **1947**, *160*, 531. https://doi.org/10.1038/160531a0.

7. Pancheri, G.; Bonolis, L. Bruno Touschek: Particle Physicist and Father of the $e^+e^-$ Collider. *arXiv* **2005**, https://arxiv.org/abs/2005.04942.

8. Thirring, W.E.; Touschek, B. On the Magnetic Moment of the Neutron. *Philos. Mag.* **1951**, *42*, 244.

9. Greco, M.; Pancheri, G. (Eds.) *1987 Bruno Touschek Memorial Lectures*; Frascati Physics Series; INFN Frascati National Laboratories: Frascati, Italy, 2004; Volume XXXIII.

10. Morpurgo, G.; Touschek, B.; Radicati, L.A. On the Magnetic Moment of the Neutron. *Nuovo Cimento* **1954**, *12*, 677. https://doi.org/10.1007/BF02781835.

11. Morpurgo, G.; Touschek, B. F. On the Magnetic Moment of the Neutron. *Nuovo Cimento* **1955**, *1*, 1159.

12. Morpurgo, G.; Touschek, B. On the Magnetic Moment of the Neutron. *Nuovo Cimento* **1956**, *4*, 691. https://doi.org/10.1007/BF02747964.

13. Lüders, G. On the Equivalence of Invariance under Time Reversal and under Particle-Anti-Particle Conjugation for Relativistic Field Theories. *Z. Phys.* **1952**, *133*, 325–339.

14. Touschek, B. On the Theory of the Betatron. *Nuovo Cimento* **1954**, *12*, 281.

15. Amaldi, E.; Fabri, E.; Hoang, T.F.; et al. Experimental Results on the Decay Process $K^+ \rightarrow \mu^+ + \nu + \gamma$. *Nuovo Cimento* **1954**, *12*, 419.

16. Gatto, R. $\beta$ Decay of Hyperons. Ph.D. Thesis, University of California, Berkeley, CA, USA, 1957.

17. Gell-Mann, M. Isotopic Spin and New Unstable Particles. *Phys. Rev.* **1953**, *92*, 833.

18. Nakano, T.; Nishijima, K. Charge Independence for V-Particles. *Prog. Theor. Phys.* **1953**, *10*, 581.

19. Chamberlain, O.; Segrè, E.; Wiegand, C.; et al. Observation of Antiprotons. *Phys. Rev.* **1955**, *100*, 947.

20. Cork, B.; Lambertson, G.R.; Piccioni, O.; et al. Antineutrons Produced from Antiprotons in Charge-Exchange Collisions. *Phys. Rev.* **1956**, *104*, 1193.

21. Amaldi, E.; Castagnoli, C.; Cortini, G.; et al. Properties of Heavy Unstable Particles Produced by Cosmic Rays. *Nuovo Cimento* **1955**, *1*, 492.

22. Gatto, R. On the Vector Character of the Weak Coupling in the Decay $K^+ \rightarrow \mu^+ + \nu$. *Nuovo Cimento* **1956**, *3*, 468.

23. Grilli, G. *Maestri e Allievi Nella Fisica Italiana del Novecento*; La Goliardica Pavese: Pavia, Italy, 2008; Chapter 11, pp. 333–360.

24. Lee, T.D.; Yang, C.N. Question of Parity Conservation in Weak Interactions. *Phys. Rev.* **1956**, *104*, 254.

25. Gatto, R. Theory of Weak Interactions. *Nuovo Cimento* **1957**, *5*, 1024.

26. Gatto, R. Weak Interactions and $SU(3)$ Symmetry. *Nucl. Phys.* **1958**, *5*, 183.

27. Morpurgo, G.; Radicati, L.A.; Touschek, B. On the Magnetic Moment of the Neutron. In Proceedings of the 1954 Glasgow Conference on Nuclear and Meson Physics, IUPAP, New York, NY, USA, 13–17 July 1954.

28. Schwinger, J.S. On Gauge Invariance and Vacuum Polarization. *Phys. Rev.* **1951**, *82*, 914.

29. Bell, J.S. Time Reversal in Field Theory. *Proc. R. Soc. Lond. A* **1955**, *231*, 479.

30. Lüders, G. On the Equivalence of Invariance under Time Reversal and under Particle-Anti-Particle Conjugation for Relativistic Field Theories. *Det K. Dan. Vidensk. Selsk. Mat. Fys. Meddeleser* **1954**, *28*, 1.

31. Pauli, W. Exclusion Principle, Lorentz Group and Reflection of Space-Time and Charge. In *Niels Bohr and the Development of Physics*; McGraw-Hill Book Co.: New York, NY, USA, 1955; pp. 30–51.

32. Lüders, G.; Zumino, B. Some Consequences of TCP-Invariance. *Phys. Rev.* **1957**, *106*, 385.

33. Wu, C.S.; Ambler, E.; Hayward, R.W.; et al. Experimental Test of Parity Conservation in Beta Decay. *Phys. Rev.* **1957**, *105*, 1413.

34. Friedman, J.I.; Telegdi, V.L. Evidence for Parity Nonconservation in the Decay Chain $\pi^+ \rightarrow \mu^+ \rightarrow e^+$. *Phys. Rev.* **1957**, *106*, 1290.

35. Garwin, R.L.; Lederman, L.M.; Weinrich, M. Observations of the Failure of Conservation of Parity and Charge Conjugation in Meson Decays: the Magnetic Moment of the Free Muon. *Phys. Rev.* **1957**, *105*, 1415.

36. Cini, M.; Touschek, B. On the Theory of Synchrotron Radiation. *Nuovo Cimento* **1958**, *7*, 422. https://doi.org/10.1007/BF02747708.

37. Touschek, B. On the Radiation of a Relativistic Electron in a Circular Orbit. *Nuovo Cimento* **1958**, *8*, 181. https://doi.org/10.1007/BF02828864.

38. Touschek, B. On the Quantum Theory of Synchrotron Radiation. *Nuovo Cimento* **1957**, *5*, 754. https://doi.org/10.1007/BF02835605.

39. Pauli, W.; Touschek, B. On the Invariant Regularization in Relativistic Quantum Theory. *Nuovo Cimento* **1959**, *14*, 205.

40. Gatto, R.; Lüders, G. Charge Conjugation and the $\beta$-Decay of Hyperons. *Nuovo Cimento* **1958**, *7*, 806.

41. Marin, P. *Un Demi-Siècle D'accélérateurs de Particules*; Éditions du Dauphin: Paris, France, 2009.

42. Gatto, R. Electromagnetic Corrections to Weak Interactions. In Proceedings of the Aix-en-Provence International Conference on Elementary Particles, Gif-sur-Yvette, France, 14–20 September 1961; Volume 1, pp. 487–502.

43. Cabibbo, N.; Gatto, R. Electromagnetic Corrections to Weak Interactions: the $\beta$-Decay. *Phys. Rev. Lett.* **1960**, *4*, 313.

44. Cabibbo, N.; Gatto, R. Electromagnetic Corrections to Weak Interactions. *Phys. Rev.* **1961**, *124*, 1577.

45. Amman, F.; Bernardini, C.; Gatto, R.; et al. Proposal for a 1 GeV Electron-Positron Storage Ring. *Tech. Rep.* **1961**, LNF-61/5.

46. Mersits, U. Introduction. In *History of CERN. Launching the European Organization for Nuclear Research*; Hermann, A., Krige, J., Mersits, U., et al., Eds.; North-Holland: Amsterdam, The Netherlands, 1987; Volume 1, pp. 3–52.

47. Salvini, G. *L'elettrosincrotrone e i Laboratori di Frascati*; Nicola Zanichelli, Bologna, Italy, 1962.

48. Bonolis, L.; Bossi, F.; Pancheri, G. Bruno Touschek and the Art of Physics. *Nuovo Saggiatore* **2021**, *37*, 47.

49. Haïssinski, J. From AdA to LEP: Bruno Touschek and the Physics of Electron-Positron Colliders. *Springer Proc. Phys.* **2023**, *287*, 33.

50. Buccella, F. Personal Recollections of Bruno Touschek. In *Bruno Touschek 100 Years Memorial Symposium 2021*; Bonolis, L., Maiani, L., Pancheri, G., Eds.; Springer Nature: Cham, Switzerland, 2023; p. 301. https://doi.org/10.1007/978-3-031-23042-4_23.

51. Altarelli, G.; Buccella, F. Electromagnetic Corrections to Nonleptonic Hyperon Decays. *Nuovo Cimento* **1964**, *34*, 1337. https://doi.org/10.1007/BF02748859.

52. Bernardini, C.; Corazza, G.; Di Giugno, G.; et al. First Results from AdA (Anello di Accumulazione). *Nuovo Cimento* **1964**, *34*, 1473.

53. Margaritondo, G. Bruno Touschek: A Pioneer of Particle Accelerators. *Quad. Stor. Fis.* **2021**, *25*, 99.

54.  Di Castro, C. Reminiscences of Bruno Touschek. In *Bruno Touschek 100 Years Memorial Symposium 2021*; Bonolis, L., Maiani, L., Pancheri, G., Eds.; Springer Nature: Cham, Switzerland, 2023; pp. 303–304, https://doi.org/10.1007/978-3-031-23042-4_23.

55.  Rossi, G. Bruno Touschek and Statistical Mechanics. *Springer Proc. Phys.* **2023**, *287*, 45.

56.  Maiani, L.; Bonolis, L. Bruno Touschek: Particle Physicist and Father of the $e^+e^-$ Collider. *Eur. Phys. J. H* **2017**, *42*, 475. https://doi.org/10.1140/epjh/e2017-80052-8.

57.  Casalbuoni, R.; Dominici, D. Bruno Touschek and the Creation of AdA. *arXiv* **2018**, *arXiv:1810.06413*.

58.  Battimelli, G.; Buccella, F.; Napolitano, P. Bruno Touschek: Science and Life. *Quad. Stor. Fis.* **2019**, *22*, 145.

59.  Preparata, G. *Dai Quark ai Cristalli. Breve Storia di un Lungo Viaggio Dentro la Materia. Ediz. Ampliata*; Bibliopolis: Naples, Italy, 2020; ISBN 978-8870886658.

60.  Di Vecchia, P.; Greco, M. Electron-Positron Annihilation into Hadrons at High Energy. *Nuovo Cimento* **1967**, *50*, 319.

61.  Di Vecchia, P. Reminiscences of the Early Days of QCD. *Springer Proc. Phys.* **2023**, *287*, 239.

62.  Touschek, B.; Rossi, G. *Meccanica Statistica*; Boringhieri: Torino, Italy, 1970.

63.  Greco, M.; Rossi, G. A Phenomenological Approach to Regge-Cut Models. *Nuovo Cimento* **1967**, *50*, 168.

64.  Brown, L.M.; Calogero, F. Electromagnetic Structure of the Nucleon. *Phys. Rev. Lett.* **1960**, *4*, 315. https://doi.org/10.1103/PhysRevLett.4.315.

65.  Calogero, F.; De Stefano, M. B. Electromagnetic Structure of the Nucleon. *Phys. Rev.* **1966**, *146*, 1195.

66.  Gatto, R. Theoretical Aspects of Colliding Beam Experiments. In *International School of Physics "Enrico Fermi" Course XXXIII*; Springer: Berlin/Heidelberg, Germany, 1965; pp. 106–137, ISBN 978-3-540-37142-7. https://doi.org/10.1007/BFb0045445.

67.  Etim, G.; Pancheri, G.; Touschek, B. The $e^+e^-$ Annihilation Cross Section. *Nuovo Cimento B* **1967**, *51*, 276.

68.  Glashow, S. L.; Iliopoulos, J.; Maiani, L. Weak Interactions with Lepton-Hadron Symmetry. *Phys. Rev. D* **1970**, *2*, 1285.

69.  Heisenberg, W. Theory of Alternating-Gradient Synchrotrons. In Proceedings of the Conference on the Theory and Design of an Alternating-Gradient Proton Synchrotron, Geneva, Switzerland, 26–28 October 1953; pp. 179–180.

70.  Cabibbo, N.; Parisi, G.; Testa, M. Light-Cone and Short-Distance Behavior in Perturbation Theory. *Lett. Nuovo Cimento* **1970**, *4*, 35.

71.  Ferrara, S.; Grillo, A.F.; Parisi, G.; et al. Conformal Covariance of the Operator Product Expansion. *Nucl. Phys. B* **1972**, *49*, 77; Erratum: *Nucl. Phys. B* **1973**, *53*, 643.

72.  Gatto, R.; Preparata, G. Conformal Covariance and the Operator Product Expansion. *Nucl. Phys. B* **1973**, *67*, 362.

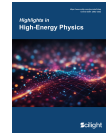

*Review*

# From ADONE's Multi-Hadron Production to the $J/\Psi$ Discovery

## Mario Greco

Department of Mathematics and Physics and INFN, University of Roma Tre, 00154 Rome, Italy; mario.greco@roma3.infn.it

**Abstract:** The physics at Frascati in the years 60's–70's is reviewed together with the $J/\Psi$ discovery.

**Keywords:** ADONE; multi-hadron production; $J/\Psi$ discovery

I arrived at the Frascati National Laboratory in January 1965, joining the ADONE Group, and stayed there for 25 years. The first electron-positron collider AdA had been successfully constructed in one year only after the famous seminar given by Bruno Touschek on March 7th, 1960. Taken to Orsay to improve the injection rate obtainable with the linear accelerator, AdA completed its cycle there [1] and the last publication [2] was at the end of 1964. On the contrary the ADONE adventure took a much longer time. The first draft of ADONE and its physics programme dated November 1960 and was written in the Bruno Touschek's notebook, where he listed a number of physical processes which could be measured, including the production of a proton-antiproton pair. The maximum total energy of the machine was then set to 3 GeV [3]. ADONE's approval was in 1963 and the first collisions were finally obtained in 1969 with the first generation experiments.

The radiative corrections for ADONE experiments were the main problem Bruno Touschek had in his mind, because of their importance in the new electron-positron collider, and indeed they had been the main field of theoretical activity of the new theory group. In addition the calculation of the double bremsstrahlung process as a monitor reaction for the luminosity was also performed using two different approaches [4]. In more detail the infrared corrections to be applied in an electron-positron collider experiment were first obtained with the help of the Bloch-Nordsieck theorem, using a statistical approach to define the probability for the four-momentum to be carried away by the electromagnetic radiation [5]. Alternatively, from a field theoretical point of view, a new finite S-Matrix was defined using a realistic definition of initial and final states, by "dressing" the charged particle's states with a phase containing the electromagnetic operator in the exponential, in order to create an undetermined number of soft photons with a Bloch-Nordsiek spectrum. The new S-Matrix was explicitly shown to be equivalent to all orders in $\alpha$ to the conventional perturbative result [6]. In other words this approach corresponds to the introduction of the idea of coherent states in QED. It is extraordinary that both approaches led exactly to the same result for the soft radiative effect, namely that the observed cross section can be written as

$$d\sigma = \frac{1}{\gamma^\beta \Gamma(1+\beta)} \left(\frac{\Delta\omega}{E}\right)^\beta d\sigma_E$$

where 2E is the total c.m. energy, $(\Delta\omega/E)$ is the relative energy resolution of the experiment, $\gamma$ is the Euler constant, $d\sigma_E$ differs from the lowest cross section $d\sigma_0$ by finite terms of $\mathcal{O}(\alpha)$, and $\beta$ is the famous Bond-factor, so named by Bruno because its numerical value at ADONE was 0.07, and more generally $\beta = \frac{4}{\pi}[\ln(2E/m) - 1/2]$.

The coherent states approach played a major role later in the description of the radiative effects in case of the production of the $J/\Psi$ and of the Z boson, as we'll discuss later. Also that was first extended to QCD in the late '70 [7] and studied further [8,9]. Many and important QCD results concerning exponentiation, resummation formulae, K-factors, transverse momentum distributions of DY pairs, W/Z and H production, have their roots in Bruno Touschek ideas on the exponentiation and resummation formulae in QED. We give here a reference list [10–16] of the papers that were written later in the Frascati-Rome area and certainly inspired by his ideas.

Let's discuss now the theoretical framework and the expectations concerning ADONE and the experimental results. At the time, the Vector Meson Dominance (VMD) model of J.J. Sakurai [17] was quite successful in describing the electromagnetic interaction of hadrons as being mediated by the vector mesons $\rho, \omega$ and $\phi$. That led

T.D. Lee, N. Kroll and B. Zumino to try to give a field theoretical approach to VMD [18]. In this framework the total hadronic annihilation cross section was expected to behave at large $s$ as

$$\sigma(s) = \left(\frac{1}{s}\right)^2$$

However departures from the simple VMD model were observed in some radiative decays of mesons and the possible existence of new vector mesons was suggested by A. Bramon and myself [19], as also predicted by dual resonance models and the Veneziano model [20]. On the other hand, the results of Deep Inelastic Scattering (DIS) experiments at SLAC, with the idea of Bjorken scaling and the Feynman parton model were naturally leading to

$$\sigma(s) = \frac{1}{s}$$

and indeed N. Cabibbo, M. Testa and G. Parisi [21] suggested that the ratio R of the hadronic to the point-like cross section would asymptotically behave as

$$R = \frac{\sigma_{had}(s)}{\sigma_{\mu\mu}} \to \sum_i Q_i^2$$

where the sum extends to all spin 1/2 elementary constituents, neglecting scalars.

As it's well known, the results of all experiments, namely the MEA Group [22], the $\gamma\gamma$ Group [23], the $\mu\pi$ Group [24], and the Bologna-CERN-Frascati Collaboration [25] showed a clear evidence of a large multihadron production with $R \approx 2$, pointing to the coloured quark model. On the other hand they also indicated evidence for a new vector meson $\rho'(1.6)$ with a dominant decay in four charged pions, which had been suggested by A. Bramon and myself [26]. The experimental data are shown in Figure 1, taken from a review paper of C. Bernardini and L. Paoluzi [27].

The ADONE results together with the request of scaling, both in DIS and $e^+e^-$ annihilation, and the Veneziano's duality idea led us to propose a new scheme where the asymptotic scaling is reached through the low energy resonances mediating the asymptotic behaviour [28]. Thus the value of R is also connected to the low energy resonances's couplings, and in the 3 coloured quark model, led us to the prediction $R \approx 2.4$. This scheme—named duality in $e^+e^-$ annihilation - was immediately shared by J. J. Sakurai [29]. Later J. Bell and collaborators also studied a potential model where the bound states could be solved analytically and verified this idea of duality [30,31]. In addition a set of $e^+e^-$ duality sum rules was derived from the canonical trace anomaly of the energy momentum tensor by E. Etim and myself [32], much earlier than the Russian sum rules of M. A. Shifman et al. [33]. The lowest order sum rule gives

$$\int_{s_0}^{\bar{s}} ds \left( \text{Im}\Pi(s) - \frac{\alpha R}{3} \right) = 0$$

where $\text{Im}\Pi(s) = s\sigma_{had}(s)/4\pi\alpha$ and clearly it relates the asymptotic value of $R$ to the low energy behaviour. One has to stress here that QCD wasn't there yet at that time. The average value of $R$ in particular, in the ADONE region, is about 2.4 as it was also confirmed by the SPEAR data at the c.m. energy just below the $J/\Psi$. On the other hand, for larger c.m. energies, the SPEAR data also showed an increasing behaviour of $R$ suggesting the presence of a new component in $\text{Im}\Pi(s)$ with threshold at about 3 GeV. That was our conclusion in ref. [32], which dates a few weeks before the $J/\Psi$ discovery.

Let's consider in detail now the $J/\Psi$ discovery, or what was called the November Revolution, from a Frascati point of perspective. As it's well known, on November 11th 1974, B. Richter and S.C.C. Ting jointly announced in Stanford the discovery of the $J/\Psi$ both at SLAC and at Brookhaven [34,35]. I had the terrific chance of arriving at SLAC the day after, with an invitation by Sid Drell to give a seminar on our duality works, on the way for a visit of a few weeks to Mexico City. Sid had been on a sabbatical leave the year before at Frascati and Rome, so we knew each other pretty well. There was a great excitement in the theory discussion room and once I was informed of the details of the discovery I realized immediately that the $J/\Psi$ could be seen possibly also at ADONE. I asked Sid to let me call Frascati, and from his confidential office—he was scientific advisor of the President of United States—I gave to Giorgio Bellettini, the director of the Laboratory, the exact position of the $J/\Psi$. The night after, the resonance was also observed at Frascati. Giorgio Bellettini communicated the results to the Phys. Rev. Lett. over the telephone and the paper was published [36] in the same issue of the American results.

As far as the theoretical interpretation of the $J/\Psi$ is concerned, hundreds of papers had been published on the argument, as it's well known. In a recent review article on this subject, Alvaro De Rujula has reported [37] the papers published on the first issue of Phys. Rev. Letters after the discovery, as shown in Figure 2.

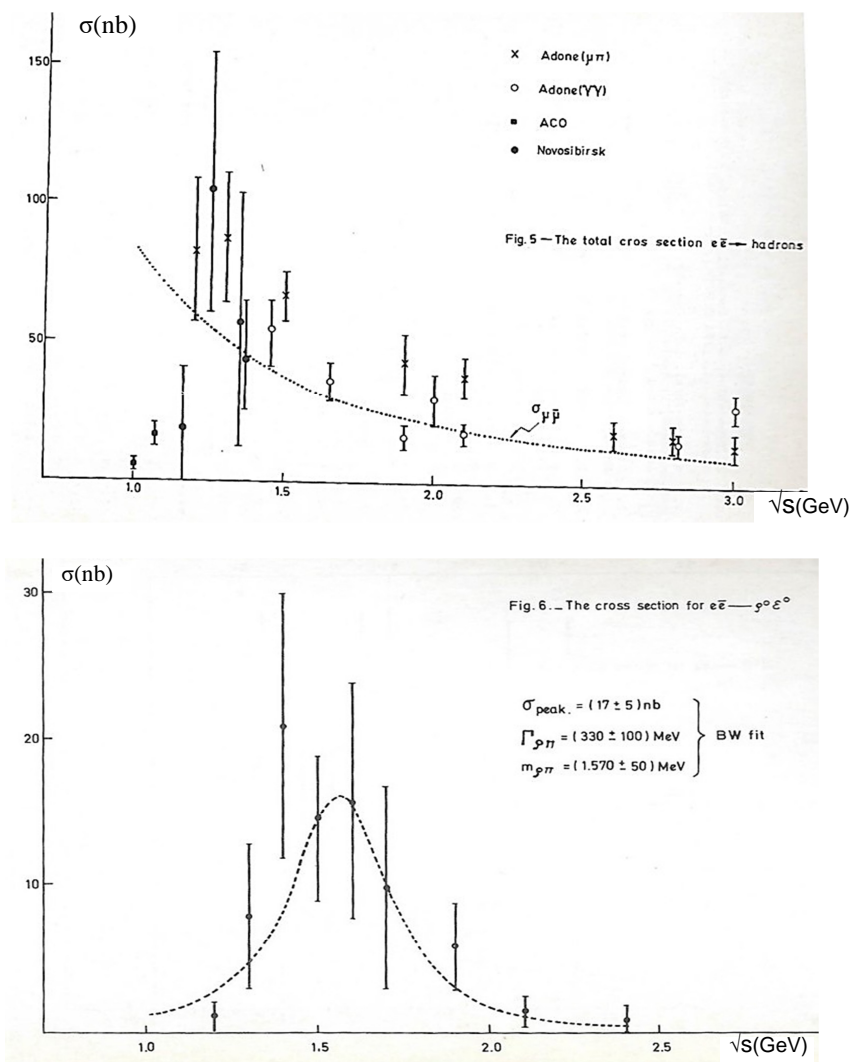

**Figure 1.** ADONE experimental results, from ref. [27] .

Of course only two of them, both from Harvard, had the right interpretation: by T. Applequist and H.D. Politzer [38], who related the reason for the very narrow width of the $J/\Psi$ to the asymptotic freedom of QCD just discovered, and by A. De Rujula and S.L. Glashow [39] because of the GIM mechanism and charm suggested earlier [40]. On the other hand Alvaro is also quoting two papers published on Lett. Nuovo Cimento, by G. Altarelli, N. Cabibbo, L. Maiani, R. Petronzio and G. Parisi [41], who had the wrong interpretation in favour of the weak boson, and C. Dominguez and myself [42], written in Mexico City a few days after I had left Stanford, who also had the right interpretation in favour of charm. Indeed I had taken with me the detailed data of the $J/\Psi$ and as soon as I found the news on a local newspaper of the subsequent discovery of the $\Psi'$, by using the duality ideas discussed above, we arrived at the conclusion that the new series of resonances was indeed composed by $c$-$\bar{c}$ pairs, and the value of $R$ would reach about 3.7 with the new charm contribution. On the other hand the limits of our paper were both a very naive assumption on the $\Psi's$ mass spectrum and no understanding of the smallness of the $J/\Psi$ width.

Forty years later, in December 2013, after Sam Ting had mentioned it in his "Bruno Touschek Memorial Lecture" in Frascati, I discovered in the archives at the CERN library that our preprint was dated 18 November, preceding by a few days those of Applequist and Politzer and of De Rujula and Glashow.

Now, fifty years later, by comparing the value of R from the Particle Data Group with all the experimental information, as shown in Figure 3, with the theoretical prediction of QCD with $\mathcal{O}(\alpha_s)$, $\mathcal{O}(\alpha_s^2)$ and $\mathcal{O}(\alpha_s^3)$ corrections included—as indicated by the continuous red line—one easily concludes that our duality predictions were well satisfied, and indeed in very good agreement with QCD.

## Theoretical interpretation of J/Ψ

**Are the New Particles Baryon-Antibaryon Nuclei?**
Alfred S. Goldhaber and Maurice Goldhaber

**Interpretation of a Narrow Resonance in e+ e- Annihilation**
Julian Schwinger

**Possible Explanation of the New Resonance in e+ e- Annihilation**
S. Borchardt, V. S. Mathur, and S. Okubo

**Model with Three Charmed Quarks** R. Michael Barnett

**Heavy Quarks and e+ e- Annihilation** Thomas Appelquist and H. David Politzer

**Is Bound Charm Found?** A. De Rújula and S. L. Glashow

**Possible Interactions of the J Particle**
H. T. Nieh, Tai Tsun Wu, and Chen Ning Yang

**Is the 3104-MeV Vector Meson the psi - Charm or the W0?**
G. Altarelli, N. Cabibbo, R. Petronzio, L. Maiani, G. Parisi

**Charm, EVDM and Narrow Resonances in** $e^+e^-$ **Annihilation**
C. A. Dominguez and M. Greco

Fig. 15. Immediate interpretations of the $J/\psi$, with their titles. PRL is Phys. Rev. Lett. **4**, Jan. 6th, 1975. The last two papers[88,89] are in Lett. Nuovo Cim.

**Figure 2.** The immediate interpretation of $J/\Psi$ from ref. [37] . PRL is Phys. Rev. Lett. 34, Jan. 16th, 1975.

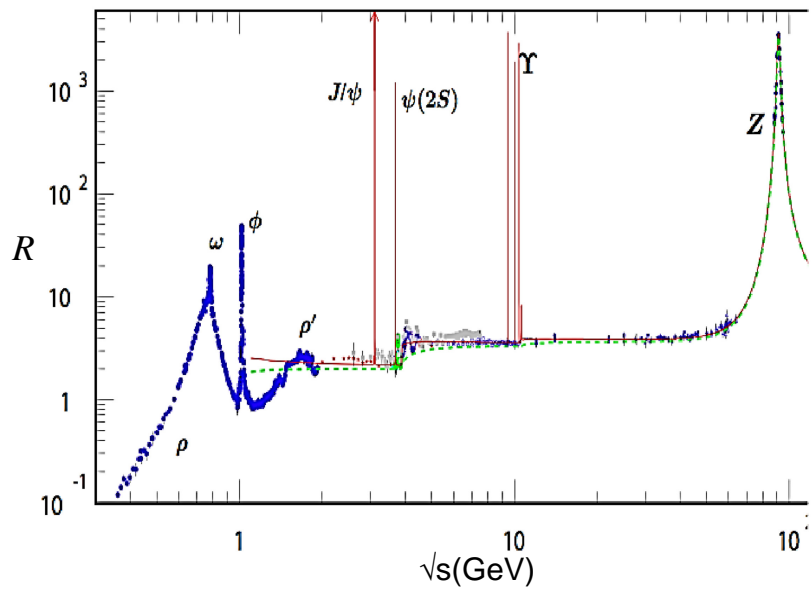

**Figure 3.** The ratio $R = \sigma_{had}(s)/\sigma_{\mu\mu}(s)$ as a function of $\sqrt{s}$, from Particle Data Group.

The problem of the radiative corrections to the $J/\Psi$ line-shape, because of the very narrow width involved, showed the crucial role played by the theoretical ideas of the early times on the infrared behaviour of QED, namely the exponentiation results and the approach of the coherent states. The detailed analysis by G. Pancheri, Y. Srivastava and myself [43,44], showed that the main infrared correction factor was of the type

$$C_{infra} \approx \left( \frac{\Gamma}{M} \right)^{\beta}$$

where $\Gamma$ and $M$ are the width and the mass of the resonance respectively, and $\beta$ the Bond-factor. The detailed result of this analysis, compared with the SLAC and Frascati data, is shown in Figure 4.

When we showed our result to Bruno Touschek, he immediately commented that the experimental errors of the Frascati data had been clearly overestimated. I should also add that the SLAC analysis of their data had been based on a paper by D.R. Yennie [45] that contained a wrong dependence on the width $\Gamma$ and the parameter $\sigma$ of the Gaussian energy distribution of the beams, with a resulting difference with respect to our analysis on the leptonic width of the $J/\Psi$. It was only in 1987, in the occasion of the first La Thuile meeting, that I convinced Burt Richter to update the SLAC radiative corrections codes with the right formulae, in perspective of the coming data on the Z boson physics at SLC. Indeed the re-analysis of all charm data at SLAC caused a change of many properties of the charm particles in the Particle Data Group in 1988, including the leptonic width of the $J/\Psi$, in better agreement with our estimate.

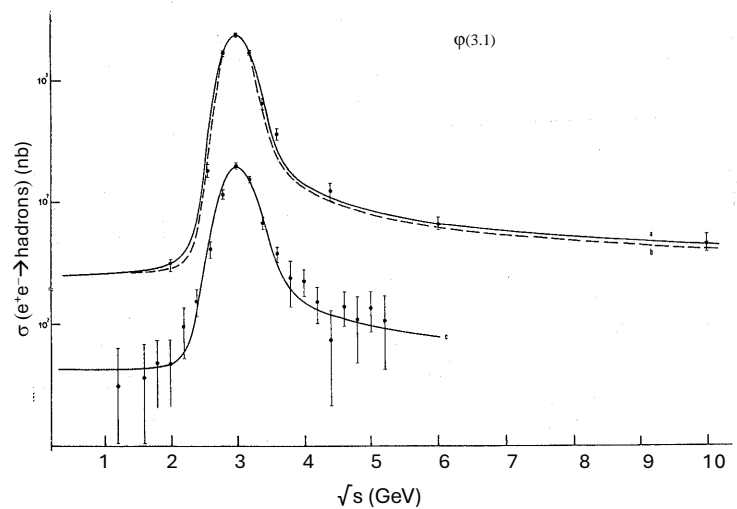

**Figure 4.** Experimental results for $J/\Psi$ production in $e^+e^-$ annihilation. The data are from SPEAR and ADONE (see text). The full lines refer to the theoretical analysis including radiative corrections of Ref. [43,44].

The above treatment of the radiative corrections for the $J/\Psi$ production was extended a few years later to study the radiative effects in the case of Z production at LEP/LHC [46]. Our work was the first study to all orders in the infrared corrections, with a complete evaluation of all finite terms of $\mathcal{O}(\alpha)$ and at the base of the later analyses of all experiments. Very recently, within the general discussion on the possibility of constructing a muon collider Higgs factory to study with great care on resonance the properties of the H, the line-shape has been studied [47,48] in the same way, as in the old times. As a result we have shown that the radiative effects put very stringent bounds on the energy spread of the beams, and make this project very tough.

To conclude, from AdA/ADONE to LEP/LHC and the future colliders, the seminal idea of Bruno Touschek has contributed with so many discoveries to the progress of the Standard Model. This certainly constitutes his main legacy. In addition some of his suggestions have also strongly contributed to the precision assessment of the theory. Finally duality ideas have been a very powerful tool for the interpretation of e+e- colliding beams results and in particle phenomenology before the advent of QCD.

**Conflicts of Interest**

The author declares no conflict of interest.

**References**

1. Bernardini, C. AdA: The First Electron-Positron Collider. *Phys. Perspect.* **2004**, *6*, 156.
2. Bernardini, C.; Corazza, G.; Giugno, G.D.; et al. Measurements of the rate of interaction between stored electrons and positrons. *Nuovo C.* **1964**, *34*, 1473.
3. Pancheri, G. *Bruno Touschek's Extraordinary Journey. From Death Rays to Antimatter*; Springer Biographies: Berlin/Heidelberg, Germany, 2022.
4. Vecchia, P.D.; Greco, M. Double photon emission in $e^\pm e^-$ collisions. *Nuovo C.* **1967**, *50*, 319.
5. Etim, E.; Pancheri, G.; Touschek, B. The infra-red radiative corrections for colliding beam (electrons and positrons) experiments. Nuovo C. **1967**, *51B*, 276.
6. Greco, M.; Rossi, G. A note on the infra-red divergence. *Nuovo C.* **1967**, *50A*, 168.
7. Greco, M.; Palumbo, F.; Pancheri, G.; et al. Coherent state approach to the infra-red behaviour of non-abelian gauge theories. *Phys. Lett.* **1978**, *77B*, 282.
8. Curci, G.; Greco, M. Mass Singularities and Coherent States in Gauge Theories. *Phys. Lett.* **1978**, *79B*, 406.
9. Catani, S.; Ciafaloni, M.; Marchesini, G. Asymptotic coherent states and colour screening. *Phys. Lett. B* **1986**, *168*, 284.
10. Parisi, G. Summing large perturbative corrections in QCD. *Phys. Lett. B* **1980**, *90*, 295.

11. Curci, G.; Greco, M. Large infra-red corrections in QCD processes. *Phys. Lett. B* **1980**, *92*, 175.

12. Pancheri, G.; Srivastava, Y. Energy-Momentum Distribution in $e^+e^-$ Annihilation. *Phys. Rev. Lett.* **1979**, *43*, 11.

13. Curci, G.; Greco, M.; Srivastava, Y. Coherent Quark-Gluon Jets. *Phys. Rev. Lett.* **1979**, *43*, 834.

14. Parisi, G.; Petronzio, R. Small transverse momentum distributions in hard processes. *Nucl. Phys. B* **1979**, *154*, 427.

15. Curci, G.; Greco, M.; Srivastava, Y. QCD Jets From Coherent States. *Nucl. Phys. B* **1979**, *159*, 451.

16. Altarelli, G.; Ellis, K.; Greco, M.; et al. Vector boson production at colliders: A theoretical reappraisal. *Nucl. Phys. B* **1984**, *246*, 12.

17. Sakurai, J.J. Vector-Meson Dominance and High-Energy Electron-Proton Inelastic Scattering. *Phys. Rev. Lett.* **1969**, *22*, 981.

18. Kroll, N.M.; Lee, T.D.; Zumino, B. Neutral Vector Mesons and the Hadronic Electromagnetic Current. *Phys. Rev.* **1967**, *157*, 1376.

19. Bramon, A.; Greco, M. Hadron production in $e^+e^-$ collisions and the existence of new vector mesons. *Lett. Nuovo C.* **1971**, *1*, 739.

20. Veneziano, G. Construction of a crossing-simmetric, Regge-behaved amplitude for linearly rising trajectories. *Nuovo C. A* **1968**, *57*, 190.

21. Cabibbo, N.; Parisi, G.; Testa, M. Deep inelastic scattering and the nature of partons. *Lett. Nuovo C.* **1970**, *4*, 35.

22. Bartoli, B.; Coluzzi, B.; Felicetti, F.; et al. Multiple particle production from $e^+e^-$ interactions at C.M. energies between 1.6 and 2 GeV. *Nuovo C. A* **1970**, *70*, 615.

23. Bacci, C.; Penso, G.; Salvini, G.; et al. Multihadronic cross sections from $e^+e^-$ annihilation at C.M. energies between 1.4 and 2.4 GeV. *Phys. Lett. B* **1972**, *38*, 551.

24. Ceradini, F.; Santonico, R.; Conversi, M.; et al., Multiplicity in hadron production by $e^+e^-$ colliding beams. *Phys. Lett. B* **1972**, *42*, 501.

25. Alles-Borelli, V.; Bernardini, M.; Bollini, D.; et al. $e^+e^-$ annihilation into two hadrons in the energy interval 1400–2400 MeV. *Phys. Lett. B* **1972**, *40*, 433.

26. Bramon, A.; Greco, M. The reaction $e^+e^- \rightarrow \pi^+\pi^-\pi^+\pi^+$ and the $\sigma'$-meson. *Lett. Nuovo C.* **1972**, *3*, 693.

27. Bernardini, C.; Paoluzi, L. In Proceedings of the 2nd International Winter Meeting on Fundamental Physics, Huesca, Spain, February 1974.

28. Bramon, A.; Etim, E.; Greco, M. A vector meson dominance approach to scale invariance. *Phys. Lett. B* **1972**, *41*, 609.

29. Sakurai, J.J. Duality in $e^+ + e^- \rightarrow$ hadrons? *Phys.Lett. B* **1973**, *46*, 207.

30. Bell, J.S.; Bertlmann, R. Testing $Q^2$ duality with non-relativistic potentials. *Z. Phys. C* **1980**, *4*, 11.

31. Bell, J.S.; Bertlmann, R. Magic moments. *Nucl. Phys. B* **1981**, *177*, 218.

32. Etim, E.; Greco, M. Duality sum rules in $e^+ + e^-$ annihilation from canonical trace anomalies. *Lett. Nuovo C.* **1975**, *12*, 91.

33. Shifman, M.A.; Vainshtein, A.I.; Zakharov, V.I. QCD and resonance physics. theoretical foundations. *Nucl. Phys. B* **1979**, *147*, 385.

34. Aubert, J.J.; Becker, U.; Biggs, P.J.; et al. Experimental Observation of a Heavy Particle $J$. *Phys. Rev. Lett.* **1974**, *33*, 1404.

35. Augustin, J.E.; Boyarski, A.M.; Breidenbach, M.; et al. Discovery of a Narrow Resonance in $e^+ + e^-$ Annihilation. *Phys. Rev. Lett.* **1974**, *33*, 1406.

36. Bacci, C.; Celio, R.B.; Berna-Rodini, M.; et al. Preliminary Result of Frascati (ADONE) on the Nature of a New 3.1-GeV Particle Produced in $e^+ + e^-$ Annihilation. *Phys. Rev. Lett.* **1974**, *33*, 1408.

37. De Rujula, A. QCD, from its inception to its stubbornly unsolved problems. *Int. J. Mod. Phys. A* **2019**, *34*, 32.

38. Applelquist, T.; Politzer, H.D. Heavy Quarks and $e^+e^-$ Annihilation. *Phys. Rev. Lett.* **1975**, *34*, 43.

39. De Rujula, A.; Glashow, S.L.; Is Bound Charm Found? *Phys. Rev. Lett.* **1975**, *34*, 46.

40. Glashow, S.L.; Iliopoulos, J.; Maiani, L. Weak Interactions with Lepton-Hadron Symmetry. *Phys. Rev. D* **1970**, *2*, 1285.

41. Altarelli, G.; Cabibbo, N.; Maiani, L.; Parisi, G.; et al. Is the 3104 MeV vector meson the $\varphi_c$ or the $W_0$? *Lett. Nuovo C.* **1974**, *11*, 14.

42. Dominguez, C.; Greco, M. Charm, EVDM and narrow resonances in $e^+e^-$ annihilation. *Lett. Nuovo C.* **1975**, *12*, 439.

43. Greco, M.; Pancheri-Srivastava, G.; Srivastava, Y. Radiative effects for resonances with applications to colliding beam processes. *Phys. Lett. B* **1975**, *56*, 367.

44. Greco, M.; Pancheri-Srivastava, G.; Srivastava, Y. Radiative corrections for colliding beam resonances. *Nucl. Phys. B* **1975**, *101*, 234.

45. Yennie, D.N. Comment on Radiative Corrections to $e^+e^- \rightarrow \psi$(3105). *Phys. Rev. Lett.* **1975**, *34*, 239.

46. Greco, M.; Pancheri, G.; Srivastava, Y. Radiative corrections to $e^+e^- \rightarrow \mu^+\mu^-$ around the $Z_0$. *Nucl. Phys. B* **1980**, *171*, 118.

47. Greco, M. On the study of the Higgs properties at a muon collider. *Mod. Phys. Lett. A* **2015**, *30*, 1530031.

48. Greco, M.; Han, T.; Liu, Z. ISR effects for resonant Higgs production at future lepton colliders. *Phys. Lett. B* **2016**, *763*, 409.

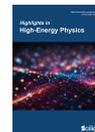

*Review*
# From Bjorken Scaling to Scaling Violations

Giorgio Parisi [1,2,3]

[1] Dipartimento di Fisica, Università degli Studi di Roma "La Sapienza", 00185 Roma, Italy; giorgio.parisi@gmail.com
[2] Istituto Nazionale di Fisica Nucleare, Sezione di Roma I, 00185 Roma, Italy
[3] Institute of Nanotechnology (NANOTEC)-CNR, Rome Unit, 00185 Roma, Italy

**Abstract:** This paper traces the historical and conceptual journey from Bjorken scaling to the discovery of scaling violations in deep inelastic scattering, culminating in the development of Quantum Chromodynamics (QCD). Beginning with the challenges faced by early strong interaction theories in the 1950s, we explore the emergence of agnostic approaches such as the bootstrap philosophy and current algebra, which sought to describe hadronic phenomena without relying on specific field theories. The pivotal role of experimental results from SLAC in the late 1960s is highlighted, leading to Bjorken's proposal of scaling in deep inelastic scattering and Feynman's parton model. We then delve into the theoretical breakthroughs of the 1970s, including Wilson's operator product expansion and the renormalization group, which provided the framework for understanding scaling violations. The discovery of asymptotic freedom in non-Abelian gauge theories by Gross, Wilczek, and Politzer marked a turning point, establishing QCD as the theory of strong interactions. Finally, we discuss the formulation of the Altarelli-Parisi equations, which elegantly describe the evolution of parton distribution functions and scaling violations, and their profound impact on the study of hard processes in particle physics. This paper not only recounts the key developments but also reflects on the interplay between theory and experiment that drove the field forward.

**Keywords:** Bjorken scaling; scaling violations; deep inelastic scattering; QCD; Altarelli-Parisi equation

## 1. Introduction

It is a well-known fact that at the beginning of '50 people were losing hope of a microscopic theory of strong interactions. In 1953, after spending nearly one year with his students on some computation in perturbative pion nucleon theory Dyson went to Fermi to ask an opinion [1]; Fermi said: "With the pseudoscalar meson theory there is no physical picture, and the forces are so strong that nothing converges. To reach your calculated results, you had to introduce arbitrary cut-off procedures that are not based either on solid physics or solid mathematics".

In desperation I asked Fermi whether he was not impressed by the agreement between our calculated numbers and his measured numbers. He replied, "How many arbitrary parameters did you use for your calculations?" I thought for a moment about our cut-off procedures and said, "Four." He said, "I remember my friend Johnny von Neumann used to say, with four parameters I can fit an elephant, and with five I can make him wiggle his trunk".

The reason was clear as stressed by Fermi: the coupling constant was so large and one could not compute more than the first and second order in the perturbative expansion. The discovery of many resonances (e.g., the $\Delta$) made clear that the situation was far from a perturbative framework. Moreover, nothing indicated that the nucleon and the pion were more fundamental than other particles.

There was no convincing suggestion for having a particular type of field theory. Moreover, if a wizard could tell which was the basic field theory, such knowledge would be considered useless as far as there was no way to exploit it: no reliable techniques existed beyond perturbation theory. As remarked by David Gross until 1973 it was not thought proper to use field theories without apologies [2].

This paper aims to recall the various steps that led to the resurrection of Lagrangian field theories.

## 1.1. Agnostic Theories

In the absence of a reliable theory, people started to study the consequences of a generic field theory without committing to a particular one. Many different keywords were used: symmetry groups, axiomatic field theory dispersion relations, S-matrix, Regge poles, superconvergence sum rules, and bootstrap, but there was no space for field theories with Lagrangians, with well-defined equations of motion.

In one case, there was a well-defined mathematical setup where one could deduce rigorous results. The basic idea was that some fields act as a creation operator of a given particle; these fields may not be fundamental they could be composite. In other words, for a particle of type $\alpha$ there is an interpolating field $\psi_\alpha(x)$ such that

$$\langle 0|\psi_\alpha(x)|\alpha\rangle \neq 0\,. \tag{1}$$

The fields are local, i.e., the commutator (or the anticommutators) are zero at space-like distances, as it should be in a Lagrangian field theory.

This approach was very successful.

- One proved that the CTP had to be exact [3].
- The Lehmann-Symanzik and Zimmermann (LSZ) formalism [4] could be used to extract the S matrix from the vacuum expectation of the product of the interpolating field.
- Mostly surprising in some kinematical range it was possible to prove the validity of the dispersion relation that relates the imaginary and real part of the scattering amplitudes [5].

Around 1960, Geoffrey Chew introduced the bootstrap philosophy, which posited that there were no elementary constituents; all particles were considered equally fundamental. According to bootstrap theory, every particle is composed of all other particles; there is a democracy among elementary particles, and none is more fundamental than any other. The age-old search for the constituent elements of matter had come to an end; there were no longer constituent elements of matter, only relationships between the various particles. It was an idea that proved to be very successful.

A book on the bootstrap approach even suggested that a detailed knowledge of quantum field theory could hinder the adoption of new ideas. This stance was somewhat paradoxical, given that dispersion relations—a cornerstone of the bootstrap—were originally derived within the context of local field theory. Ironically, the bootstrap approach played a pivotal role in the emergence of string theories, which are among the most sophisticated quantum field theories.

Murray Gell-Mann himself was not taking quarks seriously: he was considering them more as a mathematical model to implement $SU(3)$ symmetry. In 1964, he wrote [6] *We use the method of abstraction from a Lagrangian field theory model. In other words, we construct a mathematical theory of the strongly interacting particles, which may or may not have anything to do with reality, find suitable algebraic relations that hold in the model, postulate their validity, and then throw away the model. We compare this process to a method sometimes employed in French cuisine: a piece of pheasant meat is cooked between two slices of veal, which are then discarded.*

The absence of elementary point-like objects suggested that hadrons were extremely soft. This viewpoint was confirmed by the very fast decay of the proton form factor and by the exponential suppression of particle productions at large momentum transfer. Hagedorn theory [7] predicted a maximum temperature somewhat less than 200 MeV: also a very energetic hadronic fireball would emit hadrons of no more than a few hundredth MeV. The exponentially arising spectrum was later explained as the effect of the transition that leads to the creation of a quark-gluon plasma [8].

## 1.2. There Was Also a Different Viewpoint

Electromagnetism, Fermi interactions, and the V-A theories for weak interaction were based on local currents and quantum field theory. The purely leptonic electro-weak decays were described by a 4-Fermions field theory that was non-renormalizable; however, there was some hope that the introduction of heavy vector Bosons could make the theory renormalizable.

A hadronic current mediated the semi-leptonic weak interaction, and the resulting current algebra (based on local commutators) was crucial to normalizing the weak interaction vertices, which played a fundamental role in 1963 Cabibbo's theory of weak interaction [9]; indeed the normalization of the weak current was possible only after the identification of the current inside the generators of the symmetry group $SU(3)$. An angle $\theta$ among the currents was previously suggested in 1960 by Gelmann and Levy [10]: unfortunately, they have not introduced a symmetry group and concluded that There is, of course, a renormalization factor for that decay, so we cannot be sure that the low rate really fits in with such a picture. The group $SU(3)$ was introduced in 1961: the identification of the weak vector current with the current that generates the $SU(3)$ group was impossible at the time of [10].

The quantum field theory approach to physics was strongly pushed in Europe: there were many collaborations among different scientific institutions that were later formalized in the Triangular Meetings (Paris-Rome-Utrecht).

## 2. Crucial Steps to Bjorken Scaling

The path to the formulation of Bjorken scaling involved numerous pivotal developments. A key aspect was understanding the joint properties of the electromagnetic and weak currents, as well as exploring the consequences of their local commutators, particularly concerning axial current normalization.

This period saw a decade of intense research. Below, I highlight some of the most significant contributions:

- (1958) The discovery of the V-A theory for $\Delta S = 0$ semileptonic transitions (Feynman, Gell-Mann) [11]. The currents in the $\Delta S = 0$ sector resembled those of the leptonic sector.
- (1963) Cabibbo's theory [9] extended this framework to $\Delta S = 1$ semileptonic transitions, identifying the relevant currents in this sector.
- (1964) Ademollo and Gatto [12] proved that, at first order in $SU(3)$ symmetry breaking, the matrix elements of the vector currents remain unnormalized. This result validated a key approximation in Cabibbo's theory.
- (1964) Gell-Mann [6] introduced the $SU(3) \times SU(3)$ symmetry, generalizing the $SU(2) \times SU(2)$ symmetry proposed by Güraly and Radicati [13].
- (1965) Dashen and Gell-Mann [14] emphasized the importance of local current commutators, leading to the development of current algebra. While current algebra arises naturally in Lagrangian theories, no realistic Lagrangian framework existed at the time.
- (1965) In a seminal paper, Fubini, Furlan, and Rossetti [15] demonstrated that local commutators imply sum rules. These sum rules simplify in the infinite momentum frame, a concept introduced in this work. The infinite momentum frame later played a crucial role in developments such as the parton model.
- (1966) Bjorken [16] gave a comprehensive analysis of the implications of current algebra. The following year, he published a review paper [17] with the highly suggestive title: *Current Algebra at Small Distances*.
- (1968) Callan and Gross derived a sum rule for the cross-section in deep inelastic scattering [18], as a direct consequence of current commutators. This formula was later generalized by Cornwall and Norton [19], who derived additional sum rules.
- (1969) Bjorken proposed the concept of Bjorken scaling for deep inelastic scattering, building on earlier sum rules [20]. This proposal will be discussed in detail in the following section.
- (1969) In a groundbreaking paper, Feynman introduced the parton model [21]. Technically, Feynman's approach mirrored field theory, positing the existence of fundamental objects. However, he was making the implicit (and impossible) assumption that the interactions were super-normalizable, eliminating divergences. The framework was formulated in the infinite momentum frame, making it more tractable. The fundamental fields were unknown, but some consequences could still be explored. This approach provided a simplified framework, free from the complexities of renormalization and logarithmic divergences that Feynman himself had introduced in electrodynamics.
- (1969) Shortly after Feynman's proposal, Bjorken and Paschos outlined the implications of partons in a paper titled *Deep Inelastic Scattering in the Parton Model* [22]. The parton model emerged as the simplest explanation for deep inelastic scaling and the associated sum rules, which could be straightforwardly derived in the large momentum limit.

## 3. Deep Inelastic Scattering

The acceleration of the theoretical results at the end of the sixties was driven by the experimental results on deep inelastic scattering [23]. In 1966 the Electron Linear Accelerator of SLAC started to work. In 1969 the electron beam reached an energy of 17 GeV. The process studied was $electron + nucleon \rightarrow electron + hadrons$. At the leading order in quantum electrodynamics, it is equivalent to $virtual photon + nucleon \rightarrow hadrons$.

In the interesting region, the process is highly inelastic. The wonderful idea was to look only at the energy and momentum of the final hadrons without studying all other observables. However, one needed to study the energy and momentum loss of the scattered electron, so no hadron detector was necessary. There was only one electron detector at a fixed angle, that could be changed.

The only two relevant parameters were the mass squared of the virtual photon ($-q^2$) and its energy loss $\Delta E$. It was convenient to introduce the parameter $\nu \equiv \Delta EM$, $M$ being the nucleon mass. One was interested in the region where both parameters are large, away from elastic scattering and from real photon scattering. It is convenient to write: $x = q^2/\nu$. The variable $x$ is kinematically constrained to belong to the interval 0–1.

At the end, one can write the cross-section in terms of two functions of these two variables $F_1(q^2, x)$, $F_2(q^2, x)$. Looking at the dependence on the angle one could separate the two contributions.

*Bjorken Scaling*

Let us concentrate on the function $F_2(q^2, x)$, also because it gives the most relevant contribution.

Roughly speaking, Bjorken assumed that some equal-time commutators of currents and their derivatives have a non-zero vanishing element. After some computations, he found that

$$\lim_{q^2 \to \infty} \int_0^1 dx F_2(q^2, x) x^{n-1} = M_n \,,$$ (2)

(the case $n = 1$ is related to standard equal-time commutators of currents).

The quantities $M_n$ are equal to the matrix element of the commutators. If one makes the simplest and innocent-looking hypothesis that matrix elements of the commutators are non-zero, one arrives at the formula

$$\lim_{q^2 \to \infty} F_2(q^2, x) = F_2^\infty(x) \,, \qquad \int_0^1 dx F_2^\infty(x) x^{n-1} = M_n \,.$$ (3)

Finally, Bjorken and Paschos showed the function $F_2^\infty(x)$ was proportional to the fraction of charged partons carrying a fraction $x$ of the momentum of the proton in the $P = \infty$ frame, the factor of proportionality being the squared charge of the parton. The parton properties and distribution in the infinite momentum frame were observed in a conceptually simple experiment.

The asymptotic independence of $F_2(q^2, x)$ on $q^2$ was in reasonable agreement with the data. The experimentalists liked very much the proposal: of course, there was some dependence on $q^2$, however, this was small and it was supposed to disappear in the asymptotic large momenta region.

## 4. Subtle Is Field Theory

### 4.1. Lacking of Scaling in Perturbative Field Theory

The main problem to fix was the identification of the partons and their interaction. A quite natural proposal was the identification of partons with quarks, but other identifications were also possible. The troubling point was the choice of the interaction. Indeed the partons are very similar to free particles, or to particles interacting with a superrenormalizable interaction that could be neglected at high energy: in renormalizable theories, Bjorken scaling was not perturbatively correct: strong logarithmic (and maybe power) corrections were present. Unfortunately in 4 dimensions, no super-renormalizable theories with Fermions are available. The approach recalls what happened during the discovery of quantum mechanics [24]; bold (and wrong) hypotheses were first made and the progress made using the hypothesis led to the correct conclusions.

In the sixties, beyond perturbation theory, the large momentum behavior of the Green functions was a mystery. The renormalization group was relevant but the consequences of the renormalization approach were not understood. In a famous book on electrodynamics [25] Bjorken and Drell wrote that in electrodynamics if the function $\beta(e^2)$ has a zero at $e_0^2$, the bare charge is given by

$$e_0^2 = F^{-1}(\infty).$$ (4)

We are presented with a dilemma. All our arguments in this section have been based on the renormalization program in perturbation theory, which permits us to expand propagators, vertex functions, etc, in power series in both the normalized and bare charges $e$ and $e_0$. However, in the previous equation, we have come up with a result that puts a condition on the value of $e_0$ indicating that it cannot chosen arbitrarily. The behavior is completely foreign to the perturbation development and forces us to conclude that at least one of our assumptions along the way has been wrong. (...) There, conclusions based on the renormalization group arguments concerning the behavior of the theory summed at all orders are dangerous and must be viewed with due caution. So is it with all conclusions from local relativistic field theories.

Nowadays Equation (4) seems perfectly logical. For Bjorken and Drell the theory was defined in terms of its perturbative expansion, while nowadays the theory is defined by the path integral formulation and the non-convergent [26,27] (but asymptotic) perturbative expansion is just a tool that is useful in the weak coupling regime.

### 4.2. Operator Dimensions and Non-Canonical Scaling

More or less at the same time Bjorken and Drell were writing their discouraging conclusion (that echoed the same conclusion of Gell-Mann and Low), Ken Wilson had a completely different approach. His viewpoint was that strong interacting field theory exists and that one can compute its properties with due ingenuity. This viewpoint was clearly exposed in a magnificent and forgotten paper [28] of 1965, where many of the ideas of the later most famous papers are already present.

In the introduction, he writes: *The Hamiltonian formulation of quantum mechanics has been essentially abandoned in investigations of the interactions of mesons, nucleons, and strange particles. This is a pity. (...) There are two reasons why the Hamiltonian approach was discarded in the study of strong interactions. One reason was that no one knew what Hamiltonian to use, or how to obtain the correct Hamiltonian. The other reason was the problem of renormalization: the problem that whenever one tried to solve a Hamiltonian for a Lorentz-invariant theory, particle self-energies and the like were infinite.*

In this paper, he discusses the solution of a field theory model introducing a sequence of scale-dependent Hamiltonians. The renormalization group is seen as the functional relation of the Hamiltonian at one scale with the one at another scale. In this constructive approach, the Gell-Mann Low Bjorken Drell paradox fades out. The bare coupling constant is the fixed point of this transformation: nothing strange that its value is fixed.

### 4.3. Wilson's Operator Expansion

More or less simultaneously (1964) a revolutionary idea enters in the game: the so-called Wilson operator expansion. The idea is written in a one-hundred-page preprint with the title *On Products of Quantum Field Operators at Short Distances* [29] that appeared as a Cornell Report and it was never published (It was submitted to Phys. Rev.: there was a long referee report to which Wilson started to write the reply, but he never finished it). The paper was not a success. It was noticed by Zimmerman and it was at the basis of Brandt's PhD thesis [30]. Indeed it was first cited by Brandt in 1967 the second citation in 1970: now it has around 40 citations.

Wilson's short-distance operator product expansion states that in the limit of small $x$ the product of two operators can be written as

$$A(x)B(0) \to_{x \to 0} \sum_C C(0)|x|^{-d_A - d_B + d_C} \ . \tag{5}$$

The leading terms come from the operators $C$ with the lowest dimensions. The dimensions of the operators were the canonical ones in free theory where everything was clear. The bold hypothesis was that such an expansion is valid also in strong coupling field theories, where the operator dimensions may be different from the canonical ones (Similar ideas were rediscovered in the context of phase transition by Migdal and Polyakov: operator fusion).

Wilson's ideas became popular much later (1969), with the magnificent paper: *Non-Lagrangian models of current algebra* [31] that has more than 3000 citations and was an immediate success. In the introduction, he writes: *"Field theories with exact scale invariance are not physically interesting since they cannot have finite mass particles. But one can hypothesize that there exists a scale-invariant theory which becomes the theory of strong interactions when one adds mass terms to the Lagrangian The strong interactions become scale-invariant at short distances. This leads to the idea of broken scale invariance proposed by Kastrup and Mack".*

Later on, it is assumed that the strong interactions contain some arbitrary fundamental parameters just as the mass and charge of the electron are fundamental parameters in electrodynamics. However, the greater complication of strong interactions means that the parameters of strong interactions are not physical masses and coupling constants; they show up explicitly only in the short-distance behavior of strong interactions. Implicitly, they determine all of the strong interactions, but to calculate physical masses and coupling constants one has to solve the strong interactions, which is not possible at present. In physics these parameters have particular values, but the theory of strong interactions is assumed to be selfconsistent for any values of the parameters. In particular, if all the parameters are zero, it will be assumed that all partial symmetries become exact. The theory with all free parameters set equal to zero will be called the "skeleton theory".

Skeleton theory, a forgotten word, radically changed the perspective: all masses are zero, and the phenomenology of resonances disappears. In skeleton theory, the fundamental objects are the off-shell Green functions. Wilson is speaking of fields (e.g., the pion field), and their dimensions.

At the end of the paper, he discusses the vertex of the axial current with the two electromagnetic currents (the vertex is crucial for the determination of the amplitude of the process $\pi_0 \to \gamma\gamma$. It is hard to imagine that one could have a complete formula for this vertex function without having a complete solution of the hadron skeleton theory. The prospects for obtaining such a solution seem dim at present.

So the physically motivated renormalization group (that had tremendous success in the field of second-order phase transitions) and the Wilson operator expansion set the stage for a non-perturbative understanding.

New interest in the renormalization group arose in 1970 with the Callan-Symanzik equation [32,33]. The Callan-Symanzik was more explicit than the renormalization group and much easier to understand. It seemed less paradoxical than the renormalization group. It worked very well with the massive theory.

In the 1971 paper, Symanzik [34] proved in one particular case the Wilson operator expansion for the scalar $\phi^4$ theory:

$$\phi(x)\phi(0) \approx C(x)\phi(0)^2 \qquad x \to 0 \,. \tag{6}$$

where the behavior of the function $C(x)$ was controlled by an anomalous exponent $\gamma_2(g)$.

The argument was that in a Green function when the two momenta go to infinity and the others remain fixed, the large momentum behavior had anomalous terms that can be controlled analytically. If we translate these results into configuration space we obtain the previous equation. In conclusion in 1971 the Wilson expansion was taken for granted and there were no doubts about its validity.

### 4.4. The Light Cone Expansion

In 1971 Brandt and Preparata discovered [35] that Bjorken scaling was related to the so-called *Light cone expansion*. The first observation was kinematical: The value of the cross-section in deep inelastic scattering is related to configuration space to the singular behavior near the light cone $x^2 = 0$ of the function

$$\langle p|J(x)J(0)|p\rangle \,, \tag{7}$$

where $J$ is the electromagnetic current (vector indices are neglected) and $|p\rangle$ is the one proton state. Using the Wilson expansion near $x = 0$, the behavior near the light cone could be obtained by looking at many terms:

$$J(x)J(0) \to_{x^2\to 0} = \sum_{a=0,\infty} O_{\mu_1,\dots,\mu_a} x^{\mu_1} \dots x^{\mu_a} C_a(x) \,. \tag{8}$$

It is possible to show that each term in the infinite sum over $a$ corresponds to a different moment of the experimental structure-function.

Bjorken scaling follows if the lightcone singularities are the same as in free theory, i.e., if the operators that enter into the Wilson expansion have canonical dimensions. The requirement that the equal time commutators of the derivatives of the current have non-zero matrix elements is replaced by the requirement that the operator in the Wilson expansion have canonical dimensions. So the technical hypothesis on the commutators becomes a global hypothesis on the operator dimensions of the theory.

In this way, the Wilson short-distance operator product becomes deeply related to the experimentally observed approximate Bjorken scaling.

In the case of canonical dimensions, one obtains a simple form for the light cone expansion:

$$J(x)J(0) \to x^2 \to 0 \frac{O(x,0)}{x^2} \,, \tag{9}$$

where $O(x,0)$ is a bilocal operator. In free field theory a simple form of the bilocal operator is found. For example in a scalar theory:

$$O(x,0) = \varphi(x)\varphi(0) \,. \tag{10}$$

### 4.5. What Happens in an Interacting Theory?

Christ-Hasslacher-Mueller (1972) computed [36] the coupling-dependent anomalous dimensions of the operators relevant for deep inelastic scattering (the so-called twist-two operators) for a pseudoscalar theory and for a theory with a simple vector interaction.

If we neglect the dependence of the running coupling constant on the momenta, one has something like

$$M_n(q^2) \equiv \int_0^1 dx\, x^{n-1} F(x, q^2) \,; \quad M_n(q^2) = C_n \exp(\gamma_n(\alpha) \log(q^2)) \,. \tag{11}$$

The dependence on the running coupling constant can be trivially added. If one assumes that in the large momentum region, the running coupling constant $\alpha$ goes to a fixed point value $\alpha_c$, one gets:

$$M_n(q^2) = C_n (q^2)^{\gamma_n(\alpha_c)} \tag{12}$$

Canonical scaling implies that $\gamma_n(\alpha_c) = 0$. So Bjorken Scaling could be obtained if $\gamma_n(\alpha_c) = 0$.

## 5. The Quest for a Scaling Invariant Theory

### 5.1. A Scaling Invariant Theory with Canonical Dimensions Is Free

This part of the paper is more based on personal recollections, they were strongly influenced by the physicists whose papers I was reading and I was in contact with, and in particular by Ken Wilson's ideas.

If strong interactions at large momenta are scaling invariant and the operators have well-defined operator dimensions, a natural question was: can a scaling invariant non-free theory have $\gamma_n(\alpha_c) = 0$ for all $n$?

The answer (although not under the form of a theorem) is no [37–40]. So only two possibilities were present:

- The scaling invariant skeleton theory is free and we have Bjorken scaling.
- The scaling invariant theory skeleton is not free and we do not have Bjorken scaling.

Symanzik [41] noticed that the $\lambda\phi^4$ with a negative coupling constant is an example of a theory that has an asymptotically computable behavior (asymptotically free in modern language): the running coupling constant being at the leading order $\lambda(q^2) = \lambda/(1 - C\lambda\log(q^2/m^2))$, with positive $C$, so that in the large $q^2$ region $\lambda(q^2) \approx 1/(C\log(q^2/m^2))$. In this theory, Bjorken scaling is asymptotically exact without logs [42].

The theory is not realistic: it is unstable and no Fermions are present. Theories of Fermions interacting with scalar or pseudoscalar particles were not asymptotically free, as remarked by Landau in the 50's. So a strong interacting theory was missing.

### 5.2. The Illusion of a Scaling Invariant Theory for Strong Interaction

In non-asymptotically free theory the short distance theory should correspond to a fixed point of the renormalization group. If this happens, one would be very far from the perturbative region and there is no clear distinction between the fundamental field and the composite fields. For some time there was the hope that the skeleton theory could be found by solving some kind of consistency equations, some kind of bootstrap equations of a novel kind, using all the symmetry of the problem, and in particular conformal invariance [43] (The approach worked for 2-dimensional scalar theory and much later for 3-dimensional scalar theory, but at that time there was not the capacity to do these computations).

What about quarks? The quark field would have a non-zero Green function and the quark field acting on the vacuum would produce a state with a non-zero quark number. At that time quark confinement was not known, so the absence of quark production was interpreted as a high mass for the quarks: hadrons should be deep bounded states of quarks. In the scaling invariant skeleton theory bringing these quarks to zero mass would be a too-violent approximation (a more reasonable approximation could be to bring their mass to infinity). Indeed in the 1968 Wilson's paper, in the list of fields, no quark field was present. All the fields listed corresponded to physical particles, currents, and the stress tensor. So the strong interacting skeleton theory was naturally quarkless.

In such a theory Bjorken scaling is not correct. However, a careful analysis showed that strong violations of Bjorken scaling were compatible with experiments [44]. Indeed most of the good experimental data were in the region where both $q^2$ and the mass of the final hadrons were large. Indeed large $q^2$ and a large produced mass were needed to stay in the asymptotic region: given the experimental limitation that implied an $x$ neither too large nor too small. Some operators must have canonical dimensions also in strong interacting skeleton theory: the current and the stress tensors (this leads to two sum rules of the kind that we have previously discussed). One finds that the $F(x, q^2)$ should decrease near $x = 1$ and increase at small $x$. Obviously, in the central $x$ the function should be nearly constant. This was known from 1973, where it was also remarked that strong violations of Bjorken scaling where possible and this field theory motivated approach may be an explanation of the fact that the "scaling limits" seem to be reached from above at large $x$, and from below at small $x$ [44]. In that same paper, it was noticed that, if only one operator for given quantum numbers contributes, one could write equations of the form

$$\frac{\partial F(x, q^2)}{\partial \log(q^2)} = \int_x^1 \frac{dy}{y} F(y, q^2) K(x/y) \,, \tag{13}$$

The kernel $K$ depends on the value of the anomalous dimensions of the operators.

This was a simple rewriting of the formula for the moments. It is useful for phenomenological analysis because the value of the function at a small $x$ does not enter the formula (in the moment's formulation we have to integrate over all range of $x$). However, its simple meaning in terms of effective parton distribution was not contained in that paper. The idea of evolving parton densities appeared (as far as I know) firstly [45,46].

### 5.3. Asymptotic Freedom Was Discovered

All these dreams of a strong interacting theory faded away when the negative sign of the beta function of the renormalization group was discovered and a viable model of strong interaction theories with asymptotic freedom was proposed.

The negative sign for the beta function was discovered independently three times.

- In 1969 Iosif Khriplovich presented a beautiful computation that could be done on the back of the envelope [47]. He computed the vacuum energy $E$ induced by a constant gauge field. $E$ is trivially $F^2$ at the tree level. At one loop level, he computed the shift in the vacuum energy by studying the quantization of free particles in the field (Landau levels). The formulae were well known and one had to take care that gluons have spin one. At the end, he got

$$E = F^2(1 - A\alpha \log(q^2/\mu^2)).\qquad(14)$$

From this formula, the value of the beta function could be directly obtained. The computation was done with on-shell particles, so there was no need to use the Fadeev-Popov ghosts.

- At a conference in Marseilles in the summer of 1972, Gerard 't Hooft, announced that he had calculated the sign of the beta function in Yang-Mills and that the result was negative! This grand announcement was met with indifference. There were few people present, and even they did not pay much attention. The only person really capable of understanding 't Hooft's result was Symanzik, who urged him to write an article on the subject. 't Hooft had begun work on some extremely difficult calculations on quantum gravity. For him, the beta function calculation was hardly more than an exercise, and he did not have time to write it up for publication.

- In 1973, as everybody knows, by Gross and Wilczek [48], and Politzer [49].

Interestingly, the results on the negative sign of the coupling theories were noticed by most scientists in the reverse order of their discovery. However, only the last papers make the connection with colored gluons and show that asymptotic freedom is valid for Quantum Chromodynamics. Indeed only these authors did the computation aiming to apply it to strong interaction theory.

Let me add a personal touch to the story [50]. I happened to be good friends with Symanzik. In November 1972 I went to visit him for two weeks in Hamburg. Surprisingly, he did not talk to me about 't Hooft's result. Symanzik wanted the result to be communicated to the world directly by 't Hooft in a written form. It was only in February 1973 that I learned from Symanzik about 't Hooft's result. I had just moved to CERN in Geneva for two months, and since 't Hooft was there we met.

We needed to identify the gauge theory of strong interactions and verify that the beta function was negative. It sounded easy; in 1972 [51] it was proposed that quarks existed in three different 'colors' interacting by exchanging colored gluons: essentially the Yang-Mills theory studied by 't Hooft. I knew this theory perfectly, but the arguments were based on the naive parton model, a free skeleton theory. I had put my money on the opposite hypothesis—a non-free skeleton theory—and very presumptuously I had put down this proposal as too naïve. Then I shelved it.

Looking back on it now, the conversation with 't Hooft was surreal. We discussed only the possibility that the gauge theory was the flavor $SU(3)$. This was nonsense and we immediately realized it. After half an hour of discussion we concluded that we could not construct a model for the strong interactions were the negative sign of the beta function. We did not give a moment's thought to the colored gluons. We were incredibly blind, for which I take full responsibility because I knew by heart the experimental work and the various models proposed in the literature. Unfortunately, I was not fast enough to realize that if the skeleton theory was free, all the conclusions coming from the assumption of a non-free skeleton theory should be canceled.

## 6. The Final Expression for Scaling Violations

### 6.1. The First Formulae for Scaling Violations

In 1973 Georgi and Politzer [52], Gross, and Wilczek [53] computed the Christ-Hasslacher-Mueller formulae for QCD. One gets a simple result in the case of the non-singlet contribution: taking care of the running coupling constant $\alpha(q^2)$, the result is:

$$\frac{\partial M_n(q^2)}{\partial \log(q^2)} = \gamma_n(\alpha(q^2)),\qquad(15)$$

Using the same chain arguments of arguments of my previous paper, I finally got [54]:

$$\frac{\partial F(x, q^2)}{\partial \log(q^2)} = \int_x^1 \frac{dy}{y} F(x, q^2) K(x/y, \alpha(q^2)),$$ (16)

where kernel $K(z, \alpha(q^2))$ (that was not interpreted as the quark fragmentation function) is the inverse Mellin transform of the anomalous dimensions $\gamma_n(\alpha(q^2))$. Using the first order in $\alpha$ for $\gamma_n(\alpha)$, one finally gets for the non-singlet contribution:

$$K(z, \alpha(q^2)) = \frac{8}{3} \frac{\alpha(q^2)}{4\pi} \left( \frac{1+z^2}{(1-z)_+} + \frac{3}{2}\delta(z-1) \right)$$ (17)

The extraction of the value of $\alpha$ using the formulae for the moments was done in 1976 [55]. Using the moment formulation and the Mellin transform, we computed the violations of scaling without any physical interpretation of the formulae (Parisi Petronzio 1976) and we got $\alpha_s = 0.4$. Curve I of Figure 1 was our prediction for compared to the experimental data. Curve II was obtained by retaining only the octet operator in the operator expansion. This was written before Altarelli-Parisi.

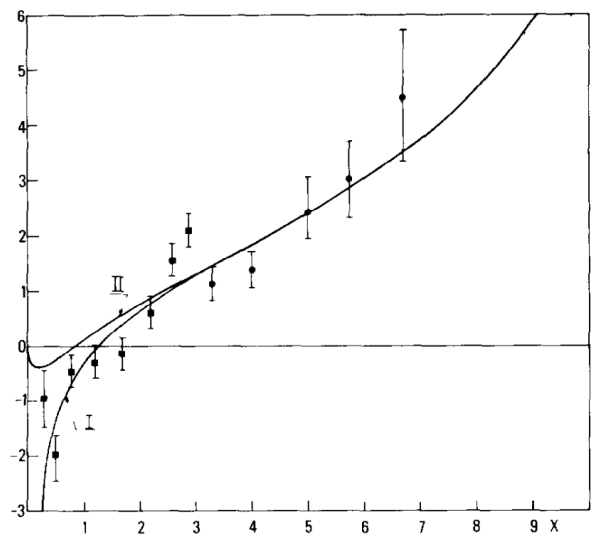

**Figure 1.** The prediction of Parisi and Petronzio [43] compared with the experimental data of SLAC [56] for the plot of $\frac{d \ln F(x,q2)}{d \ln(q^2)}$ for the proton versus $x$. Curve I is their prediction for the logarithmic derivative of the proton structure function compared with experimental data. Curve II is obtained by retaining only the octet operators in the operator expansion. The difference comes from the contribution of the gluons which is obtained by an educated guess.

### 6.2. The Altarelli Parisi Equations

The problem of logarithmic corrections in the perturbative expansion was quite old. There was a different case where the physical interpretation was quite clear: quantum electrodynamics, where electromagnetic radiative corrections are small but not negligible. In QED one often finds that radiative corrections are proportional to $\alpha \log(E/m_e)$ or $(\alpha \log(E/m_e))^k$. Various techniques can be used. The most famous approach is the *Equivalent photon approximation in quantum electrodynamics* (due to Weizsaker-Williams).

Nicola Cabibbo was strongly interested in computing these effects for applications to the experimental analysis of events collected in $e^+e^-$ collisions. Cabibbo and Rocca in 1974 wrote [57] formulae like:

$$P_{e \to e\gamma}(\eta) = \frac{\alpha}{\pi} \frac{1 + (1-\eta)^2}{\eta} \log(E/m_e); \quad P_{\gamma \to e^+e^-}(\epsilon) = \frac{\alpha}{2\pi}(1 + (1-2\epsilon)^2) \log(E/m_e),$$ (18)

where $\eta$ is the fraction of longitudinal momentum carried by the photon and $\epsilon$ is the fraction of longitudinal momentum carried by the electron.

These probabilities can be combined. They computed the probability of finding inside a $\gamma$ a triplet $\gamma, e^+, e^-$: it is proportional to $P_{\gamma \to e^+e^-} - P_{e \to e\gamma} \log(E/m_e)^2$. Nicola and I used these ideas (unpublished) to estimate the cross section for the process $e^+e^- \to 2e^+2e^-$ that was a candidate to be an important background in $e^+e^-$ colliding beam experiments.

Influenced by this paper, I extended the formulae for valence quarks to the gluons and sea-quarks [58,59]. In this case, one introduces effective quark and gluon parton distributions $N(x, \log(q^2))$:

$$\frac{dN_{q_i}(x,\log(q^2))}{d\log(q^2)} = \frac{\alpha}{4\pi} \int_x^1 \frac{dy}{y} \left[ p_{qq}(x/y) N_{q_i}(x,\log(q^2)) + p_{qg}(x/y) N_g(x,\log(q^2)) \right] ,$$

$$\frac{dN_g(x,L)}{d\log(q^2)} = \frac{\alpha}{4\pi} \int_x^1 \frac{dy}{y} \left[ p_{gg}(x/y) N_g(x,\log(q^2)) + p_{qg}(x/y) \sum_i N_{q_i}(x,\log(q^2)) \right] . \qquad (19)$$

The equations are formally the same as the Altarelli-Parisi (apart from a few typos), however, the derivation was completely different. The proof was formulated in the language of renormalization group equations for the coefficient functions of the local operators which appear in the light cone expansion for the product of two currents. Therefore the equations were useless for other hard processes. Most of the relations coming from the parton model were out of reach of the operator product expansion.

The physical derivation of the formulae using the ideas of effective parton distribution appeared in my paper with Guido; this paper stemmed from a proposal from Guido, i.e., to make previously-obtained results on scale violations clearer and more exploitable.

The motivations of the paper were clearly stated in the introduction. In this paper, we show that an alternative derivation of all results of current interest for the $Q^2$ behavior of deep inelastic structure functions is possible. In this approach all stages of the calculation refer to parton concepts and offer a very illuminating physical interpretation of the scaling violations. (...) This method can be described as an appropriate generalization of the equivalent photon approximation in quantum electrodynamics (Weizsaker-Williams .... Cabibbo-Rocca).

Indeed in the spring '77 Guido suggested that it would be pedagogically useful to derive all the equations for scaling violations using the same techniques of Cabibbo-Rocca; no loops were involved: only the evaluation of the vertices in the infinite momentum frame. Most of the formulae we needed were written there: only the gluon splitting into two gluons function was missing.

The final master (We used the wording master equations because they were classical probabilistic equations derived in a quantum setting) equations were essentially those of Equation (19). The computations were particularly transparent and simple when we extended them to the case of polarized partons using the helicity formulation in the infinite momentum frame.

The paper was a well-done cocktail of renormalization group results, parton model, and perturbation theory in an infinite momentum frame: easy to drink and to swallow. It was really pedagogic: it contained all the logical steps and the final receipt was quite easy to follow [60].

### 6.3. Aftermath

I am convinced that the most important result of the paper was not the construction of a practical way to compute scaling violations in deep inelastic scattering. The crucial point was to shift the focus from Wilson operator expansion to the effective number of partons that was dependent on the resolution, i.e., $q^2$. It was more than a computation: it was a change in the language we use. The appropriate choice of language is one of the most important scientific tools.

For example, the Drell-Yan process (i.e., $pp \to \mu^+\mu^- + \cdots$) could be not studied by a Wilson operator expansion: the Brandt-Preparata analysis did not work in this case. Similar conclusions are also true for the jet production in hadronic collisions and for the computation of the cross-section for large transverse momenta that were related to the hard scattering of partons.

However, it was now possible to study these processes using the new language: we had to factorize the amplitude for the process into a part containing the effective parton distribution at the relevant energy and into a part containing the hard scattering that could be treated in perturbation theory in the running coupling constant. Moreover, in the presence of energy-dependent effective parton distributions, the parton model predictions were ambiguous at order $\alpha_{QCD}$ and next to the leading order corrections (NLO) have to be computed to fix the predictions.

The solution to all these problems was at hand after our paper. This opportunity was immediately taken by Guido. All the following papers are published in '78 just after our '77 paper:

• Leptoproduction and Drell-Yan processes beyond the leading approximation in chromodynamics [61].

• Transverse momentum of jets in electroproduction from quantum chromodynamics [62].

• Transverse momentum in Drell-Yan processes [63].

• Processes involving fragmentation functions beyond the leading order in QCD [64].

In all these four problems the matching of the experimental data with the theoretical predictions allowed an independent measure of the QCD running coupling constant. At that time it was possible to achieve this goal in the first three cases. The compatibility of these independent determinations of $\alpha_{QCD}$ with the value coming from deep inelastic scattering (and from $\psi$ decay) was instrumental in convincing physicists that QCD was the correct theory.

**Conflicts of Interest**

The author declares no conflict of interest.

**References**

1. Dyson, F. A meeting with Enrico Fermi. *Nature* **2004**, *427*, 297–297.
2. Gross, D.J. Nobel lecture: The discovery of asymptotic freedom and the emergence of QCD. *Rev. Mod. Phys.* **2005**, *77*, 837–849.
3. Streater, R.F.; Wightman, A.S. *PCT, Spin and Statistics, and All That*; Princeton University Press: Benjamin, WA, USA, 1964.
4. Lehmann, H.; Symanzik, K.; Zimmermann, W. On the formulation of quantized field theories—II. *Nuovo Cim.* **1957**, *6*, 319–333.
5. Mandelstam, S. Determination of the pion-nucleon scattering amplitude from dispersion relations and unitarity. General theory. *Phys. Rev.* **1958**, *112*, 1344.
6. Gell-Mann, M. The symmetry group of vector and axial vector currents. *Phys. Phys. Fiz.* **1964**, *1*, 63.
7. Hagedorn, R. Statistical thermodynamics of strong interactions at high-energies. *Nuovo Cim. Suppl.* **1965**, *3*, 147–186.
8. Cabibbo, N.; Parisi, G. Exponential hadronic spectrum and quark liberation. *Phys. Lett. B* **1975**, *59*, 67–69.
9. Cabibbo, N. Unitary symmetry and leptonic decays. *Phys. Rev. Lett.* **1963**, *10*, 531.
10. Gell-Mann, M.; Lévy, M. The axial vector current in beta decay. *Il Nuovo C.* **1960**, *16*, 705–726.
11. Feynman, R.P.; Gell-Mann, M. Theory of the Fermi interaction. *Phys. Rev.* **1958**, *109*, 193.
12. Ademollo, M.; Gatto, R. Nonrenormalization theorem for the strangeness-violating vector currents. *Phys. Rev. Lett.* **1964**, *13*, 264.
13. Gürsey, F.; Radicati, L.A. Spin and unitary spin independence of strong interactions. *Phys. Rev. Lett.* **1964**, *13*, 173.
14. Dashen, R.F.; Gell-Mann, M. *Approximate Symmetry and the Algebra of Current Components*; California Institute of Technology: Pasadena, CA, USA, 1965.
15. Fubini, S.; Furlan, G.; Rossetti, C. A dispersion theory of symmetry breaking. *Il Nuovo C. A* **1965**, *40*, 1171–1193.
16. Bjorken, J.D. Applications of the chiral $U(6) \otimes U(6)$ algebra of current densities. *Phys. Rev.* **1966**, *148*, 1467.
17. Bjorken, J.D. Current algebra at small distances. *Conf. Proc. C* **1967**, *670717*, 55–81.
18. Callan, C.G., Jr.; Gross, D.J. Crucial test of a theory of currents. *Phys. Rev. Lett.* **1968**, *21*, 311.
19. Cornwall, J.M.; Norton, R.E. Current commutators and electron scattering at high momentum transfer. *Phys. Rev.* **1969**, *177*, 2584.
20. Bjorken, J.D. Asymptotic sum rules at infinite momentum. *Phys. Rev.* **1969**, *179*, 1547.
21. Feynman, R.P. Very high-energy collisions of hadrons. *Phys. Rev. Lett.* **1969**, *23*, 1415.
22. Bjorken, J.D.; Paschos, E.A. Inelastic Electron-Proton and $\gamma$-Proton Scattering and the Structure of the Nucleon *Phys. Rev.* **1969**, *185*, 1975.
23. Breidenbach, M.; Friedman, J.I.; Kendall, H.W.; et al. Observed behavior of highly inelastic electron-proton scattering. *Phys. Rev. Lett.* **1969**, *23*, 935.
24. Parisi, G. Planck's Legacy to Statistical Mechanics. *arXiv* **2001**, arXiv:cond-mat/0101293.
25. Bjorken, J.D.; Drell, S.D. *Relativistic Quantum Mechanics*; American Institute of Physics: College Park, MA, USA, 1965.
26. Itzykson, C.; Parisi, G.; Zuber, J.B. Asymptotic estimates in quantum electrodynamics. *Phys. Rev. D* **1977**, *16*, 996.
27. Hooft, G. Can we make sense out of Quantum Chromodynamics? In *The Whys of Subnuclear Physics*; Springer: Berlin, Germany, 1979; pp. 943–982.
28. Wilson, K.G. Model Hamiltonians for local quantum field theory. *Phys. Rev.* **1965**, *140*, B445.
29. Wilson, K.G. On products of quantum field operators at short distances. In *Cornell Report*; Cornell University: Ithaca, NY, USA, 1964.
30. Brandt, R.A. Derivation of renormalized relativistic perturbation theory from finite local field equations. *Ann. Phys.* **1967**, *44*, 221–265.
31. Wilson, K.G. Non-Lagrangian models of current algebra. *Phys. Rev.* **1969**, *179*, 1499.
32. Callan, C.G., Jr. Broken scale invariance in scalar field theory. *Phys. Rev. D* **1970**, *2*, 1541.
33. Symanzik, K. Small distance behaviour in field theory and power counting. *Commun. Math. Phys.* **1970**, *18*, 227–246.
34. Symanzik, K. Small-distance-behaviour analysis and Wilson expansions. *Commun. Math. Phys.* **1971**, *23*, 49–86.
35. Brandt, R.A.; Preparata, G. Operator product expansions near the light cone. *Nucl. Phys. B* **1971**, *27*, 541–567.
36. Christ, N.; Hasslacher, B.; Mueller, A.H. Light-cone behavior of perturbation theory. *Phys. Rev. D* **1972**, *6*, 3543.
37. Ferrara, S.; Grillo, A.F.; Parisi, G.; et al. Canonical scaling and conformal invariance. *Phys. Lett. B* **1972**, *38*, 333–334.
38. Parisi, G. *Serious Difficulties with Canonical Dimensions*; Comitato Nazionale per l'Energia Nucleare: Frascati, Italy, 1972.
39. Parisi, G. Bjorken scaling and the parton model. *Phys. Lett. B* **1972**, *42*, 114–116.

40. Parisi, G. How to measure the dimension of the parton field. *Nucl. Phys. B* **1973**, *59*, 641–646.

41. Symanzik, K. A field theory with computable large-momenta behaviour. *Lett. Nuovo Cim.* **1973**, *6*, 77–80.

42. Parisi, G. Deep inelastic scattering in a field theory with computable large-momenta behaviour. *Lett. Nuovo Cim.* **1973**, *7*, 84–88.

43. Ferrara, S.; Grillo, A.F.; Parisi, G.; et al. Covariant expansion of the conformal four-point function. *Nucl. Phys. B* **1972**, *49*, 77–98.

44. Parisi, G. Experimental limits on the values of anomalous dimensions. *Phys. Lett. B* **1973**, *43*, 207–208.

45. Polyakov, A. A similarity hypothesis in the strong interactions. 1. Multiple hadron production in e$^+$ e$^-$ annihilation. *Sov. Phys. JETP* **1971**, *32*, 296.

46. Kogut, J.; Susskind, L. Parton models and asymptotic freedom. *Phys. Rev. D* **1974**, *9*, 3391–3399.

47. Khriplovich, I.B. Green's functions in theories with non-abelian gauge group. *Sov. J. Nucl. Phys.* **1969**, *10*, 409.

48. Gross, D.J.; Wilczek, F. Ultraviolet behavior of non-abelian gauge theories. *Phys. Rev. Lett.* **1973**, *30*, 1343.

49. Politzer, H.D. Reliable perturbative results for strong interactions? *Phys. Rev. Lett.* **1973**, *30*, 1346.

50. Parisi, G. *In a Flight of Starlings: The Wonders of Complex Systems*; Penguin: London, UK, 2023.

51. Fritzsch, H.; Gell-Mann, M.; Leutwyler, H. Advantages of the color octet gluon picture. *Phys. Lett. B* **1973**, *47*, 365–368.

52. Georgi, H.; Politzer, H.D. Electroproduction scaling in an asymptotically free theory of strong interactions. *Phys. Rev. D* **1974**, *9*, 416.

53. Gross, D.J.; Wilczek, F. Asymptotically free gauge theories. II. *Phys. Rev. D* **1974**, *9*, 980.

54. Parisi, G. Detailed predictions for the pn structure functions in theories with computable large momenta behaviour. *Phys. Lett. B* **1974**, *50*, 367–368.

55. Parisi, G.; Petronzio, R. On the breaking of Bjorken scaling. *Phys. Lett. B* **1976**, *62*, 331–334.

56. Riordan, E.M.; Bodek, A.; Breidenbach, M.; et al. Tests of scaling of the proton electromagnetic structure functions. *Phys. Lett. B* **1974**, *52*, 249–252.

57. Cabibbo, N.; Rocca, M. *The ρ Bremsstrahlung*; CERN Report TH: Geneva, Switzerland, 1872.

58. Parisi, G. An Introduction to Scaling Violations. In Proceedings of the 11th Rencontres de Moriond, Flaine, France, 28 February–12 March 1976.

59. Parisi, G. *Field Theory, Disorder and Simulations*; World Scientific: Singapore, 1992.

60. Forte, S. Asymptotic Freedom in Parton Language: the Birth of Perturbative QCD. *arXiv* **2025**, arXiv:2501.13158.

61. Altarelli, G.; Ellis, R.K.; Martinelli, G. Leptoproduction and Drell-Yan processes beyond the leading approximation in chromodynamics. *Nucl. Phys. B* **1978**, *143*, 521–545.

62. Altarelli, G.; Martinelli, G. Transverse momentum of jets in electroproduction from quantum chromodynamics. *Phys. Lett. B* **1978**, *76*, 89–94.

63. Altarelli, G.; Parisi, G.; Petronzio, R. Transverse momentum in drell-yan processes. *Phys. Lett. B* **1978**, *76*, 351–355.

64. Altarelli, G.; Ellis, R.K.; Martinelli, G.; et al. Processes involving fragmentation functions beyond the leading order in QCD. *Nucl. Phys. B* **1979**, *160*, 301–329.